\documentstyle[twoside,11pt]{report}

\def\ed{\end{document}}

\begin{document}

%\input macros.tex

% EQUATION NUMBERING STUFF:
%
% \def\inc#1{\hbox{\global \advance#1 1}}
% \countdef\eqnr=1    \countdef\refnr=2
% \eqnr=1             \refnr=1
% \def\nexteq{\inc{\eqnr}}    
% \def\enr{\number\eqnr\nexteq}   
% \def\eq{\eqno(\enr)}   

%
% REFERENCE STUFF:
%
\def\pp{\noindent\parshape 2 0truecm 15.6truecm 2truecm 13.6truecm}
\def\rf#1;#2;#3;#4 {\par\pp#1, {\it #2}, {\bf #3}, #4. \par}
\def\rn{\pp}

\def\streck{\noalign{\vskip2pt\hrule\vskip2pt}}
\def\streck{\noalign{\vskip2pt\hrule\vskip2pt}}

\def\ie{{\frenchspacing\it i.e.}}
\def\eg{{\frenchspacing\it e.g.}}
\def\etc{{\frenchspacing\it etc.}}
\def\etal{{\frenchspacing\it et al.}}

\def\tt#1{\times10^{#1}}
\def\aet#1#2{\approx #1 \tt{#2}}
\def\where{,\quad\hbox{where}}

% UNITS:
\def\Ms{M_{\odot}}     
\def\Ls{L_{\odot}}
\def\K{{\rm K}}
\def\s{{\rm s}}
\def\m{{\rm m}}
\def\sec{{\rm s}}
\def\cm{{\rm cm}}
\def\g{{\rm g}}
\def\ergs{{\rm erg}}
\def\erg{{\rm erg}}
\def\eV{{\rm eV}}
\def\Mpc{{\rm Mpc}}
\def\kpc{{\rm kpc}}
\def\pc{{\rm pc}}
\def\AU{{\rm AU}}
\def\Gyr{{\rm Gyr}}

\def\expec#1{\langle#1\rangle}

%\simlt and \simgt produce > and < signs with twiddle underneath
\def\spose#1{\hbox to 0pt{#1\hss}}
\def\simlt{\mathrel{\spose{\lower 3pt\hbox{$\mathchar"218$}}
     \raise 2.0pt\hbox{$\mathchar"13C$}}}
\def\simgt{\mathrel{\spose{\lower 3pt\hbox{$\mathchar"218$}}
     \raise 2.0pt\hbox{$\mathchar"13E$}}}
%\simpropto produces \propto with twiddle underneath
\def\simpropto{\mathrel{\spose{\lower 3pt\hbox{$\mathchar"218$}}
     \raise 2.0pt\hbox{$\propto$}}}

\def\Ylm{Y_{lm}}
\def\nh{\widehat{n}}
\def\imii{\int_{-\infty}^{\infty}}
\def\izi{\int_0^{\infty}}
\def\izp{\int_0^{\pi}}
\def\crr{\cr\noalign{\vskip2pt}}
\def\crl{\\ \noalign{\vskip2pt}}
\def\Wc{W_{cobe}}
\def\Wsp{W_{sp}}
\def\Wmax{W_{max}}
\def\Wlp{W_{lp}}
\def\clp{c_{lp}} \def\csp{c_{sp}}  \def\ccobe{c_{c}}
\def\slp{s_{lp}} \def\ssp{s_{sp}}  \def\scobe{s_{c}}
\def\slph{\hat{s}_{lp}} \def\ssph{\hat{s}_{sp}}  
\def\scobeh{\hat{s}_{c}}
\def\f{f}
\def\flp{\f_{lp}} \def\fsp{\f_{sp}}  \def\fcobe{\f_{c}}
\def\vlp{V_{lp}} \def\vsp{V_{sp}}
\def\sigc{\sigma_{c}}
\def\sigsp{\sigma_{s}}
\def\sigmax{\sigma_{m}}
\def\asp{\alpha_{s}}
\def\amax{\alpha_{m}}
\def\Rc{R_{c}}
\def\oh{{1\over 2}}
\def\summ{\sum_{m=-l}^l}
\def\jbar{\bar j}

\def\sc{\sigma_c}
\def\st{\sigma_t}
\def\Rc{R_{c}}
\def\oh{{1\over 2}}
\def\crr{\cr\noalign{\vskip 4pt}}
\def\t{\tau_0}
\def\fuzz{\epsilon a}
\def\sumi{\sum_{i=1}^n}
\def\sumj{\sum_{j=1}^n}
\def\intk{\int_0^{\infty}}
\def\vl{{\bf\lambda}}
\def\vp{{\bf p}}   \def\vx{{\bf x}}  \def\vk{{\bf k}}
\def\va{{\bf a}} \def\vb{{\bf b}} \def\vc{{\bf c}}
\def\vv{{\bf v}} \def\ve{{\bf \epsilon}}
\def\x{\eta}
\def\xh{\hat\x}
\def\s{s}
\def\sh{\hat s}
\def\c{c}
\def\fx{f_{\x}}  \def\Fx{F_{\x}}
\def\Qrms{Q_{rms,ps}}
\def\kms{{\,\rm km/s}} \def\mK{{\,\rm \mu K}}
\def\N{n_{\gamma}}
\def\elp{\eta_l}
\def\epp{\eta_p}
\def\elph{\hat\eta_l}
\def\epph{\hat\eta_p}

% RAY TRACING STUFF:

\def\etal{{\it et al.}}
\def\ie{{\it i.e.}}
\def\izi{\int_0^{\infty}}
\def\ioi{\int_1^{\infty}}
\def\imoo{\int_{-1}^1}
\def\izz{\int_0^z}
\def\zr{z_{rec}}       \def\Pr{P_{rec}} 
\def\izzr{\int_0^{\zr}}
\def\expec#1{\langle#1\rangle}
\def\vr{{\bf r}}  \def\vx{{\bf x}} 
\def\vk{{\bf k}}  \def\vq{{\bf q}}
\def\vn{{\bf\hat n}}
\def\vxzero{\vx^{(0)}}
\def\vxone{\vx^{(1)}}
\def\crr{\cr\noalign{\vskip 4pt}}
\def\iigm{\chi_{IGM}}
\def\rydberg{E_0}
\def\Ob{\Omega_b}
\def\K{{\rm K}} 
\def\t{\tau}        % Conformal time
\def\Dt{\Delta\t} 
\def\ts{\t_s}     % Conformal scattering time
\def\rad{\sqrt{u^2+v^2}}
\def\fth{f_{\theta}}  \def\fph{f_{\phi}}
\def\ft{f_{\t}}    
\def\fx{f_{\vx}}
\def\fxzero{f_{\vx}^{(0)}}
\def\fxone{f_{\vx}^{(1)}}
\def\vol{{\partial(\ts,\theta,\phi)\over\partial(u,v,w)}}
\def\arcosh{\cosh^{-1}}
\def\arsinh{{\sinh^{-1}}}
\def\artanh{{\tanh^{-1}}}
\def\prob{g}
\def\sqm{\sqrt{1-\mu^2}}
\def\Gh{\widehat{G}}
\def\x{\Delta}
\def\Dx{\Delta\vx}
\def\xh{\hat\x}  \def\xhr{\xh_{rec}}
\def\xr{\x_{rec}}
\def\nh{\hat{\bf q}}  \def\qh{\hat{\bf q}}
\def\dh{\hat{\delta}} \def\dhr{\dh_{rec}}
\def\fh{\hat f}
\def\expec#1{\langle#1\rangle}
\def\jz#1{{\sin#1\over #1}}
\def\tc{\tau_c}
\def\ootpc{{1\over (2\pi)^3}}
\def\Oz{\Omega_0}
\def\Lz{\lambda_0}

\def\vg{{\bf g}}
\def\ootpc{{1\over (2\pi)^3}}
\def\rhobar{\bar{\rho}}
\def\deltahat{\widehat{\delta}}
\def\what{\widehat{w}}
\def\vu{\bf u}
\def\zvir{z_{vir}}
\def\erf{{\rm erf}}
\def\erfc{{\rm erfc}}

\def\dent {\hspace{6mm}}
\def \em  {\hskip 0.08 em}
\def \en  {\hskip -0.08 em}
\def\nn {\nonumber}
\def \o {\mbox{}}

\def\sR {{\sf R}}
\def\cL {{\cal L}}
\def \txs  {\textstyle}
\def \2  {{\txs {1 \over \, 2 \, }}}
\def \4  {{\txs {1 \over \, 4 \, }}}

% From reionization.tex: 

\def\ps{P_s}
\def\v{\chi}     % Volume fraction ionized
\def\ieq{\chi_{eq}}
\def\veq{\chi_{eq}} 
\def\fg{f_g}
\def\phot{\eta}
\def\fuv{f_{uv}}
\def\fupp{f_{uvpp}}
\def\euv{\expec{E_{uv}}}
\def\euvfid{E_{13.6}}
\def\sigfid{\sigma_{20}}
\def\uvsigfid{\sigma_{18}}
\def\uvsig{\sigma_{uv}}
\def\Lrec{\Lambda_{rec}}  
\def\Lci{\Lambda_{ci}}
\def\Lpi{\Lambda_{pi}}
\def\lrec{\lambda_{rec}}  
\def\lci{\lambda_{ci}}
\def\lpi{\lambda_{pi}}
\def\lcomp{\lambda_{comp}} 
\def\fnet{f_{net}}
\def\fs{f_s}
\def\fH{f_H}
\def\fmet{f_{burn}}
\def\fbh{f_{bh}}
\def\facc{f_{acc}}
\def\fuv{f_{uv}}
\def\fion{f_{ion}}
\def\fesc{f_{esc}}
\def\szmc{\sigma(M_c,0)}
\def\zvir{z_{vir}}
\def\zion{z_{ion}}
\def\opzopzv{\left({1+z\over 1+\zvir}\right)}
\def\phit{\phi_2(T)}
\def\Tcbr{T_{cbr}}
\def\Tpi{T^*}
\def\ktmc{\left({kT\over m_e c^2}\right)}
\def\recfac{\eta_{rec}}
\def\soz{\sqrt{1+\Oz z}}

% From openreion.tex: 
 
\def\boost{B}
% \def\taut{\tau}

% From y.tex:

\def\Te{T_e}  \def\Tp{T_{\gamma}}  \def\DT{\Delta T}
\def\hce{{\cal H}_{ce}}
\def\hcomp{{\cal H}_{comp}}
\def\upp{\zeta}
 
% From gp.tex: 

\def\Rd{\dot R} \def\pd{\dot p}  \def\rdot{\dot \rho}
\def\Rdd{\ddot R} \def\md{\dot m}
\def\Od{\Omega_d}
\def\rb{\rho_b}     \def\rd{\rho_d}   \def\rc{\rho_c}
\def\ri{\rho_i}     \def\rs{\rho_s}
\def\sr3{\sqrt 3}
\def\poft{{\pi\over 4 \sr3}}
\def\poet{{\pi\over 8 \sr3}}
\def\ftp{{4\over 3} \pi}
\def\rdr{{\Rd \over R}}
\def\opt{(1+\tau)}  \def\opzo{(1+z_1)}
\def\fsn{f_{sn}}  \def\eheat{f_{heat}}
\def\fm{f_m}
\def\tb{t_{burn}}
\def\taub{\tau_{burn}}
\def\Lsn{L_{sn}} \def\Lc{L_{comp}} \def\Lh{L_{diss}}
\def\Lb{L_{brems}} \def\Li{L_{ion}}
\def\l{\ell} \def\h{\eta}
\def\ls{\ell_{sn}} \def\lc{\ell_{comp}} \def\lh{\ell_{diss}}
\def\lb{\ell_{brems}} \def\li{\ell_{ion}}
\def\Et{E_t}  \def\Ek{E_k}
\def\ff{\phi}
\def\r{r}
\def\p{q}
\def\P{P}
\def\et{\varepsilon_t}  \def\ek{\varepsilon_k}  \def\e{\varepsilon}
\def\ein{\varepsilon_{in}}
\def\fcoll{f_d}
\def\eM{\left<M\right>} \def\eMtf{\left<M^{3/5}\right>}
\def\prto{\>\propto\>}
\def\ng{n_{\gamma}}  \def\ngz{n_{\gamma 0}}
\def\Tg{T_{\gamma}}  \def\Tgz{T_{\gamma 0}}
\def\Tb{T_{bubble}}
\def\slya{\sigma_{Ly\alpha}}
\def\ddz{{\partial\over\partial z}}
 
\def\ii{\chi} \def\is{\chi_s}  \def\iigm{\chi_{IGM}}
\def\Ti{T_i} \def\Ts{T_s} \def\Tigm{T_{IGM}}
\def\lion{\lambda_{ion}}
\def\Lion{\Lambda_{ion}}
\def\omdc {(1-\delta)^3}
\def\hubf{\left({\r'\over r}- {2\over 3}\h\right)}
\def\hubpar{\left(\r'-{2\over 3}\h\r\right)}
\def\linj{\ell_{inj}}  \def\finj{f_{inj}}
\def\Tigm{T_{IGM}}

% For chapter2.tex:

\def\R{{\sR}}
\def\suml{\sum_{l=0}^{\infty}}
\def\summ{\sum_{m=-l}^l}
\def\dd#1#2{{\partial #1\over\partial #2}}

\def\nhat{\widehat{n}}
\def\Dhat{\widehat{\Delta}}
\def\taut{\tau_t}

% Figures:

% \def\mycaption#1{\bigskip\noindent\parshape 1 1.8truecm 14.35truecm{#1}}
\def\mycaption#1{\bigskip\noindent\parshape 1 2.3truecm 11.45truecm{#1}}
\def\fheight{12cm}
\def\fwidth{17cm}

\def\notdone{\bigskip{\bf THIS PART IS NOT DONE YET.}\bigskip} 

\def\nothing{\noindent\centerline{\,}} 
\def\mysection#1{\newpage\noindent\centerline{\,}\vskip2.2truecm{\Huge\bf#1}\vskip1.4truecm} 
\def\mysectionn#1{\newpage\noindent\centerline{\,}\vskip1.2truecm{\Huge\bf#1}\vskip1.4truecm} 
\def\mysectionnn#1{\newpage{\Huge\bf#1}\vskip0.4truecm} 

\def\beq#1{\begin{equation}\label{#1}}
\def\eeq{\end{equation}}
\def\beqa#1{\begin{eqnarray}\label{#1}}
\def\eeqa{\end{eqnarray}}
\def\eq#1{equation~(\ref{#1})}
\def\eqnum#1{~(\ref{#1})}

\def\bfig{\begin{figure}[h] \centerline{\hbox{}}\vfill}
\def\efig{\end{figure}\vfill\newpage}

\def\ns{\vskip-0.2truecm}
\def\double{\baselineskip17pt}
\ifx\undefined\psfig\else\endinput\fi

%
% from a suggestion by eijkhout@csrd.uiuc.edu to allow
% loading as a style file:
\edef\psfigRestoreAt{\catcode`@=\number\catcode`@\relax}
\catcode`\@=11\relax
\newwrite\@unused
\def\ps@typeout#1{{\let\protect\string\immediate\write\@unused{#1}}}
\ps@typeout{psfig/tex 1.8}

%% Here's how you define your figure path.  Should be set up with null
%% default and a user useable definition.

\def\figurepath{./}
\def\psfigurepath#1{\edef\figurepath{#1}}

%
% @psdo control structure -- similar to Latex @for.
% I redefined these with different names so that psfig can
% be used with TeX as well as LaTeX, and so that it will not 
% be vunerable to future changes in LaTeX's internal
% control structure,
%
\def\@nnil{\@nil}
\def\@empty{}
\def\@psdonoop#1\@@#2#3{}
\def\@psdo#1:=#2\do#3{\edef\@psdotmp{#2}\ifx\@psdotmp\@empty \else
    \expandafter\@psdoloop#2,\@nil,\@nil\@@#1{#3}\fi}
\def\@psdoloop#1,#2,#3\@@#4#5{\def#4{#1}\ifx #4\@nnil \else
       #5\def#4{#2}\ifx #4\@nnil \else#5\@ipsdoloop #3\@@#4{#5}\fi\fi}
\def\@ipsdoloop#1,#2\@@#3#4{\def#3{#1}\ifx #3\@nnil 
       \let\@nextwhile=\@psdonoop \else
      #4\relax\let\@nextwhile=\@ipsdoloop\fi\@nextwhile#2\@@#3{#4}}
\def\@tpsdo#1:=#2\do#3{\xdef\@psdotmp{#2}\ifx\@psdotmp\@empty \else
    \@tpsdoloop#2\@nil\@nil\@@#1{#3}\fi}
\def\@tpsdoloop#1#2\@@#3#4{\def#3{#1}\ifx #3\@nnil 
       \let\@nextwhile=\@psdonoop \else
      #4\relax\let\@nextwhile=\@tpsdoloop\fi\@nextwhile#2\@@#3{#4}}
% 
% \fbox is defined in latex.tex; so if \fbox is undefined, assume that
% we are not in LaTeX.
% Perhaps this could be done better???
\ifx\undefined\fbox
% \fbox code from modified slightly from LaTeX
\newdimen\fboxrule
\newdimen\fboxsep
\newdimen\ps@tempdima
\newbox\ps@tempboxa
\fboxsep = 3pt
\fboxrule = .4pt
\long\def\fbox#1{\leavevmode\setbox\ps@tempboxa\hbox{#1}\ps@tempdima\fboxrule
    \advance\ps@tempdima \fboxsep \advance\ps@tempdima \dp\ps@tempboxa
   \hbox{\lower \ps@tempdima\hbox
  {\vbox{\hrule height \fboxrule
          \hbox{\vrule width \fboxrule \hskip\fboxsep
          \vbox{\vskip\fboxsep \box\ps@tempboxa\vskip\fboxsep}\hskip 
                 \fboxsep\vrule width \fboxrule}
                 \hrule height \fboxrule}}}}
\fi
%
%%%%%%%%%%%%%%%%%%%%%%%%%%%%%%%%%%%%%%%%%%%%%%%%%%%%%%%%%%%%%%%%%%%
% file reading stuff from epsf.tex
%   EPSF.TEX macro file:
%   Written by Tomas Rokicki of Radical Eye Software, 29 Mar 1989.
%   Revised by Don Knuth, 3 Jan 1990.
%   Revised by Tomas Rokicki to accept bounding boxes with no
%      space after the colon, 18 Jul 1990.
%   Portions modified/removed for use in PSFIG package by
%      J. Daniel Smith, 9 October 1990.
%
\newread\ps@stream
\newif\ifnot@eof       % continue looking for the bounding box?
\newif\if@noisy        % report what you're making?
\newif\if@atend        % %%BoundingBox: has (at end) specification
\newif\if@psfile       % does this look like a PostScript file?
%
% PostScript files should start with `%!'
%
{\catcode`\%=12\global\gdef\epsf@start{%!}}
\def\epsf@PS{PS}
\def\epsf@getbb#1{%
%
%   The first thing we need to do is to open the
%   PostScript file, if possible.
%
\openin\ps@stream=#1
\ifeof\ps@stream\ps@typeout{Error, File #1 not found}\else
%
%   Okay, we got it. Now we'll scan lines until we find one that doesn't
%   start with %. We're looking for the bounding box comment.
%
   {\not@eoftrue \chardef\other=12
    \def\do##1{\catcode`##1=\other}\dospecials \catcode`\ =10
    \loop
       \if@psfile
	  \read\ps@stream to \epsf@fileline
       \else{
	  \obeyspaces
          \read\ps@stream to \epsf@tmp\global\let\epsf@fileline\epsf@tmp}
       \fi
       \ifeof\ps@stream\not@eoffalse\else
%
%   Check the first line for `%!'.  Issue a warning message if its not
%   there, since the file might not be a PostScript file.
%
       \if@psfile\else
       \expandafter\epsf@test\epsf@fileline:. \\%
       \fi
%
%   We check to see if the first character is a % sign;
%   if so, we look further and stop only if the line begins with
%   `%%BoundingBox:' and the `(atend)' specification was not found.
%   That is, the only way to stop is when the end of file is reached,
%   or a `%%BoundingBox: llx lly urx ury' line is found.
%
          \expandafter\epsf@aux\epsf@fileline:. \\%
       \fi
   \ifnot@eof\repeat
   }\closein\ps@stream\fi}%
%
% This tests if the file we are reading looks like a PostScript file.
%
\long\def\epsf@test#1#2#3:#4\\{\def\epsf@testit{#1#2}
			\ifx\epsf@testit\epsf@start\else
\ps@typeout{Warning! File does not start with `\epsf@start'.  It may not be a PostScript file.}
			\fi
			\@psfiletrue} % don't test after 1st line
%
%   We still need to define the tricky \epsf@aux macro. This requires
%   a couple of magic constants for comparison purposes.
%
{\catcode`\%=12\global\let\epsf@percent=%\global\def\epsf@bblit{%BoundingBox}}
%
%
%   So we're ready to check for `%BoundingBox:' and to grab the
%   values if they are found.  We continue searching if `(at end)'
%   was found after the `%BoundingBox:'.
%
\long\def\epsf@aux#1#2:#3\\{\ifx#1\epsf@percent
   \def\epsf@testit{#2}\ifx\epsf@testit\epsf@bblit
	\@atendfalse
        \epsf@atend #3 . \\%
	\if@atend	
	   \if@verbose{
		\ps@typeout{psfig: found `(atend)'; continuing search}
	   }\fi
        \else
        \epsf@grab #3 . . . \\%
        \not@eoffalse
        \global\no@bbfalse
        \fi
   \fi\fi}%
%
%   Here we grab the values and stuff them in the appropriate definitions.
%
\def\epsf@grab #1 #2 #3 #4 #5\\{%
   \global\def\epsf@llx{#1}\ifx\epsf@llx\empty
      \epsf@grab #2 #3 #4 #5 .\\\else
   \global\def\epsf@lly{#2}%
   \global\def\epsf@urx{#3}\global\def\epsf@ury{#4}\fi}%
%
% Determine if the stuff following the %%BoundingBox is `(atend)'
% J. Daniel Smith.  Copied from \epsf@grab above.
%
\def\epsf@atendlit{(atend)} 
\def\epsf@atend #1 #2 #3\\{%
   \def\epsf@tmp{#1}\ifx\epsf@tmp\empty
      \epsf@atend #2 #3 .\\\else
   \ifx\epsf@tmp\epsf@atendlit\@atendtrue\fi\fi}

% End of file reading stuff from epsf.tex
%%%%%%%%%%%%%%%%%%%%%%%%%%%%%%%%%%%%%%%%%%%%%%%%%%%%%%%%%%%%%%%%%%%

%%%%%%%%%%%%%%%%%%%%%%%%%%%%%%%%%%%%%%%%%%%%%%%%%%%%%%%%%%%%%%%%%%%
% trigonometry stuff from "trig.tex"
\chardef\letter = 11
\chardef\other = 12

\newif \ifdebug %%% turn me on to see TeX hard at work ...
\newif\ifc@mpute %%% don't need to compute some values
\c@mputetrue % but assume that we do

\let\then = \relax
\def\r@dian{pt }
\let\r@dians = \r@dian
\let\dimensionless@nit = \r@dian
\let\dimensionless@nits = \dimensionless@nit
\def\internal@nit{sp }
\let\internal@nits = \internal@nit
\newif\ifstillc@nverging
\def \Mess@ge #1{\ifdebug \then \message {#1} \fi}

{ %%% Things that need abnormal catcodes %%%
	\catcode `\@ = \letter
	\gdef \nodimen {\expandafter \n@dimen \the \dimen}
	\gdef \term #1 #2 #3%
	       {\edef \t@ {\the #1}%%% freeze parameter 1 (count, by value)
		\edef \t@@ {\expandafter \n@dimen \the #2\r@dian}%
				   %%% freeze parameter 2 (dimen, by value)
		\t@rm {\t@} {\t@@} {#3}%
	       }
	\gdef \t@rm #1 #2 #3%
	       {{%
		\count 0 = 0
		\dimen 0 = 1 \dimensionless@nit
		\dimen 2 = #2\relax
		\Mess@ge {Calculating term #1 of \nodimen 2}%
		\loop
		\ifnum	\count 0 < #1
		\then	\advance \count 0 by 1
			\Mess@ge {Iteration \the \count 0 \space}%
			\Multiply \dimen 0 by {\dimen 2}%
			\Mess@ge {After multiplication, term = \nodimen 0}%
			\Divide \dimen 0 by {\count 0}%
			\Mess@ge {After division, term = \nodimen 0}%
		\repeat
		\Mess@ge {Final value for term #1 of 
				\nodimen 2 \space is \nodimen 0}%
		\xdef \Term {#3 = \nodimen 0 \r@dians}%
		\aftergroup \Term
	       }}
	\catcode `\p = \other
	\catcode `\t = \other
	\gdef \n@dimen #1pt{#1} %%% throw away the ``pt''
}

\def \Divide #1by #2{\divide #1 by #2} %%% just a synonym

\def \Multiply #1by #2%%% allows division of a dimen by a dimen
       {{%%% should really freeze parameter 2 (dimen, passed by value)
	\count 0 = #1\relax
	\count 2 = #2\relax
	\count 4 = 65536
	\Mess@ge {Before scaling, count 0 = \the \count 0 \space and
			count 2 = \the \count 2}%
	\ifnum	\count 0 > 32767 %%% do our best to avoid overflow
	\then	\divide \count 0 by 4
		\divide \count 4 by 4
	\else	\ifnum	\count 0 < -32767
		\then	\divide \count 0 by 4
			\divide \count 4 by 4
		\else
		\fi
	\fi
	\ifnum	\count 2 > 32767 %%% while retaining reasonable accuracy
	\then	\divide \count 2 by 4
		\divide \count 4 by 4
	\else	\ifnum	\count 2 < -32767
		\then	\divide \count 2 by 4
			\divide \count 4 by 4
		\else
		\fi
	\fi
	\multiply \count 0 by \count 2
	\divide \count 0 by \count 4
	\xdef \product {#1 = \the \count 0 \internal@nits}%
	\aftergroup \product
       }}

\def\r@duce{\ifdim\dimen0 > 90\r@dian \then   % sin(x+90) = sin(180-x)
		\multiply\dimen0 by -1
		\advance\dimen0 by 180\r@dian
		\r@duce
	    \else \ifdim\dimen0 < -90\r@dian \then  % sin(-x) = sin(360+x)
		\advance\dimen0 by 360\r@dian
		\r@duce
		\fi
	    \fi}

\def\Sine#1%
       {{%
	\dimen 0 = #1 \r@dian
	\r@duce
	\ifdim\dimen0 = -90\r@dian \then
	   \dimen4 = -1\r@dian
	   \c@mputefalse
	\fi
	\ifdim\dimen0 = 90\r@dian \then
	   \dimen4 = 1\r@dian
	   \c@mputefalse
	\fi
	\ifdim\dimen0 = 0\r@dian \then
	   \dimen4 = 0\r@dian
	   \c@mputefalse
	\fi
	\ifc@mpute \then
        	% convert degrees to radians
		\divide\dimen0 by 180
		\dimen0=3.141592654\dimen0
		\dimen 2 = 3.1415926535897963\r@dian %%% a well-known constant
		\divide\dimen 2 by 2 %%% we only deal with -pi/2 : pi/2
		\Mess@ge {Sin: calculating Sin of \nodimen 0}%
		\count 0 = 1 %%% see power-series expansion for sine
		\dimen 2 = 1 \r@dian %%% ditto
		\dimen 4 = 0 \r@dian %%% ditto
		\loop
			\ifnum	\dimen 2 = 0 %%% then we've done
			\then	\stillc@nvergingfalse 
			\else	\stillc@nvergingtrue
			\fi
			\ifstillc@nverging %%% then calculate next term
			\then	\term {\count 0} {\dimen 0} {\dimen 2}%
				\advance \count 0 by 2
				\count 2 = \count 0
				\divide \count 2 by 2
				\ifodd	\count 2 %%% signs alternate
				\then	\advance \dimen 4 by \dimen 2
				\else	\advance \dimen 4 by -\dimen 2
				\fi
		\repeat
	\fi		
			\xdef \sine {\nodimen 4}%
       }}

% Now the Cosine can be calculated easily by calling \Sine
\def\Cosine#1{\ifx\sine\UnDefined\edef\Savesine{\relax}\else
		             \edef\Savesine{\sine}\fi
	{\dimen0=#1\r@dian\advance\dimen0 by 90\r@dian
	 \Sine{\nodimen 0}
	 \xdef\cosine{\sine}
	 \xdef\sine{\Savesine}}}	      
% end of trig stuff
%%%%%%%%%%%%%%%%%%%%%%%%%%%%%%%%%%%%%%%%%%%%%%%%%%%%%%%%%%%%%%%%%%%%

\def\psdraft{
	\def\@psdraft{0}
	%\ps@typeout{draft level now is \@psdraft \space . }
}
\def\psfull{
	\def\@psdraft{100}
	%\ps@typeout{draft level now is \@psdraft \space . }
}

\psfull

\newif\if@scalefirst
\def\psscalefirst{\@scalefirsttrue}
\def\psrotatefirst{\@scalefirstfalse}
\psrotatefirst

\newif\if@draftbox
\def\psnodraftbox{
	\@draftboxfalse
}
\def\psdraftbox{
	\@draftboxtrue
}
\@draftboxtrue

\newif\if@prologfile
\newif\if@postlogfile
\def\pssilent{
	\@noisyfalse
}
\def\psnoisy{
	\@noisytrue
}
\psnoisy
%%% These are for the option list.
%%% A specification of the form a = b maps to calling \@p@@sa{b}
\newif\if@bbllx
\newif\if@bblly
\newif\if@bburx
\newif\if@bbury
\newif\if@height
\newif\if@width
\newif\if@rheight
\newif\if@rwidth
\newif\if@angle
\newif\if@clip
\newif\if@verbose
\def\@p@@sclip#1{\@cliptrue}

\newif\if@decmpr

%%% GDH 7/26/87 -- changed so that it first looks in the local directory,
%%% then in a specified global directory for the ps file.
%%% RPR 6/25/91 -- changed so that it defaults to user-supplied name if
%%% boundingbox info is specified, assuming graphic will be created by
%%% print time.
%%% TJD 10/19/91 -- added bbfile vs. file distinction, and @decmpr flag

\def\@p@@sfigure#1{\def\@p@sfile{null}\def\@p@sbbfile{null}
	        \openin1=#1.bb
		\ifeof1\closein1
	        	\openin1=\figurepath#1.bb
			\ifeof1\closein1
			        \openin1=#1
				\ifeof1\closein1%
				       \openin1=\figurepath#1
					\ifeof1
					   \ps@typeout{Error, File #1 not found}
						\if@bbllx\if@bblly
				   		\if@bburx\if@bbury
			      				\def\@p@sfile{#1}%
			      				\def\@p@sbbfile{#1}%
							\@decmprfalse
				  	   	\fi\fi\fi\fi
					\else\closein1
				    		\def\@p@sfile{\figurepath#1}%
				    		\def\@p@sbbfile{\figurepath#1}%
						\@decmprfalse
	                       		\fi%
			 	\else\closein1%
					\def\@p@sfile{#1}
					\def\@p@sbbfile{#1}
					\@decmprfalse
			 	\fi
			\else
				\def\@p@sfile{\figurepath#1}
				\def\@p@sbbfile{\figurepath#1.bb}
				\@decmprtrue
			\fi
		\else
			\def\@p@sfile{#1}
			\def\@p@sbbfile{#1.bb}
			\@decmprtrue
		\fi}

\def\@p@@sfile#1{\@p@@sfigure{#1}}

\def\@p@@sbbllx#1{
		%\ps@typeout{bbllx is #1}
		\@bbllxtrue
		\dimen100=#1
		\edef\@p@sbbllx{\number\dimen100}
}
\def\@p@@sbblly#1{
		%\ps@typeout{bblly is #1}
		\@bbllytrue
		\dimen100=#1
		\edef\@p@sbblly{\number\dimen100}
}
\def\@p@@sbburx#1{
		%\ps@typeout{bburx is #1}
		\@bburxtrue
		\dimen100=#1
		\edef\@p@sbburx{\number\dimen100}
}
\def\@p@@sbbury#1{
		%\ps@typeout{bbury is #1}
		\@bburytrue
		\dimen100=#1
		\edef\@p@sbbury{\number\dimen100}
}
\def\@p@@sheight#1{
		\@heighttrue
		\dimen100=#1
   		\edef\@p@sheight{\number\dimen100}
		%\ps@typeout{Height is \@p@sheight}
}
\def\@p@@swidth#1{
		%\ps@typeout{Width is #1}
		\@widthtrue
		\dimen100=#1
		\edef\@p@swidth{\number\dimen100}
}
\def\@p@@srheight#1{
		%\ps@typeout{Reserved height is #1}
		\@rheighttrue
		\dimen100=#1
		\edef\@p@srheight{\number\dimen100}
}
\def\@p@@srwidth#1{
		%\ps@typeout{Reserved width is #1}
		\@rwidthtrue
		\dimen100=#1
		\edef\@p@srwidth{\number\dimen100}
}
\def\@p@@sangle#1{
		%\ps@typeout{Rotation is #1}
		\@angletrue
%		\dimen100=#1
		\edef\@p@sangle{#1} %\number\dimen100}
}
\def\@p@@ssilent#1{ 
		\@verbosefalse
}
\def\@p@@sprolog#1{\@prologfiletrue\def\@prologfileval{#1}}
\def\@p@@spostlog#1{\@postlogfiletrue\def\@postlogfileval{#1}}
\def\@cs@name#1{\csname #1\endcsname}
\def\@setparms#1=#2,{\@cs@name{@p@@s#1}{#2}}
%
% initialize the defaults (size the size of the figure)
%
\def\ps@init@parms{
		\@bbllxfalse \@bbllyfalse
		\@bburxfalse \@bburyfalse
		\@heightfalse \@widthfalse
		\@rheightfalse \@rwidthfalse
		\def\@p@sbbllx{}\def\@p@sbblly{}
		\def\@p@sbburx{}\def\@p@sbbury{}
		\def\@p@sheight{}\def\@p@swidth{}
		\def\@p@srheight{}\def\@p@srwidth{}
		\def\@p@sangle{0}
		\def\@p@sfile{} \def\@p@sbbfile{}
		\def\@p@scost{10}
		\def\@sc{}
		\@prologfilefalse
		\@postlogfilefalse
		\@clipfalse
		\if@noisy
			\@verbosetrue
		\else
			\@verbosefalse
		\fi
}
%
% Go through the options setting things up.
%
\def\parse@ps@parms#1{
	 	\@psdo\@psfiga:=#1\do
		   {\expandafter\@setparms\@psfiga,}}
%
% Compute bb height and width
%
\newif\ifno@bb
\def\bb@missing{
	\if@verbose{
		\ps@typeout{psfig: searching \@p@sbbfile \space  for bounding box}
	}\fi
	\no@bbtrue
	\epsf@getbb{\@p@sbbfile}
        \ifno@bb \else \bb@cull\epsf@llx\epsf@lly\epsf@urx\epsf@ury\fi
}	
\def\bb@cull#1#2#3#4{
	\dimen100=#1 bp\edef\@p@sbbllx{\number\dimen100}
	\dimen100=#2 bp\edef\@p@sbblly{\number\dimen100}
	\dimen100=#3 bp\edef\@p@sbburx{\number\dimen100}
	\dimen100=#4 bp\edef\@p@sbbury{\number\dimen100}
	\no@bbfalse
}
% rotate point (#1,#2) about (0,0).
% The sine and cosine of the angle are already stored in \sine and
% \cosine.  The result is placed in (\p@intvaluex, \p@intvaluey).
\newdimen\p@intvaluex
\newdimen\p@intvaluey
\def\rotate@#1#2{{\dimen0=#1 sp\dimen1=#2 sp
%            	calculate x' = x \cos\theta - y \sin\theta
		  \global\p@intvaluex=\cosine\dimen0
		  \dimen3=\sine\dimen1
		  \global\advance\p@intvaluex by -\dimen3
% 		calculate y' = x \sin\theta + y \cos\theta
		  \global\p@intvaluey=\sine\dimen0
		  \dimen3=\cosine\dimen1
		  \global\advance\p@intvaluey by \dimen3
		  }}
\def\compute@bb{
		\no@bbfalse
		\if@bbllx \else \no@bbtrue \fi
		\if@bblly \else \no@bbtrue \fi
		\if@bburx \else \no@bbtrue \fi
		\if@bbury \else \no@bbtrue \fi
		\ifno@bb \bb@missing \fi
		\ifno@bb \ps@typeout{FATAL ERROR: no bb supplied or found}
			\no-bb-error
		\fi
		%
%\ps@typeout{BB: \@p@sbbllx, \@p@sbblly, \@p@sbburx, \@p@sbbury} 
%
% store height/width of original (unrotated) bounding box
		\count203=\@p@sbburx
		\count204=\@p@sbbury
		\advance\count203 by -\@p@sbbllx
		\advance\count204 by -\@p@sbblly
		\edef\ps@bbw{\number\count203}
		\edef\ps@bbh{\number\count204}
		%\ps@typeout{ psbbh = \ps@bbh, psbbw = \ps@bbw }
		\if@angle 
			\Sine{\@p@sangle}\Cosine{\@p@sangle}
	        	{\dimen100=\maxdimen\xdef\r@p@sbbllx{\number\dimen100}
					    \xdef\r@p@sbblly{\number\dimen100}
			                    \xdef\r@p@sbburx{-\number\dimen100}
					    \xdef\r@p@sbbury{-\number\dimen100}}
%
% Need to rotate all four points and take the X-Y extremes of the new
% points as the new bounding box.
                        \def\minmaxtest{
			   \ifnum\number\p@intvaluex<\r@p@sbbllx
			      \xdef\r@p@sbbllx{\number\p@intvaluex}\fi
			   \ifnum\number\p@intvaluex>\r@p@sbburx
			      \xdef\r@p@sbburx{\number\p@intvaluex}\fi
			   \ifnum\number\p@intvaluey<\r@p@sbblly
			      \xdef\r@p@sbblly{\number\p@intvaluey}\fi
			   \ifnum\number\p@intvaluey>\r@p@sbbury
			      \xdef\r@p@sbbury{\number\p@intvaluey}\fi
			   }
%			lower left
			\rotate@{\@p@sbbllx}{\@p@sbblly}
			\minmaxtest
%			upper left
			\rotate@{\@p@sbbllx}{\@p@sbbury}
			\minmaxtest
%			lower right
			\rotate@{\@p@sbburx}{\@p@sbblly}
			\minmaxtest
%			upper right
			\rotate@{\@p@sbburx}{\@p@sbbury}
			\minmaxtest
			\edef\@p@sbbllx{\r@p@sbbllx}\edef\@p@sbblly{\r@p@sbblly}
			\edef\@p@sbburx{\r@p@sbburx}\edef\@p@sbbury{\r@p@sbbury}
%\ps@typeout{rotated BB: \r@p@sbbllx, \r@p@sbblly, \r@p@sbburx, \r@p@sbbury}
		\fi
		\count203=\@p@sbburx
		\count204=\@p@sbbury
		\advance\count203 by -\@p@sbbllx
		\advance\count204 by -\@p@sbblly
		\edef\@bbw{\number\count203}
		\edef\@bbh{\number\count204}
		%\ps@typeout{ bbh = \@bbh, bbw = \@bbw }
}
%
% \in@hundreds performs #1 * (#2 / #3) correct to the hundreds,
%	then leaves the result in @result
%
\def\in@hundreds#1#2#3{\count240=#2 \count241=#3
		     \count100=\count240	% 100 is first digit #2/#3
		     \divide\count100 by \count241
		     \count101=\count100
		     \multiply\count101 by \count241
		     \advance\count240 by -\count101
		     \multiply\count240 by 10
		     \count101=\count240	%101 is second digit of #2/#3
		     \divide\count101 by \count241
		     \count102=\count101
		     \multiply\count102 by \count241
		     \advance\count240 by -\count102
		     \multiply\count240 by 10
		     \count102=\count240	% 102 is the third digit
		     \divide\count102 by \count241
		     \count200=#1\count205=0
		     \count201=\count200
			\multiply\count201 by \count100
		 	\advance\count205 by \count201
		     \count201=\count200
			\divide\count201 by 10
			\multiply\count201 by \count101
			\advance\count205 by \count201
		     \count201=\count200
			\divide\count201 by 100
			\multiply\count201 by \count102
			\advance\count205 by \count201
		     \edef\@result{\number\count205}
}
\def\compute@wfromh{
		% computing : width = height * (bbw / bbh)
		\in@hundreds{\@p@sheight}{\@bbw}{\@bbh}
		%\ps@typeout{ \@p@sheight * \@bbw / \@bbh, = \@result }
		\edef\@p@swidth{\@result}
		%\ps@typeout{w from h: width is \@p@swidth}
}
\def\compute@hfromw{
		% computing : height = width * (bbh / bbw)
	        \in@hundreds{\@p@swidth}{\@bbh}{\@bbw}
		%\ps@typeout{ \@p@swidth * \@bbh / \@bbw = \@result }
		\edef\@p@sheight{\@result}
		%\ps@typeout{h from w : height is \@p@sheight}
}
\def\compute@handw{
		\if@height 
			\if@width
			\else
				\compute@wfromh
			\fi
		\else 
			\if@width
				\compute@hfromw
			\else
				\edef\@p@sheight{\@bbh}
				\edef\@p@swidth{\@bbw}
			\fi
		\fi
}
\def\compute@resv{
		\if@rheight \else \edef\@p@srheight{\@p@sheight} \fi
		\if@rwidth \else \edef\@p@srwidth{\@p@swidth} \fi
		%\ps@typeout{rheight = \@p@srheight, rwidth = \@p@srwidth}
}
%		
% Compute any missing values
\def\compute@sizes{
	\compute@bb
	\if@scalefirst\if@angle
% at this point the bounding box has been adjsuted correctly for
% rotation.  PSFIG does all of its scaling using \@bbh and \@bbw.  If
% a width= or height= was specified along with \psscalefirst, then the
% width=/height= value needs to be adjusted to match the new (rotated)
% bounding box size (specifed in \@bbw and \@bbh).
%    \ps@bbw       width=
%    -------  =  ---------- 
%    \@bbw       new width=
% so `new width=' = (width= * \@bbw) / \ps@bbw; where \ps@bbw is the
% width of the original (unrotated) bounding box.
	\if@width
	   \in@hundreds{\@p@swidth}{\@bbw}{\ps@bbw}
	   \edef\@p@swidth{\@result}
	\fi
	\if@height
	   \in@hundreds{\@p@sheight}{\@bbh}{\ps@bbh}
	   \edef\@p@sheight{\@result}
	\fi
	\fi\fi
	\compute@handw
	\compute@resv}

%
% \psfig
% usage : \psfig{file=, height=, width=, bbllx=, bblly=, bburx=, bbury=,
%			rheight=, rwidth=, clip=}
%
% "clip=" is a switch and takes no value, but the `=' must be present.
\def\psfig#1{\vbox {
	% do a zero width hard space so that a single
	% \psfig in a centering enviornment will behave nicely
	%{\setbox0=\hbox{\ }\ \hskip-\wd0}
	%
	\ps@init@parms
	\parse@ps@parms{#1}
	\compute@sizes
	\ifnum\@p@scost<\@psdraft{
		\special{ps::[begin] 	\@p@swidth \space \@p@sheight \space
				\@p@sbbllx \space \@p@sbblly \space
				\@p@sbburx \space \@p@sbbury \space
				startTexFig \space }
		\if@angle
			\special {ps:: \@p@sangle \space rotate \space} 
		\fi
		\if@clip{
			\if@verbose{
				\ps@typeout{(clip)}
			}\fi
			\special{ps:: doclip \space }
		}\fi
		\if@prologfile
		    \special{ps: plotfile \@prologfileval \space } \fi
		\if@decmpr{
			\if@verbose{
				\ps@typeout{psfig: including \@p@sfile.Z \space }
			}\fi
			\special{ps: plotfile "`zcat \@p@sfile.Z" \space }
		}\else{
			\if@verbose{
				\ps@typeout{psfig: including \@p@sfile \space }
			}\fi
			\special{ps: plotfile \@p@sfile \space }
		}\fi
		\if@postlogfile
		    \special{ps: plotfile \@postlogfileval \space } \fi
		\special{ps::[end] endTexFig \space }
		% Create the vbox to reserve the space for the figure
		\vbox to \@p@srheight true sp{
			\hbox to \@p@srwidth true sp{
				\hss
			}
		\vss
		}
	}\else{
		% draft figure, just reserve the space and print the
		% path name.
		\if@draftbox{		
			% Verbose draft: print file name in box
			\hbox{\frame{\vbox to \@p@srheight true sp{
			\vss
			\hbox to \@p@srwidth true sp{ \hss \@p@sfile \hss }
			\vss
			}}}
		}\else{
			% Non-verbose draft
			\vbox to \@p@srheight true sp{
			\vss
			\hbox to \@p@srwidth true sp{\hss}
			\vss
			}
		}\fi

	}\fi
}}
\psfigRestoreAt

%\input chap1figs.tex
%\input chap2figs.tex
%\end{document}
%%%%%%%%%%%%%%%%%%%%%%%%%%%%%

\renewcommand{\thepage}{\roman{page}}
\begin{titlepage}   % Count as [i], not numbered.
\setcounter{page}{1}

\begin{center}

{\Huge\bf Probes of the Early Universe}
\bigskip\bigskip

by

\bigskip\bigskip

{\Large\bf Max Erik Tegmark}

\bigskip\bigskip

B.A., Stockholm School of Economics, 1989

B.A., Royal Institute of Technology, Stockholm, 1990

M.A., University of California at Berkeley, 1992

\bigskip
\bigskip
\bigskip
A dissertation submitted in satisfaction of 
the final requirement for the degree of
\bigskip

{\Large\bf Doctor of Philosophy}
\bigskip

in
\bigskip

{\Large\bf Physics}
\bigskip

in the

Graduate Division

of the
\bigskip

{\Large\bf University of California at Berkeley}
\end{center}

\bigskip
\bigskip\bigskip\bigskip\bigskip
\leftline{\bf\hglue4.6cm Committee in charge:}
\smallskip
\leftline{\hglue4.6cm Professor Joseph Silk, Chair}
\leftline{\hglue4.6cm Professor Bernard Sadoulet}
\leftline{\hglue4.6cmProfessor Hyron Spinrad}
\bigskip
\bigskip\bigskip\bigskip
\centerline{\bf April 1994}

\end{titlepage}

%%%%%%%%%%%%%%%%%%%%%%%%%%%%%%%%%%%%%%%%%%%%%%%%%%%%%%

\pagestyle{empty}
\setcounter{page}{2}
\cleardoublepage

%approval page (count as [ii];  not numbered)
\centerline{\,} % Stupid, but needed for latex not to ignore the 
 	        % vskip on the following line...
\vskip0.2truein
\centerline{The dissertation of Max Tegmark is approved:}
\vskip1.6truein
\centerline{\hfil\leaders\hrule \hskip 4truein\hfil}
\centerline{\hfil\quad Chair\hskip2truein\hskip6mm \hss Date\quad\hfil}
\bigskip
\bigskip
\medskip
\centerline{\hfil\leaders\hrule \hskip 4truein\hfil}
\centerline{\hfil \hskip 2.75truein \hss Date\quad\hfil}
\bigskip
\bigskip
\medskip
\centerline{\hfil\leaders\hrule \hskip 4truein\hfil}
\centerline{\hfil \hskip 2.75truein \hss Date\quad\hfil}
\vskip 1.5 truein
\centerline{University of California at Berkeley}
\bigskip
\bigskip
\centerline{April 1994}
\vfill
\cleardoublepage
%\newpage
%%%%%%%%%%%%%%%%%%%%%%%%%%%%%%%%%%%%%%%%%%%%%%%%%%%%%%

%copyright page (or blank) (not numbered)
\centerline{\,} % Stupid, but needed for latex not to ignore the 
 	        % vskip on the following line...
\vskip2.5truein
\centerline{Probes of the Early Universe}
\bigskip
\centerline{\copyright 1994}
\bigskip
\centerline{by}
\bigskip
\centerline{Max Tegmark}
\vfill\
%\newpage
\cleardoublepage
%%%%%%%%%%%%%%%%%%%%%%%%%%%%%

% \pagenumbering{arabic}    
\pagestyle{plain}

% \mysectionn{Abstract}
% \addcontentsline{toc}{section}{Abstract}
\nothing\vskip2cm

{
\double
\begin{center}

{\Large\bf Abstract}
\bigskip

Probes of the Early Universe

by

Max Erik Tegmark

Doctor of Philosophy in Physics

University of California at Berkeley

Professor Joseph Silk, Chair

\bigskip
\bigskip

\end{center}

One of the main challenges in cosmology is to quantify how small 
density fluctuations at the recombination epoch $z\approx 10^3$ 
evolved into the galaxies and the large-scale structure we 
observe in the universe today. 
This thesis discusses ways of probing the intermediate 
epoch, focusing on the thermal history.
The main emphasis is on the role played by 
non-linear feedback, where a small fraction of matter forming 
luminous objects can inject enough energy into the
inter-galactic medium (IGM) to 
radically alter subsequent events.
The main conclusions are:

\begin{itemize}

\item
Early structures
corresponding to rare Gaussian peaks in a cold dark matter (CDM)
model can
photoionize the IGM
early enough to appreciably smooth out fluctuations in 
the cosmic microwave background radiation (CBR), 
provided that these early structures are quite small, 
no more massive than about $10^8 M_{\odot}$. 

\item
Typical parameter values predict that reionization
occurs around $z=50$, thereby reducing
fluctuations on degree scales
while leaving the larger angular scales probed by COBE relatively 
unaffected.

\item
For non-standard CDM incorporating mixed dark matter,
vacuum density, a tilted primordial power spectrum or
decaying $\tau$ neutrinos, early reionization
is not likely to play a significant role.

\item
For CDM models with $\Omega<1$ and $\Lambda=0$, 
the extent of this suppression is
quite insensitive to  $\Omega_0$,
as opposing effects partially cancel.

\item

It is still not certain that the
universe underwent a neutral phase, despite the new COBE FIRAS
limit $y <2.5\times 10^{-5}$ on
Compton $y$-distortions of the cosmic microwave background.

\item
The observed absence of a
Gunn-Peterson trough in the spectra of high-redshift
quasars can be explained without photoionization, in 
in a scenario in which supernova-driven winds
from early galaxies reionize the IGM by $z=5$.

\item
It is possible to place 
constraints on cosmological models that are independent of the shape
of the primordial power 
spectrum
--- the only assumption being that the random fields are Gaussian.
As an example of an application, 
the recent measurement of 
bulk flows of galaxies by Lauer and Postman 
is shown to be inconsistent with the CBR experiment SP91
at a 95\% confidence level
regardless of the shape of the power spectrum.

\end{itemize}

}  
\vskip1.6truein
\centerline{\hfil\leaders\hrule \hskip 4truein\hfil}
\centerline{\hfil\quad Prof. Irwin Shapiro\hskip 2truein \hss Date\quad\hfil}

\cleardoublepage
%\newpage

%%%%%%%%%%%%%%%%%%%%%%%%%%%%%
% DEDICATION PAGE

\pagestyle{plain}

\begin{center}

\centerline{\,} % Stupid, but needed for latex not to ignore the 
 	        % vskip on the following line...
\vskip2.2truein
{\Large To Richard Feynman,}
\bigskip

1918-1988,
\bigskip

who was the reason I decided to switch to physics

\end{center}

\newpage

%%%%%%%%%%%%%%%%%%%%%%%%%%%%%

\tableofcontents

%%%%%%%%%%%%%%%%%%%%%%%%%%%%%

\listoffigures
\addcontentsline{toc}{section}{List of Figures}

%%%%%%%%%%%%%%%%%%%%%%%%%%%%%

\listoftables
\addcontentsline{toc}{section}{List of Tables}

%%%%%%%%%%%%%%%%%%%%%%%%%%%%%

\mysectionn{List of Abbreviations}
\addcontentsline{toc}{section}{List of Abbreviations}

\begin{description}

\item[BBKS] Bardeen, Bond, Kaiser \& Szalay (1986), a reference in Chapter 6.

\item[BDM] Baryonic Dark Matter

\item[CBR] Cosmic (Microwave) Background Radiation

\item[CDM] Cold Dark Matter

\item[COBE] Cosmic Background Explorer

\item[DMR] Differential Microwave Radiometer, an instrument on board the 
COBE satellite that measures the difference in CBR temperature between 
different parts of the sky 

\item[FIRAS] Far Infra-Red Absolute Spectrometer, an instrument on board the 
COBE satellite that measures the CBR spectrum

\item[FRW] The Friedmann-Robertson-Walker metric, sometimes 
referred to as the Friedmann-Lema\^{\i}tre metric

\def\i{x}

\item[GO] Gnedin \& Ostriker (1993), a reference in Chapter 5

\item[GR] General Relativity

\item[HDM] Hot Dark Matter

\item[IGM] Intergalactic Medium

\item[IRAS] Infra-Red Astronomical Satellite

\item[$\Lambda$CDM] Cold  Dark Matter 
with a non-zero cosmological constant $\Lambda$

\item[LP] Lauer \& Postman, a reference in Chapter 7

\item[MACHO] Massive Compact Halo Object, a dark matter candidate

\item[MDM] Mixed Dark Matter, a mixture of HDM and CDM

\item[SP91] South Pole 1991 CBR experiment referenced in chapters 3, 4 and 7

\item[QSO] Quasi-Stellar Object --- quasars are a subset of these 

\item[$\tau$CDM] Cold Dark Matter with a decaying $\tau$ neutrino

\item[UV] Ultraviolet

\item[WIMP] Weakly Interacting Massive Particle, 
a dark matter candidate

\end{description}

\newpage
%%%%%%%%%%%%%%%%%%%%%%%%%%%%%

\mysection{Preface}
\addcontentsline{toc}{section}{Preface}
\baselineskip16pt

As this thesis is laden with cosmology and astrophysics jargon, 
it is recommended that the 
non-cosmologist reader flick through Chapter 2
before attempting to read anything else. 
This chapter is 
by no means intended to be a comprehensive 
treatise on the foundations of cosmology. 
It is merely an attempt to provide the basic knowledge
necessary for a reader with a general physics background to 
be able to follow the other chapters.
Cosmologists are advised to skip directly from the introduction 
to Chapter 3. 

During my graduate studies here at Berkeley, I also did some non-cosmology 
work (Tegmark 1993, Tegmark \& Yeh 1994, Tegmark \& Shapiro 1994, 
Shapiro \& Tegmark 1994), but that research will not be 
discussed in this thesis.

\newpage

%%%%%%%%%%%%%%%%%%%%%%%%%%%%%
%\input berkeley_ack.tex
%%%%%%%%%%%%%%%%%%%%%%%%%%%%%

\mysectionn{Acknowledgements...}
\addcontentsline{toc}{section}{Acknowledgements}
\baselineskip15pt
\ns\ns

\begin{verse}

I wish to thank

my advisor Joe Silk
\begin{quote}
for the inspiration, support and guidance,\\
for the infinite degrees of freedom, and\\
for proving that success in science is not irreconcilable with having a life,
\end{quote}
 
my Mom,\\
\begin{quote}
one foxy lady,\\
thanks to whom science is still less than half of my life,
\end{quote}
 
my Dad,\\
\begin{quote}
who encouraged my curiosity and shared my fascination\\
and who thinks that the universe was created as a source of 
interesting math problems, 
\end{quote}

my brother Per 
\begin{quote}
--- iiiiiiih..., 
\end{quote}
 
Justin Bendich, Ted Bunn, Wayne Hu, Bill Poirier and Leehwa Yeh, 
\begin{quote}
with whom I've so enjoyed sharing my bizarre physics interests, 
be it over French Alps
or homebrewn beer, 
\end{quote}
 
teachers like Geoffrey Chew, Andy Elby, Alex Filippenko,\\
 Lawrence Hall and
Bruno Zumino, 
who inspired me,
 
% my various co-authors, 
% who allowed me to incorporate joint work into this thesis,
 
the Cyberscope and the Simsalabimiers, 
who stole away so much of my time,
 
the friendly staff in the physics and astronomy departments 
here at Berkeley,\\
especially Anne Takizawa and Donna Sakima,\\ 
who saved me every time I was dropped
from the rolls of the university,
 
and last but not least, the people close to me here in 
Berkeley,\\ 
who have put up with
this crazy cosmologist and made these four years\\ 
perhaps the happiest of my life.

\end{verse}

%%%%%%%%%%%%%%%%%%%%%%%%%%%%%

\mysection{...and More Acknowledgements}
\addcontentsline{toc}{section}{...and More Acknowledgements}
\double

The research presented in Chapters 3 through 7 was done in 
collaborations, all but Chapter~7 with my 
thesis advisor Joe Silk, 
Chapter 3 with Alain 
Blanchard, Chapter 6 with August Evrard, and Chapter 7 with 
Emory Bunn and Wayne Hu. 
%%%%% CUT HERE FOR BERKELEY VERSION%%%%%
%%%%%END BERKELEY CUT%%%%%
In all cases except Chapter~\ref{pindepchapter}, I made  
the calculations, produced the figures and wrote the
bulk of the text. In ``Power spectrum independent constraints on cosmological
models",
Emory Bunn performed the Monte Carlo analysis 
for modeling the COBE experiment and
made the fits described by 
equations\eqnum{COBEfitEq1} and\eqnum{COBEfitEq2}.
He also calculated all the window
functions analytically and numerically, wrote 
Appendix B and wrote the first version of our
FORTRAN data analysis code. Wayne Hu originally 
spawned the idea of applying power
spectrum independent constraints to bulk flows, 
and independently verified the
calculations. 
 
For Chapter~\ref{primerchapter}, I gratefully 
acknowledge Naoshi Sugiyama for letting me use 
power spectra generated by his Boltzmann code in Figure~\ref{powerplot}
and Martin White for letting me use window function data from 
White, Scott \& Silk (1994) in Figure~\ref{clplot}. 

For Chapters~\ref{reionchapter} and~\ref{ychapter}, I gratefully 
acknowledge William Vacca for 
providing stellar spectra used to calculate an entry in
Table~\ref{reionfig3} and Table~\ref{yfig1}. 

In addition, I wish to thank 
Emory Bunn, 
Angelica de Oliveira Costa,
Wayne Hu, 
Andreas Reisenegger, 
David Schlegel, 
Douglas Scott,
Harold Shapiro, 
Joseph Silk,
Charles Steidel,
Michael Strauss, 
and
Martin White,
for many useful comments 
and/or for help with proofreading the manuscript.

%%%%%%%%%%%%%%%%%%%%%%%%%%%%%
%%%%%%%%%%%%%%%%%%%%%%%%%%%%%

\cleardoublepage
%\newpage

\pagenumbering{arabic}    
\pagestyle{headings}
\setcounter{secnumdepth}{2}

\chapter{Introduction}
\label{introchapter}

One of the main challenges in modern cosmology is to quantify
how small density fluctuations at the recombination epoch, 
about half a million years after 
the Big Bang, evolved into the galaxies and the large-scale
structure we observe in the universe today, some $10^{10}$
years later. This thesis 
focuses on ways of probing the interesting intermediate epoch.
The main emphasis is on the role played by non-linear feedback, 
where a small fraction of matter forming luminous
objects such as stars or QSO's can inject enough energy into their
surroundings to radically alter subsequent events. 
Specific questions addressed include the following:

\begin{itemize}

\item 
Which cosmological models predict that the universe will become
photoionized early enough to measurably affect spatial fluctuations in the
cosmic microwave background radiation (CBR)?

\item 
Which cosmological models are ruled out by the constraints
from the COBE FIRAS experiment on the spectral distortions of the CBR?

\item 
Which cosmological models satisfy the Gunn-Peterson test, {\ie} predict
the intergalactic medium (IGM) to become highly ionized by $z=5$?

\item 
Is reionization required to make bulk flows consistent with degree-scale
anisotropies?

\end{itemize}

\noindent
Thus this thesis provides a 
quantitative link between cosmological models and extragalactic 
observational astronomy, showing how current observations 
constrain the allowed parameter space of cosmological models.
\bigskip

The next two sections of this introduction are aimed at placing the work in 
a larger context, and the final section gives a 
brief description of how the thesis is organized into chapters. 
As mentioned in the preface, the non-cosmologist is advised to flick through
Chapter~\ref{primerchapter} 
first, before attempting to read further. 

\section{Cosmology towards a New Millennium}

Like the universe, the field of cosmology is expanding.
Thanks to the rapid technological developments of the last few decades,
we humans now have better eyes than ever 
before with which to scrutinize our
surroundings, both from the ground and from various spacecraft. 

On the ground, the recently inaugurated ten-meter Keck telescope
on Hawaii provides record-breaking light-gathering power in optical
wavelengths. Interferometry between radio telescopes on separate continents
provides angular resolutions that our ancestors could only have dreamed of.
Semiconductor technology has replaced photographic plates by much
more sensitive electronic detectors (CCDs), and also, in the form of computers,
enabled  previously unheard of capacity for data processing. A notable
example is the MACHO project, looking for dark matter 
gravitational microlensing events in the halo of our galaxy, a search
that involves looking at some one million stars every night
(Alcock {\etal} 1993). 
Another example is the Sloan Digital Sky Survey, whose
goal is to measure the redshifts of a million galaxies 
before the end of the millennium. To place this in 
perspective, when Slipher
obtained the the 18 data points that Hubble used to support his famous
expansion law (Hubble 1929), he required weeks 
of observations with a by contemporary standards good
telescope
in order to obtain the redshift of a single nearby galaxy.

In space, scores
of detectors have been observing the cosmos unhindered by the astronomer's worst
enemy: 
the earth's atmosphere. After being repaired, the Hubble Space Telescope
now obtains optical images with ten times better angular resolution than
obtainable from the ground. The Cosmic Background Explorer satellite
(COBE) has made an all-sky map at millimeter wavelengths
(Smoot {\etal} 1992). The IRAS and ROSAT
satellites have done the same in the infrared and x-ray wavelength bands,
respectively, and a number of satellites have explored gamma rays and the
ultraviolet  as well.  Many of these wavelengths are almost completely absorbed
by the atmosphere, and thus inaccessible to ground-based experiments. 
An overview of our knowledge of the cosmological photon spectrum in all
wavelength bands from $10^{-24}\cm$ to $10^6\cm$ is given by 
Ressell \& Turner (1990).

With all this data pouring in, it is a daunting task for theorists to keep up
and try to make sense of it all. 
Although it cannot be overemphasized how much
remains to be done, the advances (interrupted by occasional 
false clues) of the last few
decades have been quite astonishing by any standards. 
As a more detailed example,
let us take something that is directly relevant to the material in this thesis:
the cosmic microwave background radiation (CBR). 
Thirty years ago, we did not even know that it existed. 
Confirming theoretical
predictions of Gamow and others, it was 
serendipitously discovered by Penzias and
Wilson in 1964 (Penzias \& Wilson 1965) and given its 
cosmological interpretation by
Dicke, Peebles, Roll \& Wilkinson (1965). 
In Figure~\ref{yplot}, the best 
limit on the Compton $y$-parameter is plotted as a
function of year. This parameter is a measure of how much the spectrum deviates
from a Planck spectrum, and when it was 
first defined (Zel'dovich \& Sunyaev
1969), one was still arguing
as to whether the CBR had a Planck spectrum or not, so the 
constraints on $y$ were of order unity. About a decade later, the limit had
dropped to about 0.25 (Woody \& Richards 1981). Almost another decade
later, it dropped to around 0.1 thanks to a Berkeley-Nagoya rocket experiment
(Matsumoto {\etal} 1988).
Here a fairly common cosmo-sociological effect occurred: 
an apparent deviation
from a Planck spectrum, now sarcastically referred to as ``the Berkeley Bump",
led to scores of theoretical models 
explaining it before it was officially
admitted to be merely an artifact caused by problems with the detector.
In 1990, the FIRAS experiment aboard the COBE 
satellite suddenly pushed the limit
down by a full two orders of magnitude (Mather {\etal} 1990). And the
exponential progress continued, the current limit being 
$2.5\tt{-5}$ (Mather {\etal} 1994). 

This mind-boggling progress with CBR spectral 
distortions has been paralleled by
improved measurements of the spatial distortions.
The 
fairly isotropic radiation field was observed to have a dipole
anisotropy of order $10^{-3}$ (Conklin 1969; Henry 1971;
Smoot {\etal} 1977), 
which is normally interpreted as a Doppler shift
due to our motion  relative to the CBR rest frame. 
A few years ago, the COBE DMR
experiment detected smaller scale fluctuations as well 
(Smoot {\etal} 1992), at
a level of about $10^{-5}$, which are 
believed to be of cosmological origin. 
Since this time period, there has been an
enormous surge in balloon-based and ground-based CBR experiments that probe
smaller angular scales, 
and now in the spring of 1994, new results are coming in at
a rate of almost one per month. 

It is precisely this type of data, in combination with 
improved constraints on the matter power spectrum 
from galaxy surveys, that
is allowing us to start pinning down 
the thermal history of the 
IGM since recombination, 
the main topic of this thesis.

\section{Reheating}

It is now widely accepted that the universe underwent a reheating phase at
some point after the standard recombination epoch at redshift 
$z\approx 10^3$. 
The absence of a Gunn-Peterson trough in the spectra of
high-redshift quasars has provided strong
evidence for the reheating occurring at a redshift $z>5$, since it indicates
that the intergalactic medium (IGM) was highly ionized at lower
redshifts (Gunn \& Peterson 1965; Steidel \& Sargent 1987; Webb
{\etal} 1992).  The smallest baryonic objects to go non-linear in a standard cold
dark matter (CDM) model are expected to reionize the IGM at a redshift somewhere
in the range $20 < z < 100$ (Couchman \& Rees 1986).
In the most recent models with baryonic dark matter, reheating and
reionization are 
predicted to occur at an even higher redshift, typically in the
range $100 < z < 1000$ (Peebles 1987; Gnedin \& Ostriker 1992;
Cen {\etal} 1992; Cen {\etal} 1993).

A reheating epoch would have at least two interesting classes of effects that may
be measurable today: effects on subsequent structure formation and effects on
the cosmic microwave background radiation (CBR).

Subsequent structure formation would be affected in a number of ways.
First of all, the heating of the IGM up to a higher adiabat would raise the
Jeans mass, thus suppressing the formation of small objects. For instance,
an IGM temperature of $10^5\K$ at a redshift of a few would
suppress the formation of galaxies of mass below $10^{10}\Ms$, thus 
alleviating the ubiquitous problem of theories overpredicting the abundance of
faint galaxies (Blanchard {\etal} 1992).
% Kauffmann, G.; White, S.D.M.; Guiderdoni, B.
% The formulation and evolution of galaxies within merging dark matter haloes.
% Monthly Notices of the Royal Astronomical Society, 1 Sept. 1993, vol.264, (no.1):201-18.
Secondly, if the objects that reheat the IGM
also enrich it with heavy elements, the ability of gas to cool would be
greatly enhanced in the temperature range $10^4\K<T<10^7\K$,
presumably facilitating future structure formation. 

The CBR would be affected in at least three ways:

\begin{enumerate}

\item
Hot ionized IGM would cause spectral distortions 
(Kompaneets 1957; Zel'dovich \& Sunyaev 1969; 
Bartlett \& Stebbins 1991) which might
violate the stringent limits on the the Compton $y$-parameter (Mather {\etal}
1994). 
% This is a problem mainly for BDM models (Tegmark \& Silk 1993).

\item
Spatial fluctuations on angular scales below a few degrees 
might be suppressed,
while fluctuations on larger scales would remain fairly unaffected. This effect
is particularly interesting since some degree-scale microwave background
experiments carried out at the South Pole 
(Meinhold \& Lubin 1991; Shuster {\etal} 1993)
and at balloon altitudes
(Devlin {\etal} 1992; Meinhold {\etal} 1993;
Shuster {\etal} 1993)
appear to detect fluctuation amplitudes lower than those expected in 
a CDM model normalized to the the COBE
DMR experiment 
(Smoot {\etal} 1992),
which probes larger angular scales.
The interpretation of this remains unclear, however, as 
many other experiments ({\eg} de Bernardis {\etal} 1993;
Dragovan {\etal} 1993; Cheng {\etal} 1993; 
Gundersen {\etal} 1993)
have detected higher levels of fluctuations.

\item
New spatial fluctuations 
will be generated on smaller angular scales, through the so called 
Vishniac effect (Vishniac 1987, Hu, Scott \& Silk 1994). 
The current upper limit on CBR fluctuations on arcminute scales
({\eg} Subrahmanyan {\etal} 1993) places constraints 
on some reheating scenarios. 

\end{enumerate}

\noindent
With the recent surge in CBR experiments and the many new numerical,
theoretical and observational results on structure formation, 
the thermal history of the universe is now coming
within better reach of our experimental probes. 
In view of this, it is
very timely to theoretically investigate the nature of the reheating epoch in
greater detail.
It is also important to use observational data to place firm constraints
on the thermal history of the universe, constraints that do not require any
assumptions about uncertain entities like the primordial power spectrum.

\section{A Sneak Preview}

In this thesis, we will do the following:

\bigskip

In 
Chapter~\ref{reionchapter}, 
early photoionization of the intergalactic medium is discussed in a
nearly model-independent way, in order to 
investigate whether early structures
corresponding to rare Gaussian peaks in a CDM model can
photoionize the intergalactic medium early enough
to appreciably smooth out the microwave background fluctuations.
We conclude that this is indeed possible for a broad range
of CDM normalizations and is almost inevitable for unbiased CDM, 
provided that the
bulk of these early structures are quite small, no more massive than about
$10^8 M_{\odot}$. Typical parameter values predict that reionization
occurs around $z=50$, thereby suppressing fluctuations on degree scales
while leaving the larger angular scales probed by COBE relatively 
unaffected.
However, for non-standard CDM, incorporating mixed dark matter,
vacuum density or a tilted primordial power spectrum, early reionization
plays no significant role.

\bigskip

In 
Chapter~\ref{openreionchapter}, 
the results of 
Chapter~\ref{reionchapter} 
are generalized to CDM models where
$\Omega < 1$ and the cosmological constant $\Lambda>0$.
Such models often require early reionization to suppress degree-scale
anisotropies in order to be consistent with  experimental data. It is found that
if the cosmological constant 
$\Lambda = 0$, the extent of this suppression is
quite insensitive to  $\Omega_0$,
as opposing effects partially cancel.
Given a $\sigma_8$-normalization today, 
the loss of small-scale power associated
with a lower $\Omega_0$ is partially canceled by higher optical depth from
longer lookback times and by structures forming at higher
redshifts as the universe becomes curvature dominated at $z\approx
\Omega_0^{-1}$.
The maximum angular scale on which fluctuations are
suppressed decreases when $\Omega_0$ is lowered,
but this effect is also rather weak and unlikely to be
measurable in the near future.
For flat models, on the other hand, where $\lambda_0 = 1 - \Omega_0$, 
the negative effects of lowering $\Omega_0$ dominate, 
and early reionization is not likely to play a
significant role if $\Omega_0\ll 1$. 
The same goes for CDM models
where the shape parameter $\Gamma$ is lowered by increasing the 
number of relativistic particle species.

\bigskip

In 
Chapter~\ref{ychapter}, 
we find that 
it is still not certain that the
universe underwent a neutral phase, despite the new COBE FIRAS
limit $y <2.5\times 10^{-5}$ on
Compton $y$-distortions of the cosmic microwave background.
Although scenarios where the very early ($z\sim 1,000$) ionization is
thermal (caused by IGM temperatures exceeding $10^4$K) are clearly
ruled out, there is a significant loophole for cosmologies with
typical CDM parameters if the dominant ionization mechanism is
photoionization.
If the ionizing radiation has a typical quasar spectrum,
then the $y$-constraint implies roughly
$h^{4/3}\Ob \Omega_0^{-0.28}<0.06$ 
for fully ionized models.
This means that baryonic dark matter (BDM)
models with $\Omega_0\approx 0.15$ and reionization
at  $z\approx 1,000$ are strongly constrained even in this very
conservative case, and can survive the $y$ test only if
most of the baryons form BDM around the reionization epoch.

\bigskip

In 
Chapter~\ref{gpchapter}, 
a model is presented in which supernova-driven winds
from early galaxies reionize the IGM by $z=5$.
This scenario can explain the observed absence of a
Gunn-Peterson trough in the spectra of high-redshift
quasars
providing that the bulk of these early galaxies are quite small,
no more massive than about $10^8 M_{\odot}$. It also
predicts that
most of the IGM was enriched to at least $10\%$ of current
metal content by $z=5$ and perhaps as early as $z=15$.
The existence of such
early mini-galaxies violates no spectral constraints
and is
consistent with a pure CDM model with $b\leq 2$. Since the final
radius of a typical ionized bubble is only around 100 kpc, the
induced modification of the galaxy autocorrelation function is
negligible, as is the induced angular smoothing of the CBR.
Some of the gas swept up by shells may be observable as
pressure-supported Lyman-alpha forest clouds.

\bigskip

In 
Chapter~\ref{pindepchapter}, 
a formalism is presented that allows cosmological
experiments to be tested for consistency, and allows a simple frequentist
interpretation of the resulting significance levels.
As an example of an application, this formalism is used to place 
constraints on bulk flows of galaxies using the
results of the microwave background anisotropy experiments COBE and SP91,
and a few simplifying approximations about the experimental 
window functions.
It is found that if taken at face value, with the quoted errors, 
the recent detection by Lauer and Postman (1994) of a bulk
flow of 689 km/s on scales of 150$h^{-1}$Mpc
is inconsistent with SP91 at a 95\% confidence level 
within the framework of a Cold Dark Matter (CDM)
model.
The same consistency test is also used to place constraints that are
completely 
model-independent, in the sense that they hold for any 
power spectrum whatsoever
--- the only assumption being that the random fields are Gaussian.
It is shown that the resulting infinite-dimensional optimization 
problem reduces
to a set of coupled non-linear equations that can readily be solved
numerically.
Applying this technique to the above-mentioned example, we find that the
Lauer and Postman result is inconsistent with SP91
even if no
assumptions whatsoever are made about the power spectrum.

\bigskip

These chapters are all self-contained and fairly independent
(with the exception of 3 and 4), so they can be read in any order.

\def\fheight{10.3cm} \def\fwidth{14.5cm}

\newpage
\bfig
\psfig{figure=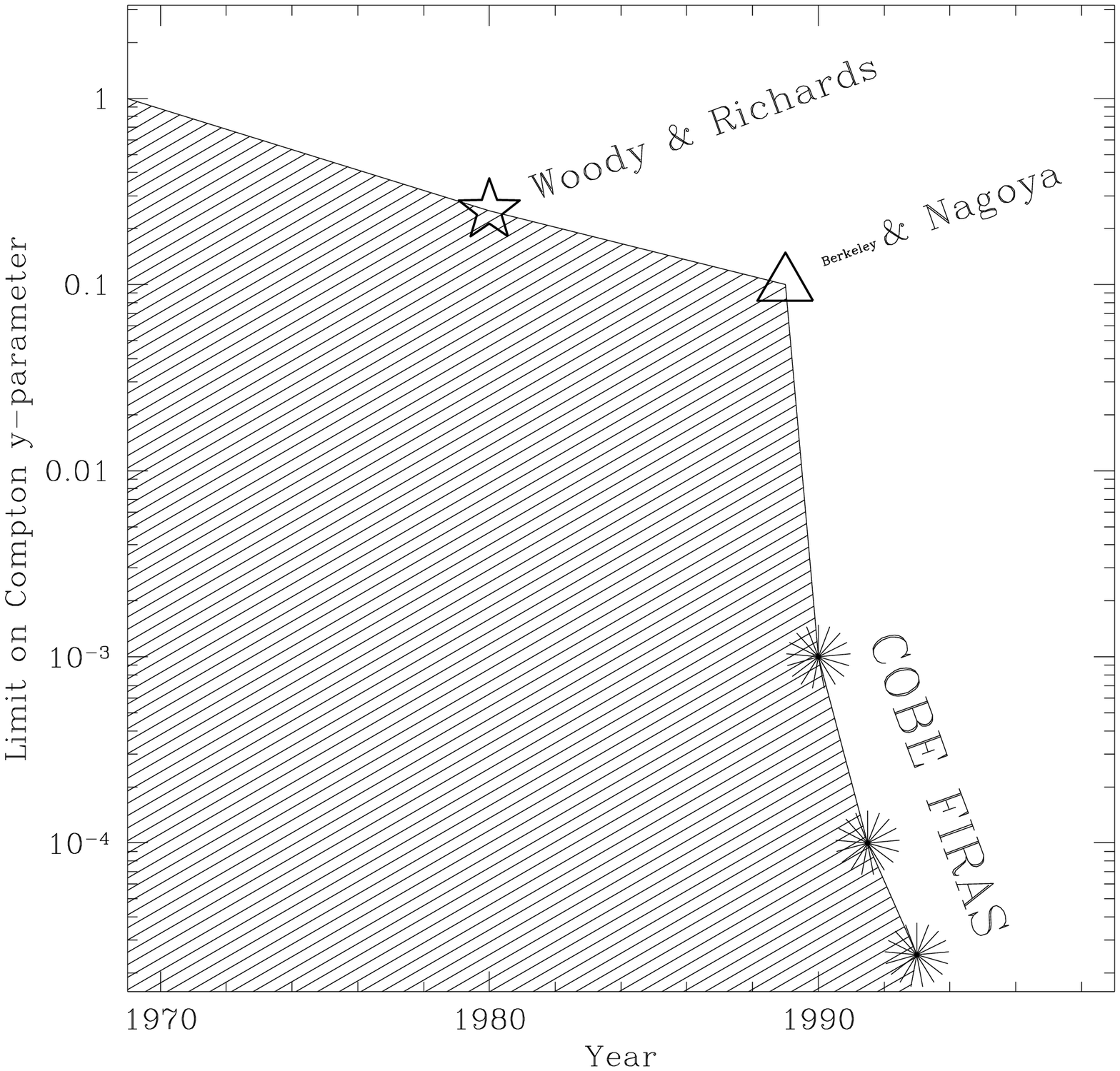,width=\fwidth,height=\fheight}
\caption{Limits on the Compton $y$-parameter.}
\label{yplot}
 
\mycaption{The observational upper limit on the 
Compton $y$-parameter is plotted as
a function of year.
}
\efig

\cleardoublepage
%\input chapter2.tex
% \countdef\ClassicalFieldEqs=176
% \countdef\JeansEq=177

\chapter{A Cosmology Primer}
\label{primerchapter}

This chapter is by no means intended to be a thorough and
systematic treatise on the basics of cosmology. 
It is merely an attempt to provide the bare-bones knowledge
necessary for the reader with a general physics background to 
be able to follow the other chapters.
Cosmologists are advised to skip directly to 
Chapter 3.

\section{Where to Read More}

For the reader interested in the basics of modern 
cosmology, there
are many excellent books to chose from.
Two early classics are 
\begin{itemize}
\item {\it Physical Cosmology} by Peebles (1971) and 
\item {\it Gravitation and Cosmology} by Weinberg (1972).
\end{itemize}
More recent texts include 
\begin{itemize}
\item {\it The Early Universe} by Kolb and Turner (1990),
\item {\it Modern Cosmology} by Dolgov, Sazhin and Zel'dovich (1990),
\item {\it Introduction to Cosmology} by Narlikar (1993)
\item {\it Structure Formation in the Universe} by Padmanabhan (1993) and 
\item {\it Principles of Physical Cosmology} by Peebles (1993).
\end{itemize}
There is also a large number of good review articles, for instance
in conference proceedings and printed lecture series.
A very incomplete list includes the following:
\begin{itemize}

\item
{\it Vatican Study Week: Large Scale Motions in the Universe},
% \footnote{Yes, the pope {\it
% does} accept the Big Bang now,  as long as we don't try to push physics beyond
% the Planck epoch...},
edited by Rubin \& Coyne (1988),

\item
{\it Development of Large-Scale Structure in the Universe} by 
Ostriker (1991),

\item
{\it Physics of the Early Universe}, edited by Peacock, Heavens and Davies 1990
(see especially the articles by Efstathiou and White),

\item
{\it Observational and Physical Cosmology}, edited 
by S\'anchez, Collados and Rebolo 1990 (see especially the article by Bernard
Jones),

\end{itemize}

\noindent
The detailed reference information for these publications 
is in the reference section at the end.
Essentially all the results in the rest of this chapter can be found in books
such as these, so very few explicit references will be given.

\section{So what is cosmology all about?}

Cosmologists study the history of the universe, from its birth some $10^{10}$
years ago up until today. 
Since the discovery of the cosmic microwave background radiation,
the Big Bang model has been generally accepted by the scientific community.
Although we still know very little about what happed before 
the universe was about a hundredth of a 
second old, the Big Bang model is widely
believed to give a correct basic picture of what happened after that. 
Here are some of the cornerstones of this standard cosmological model:

\begin{enumerate}

\item The universe is homogeneous and isotropic on large scales.

\item The universe is expanding.

\item The universe used to be dense and hot.

\item The universe is full of microwaves.

\item Structures formed through gravitational instability.

\item The universe contains dark matter.

\end{enumerate}

\noindent
We will now discuss these issues one by one.

\smallskip

1. By large-scale homogeneity is meant that if we average over sufficiently large
volumes, the density of the universe is roughly the same everywhere.
This of course implies that the density is isotropic as 
well,
{\ie} that it appears the same in whatever direction we choose to 
look\footnote{Note that homogeneity implies isotropy only for
scalar quantities (such as the density), not vectorial 
or tensorial quantities such as the velocity or the metric.}.
Not only the density, but all other fields as well (the velocity field, 
the metric, {\it etc.}) are assumed to be homogeneous and isotropic on large scales. 
There is now strong observational support for 1.
Apart from this, we
also tend to believe in homogeneity and isotropy because this greatly simplifies our
calculations...

\smallskip

2. All distant galaxies that we see in the sky are moving away from
us, and the further away they are, the faster they tend to be receding. 
The homogeneity implies that on large scales, everything is indeed receding from 
everything else. In terms of classical physics, this is analogous to the way
that all the raisins in an expanding raisin cake recede from one another, and
just as in this analogy, the speed with which to objects are receding
from each other is proportional to the distance between them.
The proportionality constant $H$ is called the Hubble constant. 
Its value is such that objects separated by more than about $10^{10}$ light-years
are receding from each other faster than the speed of light. Doesn't
this violate
special relativity? Yes it does, but it's OK in general relativity... 
As will be discussed in the next section, 
rather than picturing the matter expanding through space at enormous
speeds, a more natural way to think of this is
to imagine the the matter as merely sitting still while space itself is
expanding. The mathematical description of the expanding universe is called the
Friedmann-Robertson-Walker (FRW) metric, 
and will be introduced in the next section. 

\smallskip

3. Extrapolating this expansion backwards in time, one concludes that the
universe used to be much denser and hotter in the past. Indeed, the FRW metric
tells us that some $10^{10}$ years 
ago\footnote{An amusing fact is that when Edwin Hubble first
attempted to measure the constant that now bears his name, 
a false assumption led him to calculate a value more than five times larger
than the current estimates. This implied that the universe was ``only" about
2,000,000,000 years old. 
Since radioactive dating methods had showed certain meteorites
to be about twice as old as that, the Big Bang theory came under
fire from people wondering how 
rocks could be older than the universe itself.
This age discrepancy was one of the reasons that 
Hoyle and others developed a competing (and now dead) theory, 
the so called steady state theory,
in which there is no Big Bang --- and indeed the reason that Hoyle 
condescendingly coined the term ``Big Bang".
}, 
the universe was denser and hotter than any object we have ever been able to study.
Since we have no experimental knowledge about how matter behaves when it
is so hot and dense, 
there was an early phase of the universe about which we cannot predict anything
without making speculative extrapolations. After the universe was about a
hundredth of a second old, however, the temperatures were no greater than those
that we can study in stars and in laboratories, so the physical processes that
operated after that are fairly well understood and can be used to make detailed
quantitative predictions. 
One such prediction is that most of the mass of the universe will be
in the form of hydrogen, and about $25\%$ will be helium.
The fact that the observed helium fraction is indeed around $25\%$ 
is one of the great triumphs of the Big Bang model. 

\smallskip

4. Another such prediction is that the universe is full of microwaves.
These
microwaves have roughly the same wavelength as those that heat our breakfast
hot dogs, but their density is much
lower and their spectral distribution 
is almost exactly that which would be emitted by a black object whose
temperature was a few degrees Kelvin, {\ie} just a few degrees above absolute zero. 
The discovery of this cosmic microwave background radiation in 1964 was another 
great triumph of the Big Bang model, arguably the greatest of them all.

\smallskip

5. By studying these microwaves, we can infer that the density of the universe
was almost perfectly uniform about half a million years after the Bang. 
Today, it is full of dense clumps, ranging in size from superclusters of
galaxies to the planet we live on, not to mention the author. 
The main mechanism responsible for turning the primordial ordered state into
such a mess is believed to be gravitational instability, whereby the gravity
from regions slightly denser than average attracted surrounding matter,
gradually creating ever larger lumps. 
This phenomenon is usually modeled as perturbations to the
FRW metric, and will be discussed in 
Section~\ref{perturbationsec}.

\smallskip

6. By studying the rotation of different parts of our galaxy, we can calculate
how much mass there is at various distances from the galactic center. The
surprising conclusion is that there is much more mass present than can be
accounted for by the matter we see. The mysterious invisible substance (or
substances) is referred to as {\it dark matter}. 
There are also other indications of dark matter. 
Figuring out its nature and its density is one 
of the hottest problems in modern cosmology 
(Sadoulet \& Cronin 1991), as 
its gravitational presence affects the 
formation of structure mentioned in 5. 

\medskip

The only parts of cosmology and astrophysics 
with which the reader should be familiar to
be able to follow the following chapters are 
\begin{itemize}

\item 
The standard model of space and time in an expanding universe,
including
quantities such as the redshift $z$, the density parameter 
$\Omega$ and the cosmological constant $\Lambda$

\item
The basics of structure formation, including quantities like 
the random field $\delta(\vx)$ and the matter power spectrum $P(k)$

\item 
The microwave background, including quantities such as the
the radiation power spectrum and 
window functions

\item 
Rudimentary hydrogen chemistry

\item 
Some basic cosmology jargon, such as units like Mpc 
and fudge factors like $h$

\end{itemize}

These will be the topics of the following five sections.

\newpage
\section{Bare Bones General Relativity and the FRW Metric}

In this section, we will say a few words about general relativity and the 
large-scale structure of space-time, as well 
as list a number of handy 
formulas that will be used in subsequent chapters. 

\subsection{General Relativity}

\begin{quotation}

Journalist: 
{\it ``Professor Eddington, is it really true that only three people in
the world understand Einstein's theory of general relativity?"}

Eddington: {\it ``Who is the third?" }

\end{quotation}
\smallskip

It would obviously be absurd to try to squeeze a real presentation of general
relativity, one of the deepest physical theories ever to be invented by man,
into a subsection of a Ph.D. thesis. In fact, I have been asked how so
many cosmologists can use a theory as mathematically complex as 
``GR" in their
work without even knowing basic differential geometry.
I think that the answer is that some aspects of GR are quite
intuitive, so that one can acquire a working feeling for what is going on even
without being able to tell a Lie derivative from a Killing field. 
In addition to this, the FRW metric is so much simpler than the most generic
solutions that what happens on distance scales much smaller than 
$10^{10}$ light-years can often be adequately described by 
classical physics plus some geometric considerations. 
In fact, in some ways GR resembles classical physics more than it 
resembles special relativity. 

\subsection{The raw equations}

So how do you calculate things with GR?
What is the basic mathematical structure of the theory?
The basic objects are some fields that live on a four-dimensional manifold, 
and satisfy a bunch of messy nonlinear second order partial differential
equations.  The {\it Einstein field equations} 
are\footnote{A number of different notational conventions can
be found in the literature. Here we use the sign convention
$(+\,-\,-\,-)$ for the metric. In some texts, the signs
of the last two terms in the definition of 
of $R^{\alpha}_{\mu\nu\beta}$ are reversed, which corresponds
to a different convention for the index placement.}

$$R_{\mu\nu} - {1\over 2}R g_{\mu\nu} = 8\pi G T_{\mu\nu},$$
where
$$R \equiv g^{\mu\nu}R_{\mu\nu},$$
$$R_{\mu\nu}\equiv R^{\alpha}_{\mu\alpha\nu},$$
$$R^{\alpha}_{\mu\nu\beta} = \Gamma^{\alpha}_{\nu\beta,\mu} -
\Gamma^{\alpha}_{\mu\beta,\nu}
 + \Gamma^{\gamma}_{\mu\beta}\Gamma^{\alpha}_{\nu\gamma} -
\Gamma^{\gamma}_{\nu\beta}\Gamma^{\alpha}_{\mu\gamma},$$
$$\Gamma^{\alpha}_{\mu\nu} = 
{1\over 2} g^{\alpha\sigma}
\left(g_{\sigma\mu,\nu} + g_{\sigma\nu,\mu} - g_{\mu\nu,\sigma}\right),
$$
and $g^{\mu\nu}$ is the matrix inverse of $g_{\mu\nu}$, {\ie}
$$g^{\mu\alpha}g_{\alpha\nu} = \delta^{\mu}_{\nu}.$$
Here repeated indices are to be summed over from 0 to 3, commas denote
derivatives as in the standard tensor notation, and $G$ is the gravitational
constant. Throughout this section, we will use units where the speed of light
$c=1$.  In the Einstein field equations, the dependent variables
are the two  tensors $g_{\mu\nu}$ and $T_{\mu\nu}$. They are both symmetric, and
thus contain ten independent components each.  $g$ is called the 
{\it metric tensor}, and describes the structure of spacetime at each 
spacetime point $x^{\mu}$. 
$T$ is called the {\it stress-energy tensor}, and describes the state of the 
matter (what is {\it in} space) at each point. The
quantities $\Gamma^{\alpha}_{\mu\nu}$, $R^{\alpha}_{\mu\nu\beta}$ and 
$R_{\mu\nu}$ are named after Christoffel, Riemann and Ricci, respectively, and
will not be further discussed here.

\subsection{The FRW metric}

\begin{quote}
\indent
{\it Truth is much too complicated to allow anything but 
approximations.}

\centerline{\sl John von Neumann}
\end{quote}

The differential quantity 
$$ds^2 = g_{\mu\nu}dx^{\mu}dx^{\nu}$$
is called the {\it line element}, and for the
Friedmann-Robertson-Walker (FRW) metric takes the simple form
$$ds^2 = dt^2 - a(t)^2\left[{dr^2\over 1-kr^2} + 
r^2 d\theta^2 + r^2\sin^2\theta\>d\varphi^2\right],$$
where $k$ is a constant that is either zero or $\pm 1$, and 
$a$ is a function to be determined later.
For reasons that will soon become clear, we have renamed the
independent variables $$x^{\mu} = (t,\vr),$$
and changed to spherical coordinates
$$\vr = (x_1,x_2,x_3) = 
(\sin\theta\cos\varphi,\sin\theta\sin\varphi,\cos\theta)r.$$

\subsection{Interpreting the FRW coordinates}

The Newtonian equation of motion for a point particle moving along a
trajectory $\vx(t)$ in a gravitational field $\vg$ is 
$$\ddot x_i = g_i,$$
where $i=1,2,3$.
The analog of this in General Relativity 
for a point particle moving along a
trajectory $x^{\mu}(t)$ in spacetime, $t$ now being any parameter, is the
geodesic equation of motion 
$$\ddot x^{\mu} = \Gamma^{\mu}_{\nu\lambda}\dot x^{\nu}\dot x^{\lambda}.$$
For the FRW metric, a cumbersome but straightforward calculation shows that
$$x^{\mu} = (t,\vr_0),$$
is a solution to the geodesic equation for any constant $\vr_0$.
This means that if we interpret $t$ 
as time\footnote{This is a very fortunate feature 
of the FRW metric, since for generic metrics,
it turns out to be impossible to define a global time that observers in
different parts of the universe can agree on.
}
and call the vector $\vr$ 
the {\it comoving spatial coordinates}, then
the FRW metric can be interpreted as follows:
an object at a comoving position $\vr =\vr_0$ that is at rest (defined as 
$\dot\vr=0$) will remain at rest. The time $t$ is simply the time 
that such an object would measure if it were a 
clock.
By definition, the physical distance between two comoving points 
$\vr_1$ and $\vr_2$ at some time $t$ equals
$$\int_1^2 \sqrt{-ds^2},$$
where the integral is to be taken along the {\it geodesic}, the curve from 
$(t,\vr_1)$ to $(t,\vr_2)$ that minimizes this quantity. 
A quick look at the expression for $ds^2$ above tells us that this distance will
be proportional to $a(t)$. Thus objects that are at rest relative to the
comoving coordinates nonetheless move relative to each other, the
time-dependence of all distances scaling as the function $a(t)$, which is named
the {\it scale factor}. 
When our universe is modeled by the FRW metric, 
distant galaxies are assumed to 
stay at fixed comoving coordinates, and the 
universe expanding simply corresponds to $a(t)$ increasing with time.
The Hubble constant is evidently given by 
$$H\equiv {\dot a\over a}.$$
(Note that calling it a ``constant" is a misnomer, as 
it generally changes with time.)

\subsection{Curvature}

Let us for a moment keep $t$ fixed and 
say a few words about the constant $k$. 
If $k=0$, then the
geodesic (shortest curve) between two points $\vr_1$ and $\vr_2$ in ${\bf R}^3$
will be the conventional straight line. This is not the case when $k\neq 0$.  If
we chose three points $\vr_1$, $\vr_2$ and  $\vr_3$ and connect them in a
triangle by the three geodesics, the sum of the three angles will be greater that
$180^{\circ}$ if $k=1$ (in which case space is said to be {\it positively
curved}), equal to $180^{\circ}$ if $k=0$ (in which case
space is said to be {\it flat}),
and less than $180^{\circ}$ if $k=-1$ (in which case
space is said to be {\it negatively curved} or 
{\it hyperbolic}).
If $k\leq 0$, space is infinite at all times\footnote{Except in unusual topologies
that will not be discussed here},  
the 
coordinate $\vr$ is allowed to range over the entire space ${\bf R}^3$,
and 
the universe is said to be {\it open}.
If $k=1$, however, space has the topology of the 3-sphere, 
the universe is said to be {closed}, and its
volume is finite for any $t$. In this case, the comoving spatial
coordinates are
restricted to the range $|\vr|<1$. (This coordinate patch covers exactly half of
the three-sphere.)

\subsection{Redshift}

Let us now turn to the time-dependence. 
One can show that the momentum of all free particles decreases as the inverse of
the scale factor $a$ as the FRW universe expands. 
Thus massive particles gradually slow down and become comoving, ``go with the
flow", ending up at rest with respect to the comoving coordinates. 
For photons, which cannot slow down, the momentum loss corresponds to a lowering of
their frequency, {\ie} they become redshifted. 
Thus the wavelength of a given photon is proportional to $a$,
and we can whimsically think of the corresponding electromagnetic waves as being
``stretched out" at the same rate that space is being stretched out. 
Another way to interpret this redshifting is as a Doppler shift, since due to
the expansion of the universe, the distant objects emitting the photons that
we see are receding from us.

If we observe spectral lines in the spectrum of a distant object, they will
all be shifted towards longer wavelengths by the same factor. Calling this
factor $(1+z)$, the {\it redshift} of the object is defined to be the number $z$.
For objects receding by a velocity $v\ll c$, we have to a good approximation
that  
$$z \approx {v\over c}.$$
If we assume that the recession velocity $v$ is mainly due to the expansion of
the universe, we can immediately estimate the distance $d$ to the object as
$$d \approx {c\over H}z \approx 10^{28}\cm \times z.$$
(If $z$ is of order unity or larger, this formula is replaced by a more
complicated one.)

In cosmology, redshift is often used as a substitute for time. Indeed, the
horizontal axis on most of the plots in the subsequent chapters is labeled $z$,
not $t$. 
This is mainly because $z$ corresponds much more directly to what
astronomers actually measure. 
When we look out into space, we are looking into the past. 
If we observe a distant quasar, we know that we are observing events that
occurred at some earlier time $t$, but we have no direct
and accurate way of measuring $t$, or how long ago the quasar emitted the
light that we see. However, we can measure its redshift simply by
observing its spectrum and measuring where the spectral lines are. 
Thus ``at redshift $z$" in cosmology jargon means ``at the time 
when photons would have had to be emitted to have a redshift $z$ by now."
There is of course an explicit mathematical relationship between the two
variables $t$ and $z$, and this will be given later in this chapter.
$z=\infty$ corresponds to $t=0$ (the Big Bang), and
$z=0$ corresponds to today.

\subsection{The Friedmann equation}

So far, we have only discussed properties of the FRW metric. For it to
correspond to something physical, it has to satisfy the Einstein field equations.
If we take the stress-energy tensor $T_{\mu\nu}$ that corresponds to comoving 
matter of uniform density $\rho(t)$, substitute 
it into the Einstein field equations,
and grind through a not particularly amusing calculation, it is found that the FRW
metric is indeed a solution providing that the scale factor $a$ satisfies the
ordinary differential equation
$$\left({\dot a\over a}\right)^2 = {8\pi G\over 3}\rho - {k\over a^2}.$$
The solution to this equation, known as the 
{\it Friedmann Equation}\footnote{This is an example where general relativity 
is in a sense more ``classical" than
special relativity. The Friedmann equation is identical to that
describing the time-evolution of the radius $a$ of a self-gravitating cloud of 
uniform, pressureless
dust in classical physics. Note that $c$, the speed of light, appears
nowhere in the equation, so with this classical interpretation, 
nothing prevents the dust particles from moving faster than the speed of light. 
},
clearly
depends both on the value of $k$ and on the equation of state, the way in which
the density $\rho$ depends on $a$. 
If the density is dominated by nonrelativistic matter
(particles whose kinetic energy is much less than their rest mass),
just ``matter" for short, it will
obviously drop as the inverse of the volume as the universe expands,
giving 
$$\rho \propto a^{-3}.$$
This has been the case for quite a while, for redshifts $z\ll 10^4$.
If the density is dominated by photons (or other highly relativistic particles),
which become redshifted as discussed above, 
$$\rho \propto a^{-4}$$
since the energy of each photon
dwindles as $1/a$ as the universe expands. 
This was indeed the case very early on. 
A highly speculative third type of mass is {\it vacuum energy}, which if it exists
has the property that its density stays constant, independent of the scale
factor. The presence of such matter is equivalent to the presence of
an infamous
``cosmological constant" $\Lambda$ in the Einstein field equations.

Let us define the {\it critical density} 
$$\rho_{crit}\equiv {3H^2\over 8\pi G}$$
and the dimensionless variable 
$$\Omega\equiv {\rho\over\rho_{crit}}.$$
Then for the matter-dominated case (to which we
limit ourselves 
in the remainder of this section),
the Friedmann equation becomes
$${\dot a\over a} = H_0\left[\Oz \left({a\over a_0}\right)^{-3} + \Lz 
- (\Oz + \Lz - 1)\left({a\over a_0}\right)^{-2}\right]^{1/2},$$
where if $k\neq 0$, we chose the scale factor today to be
$$a_0\equiv \left({k\over\Oz-1}\right)^{1/2}H_0^{-1}
\approx {3000 h^{-1}\Mpc\over |\Oz-1|^{1/2}},$$
which 
can be interpreted as the current radius of curvature of the universe. 
If $\Oz=1$, then $a_0$ has no physical meaning and can be chosen 
arbitrarily --- it will always cancel out when some physical quantity is
computed.
(Here and throughout cosmology, the subscript zero on a variable denotes its value
today, at $z=0$. For instance, $H_0$ is the value of the Hubble constant today.) 
$\Lz$ is the above-mentioned cosmological constant, believed by many to be zero.
The constant $k$ was eliminated by using the fact that  
$$k = H_0^2(\Lz + \Oz - 1).$$
This relation shows that the universe is flat ($k=0$) if 
$\Lz + \Oz = 1$, and closed or open depending on whether 
$\Lz + \Oz$ exceeds unity or not.
In the standard cold dark matter (CDM) scenario, one assumes that
$\Oz=1$ and $\Lz=0$.
For this simple case, the Friedmann equation has the solution
$$a(t)\propto t^{2/3},$$
and $\Omega=1$ at all times. 
If the universe is open, it is clear that 
unless $\Lz<0$, 
we will always have $\dot a > 0$, and the expansion will continue forever.
If the universe is closed and $\Lz\leq 0$, the  
expansion eventually ceases and the universe contracts back together again,
ending with a hot Big Bang in reverse, a big crunch. 
For the $\Lz=0$ case, the graph of the function $a(t)$ is a cycloid, and 
the Friedmann equation above describes only the increasing half of the curve.
Closed universes expand forever if
(Glanfield 1966)
% My formula $$\Lz > {4\over 27} {(\Oz+\Lz-1)^3\over\Oz}.$$
% seems to be wrong. Why?
$$\Lz \geq 4\Oz\cos^3\left[{1\over 3}\arccos\left(\Oz^{-1}-1\right)
+ {4\pi\over 3}\right].$$
An recent review of properties of the Friedmann equation is given by 
Carroll {\etal} (1992).

\subsection{Some handy formulas for getting by in curved space}

This section contains a collection of useful formulas, many of which are used in
the following chapters. It is intended as more of a reference section, and
can be safely skipped by the general reader.
 
\subsubsection{Friedmann solutions, conformal time}
 
Since $a_0/a = 1+z$, the
Friedmann equation (neglecting radiation) can also be written
$$H = H_0\sqrt{\Lz + (1+z)^2(1-\Lz+\Oz z)}$$
or, alternatively, 
\beq{dzdtEq}
-{dz\over dt} = H_0(1+z)\sqrt{\Lz + (1+z)^2(1-\Lz+\Oz z)}.
\eeq
In the rest of this section, we will set $\Lz=0$ unless 
the contrary is explicitly stated. This means that the    
Friedmann equation can be solved analytically,
the $\Omega>1$ solution being the ubiquitous cycloid.

\noindent
It is often convenient to introduce a new time coordinate $\tau$ called
{\it
conformal time}, defined by 
$$\t(t) \equiv \int_0^t {dt'\over a(t')}.$$
In terms of the conformal time, the solutions $a(t)$ to
the Friedmann equation can
be written in the parametric form
 
$$H_0 t = \cases{ {\Omega_0\over 2(1-\Omega_0)^{3/2}}
(\sinh\t - \t)&for $\Omega_0<1$,\crr
{2\over 3}\left({\t\over\t_0}\right)^3&for $\Omega_0=1$,\crr
{\Omega_0\over 2(\Omega_0-1)^{3/2}}(\t-\sin\t)&for $\Omega_0>$1,
}$$
$${a\over a_0} = {1\over 1+z} =
\cases{
{\Omega_0(\cosh\t-1) \over 2(1-\Omega_0)}&for $\Omega_0<1$,
\crr 
\left({\t\over \t_0}\right)^2&for $\Omega_0=1$,\crr
{\Omega_0(1-\cos\t) \over 2(\Omega_0-1)}&for $\Omega_0>1$.
}$$
When $\Oz\neq 1$, $\t$ is often called the {\it development angle}.
When $\Oz = 1$, $\t_0 = 2/(H_0 a_0)$, where the constant
$a_0$ can be chosen arbitrarily. Again, this
is because the radius of curvature, which fixed $a_0$ when $\Oz\neq
1$, is infinite when the universe is flat.
In all cases, $\t=0$ corresponds to the Big
Bang.
Inverting these last expressions, we obtain
$$\t \equiv
\cases{
2\arsinh\sqrt{1-\Omega_0\over\Omega_0(1+z)}
&for $\Omega_0<1$,\crr
{\t_0\over\sqrt{1+z}}&for $\Omega_0=1$,\crr
2\arcsin\sqrt{\Omega_0-1\over\Omega_0(1+z)}&for $\Omega_0>1$.
}$$
Thus the conformal time today is
$$\t_0 =
\cases{
2\arsinh\sqrt{\Omega_0^{-1} - 1}
&for $\Omega_0<1$,\crr
2\arcsin\sqrt{1 - \Omega_0^{-1}}&for $\Omega_0>1$.
}$$
Although one cannot give an explicit expression
for $a$ in terms of $t$ in general,
$${a\over a_0} \approx \Oz^{1/3} \left({3\over 2}H_0 t\right)^{2/3}$$
for $t\ll H_0^{-1}$.

\subsubsection{Causality, horizons, angles}
 
A useful property of $\t$ is that during a conformal time interval
$d\t$, a light ray always travels the same comoving distance $d\t$.
If $\Omega=1$ so that space is flat, the situation becomes especially simple
when viewed in $(\t,\vx)$-space:
all light rays (null geodesics) 
simply move in straight lines, and always make 45
degree angles with the $\t$ axis.
Thus after changing variables to comoving 
spatial coordinates and conformal time,
we can use all our intuition about light propagation and causality in good old
non-expanding Euclidean space. This is illustrated in 
Figure~\ref{lightconeplot}.
 
Parts of the universe receding from us faster than the speed of light are 
not in 
causal contact with us.
No signals have ever been able to reach a comoving object in the universe
from objects that are outside of its backward
light cone. Such objects are said to be outside of its {\it horizon}.
Thus defining the
{\it horizon radius} as $a\t$ and combining some of the previous expressions,
we obtain
$$r_h \equiv a\tau = \cases{
{H_0^{-1}\over (1+z)\sqrt{1-\Oz} }
\cosh^{-1}\left[1+{2(1-\Oz)\over\Oz(1+z)}\right]
&if $\Oz<1$,\crr
{2 H_0^{-1}\over (1+z)^{3/2}}
&if $\Oz=1$,\crr
{H_0^{-1}\over (1+z)\sqrt{\Oz-1} }
\cos^{-1}\left[1-{2(\Oz-1)\over\Oz(1+z)}\right]
&if $\Oz>1$.\crr}
$$
If an object at redshift $z$ has a proper diameter $D$ (perpendicular to us),
then one can show that it will subtend an angle on the sky given by
$$2 \tan{\theta\over 2} = \cases{
{\Oz^2(1+z)\sinh\left[{1\over 2}(1+z)\sqrt{1-\Oz} H_0 D\right]
\over
\sqrt{1-\Oz}\left[\Oz z-(2-\Oz)(\sqrt{1+\Oz z}-1)\right]}
&if $\Oz<1$,\crr
{(1+z)^{3/2}H_0 D\over 2(\sqrt{1+z}-1)}
&if $\Oz=1$,\crr
{\Oz^2(1+z)\sin\left[{1\over 2}(1+z)\sqrt{\Oz-1} H_0 D\right]
\over
\sqrt{\Oz-1}\left[\Oz z-(2-\Oz)(\sqrt{1+\Oz z}-1)\right]}
&if $\Oz>1$.\crr
}$$
Thus
if we look at an object at a redshift $z$, its
horizon radius subtends an angle $\theta$ in the sky that is given by
$$
2 \tan{\theta\over 2} = 
{\Oz^{3/2}\sqrt{1+z}\over
\Oz z-(2-\Oz)(\sqrt{1+\Oz z}-1)},
$$
for all values of $\Oz$. For $\Oz=1$, this takes the particularly
simple form
$$
2 \tan{\theta\over 2} = {1\over\sqrt{1+z} - 1},
$$
whereas
for large $z$, it reduces to
$$\theta\approx\sqrt{\Oz\over z}.$$
This last expression is a rather poor approximation 
for 
for typical values of $\Oz$, unless $z\gg 100$.

\subsubsection{Age of the universe}
 
The current age of the universe, $t_0$, is given by
$$t_0 = \int dt = 
\int_0^{\infty} (aH)^{-1}(z)\left|{da\over dz}\right|\,dz
 = \int_0^{\infty}{dz\over (1+z)H(z)}.$$
Performing this integral for
$\Lz=0$ gives the age of the universe as
$$H_0 t_0 =
\cases{
{1\over1-\Oz} - {\Oz\over 2(1-\Oz)^{3/2}}
\cosh^{-1}\left({2\over\Oz}-1\right)
&for $\Omega_0<1$,\crr
{2\over 3}&for $\Omega_0=1$,\crr
{\Oz\over 2(\Oz-1)^{3/2}} \cos^{-1}\left({2\over\Oz}-1\right)
- {1\over\Oz-1}
&for $\Omega_0>1$.
}$$
For the flat case where $k=0$ and $\Oz = 1-\Lz$, the result is
$$H_0 t_0 = {1\over\Lz^{1/2}}\ln\left[
{1+\Lz^{1/2}\over(1-\Lz)^{1/2}}\right].$$

% \subsubsection{How $\Omega$ evolves}
% (There are errors in Jones's article.)

\newpage
\section{Perturbing the Universe}
\label{perturbationsec}

In this section, we will give the rudiments of the theory of structure
formation.
First we discuss how small ripples in the FRW metric grow larger as the universe
expands. We then briefly mention the theory of Gaussian random fields and the so
called Press-Schechter model for galaxy formation.

\subsection{Linear perturbation theory}

\begin{quote}
\indent
{\it Let us assume that demand is an arbitrary function of price;
$D = aP + b$.}

\centerline{\sl Anonymous economics professor}
\end{quote}

In this section, everything will indeed be linear. 
As mentioned in the previous section, it is generally futile
to look for analytic solutions to the Einstein field equations of GR.
Fortunately, the actual metric is fairly uniform on large scales, indicating that
the relative deviations from the FRW metric are rather small. 
For this reason, cosmologists attack the problem of structure formation with 
linear perturbations. One perturbs
the FRW metric, throws away all terms of second order and higher, and after
Fourier transforming ends up with a simple equation for the time-evolution of
the various perturbation modes.

As mentioned in a footnote a few pages back, the Friedmann equation can be
derived using classical physics alone. This is no coincidence. A result known
as Birkhoff's theorem states that if we take an FRW metric and remove all
matter from a spherical volume, then space will be flat inside of this sphere,
{\ie} all gravitational effects from the matter outside cancel out.
If we put the uniformly expanding 
matter back into the sphere, then its gravitational behavior will be
correctly described by classical physics.
As long as 
the gravitational field is weak
(which it is in the present problem, as opposed to say near a black hole)
and the
sphere is much smaller than the horizon scale $10^{10}$ light years, it is easy to
show that classical physics will apply even if the matter is not uniform.
With this motivation, let us write down the classical equations governing a gas
of density $\rho$, velocity $\vv$ and pressure $p$ and the corresponding
gravitational potential $\phi$. These are the continuity equation, the Euler
equation of motion and the Poisson equation of classical gravity, respectively:
\beq{ClassicalFieldEqs}
\cases{
\dot{\rho} + \nabla\cdot(\rho\vv) = 0\crr
\dot{\vv} + (\vv\cdot\nabla)\vv = -\nabla(\phi + {p\over\rho})\crr
\nabla^2\phi = 4\pi G\rho
}
\eeq
A simple solution to these equations is 
$$\cases{
\rho_0(t,\vr) = {\rho_0\over a^3},\crr
\vv_0(t,\vr) = {\dot a\over a}\vr,\crr
\phi_0(t,\vr) = {2\pi G\rho_0\over 3} r^2,
}$$
where $a$ is a function of time that satisfies the Friedmann equation, and 
since the density is independent of $\vr$, this classical
solution clearly corresponds to the unperturbed FRW solution in GR.
From here on, let us write these fields as functions of the 
{\it comoving} position $\vx$ rather than the physical position 
$\vr = a(t)\vx$. 
Let us now expand the fields as 
$$\cases{
\rho(t,\vx) = \rho_0(t)[1+\delta(t,\vx)],\crr
\vv(t,\vx) = \vv_0(t,\vx) + \vv_1(t,\vx),\crr
\phi(t,\vx) = \phi_0(\vx) + \phi_1(t,\vx),
}$$
and assume that $|\delta|$, $|\vv_1|$, $\phi_1\ll1$.
Substituting this into \eq{ClassicalFieldEqs}, dropping all non-linear
terms, Fourier transforming everything with respect to $\vx$, and doing some
algebra, one obtains the second order ordinary differential equation
\beq{JeansEq}
\ddot{\deltahat}(\vk) + 2{\dot a\over a}\dot{\deltahat}(\vk)
+ \left({v_s^2 |\vk|^2\over a^2} - 4\pi G\rho_0\right)\deltahat(\vk) = 0.
\eeq
Here the equation of state for the gas enters only through 
the {\it sound speed},
$$v_s \equiv \left({\partial p\over\partial\rho}\right)^{1/2}_{adiab}
\approx\sqrt{5 kT\over 3 m_p},$$
where the last expression applies 
if the gas consists mainly of atomic hydrogen 
($m_p$ denotes the proton mass).
If we substitute an expanding universe solution for $a$, 
\eq{JeansEq} tells us a number of interesting things.

The coefficient of $\deltahat$ has two contributions with opposite signs: 
the destabilizing effect of gravity competes with the stabilizing effect of gas
pressure. On small scales (corresponding to 
large $k$), pressure dominates and 
perturbations do not grow, merely oscillate.
On large scales, on the other hand, self-gravity dominates, and 
fluctuations grow monotonically 
until $\delta$ becomes of order unity and the entire linear approximation 
breaks down. 
The border between these two regimes is when
the perturbation has a physical
wavelength 
$$\lambda \equiv {2\pi a\over|\vk|},$$
called the {\it Jeans wavelength}
$$\lambda_J\equiv v_s\sqrt{\pi\over G\rho_0},$$
for which this coefficient vanishes. 

The coefficient of $\dot{\deltahat}$, twice the Hubble constant $H$,
acts like a kind of friction: in the absence of this term, we could get
exponentially growing modes, but as will be seen, the growth is never that rapid
in the presence of this term, {\ie} in an expanding universe. 

If the density field is smoothed on a scale
$\lambda\gg\lambda_J$, then we can neglect the $k$-dependence and
trivially Fourier transform back to real space. 
Thus all contributing modes grow at the same rate, and a 
perturbation retains its (comoving) shape as the universe expands, 
simply increasing in amplitude.
For the simple case where $\Omega=1$,
$\Lz=0$ and $a\propto t^{2/3}$, we obtain 
$$\ddot{\delta} + {4\over 3 t}\dot{\delta} - {2\over 3 t^2}\delta = 0,$$
which has the most general solution 
$$\delta(t,\vx) = 
A(\vx)\delta(t_0,\vx)\left({t\over t_0}\right)^{2/3} + 
B(\vx)\delta(t_0,\vx)\left({t\over t_0}\right)^{-1}$$ 
for some time-independent functions $A$ and $B$.
Ignoring the physically uninteresting decaying mode, 
and writing $\delta$ as a
function of redshift rather than time,
we thus have the extremely simple result
$$\delta(z,\vx) \propto  {\delta_0(\vx)\over 1+z},$$
where $\delta_0$ is the linearly extrapolated field of density
perturbations today.

\subsection{Random fields}

The mathematical theory of random fields (sometimes known as three-dimensional
stochastic processes) is a very useful tool when analyzing cosmological structure
formation. A {\it random field} is simply an infinite-dimensional random variable,
such that each realization of it is a field on some space. In cosmology
applications, these fields tend to be $\delta$, $\vv$ or $\phi$, and the space is
physical space at some fixed time $t$. 
As every quantum field theorist knows, it is a nightmare to try to
define a nice measure on an infinite-dimensional space, so 
random fields are defined by
specifying the joint probability distribution of their values at any 
$n$ points, $n=1, 2, 3, ...$, thus circumventing the need to define
a probability distribution on the space of all fields.
Hence,
to define a random field $\delta$, one must
specify the 1-point distribution (the 1-dimensional probability distribution of
$\delta(\vx_1)$ for all $\vx_1$), the 2-point-distribution
(the 2-dimensional probability distribution of the vector
$[\delta(\vx_1),\delta(\vx_2)]$ for
all  $\vx_1$ and $\vx_2$), {\it etc}. 
In cosmology, the random fields are always assumed to be translationally and
rotationally invariant, {\ie} homogeneous and isotropic. Hence the 1-point
distribution is independent of $\vx_1$, and the 2-point distribution will depend
only on the scalar quantity $x\equiv|\vx_2 - \vx_1|$. 

\subsubsection{Ergodicity}

A random field $\delta$ is said to be {\it ergodic} if one can use ensemble
averaging and spatial averaging interchangeably. 
The ensemble average of a random field $\delta$ 
at a
point, denoted $\expec{\delta(\vx)}$, is simply the expectation value of 
the random variable $\delta(\vx)$.
Thus for an ergodic field, 
$$\expec{\delta(\vx_1)} = 
\lim_{R\to\infty}
\left({4\over 3}\pi R^3\right)^{-1}\int_{|\vx|<R}\delta_*(\vx) d^3x$$
holds for all points $\vx_1$ and for 
all realizations $\delta_*(\vx)$ (except for
a set of probability measure zero).  
Ensemble averages are completely inaccessible to us, since 
we have only one universe to look at, namely the particular realization that we
happen to live in. 
So as cosmologists, we are quite happy if we have ergodicity, since this means
that we can measure these elusive ensemble averages by simply averaging over
large volumes instead.  

\subsubsection{Gaussianity}

A random field is said to be {\it Gaussian} if 
all the above-mentioned probability
distributions are Gaussian. 
This is a 
very popular assumption in cosmology,
partly because, as we will see, it greatly simplifies matters.
A first nice feature of this assumption is that 
all homogeneous and isotropic Gaussian random fields are
ergodic\footnote{Note that this is true only for random fields that live on infinite spaces
such as $\R^n$, and does not hold for fields on compact manifolds such as the sphere 
$S^2$. Thus the field of microwave background anisotropies 
(defined further on) is not ergodic, so that even if we could reduce our experimental 
errors to zero, we could still never measure any ensemble-averages with perfect accuracy.
This phenomenon is known as ``cosmic variance". It stems from the fact that
in the spatial average above, one cannot average over an infinite volume ({\ie} let
$R\to\infty$), since the volume of the compact manifold is finite.}.
Taking the spatial average of the definition of $\delta$, for instance, 
ergodicity implies that 
$$\expec{\delta(\vx)} = 0.$$
Let us define the r.m.s. fluctuations
$$\sigma\equiv\expec{\delta(\vx)^2}^{1/2},$$
and the {\it correlation function} as
$$\xi(x)\equiv{\expec{\delta(\vx_2)\delta(\vx_1)}\over\sigma^2}.$$
(Note that because of the homogeneity and isotropy, $\sigma$ is independent of
$\vx$ and the correlation function depends only on $x\equiv|\vx_2 -
\vx_1|$.)                                                       
Since the $n$-point distribution is Gaussian, it is defined by its mean vector
$\expec{\delta(\vx_n)}$
(which is identically zero) and its covariance matrix 
$$C_{mn}\equiv \expec{\delta(\vx_m)\delta(\vx_n)} = 
\sigma^2\xi(|\vx_m - \vx_n|).$$ 
Thus the Gaussian
random field $\delta$ has the extremely useful property that it is is entirely
specified by $\sigma$ and its correlation function.

\subsubsection{The power spectrum}

If we Fourier expand $\delta$ as
$$\delta(\vx) = \ootpc\int e^{i\vk\cdot\vx}\deltahat(\vk) d^3k,$$
we see that if its Fourier transform 
$$\deltahat(\vk) = \int e^{-i\vk\cdot\vx} \delta(\vx)d^3x,$$
is a Gaussian random variable for any $k$, then 
$\delta$ will automatically be a Gaussian random field (since a sum of Gaussians
is always Gaussian).
Cosmologists like to postulate that the complex numbers 
$\deltahat(\vk)$ have {\it random phases}, which implies that they are all
uncorrelated. One postulates that 
$$\expec{\deltahat(\vk)^*\deltahat(\vk')} = 
(2\pi)^3\delta(\vk - \vk') P(\vk),$$
where the function $P(\vk)$ is called the {\it power spectrum} and $\delta$ is the
Dirac delta function (which will hopefully not be confused with the random field
$\delta$).
This implies that even if $\deltahat(\vk)$ does not have a Gaussian distribution, 
the random field $\delta$, being an infinite sum of independent random
variables, will still be Gaussian by the Central Limit Theorem for many
well-behaved power spectra.
It is straightforward to show that the power spectrum is simply $\sigma^2$
times the three-dimensional Fourier transform of the correlation function,
{\ie} 
$$P(k) = 4\pi\sigma^2\int\left({\sin kr\over kr}\right)\xi(r) r^2dr.$$
Note that $P$ depends on $\vk$ only through its magnitude
$k = |\vk|$, because of the isotropy assumption.

\subsubsection{Window functions}

If we smooth $\delta$ by defining the weighted average
$$\delta_w(\vx) \equiv \int\delta(\vr')w(\vr'-\vr) d^3r',$$
$w$ being some weight function, then 
$$\sigma^2_w\equiv\expec{|\delta_w(\vx)|^2} = 
\ootpc\int|\what(k)|^2 P(k) d^3k.$$
This is quite a typical pattern: the variance of some physical quantity is given
by an integral of the power spectrum against some function, referred to as a {\it
window function}. In this particular example, where for an appropriately chosen
$w$ the physical quantity can be interpreted as the mass within a volume of some
given shape, the window function is $|\what|^2$. As we will see in 
Chapter~\ref{pindepchapter},
basically all cosmological experiments that measure a single number can be
described by some window function, regardless of whether they involve
mass distribution, bulk flows of galaxies or microwave background fluctuations.

\subsection{The Press-Schechter recipe}

In this section, we will describe a rather simple-minded model of structure
formation that agrees remarkably well with observational data.

\subsubsection{The top hat solution}

Early on, while $z\gg\Oz^{-1}$, space is approximately flat and 
the Friedmann equation has the approximate solution
$$a(t)\propto t^{2/3}$$
regardless of the values of $\Oz$ and $\Lz$. 
If an $\Omega=1$ universe has a completely uniform density
$\rho$ except for a ``top hat" overdensity, a spherical region 
where the density at a given time
is some constant $\rho'>\rho$, then this top hat region will gradually begin to
expand slower than the rest of the universe, stop expanding and recollapse to a
point. By Birkhoff's theorem, the radius of this region will evolve according to
the Friedmann equation, but with $\Omega>1$. It is easy to show that the
overdensity 
$$\delta\equiv{\rho'\over\rho} - 1$$
evolves as 
\beq{TopHatEq1}
(1+\delta) = 
{9\over 2}{(\alpha - \sin\alpha)^2\over(1-\cos\alpha)^3}
= 1 + {3\over 20}\alpha^2 + O(\alpha^3),
\eeq
where the parameter $\alpha$, the ``development angle", 
is related to the redshift
through
\beq{TopHatEq2}
{1+\zvir\over 1+z} = \left({\alpha - \sin\alpha\over 2\pi}\right)^{2/3}
= {\alpha^2\over (12\pi)^{2/3}} + O(\alpha^{8/3}).
\eeq
Here $\zvir$ is the redshift at which the top hat would collapse to a point. In
reality, an overdense region would of course not collapse to a point
(and form a black hole).
Since it would
not be perfectly spherically symmetric,  
dark matter particles would
mostly miss each other as they whizzed past the central region and out again on
the other side, eventually settling down in some (quasi-) equilibrium 
configuration known as a virial state. 
For baryons, gas-dynamical processes become important, and pressure eventually 
slows the collapse.
Strictly speaking, virial states are not
stable over extremely long periods of time, and their density is certainly not
uniform. 
For a virialized lump, often referred to as a ``halo", 
a typical density profile peaks 
around some constant value in its core and falls off like $1/r^2$ over
some range of radii. Nonetheless, people often say that they have a ``typical"
density 
$$\rho_{vir} \approx 18\pi^2\rho_0(1+\zvir)^3,$$
which is a useful rule of thumb. 
Thus in the top-hat collapse model, 
density in the perturbed region is assumed to
evolve as in 
Figure~\ref{znplot}.
The density starts out decreasing almost as fast as the background
density $\rho$, with 
$$\delta \propto (1+z)^{-1}$$ 
early on, just as in linear
theory, but gradually stops decreasing and increases radically as $z$ approaches
$\zvir$. 
It never increases past the virial value $\approx 18\pi^2\rho_0(1+\zvir)^3$,
however, but stays at that density for all $z<\zvir$.
Unfortunately, $\alpha$ cannot be eliminated from the 
equations that relate $\delta$ and $z$,
{\ie} equations\eqnum{TopHatEq1} and\eqnum{TopHatEq2},
by using elementary functions. 
For this reason, I made the following fit to the function $\delta(z)$, which is
accurate to about $5\%$ until $z$ is within $10\%$ of $\zvir$, at which the
density is assumed to start approaching the virial value anyway:
$$\rho(z)\approx 
\rho_0(1+z)^3
\exp\left[-{1.9A\over 1-0.75
A^2}\right],$$
where 
$$A(z) \equiv {1+\zvir\over 1+z}.$$

\subsubsection{The Press-Schechter formula}

In the above top-hat solution, we have approximately 
$$\delta = {3\over 20} (12\pi)^{2/3} A\approx 1.686 A$$
early on, while $A\ll 1$.
If we naively extrapolate this far beyond its region of validity, 
we thus obtain
$$\delta = \delta_c \equiv {3\over 20} (12\pi)^{2/3}\approx 1.69$$
when $A=1$, {\ie} when $z=\zvir$. 
Press and Schechter proposed the following 
simplistic prescription for modeling structure formation:
define a smoothed density field $\delta_M$ by averaging $\delta$ over spherical
regions of such a radius that they contain a mass $M$ on the average. 
Define $\sigma(M)$ as the r.m.s. mass fluctuation in these spheres, {\ie}
$$\sigma(M) \equiv \expec{\delta_M^2}^{1/2}.$$
Now make the interpretation that a point $\vx$ is part of a virialized halo
of mass exceeding $M$ if 
$$\delta_M(\vx) \geq \delta_c\approx 1.69.$$ 
In linear theory, for $\Oz=1$ and $\Lz=0$, we clearly have 
$$\sigma(M,z) = {\sigma(M,0)\over 1+z}.$$
Since $\delta$ is a Gaussian random field, so is $\delta_M$, which means that 
$\delta_M(\vx)$ is a Gaussian random variable with zero mean and a standard
deviation $\sigma(M,z)$.
Thus with the Press-Schechter interpretation, the fraction of all mass that is in
virialized lumps of mass exceeding $M$ at a redshift $z$ is 
$$f_g(M,z) \equiv P[\delta_M(\vx) > \delta_c] =
{1\over 2}\erfc\left[{(1+z)\delta_c\over\sqrt{2}\sigma(M,0)}\right],$$
where the complementary error function is defined by
$$\erfc[x]\equiv {2\over\sqrt{\pi}}\int_x^{\infty} e^{-x^2}dx.$$
Unfortunately, the so defined 
$f_g(M,z)$ approaches ${1\over 2}$ as 
$t\to\infty$, {\ie} half of all mass remains unaccounted for. To remedy
this, Press and Schechter introduced a ``fudge factor" of 2 and
predicted that 
$$f_g(M,z) =
\erfc\left[{(1+z)\delta_c\over\sqrt{2}\sigma(M,0)}\right].$$ And lo and behold:
this is quite a good fit both to the observed mass distribution of galaxies today
and to the $f_g(M,z)$ that is observed in numerical $n$-body simulations. Although
a score of papers have been written on the topic since this formula was proposed
some twenty years ago, often attempting to justify the factor of two, not much has
happened to the basic formula except that other values have been proposed
for the constant $\delta_c$, ranging from $1.33$ to $2.00$.

%%%%%%%%%%%%%%%%%%%%%%%%%%%%%%%%%%%%%%%%%%%%%%%%%%%%%%%%%%%%%%%%%%%
\subsection{CDM and BDM}

% Give credit to Lifschitz. 
% Mention that linear theory breaks down eventually (indeed always, 
% on small enough
% scales, so you need an additional assumption that nobody has justified
% mathematically about small scales not influencing large scales), 
% ``go non-linear".

After all this talk about the power spectrum $P(k)$, what does theory predict
it to be?
There is now a plethora of rival theories on the market predicting all
sorts of power spectra that are tailor-made to fit various observational facts.
In this extremely cursory discussion, we will give only the highlights of two
models.  The most popular model 
over the last decade has probably been 
the 
cold dark matter (CDM) model with adiabatic initial fluctuations and an $n=1$
Harrison Zel'dovich initial power spectrum. We will refer to this mouthful as
``standard CDM" for short. This model is simple and elegant, as it contains
almost no free parameters, and 
roughly speaking fits 
all the observed power spectrum data to within
about a factor of two when $k$ ranges over many orders of magnitude.
A recent rising star (albeit always surrounded by 
rumors of its imminent death) is the baryonic dark matter (BDM) model with 
isocurvature initial fluctuations and a power-law initial power spectrum.
We will refer to this as just BDM for short. 

All theories of structure formation (theories that predict $P(k)$) split into
two parts: one part that gives some ``initial" or ``primordial" power spectrum 
$P_i(k)$ when the
universe was very young, and a second part that describes how this power
spectrum is altered by various physical processes
to produce the $P(k)$ we observe today.
Neglecting speculative theoretical constructs such as cosmic strings and other
so called topological defects, the latter processes 
usually do not mix different Fourier
modes as long as $\delta\ll 1$. 
Thus the second half of a theory can be specified by a single function,
the {\it transfer function} $T(k)$, defined so that
$$P(k) = T(k)^2 P_i(k).$$

\subsubsection{The initial power spectrum}

For lack of any better assumptions, cosmologists usually assume that the
primordial power spectrum was a simple power law
$$P(k)\propto k^n$$
for some constant $n$ termed the {\it spectral index}, at least over the range of
$k$-values that are accessible to our observations. In CDM, the primordial
fluctuations are usually assumed to arise from quantum
fluctuations during a speculative inflationary epoch when the universe was
less than $10^{-32}$ seconds old.
Thirteen years after the theory of inflation was invented by Alan
Guth, there are now scores of different versions of it.
Most of them predict a spectral index between 0.7 and 1.
These initial fluctuations are {\it adiabatic}, which means that 
the density varies slightly from point to point but that the ratio of photons
to baryons and dark matter remains constant initially. 
For BDM, the primordial spectrum is an Achilles' heel, as the theory contains no
physical mechanism for generating it. One simply postulates that it is a
power-law. In addition, one postulates that the initial fluctuations are
of {\it isocurvature} type, which means that the fluctuations in the photon
density cancel the baryon fluctuations, so that the total density remains 
constant. One may question the value or calculating the BDM transfer
function, since at the end of the day it is multiplied by a function that
was put in arbitrarily by hand, but peoples' opinions differ on what is elegant.
An aesthetic advantage of BDM is that it requires no new particles
besides those we know exist, whereas CDM postulates that more than $90\%$ of
the matter  is of a type that we have never detected in our laboratories.

\subsubsection{The CDM transfer function}

The standard way to compute transfer functions is to write down the
Boltzman equation for the time-evolution of the phase-space distributions of all
particle species in the model (baryons, photons, neutrinos, collisionless dark
matter, {\etc}), linearize it by throwing away all quadratic and higher order
terms, plug it into a fast computer and go for lunch. To model evolution of the
power spectrum at late times, when the approximation $\delta\ll 1$ breaks down,
so called $N$-body simulations are employed: the time-evolution of some 
$10^6$ to $10^8$ point-particles in a box is studied by numerically integrating
the Newtonian equations of motion, assuming periodic boundary conditions.  
The power spectra resulting from a few assorted models are plotted in 
Figure~\ref{powerplot}.

Fortunately, the main features of the CDM spectrum can be understood without
numerical calculations, with a bit of hand-waving, as we will now see.
If the transfer function is to be anything else than a simple power law in
$k$, some process must imprint a physical length scale on it.
Collisionless dark matter particles have no particular length scale at all
associated with them that is relevant in this context. 
However, at redshifts $z\simgt 10^4$, the main contribution to the density 
and the self-gravity of
the universe was CBR photons, not dark matter, so before this, the crucial
issue was what the photons decided to do. 
Were it not for this fact, a power-law initial power spectrum would
necessarily lead to a power-law spectrum today. In fact, the transfer function
would be simply $T(k)=1$, giving 
$$P(k)\propto k^n.$$

As the universe expands, the horizon size keeps increasing, also in comoving
coordinates. This means that the wavelength of any sinusoidal perturbation
(a distance which remains constant in comoving coordinates) will sooner or later
become smaller than the horizon scale. When this happens, the perturbation is 
said to ``enter the horizon". 
To avoid various painful GR gauge ambiguities
associated with super-horizon sized modes, 
it is convenient to specify the amplitude of 
perturbations with a given wavenumber $k$ when they enter the horizon. 
Since microphysical processes such as pressure support turn out to be
irrelevant for super-horizon sized modes, the power spectrum at horizon
entry must also have been some power law. 

Before recombination, the photons and baryons were tightly coupled and acted
as a single fluid where the sound speed was of the order of the speed of
light.  This means that the above-mentioned 
Jeans length was enormous, indeed several
times greater than the horizon size, so perturbations 
that had entered the
horizon did not grow much at all early on, indeed logarithmically at best.
As the redshifting of the photons caused the density of the universe to become
matter-dominated around $z\approx 10^4$, the perturbations in the dark
matter began to grow. And at $z\approx 10^3$,
recombination caused the Jeans length to drop drastically, down to a comoving
scale much lower than the sizes of present-day galaxies, and perturbations in
the baryons began to grow as well. Thus dark matter fluctuations that entered the
horizon long after matter-radiation equality, fluctuations whose comoving
wavelength exceeds  
$$\lambda_{eq}\approx 13(\Oz h^2)^{-1}\Mpc,$$ 
were never affected by any of this microphysics that has a length scale
associated with it, which means that 
$T(k)\approx 1$ for $k\ll\lambda_{eq}^{-1}$. Fluctuations 
with $k\gg\lambda_{eq}^{-1}$, on the other hand, entered the horizon
while dark matter was still only a minuscule contributor to the overall
density,
and did not start growing properly until after matter-radiation equality. Thus
we know that for $k\gg\lambda_{eq}^{-1}$,  $T(k)^2$ must be some different power
law $k^m$, with $m<0$. We will now see that $m=-4$.
Take $\Omega=1$ and consider a perturbation of comoving wavelength $\lambda$.
Then its physical scale is 
$$d_{\lambda} = {\lambda\over 1+z}$$
whereas the horizon scale is
$$d_H\propto t \propto (1+z)^{-3/2}.$$
Equating these two scales gives the redshift for horizon entry
$$(1+z_{ent}) \propto \lambda^{-2} \propto k^2.$$
Since this redshift, $\delta$ has grown by a factor of $(1+z_{ent})$,
so $P(k)$ has grown by a factor of $(1+z_{ent})^2\propto k^4$.
The modes with $k\gg\lambda_{eq}^{-1}$, on the other hand, have
all grown by the same factor since they entered the horizon, 
so in this part of the power spectrum, the slope is down by the sought for 
factor of $k^4$. 
In conclusion, the net result of all this is that
$$P(k)\propto
\cases{
k^n&for $k\ll\lambda_{eq}^{-1},$\crr
k^{n-4}&for $k\gg\lambda_{eq}^{-1}$,
}$$
where $\lambda_{eq}\approx 50\,\Mpc$ for $\Omega=1$ and $h=0.5$.

\subsubsection{The transfer functions for baryons and photons}

All this concerned the power spectrum of the density of dark matter.
The baryons and the photons 
are affected by two additional effects, known as
Landau damping and Silk damping, which for the case of
adiabatic fluctuations effectively wipes out their 
fluctuations on galactic scales. 
Since we know that galaxies do indeed exist, a successful theory of structure
formation must find a way around this disaster. With CDM, the trick is that the
cold dark matter is unaffected by this damping, and that gravitational
instability eventually makes the baryon fluctuations ``catch up" with the
fluctuations in the dark matter. With BDM, the trick is to utilize
isocurvature
fluctuations instead, which do not suffer Silk damping. 
We will return to the transfer function for photons in the CBR section.

\subsubsection{The complaint department}

As to which model of structure formation is correct, 
the jury is still out. We will conclude this section by 
criticizing all the models, as democratically as possible.
According to many workers, standard CDM is off by about a factor of two. This
may not be much to fuss about in a field where one is often satisfied with
getting the right answer to within an order of magnitude, but as data gets
better, concern grows. 
This factor of two can be dealt with in at least six ways:

\begin{itemize}

\item One can argue that what we see is a biased version of what is
actually there, and introduce different fudge factors called {\it bias
factors} for various types of measurements.

\item The spectral index $n$ can be lowered to around 0.7 by tweaking certain
parameters associated with inflation, 
which seems to underproduce high redshift quasars. So  do several of the other 
models listed
below.

\item Repeating what Einstein described as
``the biggest blunder of my life" (Gamow 1970, p.~44),
a non-zero cosmological constant $\Lambda$
can be introduced.

\item Yet another free parameter can be introduced by adding just the right
amount of massive neutrinos (about $30\%$) to the cosmological mixture.

\item $\lambda_{eq}$ can be increased by postulating that
the $\tau$ neutrino decays in an appropriate manner and that there
are new and unseen particles, for instance additional massless 
Nambu-Goldstone bosons.

\item $\lambda_{eq}$ can be increased by lowering the Hubble constant
to about $h=0.25$, causing raised eyebrows among those astronomers who 
claim that
$h=0.8$.

\end{itemize}

\noindent
In addition, all of these inflation-based theories 
are challenged by the fact that some CBR experiments have 
been argued to  
indicate $n\approx 1.5$ on large scales, whereas essentially all inflationary
scenarios predict $n\leq 1$.

Meanwhile, BDM is fighting a difficult battle, ensnarled by a number of tight
constraints.
The Compton $y$-parameter prevents reionization from being too complete, too hot
and too early (see 
Chapter~\ref{ychapter}),
degree-scale CBR-experiments prevent reionization from being too late,
and the observed power-spectrum constrains the thermal history as well. 
As the author is writing this, one of his colleagues is running a Boltzmann
code for BDM models which have nine free parameters, so it appears that in
dodging the many constraints, the BDM model has lost much of its 
initial elegant simplicity, and with that perhaps most of its appeal.

\newpage
\section{The Microwave Background}
% Mention that we ignore dust and Lithium Hydride.

In 1964, Arno Penzias and Robert Wilson at Bell Laboratories discovered
microwave radiation of an unknown source that they at one point
thought might have been bird droppings in their antenna horn.
Eventually, their discovery was given a more grandiose interpretation,  
which resulted in them both winning a free trip to Stockholm. 
This radiation is the topic of the present section.
In the first few subsections, where its origin is described, we will also 
make a digression about the thermal history of the universe. 

\subsection{The thermal history of the universe}

When electrons are accelerated, they emit photons. Thus if one heats a
hydrogen plasma sufficiently, the fast-moving electrons will frequently
scatter off of nuclei and other electrons, resulting in a significant number of 
photons being created and radiated away.
For cosmological applications, the two most important of these
scattering processes are thermal bremsstrahlung 
$$p^+ + e^- \mapsto p^+ + e^- + \gamma$$
and double Compton scattering 
$$e^- + \gamma \mapsto e^- + \gamma + \gamma.$$
These processes can also go in the reverse direction, so 
if such a hot plasma filled the
entire universe, some of these photons would eventually get into
collisions where they would become absorbed.
One can show that 
if the plasma is held at a constant, uniform density long enough, then 
a thermal equilibrium situation will arise where the electrons and protons have
Maxwell-Boltzman velocity distributions corresponding to some temperature $T_e$
and the photons have a Planck blackbody distribution
$$\ng(\vx,\vk) = 
{1\over e^{h\nu/k\Tg} -1},$$ 
where $\Tg=T_e$. Here the photon frequency $\nu = |\vk|c/2\pi$, 
and the wave vector $\vk$ is 
not to be confused with Boltzmann's constant $k$. Note that since 
the number density $\ng$ is
a distribution over the six-dimensional photon phase space $(\vx,\vk)$, it is
dimensionless, as the units of $\m^{-3}$ arising from  $\vx$ are canceled by the
$\m^3$ arising from $\vk$.

\subsubsection{In the beginning, there was thermal equilibrium}

This was not quite the situation in the early universe, since the expansion
kept lowering the density (and hence the reaction rates). However, it is easy
to show that there was nonetheless sufficient time for thermal equilibrium to
be attained early on. 
Thus at high redshifts (say $z\gg 10^3$), the entire state of the 
hydrogen plasma
and the photons could be described by a single number, the 
temperature $T = T_e = \Tg$.
Indeed,
if we go back to high enough
redshifts, thermal equilibrium was attained not only for electron
and proton velocities and for electromagnetic radiation, but indeed for 
all particles that we know of, as a plethora of reactions kept rapidly 
creating and destroying all sorts of particles left and right. 
For instance, when the universe was much less than a
second old, electrons and positrons where being created and destroyed so
rapidly that their densities where accurately given by the standard Boltzman
formula. And even earlier on, the same applied to heavier particles like protons
and W bosons. 

\subsubsection{Freezeout}

As the universe expanded, the reaction rates that affected the
abundance and energy distribution of any given particle species kept dropping
until that species ``decoupled" from the merry thermal mixture, or ``froze 
out" as cosmologists say\footnote{a more quantitative definition is to say that
a given reaction is frozen out when the rate $\Gamma$ at which a particle 
is subject to such reactions is so low, that less than one such event is
expected to occur while the universe doubles its age. This
happens roughly when $\Gamma = H$, the Hubble constant.}.
One by one, species
after species froze out.  If the rest mass of a particle was much greater than
the thermal energy  $kT$ when its species froze out, then tough luck: the
freezeout abundance would have been exponentially suppressed by a huge Boltzman
factor, and  almost no such particles would still be around today to bear
witness.  This was the grim fate of the $W^+$, $W^-$ and $Z$
bosons, which constitute only a minuscule fraction of the density of the
universe today. 
This is also the assumed fate of many highly speculative 
particles\footnote{Particle physicists desperate for
funding have postulated the existence of many dozens of particles never seen by
man nor beast...}
whose existence today would be in conflict with observations. 
In the other extreme, where the particle is so weakly interacting that it
freezes out while $kT$ is still larger than its rest mass, there is still
an ample abundance of such
particles today. Such weakly interacting 
massive particles, known as WIMPS, are interesting candidates for the dark
matter in the universe. 
Another classic example of freezeout survivors are the 
cosmic background neutrinos,
which are still with us today.
A most interesting intermediate case is the neutron.
Being about $0.2\%$ heavier than the proton, the neutron to proton ratio in
equilibrium is 
$${n_n\over n_p}\approx e^{-10^{10}\K/T}.$$
Thus the neutron and proton 
abundances would be 
almost identical while the temperature of the universe exceeded their energy
difference, but eventually, as $T\to 0$, the
ratio would approach zero exponentially. 
Thus we might expect to find either virtually no neutrons today or a
$50\%$ abundance. In the latter case, it turns out that virtually all the
hydrogen would have formed Helium during nucleosynthesis, and we would find
things such as water quite difficult to come by. If we were
around... Interestingly, as mentioned in the nucleosynthesis section
above, nature chose to be right in the middle: neutron freezeout occurred just
around this critical temperature, and the primordial abundance is about $12\%$.

After this digression, we will limit ourselves entirely to 
electrons and photons.

\subsubsection{Recombination and the CBR freezeout}

The freezeout picture applies to hydrogen recombination as well.
In thermal equilibrium, the fraction of hydrogen that is ionized is
$$x \approx
\left[1+2.8\times10^{-6}T_5^{-7/6}e^{1.58/T_5}\right]^{-1},$$
where $T_5\equiv T/10^5\K$,
which means that if one slowly lowers the temperature of a hot hydrogen plasma, 
it will recombine when $T\approx 15,000\K$. 
However, the recombination reaction
$$p^+ +  e^- \mapsto H$$
has already begun freezing out by then. 
Quantum effects such as stimulated emission complicate the process further, and
the end result is that recombination does not occur until much later, at a
temperature of a few thousand K, at $z\approx 10^3$.
At this point, something else of great importance happens as well: the photons
freeze out. Until then, they were constantly Thomson scattering off of free
electrons, but from this point on, a generic photon merely moves along
in a straight line (or, more precisely, a geodesic) until today. 
Another way of phrasing this is that the universe became transparent at
$z\approx 10^3$, so that in a sense we can see that far back.
These frozen out photons gradually become redshifted as was described in the GR
section, and it is easy to show that this redshifting preserves the
blackbody nature of their spectrum, simply lowering the blackbody temperature 
as
$$\Tg = (1+z)\Tgz$$
as the
the universe expands. 
It is these frozen out photons that are known as the cosmic microwave
background radiation, the CBR. 
The best measurement to date to puts their current temperature at 
$$\Tgz \approx 2.726\K\pm0.005\K,$$
and was made by the Far Infra-Red Absolute Spectrometer (FIRAS) detector on
board the celebrated COBE satellite (Mather {\etal} 1994).

\subsection{CBR fluctuations}

If one corrects for our motion relative to the comoving 
CBR rest frame\footnote{Note that whereas there are 
no preferred observers in classical physics or in generic GR metrics, which
renders concepts such as ``at rest" meaningless, this is in fact not the case
in the FRW metric. Here comoving observers have a special status, and the CBR
plays the role of the infamous ``aether" which Michelson and Morley (1887)
discredited with their famous speed-of-light experiment.}, the
observed CBR
is almost perfectly thermal and isotropic. In other words, its spectrum is
almost exactly a Planck spectrum, and the temperature $\Tgz$ is almost
exactly the same in all directions in the sky.
If it were not for the caveat ``almost", the CBR would not be so interesting,
as it would contain no more information than the single number $\Tgz$.
(In addition, a perfectly isotropic CBR would have implied that all
our theories about structure formation would be dead wrong, as we shall see
below.) It is the small deviations from thermality and isotropy that
have the potential to give us information about the early universe.

These deviations can be conveniently split into two categories:
spectral fluctuations and spatial fluctuations.
To study the former, we average the spectrum from all directions in the sky
and investigate the extent to which it deviates from a Planck spectrum.   
To study the latter, we summarize the spectrum in each direction $\vn$ by a
single number, the best fit temperature $T(\vn)$, and investigate how this
temperature varies across the sky. 
In some instances, these two types of deviations are linked
--- we will return to this case when we discuss the so called SZ effect below. 

What sort of spectral distortions might we theoretically expect?
One recurring scenario in this thesis is that the IGM was reheated and
reionized at some redshift $z < 10^3$. As long as the IGM temperatures are not
too extreme, the thermal bremsstrahlung and
double Compton processes will remain frozen out, so that the number of CBR
photons will not change. However, the photons will Thomson scatter off free
electrons, which will modify their energy slightly and change their
direction of motion radically. As we will see in the following two sections, the
former effect tends to cause spectral distortions whereas the latter effect tends
to suppress spatial
fluctuations.

\subsection{Spectral distortions}

In this subsection, we will assume that the universe is completely uniform and
isotropic.
With this assumption, 
the time-evolution of 
the photon spectrum  $\N(\vr,\vk)$ in
the presence of free electrons of temperature $T_e$ is given by the Kompaneets
equation (Kompaneets 1957)
$$\dot\N = {k T_e\over m_e c^2}\sigma_t n_e c
{1\over x^2}{\partial\over\partial x}
\left[x^4\left({\partial\N\over\partial x} + \N + \N^2\right)\right],$$
where 
$$x\equiv {h\nu\over k T_e}.$$
This equation describes the effect of Compton scattering only, and is valid 
after about $z\sim 10^5$ when double Compton scattering
and thermal Bremsstrahlung have frozen out so that the number of photons 
is conserved. 
Here $\vr$ and $\vk$ are taken to be the {\it comoving} quantities, so that 
the redshift term disappears from the equation and we simply have
$\dot\N = 0$ if $n_e = 0$.
Direct substitution shows that $\dot\N = 0$ also in the more interesting case 
where 
$$\N = {1\over e^{x+\mu} -1}.$$
This means that if the photons start out with a Planck spectrum and if
the electrons have the same temperature as the photons, then there 
will be no spectral distortions at all.
Indeed, numerical integration of this nonlinear partial differential equation
shows that for most physically reasonable initial spectra, the presence of the
free electrons causes the photons to approach such a steady state. 
However, the resulting steady state typically has a ``chemical potential" 
$\mu \neq 0$.
Thus if some process such as the decay of a mysterious relic
particle creates non-thermal photons in the early universe, the result will be
a spectrum with $\mu\neq 0$. The FIRAS experiment has placed the strong 
limit $|\mu|< 3.3\tt{-4}$ 
on the value of $\mu$ (Mather {\etal} 1994). 

At the redshifts
$z< 10^3$ that we probe in this thesis,
this convergence towards a steady state is no longer effective, being basically
frozen out.
Suppose the free electrons have a temperature $T_e\neq \Tg$.
Substituting this into the right-hand-side of the Kompaneets equation and
using the fact that the resulting spectral distortions will be quite small, we
obtain the first order approximation that the spectral distortion today will be
simply (Zel'dovich \& Sunyaev 1969)
$${\delta\N\over\N} \approx 
\left[{x^2 e^x(e^x+1)\over (e^x-1)^2} - {4x e^x\over e^x-1}\right] y
\approx\cases{-2y&for $x\ll 1$,\cr x^2y&for $x\gg1$,}$$
where 
$$y \equiv \int\left({kT_e-k\Tg\over m_e c^2}\right) n_e\st c\> dt,$$
and the integral is to be taken from the early time when photon-creating
processes froze out until today. Thus we see that to
first order, the distortion will always have the same spectral 
profile, and that
only the magnitude of the distortion depends on the thermal history of the
electrons. This magnitude, referred to as the {\it Compton y-parameter},
thus
summarizes the effect of free electrons by a single convenient number. Loosely
speaking, the $y$-distortion indicates a relative excess of high-energy photons
and a deficit of low-energy ones, which is caused by collisions in which hot
electrons transfer energy to CBR photons. 

There are two types of $y$-distortions of current observational interest. 
The first type is the average $y$-distortion over all of the sky, typically
caused by very early reionization. This is the topic of 
Chapter~\ref{ychapter}.
To date, no
such $y$-distortion has been observed, and the 
upper limit of $y < 2.5\tt{-5}$ places interesting constraints on the thermal
history of the IGM. The second type is the localized $y$-distortion caused by
hot plasma in rich clusters of galaxies. 
This is usually referred to as the Sunyaev-Zel'dovich effect, or the SZ effect
for short, after the Russians who invented the $y$-parameter 
(Zel'dovich \& Sunyaev 1969). 
The SZ-effect has indeed been observed in the direction of a number of large
galaxy clusters --- see for instance Birkinshaw \& Hughes (1994). 
The experimental uncertainties are still large; around a
factor of two. As soon as these shrink, the SZ-effect will become of great
importance, since it provides an independent way of computing the
distance to a galaxy cluster and thus can be used to determine the elusive
Hubble constant. A recent review of the issues touched upon above is given by 
Hu \& Silk (1993).

\subsection{Spatial distortions}

In this subsection, we will take the opposite approach: we will assume that 
there are no spectral distortions, so that the CBR 
that we observe from the direction
$\vn$ in the sky has a Planck spectrum 
characterized by a temperature 
$\Tg(\vn)$. It is the COBE observation of this function (Smoot {\etal} 1992) 
that
has recently been plastered on front covers of magazines across the world and
prompted various famous people to say silly things about holy grails and seeing
the face of God. 
Our discussion of the ripples in $\Tg(\vn)$, usually referred to simply as CBR
fluctuations, will be organized as follows: first, we list some handy formulas
for working with spherical harmonics. Then we review how the
above-mentioned formulas for random fields are altered when dealing with a
sphere instead of $\R^3$. After this, we discuss how the CBR fluctuations are
related to the fluctuations in the density field $\delta$.
Finally, we show how early reionization can suppress CBR fluctuations, which is
relevant to Chapters 3, 4 and 5.
A recent review of CBR fluctuations is given by White, Scott \& Silk (1994).

\subsubsection{Working with spherical harmonics}

For functions that live on a sphere, the analogue of a Fourier
expansion is an expansion in {\it spherical harmonics}.
The spherical harmonics are defined as
$$Y_{lm}(\theta,\varphi) = \sqrt{{2l+1\over 4\pi}{(l-m)!\over(l+m)!}}
P_l^m(\cos\theta) e^{im\varphi},$$
where $P_l^m$ are the associated Legendre functions, and $l$ and
$m$ are integers such that $l\geq 0$ and $|m|\leq l$.
They have the symmetry property that 
$$Y_{l,-m} = (-1)^m Y_{lm}^*,$$
where $*$ denotes complex conjugation.
These functions form a complete
orthonormal set on the unit sphere.  The orthonormality means that 
$$\int Y_{lm}^*(\theta,\varphi)Y_{l'm'}^*(\theta,\varphi)d\Omega
= \delta_{ll'}\delta_{mm'},$$
and the completeness means that we can expand any $L^2$ function
$\Delta$ as 
$$\Delta(\theta,\varphi) = 
\suml \summ a_{lm}Y_{lm}(\theta,\varphi),$$
where 
$$a_{lm} \equiv \int Y_{lm}^*(\theta,\varphi)\Delta(\theta,\varphi)d\Omega.$$
Throughout this section, we use the differential solid angle notation
$$d\Omega = \sin\theta\> d\theta d\varphi.$$
Also, we will sometimes replace $\theta$ and $\varphi$ by the unit vector 
$$\vn = (\sin\theta\cos\varphi, \sin\theta\sin\varphi,\cos\theta)$$
and write things like 
$$Y_{lm} (\theta,\varphi) = Y_{lm} (\vn).$$
With this notation, the so called addition theorem for spherical harmonics
states that 
$$\summ Y_{lm}^*(\vn)  Y_{lm}(\vn') = 
{2l + 1\over 4\pi}P_l(\vn\cdot\vn'),$$
where $P_l = P_l^0$ are the Legendre polynomials. 

For the reader who wants more intuition about spherical 
harmonics, it is good to lump together all harmonics with the same $l$ value.
For $l=0$, 1, 2, 3 and 4, these sets are referred to as the
monopole, the dipole, the quadrupole, the octupole and the 
hexadecapole, respectively.
Geometrically, when one expands a function $f(\vn)$ in spherical harmonics, 
$l=0$ picks up the constant part, $l=1$ picks up the remaining linear part, 
$l=2$ picks up the remaining quadratic part, $l=3$ picks
up the remaining cubic part, {\it etc}.
In group theory jargon, the spherical harmonics corresponding to different
$l$-values are the irreducible representations of the group of rotations of the
sphere.  This means that if one expands a rotated version of the same function
$f$, the new spherical harmonic coefficients $a'_{lm}$ will be some linear
combination of the old ones $a_{lm}$, 
$$a'_{lm'} = \summ D_{lm'm}\>a_{lm},$$
such that different $l$-values live separate lives and never mix. 
For instance, however one chooses to rotate a linear function ($l=1$), it will
always remain linear and never say pick up quadratic terms.

\subsubsection{Random fields on the sphere}
\label{RandFieldsOnSphere}

$\Tg(\vn)$, the CBR temperature that we observe in the direction
$\vn$ in the sky, is modeled as a random field. 
It is more convenient to work with the dimensionless version
$$\Delta(\vn) \equiv {\Tg(\vn)\over\expec{\Tg(\vn)}} - 1,$$
which is often denoted $\Delta T/T$ in the literature. 
This field is related to the random density field $\delta$ that we discussed
earlier on, and we will return to how they are relater further on.
All the formulas and definitions we gave for $\delta$ have spherical analogues
for $\Delta$, as we will now see.
$\Delta$ is usually assumed to be Gaussian, which follows from the
random phase assumption
$$\expec{a_{lm}^*a_{l'm'}} = \delta_{ll'}\delta_{mm'} C_l,$$
where the coefficients $C_l$ constitute the spherical version of 
the power spectrum $P(k)$. 
The {\it angular correlation function} is defined as
$$c(\theta) = \expec{\Delta(\vn)\Delta(\vn')},$$
where $\vn\cdot\vn'=\cos\theta$, and the right hand side is independent of the
actual directions by the isotropy assumption. 
Just as the spatial correlation function was the Fourier transform of the
spatial power spectrum, the angular correlation function is what might be
called a ``Legendre transform"
of the angular power spectrum $C_l$. Using the addition theorem, one readily
obtains
$$c(\theta) = {1\over 4\pi}\suml(2l+1)C_l P_l(\cos\theta).$$
And just as the r.m.s. fluctuations of the smoothed density field was given by
integrating the power spectrum against a window function $\widehat{w}$,  
the r.m.s. temperature fluctuations seen by a CBR experiment with some 
angular selection function (beam pattern) is given by 
a weighted average of the coefficients $C_l$, the weights being  
a sort of ``Legendre transform" of the beam pattern. 
Explicit examples of this are given in 
Appendix~\ref{windowappendix},
and a few selected window functions are plotted 
Figure~\ref{wlplot}.

\subsubsection{The monopole and the dipole}

The non-cosmologist reader may wonder why all sums over $l$ in that
chapter start with $l=2$ rather than with $l=0$.
This is because we have no way of measuring the monopole coefficient $c_0$ 
or the dipole coefficient $c_1$. 

If we knew the ensemble average 
$\expec{\Tg}$ (or could measure it by repeating the COBE experiment in many
different horizon volumes throughout the universe and invoking ergodicity),
then we could calculate our observed monopole 
as the difference between the 
average $\Tg$ that we observe in our sky and $\expec{\Tg}$. But we can't...
In other words, our CBR monopole is  
the difference between the average temperature in our sky and the average
temperature of the entire universe. But since we use the former as our estimate
of the latter, this gets us nowhere with the monopole.
Since direct estimates of the age of the universe are still quite uncertain,
but we know the CBR temperature to three decimal places, the CBR
in a sense merely tells us what time it is,
{\ie} how much our universe had expanded by the time this civilization of ours
turned up on the scene. 
Thus it seems highly unlikely that we will be able to use any kind of physics
to predict $\expec{\Tg}$ accurately (to one part in $10^{-5}$ or so)
in the foreseeable future.

The situation is similar with the dipole. If we knew that we were comoving
observers, {\ie} at rest with respect to the comoving FRW coordinates, then we
could measure our local CBR dipole as the dipole we actually see in the sky. 
But the most accurate way we have to determine the comoving rest frame is to 
calculate the frame in which the CBR dipole vanishes.
In other words, our CBR dipole is the
difference between the average dipole in our sky and our velocity vector.
Since we use the former as
our estimate of the latter, this gets us nowhere with the dipole either.
Thus the lowest multipole which gives us information about the power spectrum
is $l=2$, the quadrupole.

% \subsubsection{Cosmic Variance}

\subsubsection{The relationship between $\delta$ and $\Delta$}

To date, the only really accurate way to compute the CBR power spectrum 
$C_l$ is to integrate the Boltzmann equation numerically. Today it takes a few
hours of CPU time on a high-end workstation to compute the first few
thousand $C_l$-coefficients. The results for a CDM model and a BDM model 
are plotted in
Figure~\ref{clplot}, multiplied by
$l(l+1)$.
To give some intuition about the general shape of these functions, we will very
briefly mention the main physical effects at work.

In the absence of reionization, the CBR photons that reach us today are 
carrying information about the electron distribution on the 
{\it last-scattering surface}, the spherical region around us where 
the photons last Thomson scattered off of an electron
some some $10^{10}$ lightyears away.
There are three mechanisms through which adiabatic density fluctuations
$\delta$ on the last-scattering surface cause CBR anisotropies:

\begin{enumerate}

\item Fluctuations in the gravitational potential $\phi$ causing a gravitational
redshift/blueshift

\item Bulk velocities on the last-scattering surface, causing Doppler shifts

\item Fluctuations in the photon density on the last-scattering surface

\end{enumerate}

\noindent
For adiabatic fluctuations, it turns out that the third effect cancels two
thirds of the first effect. The net result in known as the Sachs-Wolfe effect, 
and for a matter power spectrum $P(k)\propto k^n$,  
the Sachs-Wolfe effect turns out to give 
$$C_l \propto {\Gamma\left(l+{n-1\over 2}\right)\over
\Gamma\left(l+{5-n\over 2}\right)},$$ 
which for the case $n=1$ reduces to simply
$$C_l \propto {1\over l(l+1)}.$$ 
The Sachs-Wolfe effect is the dominant source of anisotropies on large angular
scales $\theta\gg 1^{\circ}$, corresponding to small $l$. This is why the 
$n=1$ CDM curve in 
Figure~\ref{clplot}
is flat for small $l$.

The funny-looking bumps in the CDM curve, 
known as ``Doppler peaks" are due to a combination of
effects 2 and 3. Finally, the last scattering surface has a finite thickness,
which means that when we look in a given direction, we are in fact seeing a
mixture of photons emanating from different points along that line of sight. 
This averaging of the fluctuations at many different points 
washes out fluctuations on very small angular scales, which is why all the
curves approach zero for $l\gg 10^3$.

\subsection{How reionization suppresses fluctuations}

As mentioned in the introduction, reionization would affect the CBR in at
least three ways:

\begin{enumerate}

\item The hot electrons would cause a $y$-distortion, as discussed above.

\item The fluctuations in the density and velocity of these hot electrons
would generate new CBR anisotropies. For $\Omega=1$, due to various
cancellations,
it turns out that this so called Vishniac effect becomes important only on very
small angular scales. We will not discuss it further here ---
see Hu {\etal} 1994
for a thorough discussion.

\item{Spatial fluctuations on angular scales below a few degrees could become
suppressed, while fluctuations on larger scales such as those probed by COBE
would remain fairly unaffected.}

\end{enumerate}
This third effect can be calculated numerically by integrating the 
linearized Boltzmann equation as mentioned above. Since 
hitting [RETURN] and waiting for hours does not greatly enhance ones intuitive
understanding of the underlying physics, we will devote this section to 
deriving a simple analytic approximation which agrees fairly well with 
the numerical results
for relatively late ($z<200$) reionization.
This is based on some ray tracing calculations made by
the author. Examples of ray tracing work in the literature are
Peebles (1987) and Anninos {\etal} (1991).

\subsubsection{The ray tracing approach}

Let $n_z(\vx,\vq)$ denote the six-dimensional 
phase-space density of
photons at redshift $z$ that have comoving position $\vx$ and wave 
vector $\vq = q\qh$. 
If this function is known at our location ($\vx=0$) today ($z=0$),
it is straightforward to predict the outcome of any microwave
background experiment. 
The time evolution of this function between
early times and today is described by the 
Boltzmann equation for radiative transfer. Since this
equation is
linear in $n$ and invariant under spatial translations, we can
write 
\beq{BoltzmanKernelEq}
n_z(\vx,\vq) = 
\int G(\vx-\vx',\vq,\vq',z,z') n_{z'}(\vx',\vq')
d^3x'd^3q'
\eeq
for some integral kernel $G$.
One way to interpret this $G$ is as the probability density that
a photon that has position $\vx$ and momentum $\vq$ at redshift
$z$ was at $\vx'$ with momentum $\vq'$ at redshift $z'$.
Thus an alternative to integrating the Boltzman equation numerically is to
evaluate this probability distribution by
Monte Carlo simulations of trajectories of large numbers of photons.
As we will see, it is sometimes
possible to obtain good analytic approximations for $G$ as well.

\subsubsection{The geometrical smudging effect}

Reionization of the intergalactic medium (IGM) causes
CBR photons to Thomson scatter off of free electrons, which affects
CBR fluctuations in the three ways listed above. 
The third of these effects, to which we are
limiting our attention, is purely geometrical in nature, 
{\ie} independent of the
fluctuations in the electron temperature and velocity.
Thus ignoring the frequency change that photons experience when they
scatter, the magnitude $q$ simply stays constant
for all
photons (there redshift is taken care of by the fact that we are using
comoving position and momentum coordinates). Hence we can 
write \eq{BoltzmanKernelEq} as 
\beq{KernelEq}   
n_z(\vx,\qh,q) = 
\int G(\vx-\vx',\qh,\qh',z,z') n_{z'}\left(\vx',\qh',q\right)
d^3x'd\Omega',
\eeq
where the angular integration is over all directions $\qh'$.
Assuming Planck spectra and using the notation introduced in 
Section~\ref{RandFieldsOnSphere}, 
this implies that 
\beq{DeltaKernelEq}   
\Delta_z(\vx,\qh) = 
\int G(\vx-\vx',\qh,\qh',z,z') \Delta_{z'}\left(\vx',\qh'\right)
d^3x'd\Omega'.
\eeq

\subsubsection{The isotropy approximation}

A very useful approximation is that 
the radiation is isotropic early on, at the last scattering 
epoch $z=z_{rec}$. In this approximation, we can write 
$$\Delta_{z_{rec}}(\vx,\qh) = \Delta_{rec}(\vx),$$
since the left hand side is independent of $\qh$.
For a thin last-scattering surface, it is easy to see that 
this is the case
for the contribution from both the Sachs-Wolfe and the intrinsic density
fluctuations, whereas the Doppler term will have a dipole anisotropy. 
With the isotropy approximation,
we have 
\beq{IsotropicKernelEq}
\Delta_z(\vx,\qh) = 
\int G(\vx-\vx',\qh,z,z_{rec}) \Delta_{rec}(\vx') d^3x',
\eeq
where we have defined the angularly averaged kernel
$$G(\Delta\vx,\qh,z,z') \equiv 
\int G(\Delta\vx,\qh,\qh',z,z') d\Omega'.$$
From here on, we will suppress $z$ and $z'$, taking $z=0$ and $z'=z_{rec}$.
Fourier transforming
\eq{IsotropicKernelEq} with respect
to $\vx$ and using the convolution theorem leaves us with
\beq{FourierKernelEq}
\Dhat_0(\vk,\qh) = 
\Gh(\vk,\qh) \Dhat_{rec}(\vk).
\eeq
Thus there is no mode coupling at all, and for fixed 
$\qh$, the transfer
function is simply the Fourier transform of the probability
distribution $G(\Dx) \equiv G(\Dx,\qh)$.

\subsubsection{The transfer function $\Gh$}

Before we turn to the problem of explicitly evaluating $\Gh$, we will briefly
discuss its interpretation. 
Assuming that $\Delta$ is a Gaussian random field, 
it is easy to show that 
$$
\expec{\Dhat_0(\vk,\qh) \Dhat_0(\vk',\qh)} = 
|\Gh(\vk,\qh)|^2 \expec{\Dhat(\vk)_{rec} \Dhat_{rec}(\vk')},
$$
{\ie} that the quantity $|\Gh(\vk,\qh)|$ plays the 
role of a transfer function. 
Thus the radiation power spectrum on scales $\lambda = 2\pi/k$ 
is suppressed by a factor $|\Gh(\vk,\qh)|^2$.
Let us make a few
elementary observations
that are valid for an arbitrary probability distribution $G$.
\begin{description}

\item[Observation (I):]
$|\Gh(\vk)| \leq 1$, with equality if $\vk=0$ or 
$G(\Dx)=\delta(\Dx-\Dx_0)$.

\item[Observation (II):]
$\Gh(\vk) \to 0$ as $|\vk| \to \infty$ if $G$ is an integrable
function (as opposed to say a tempered distribution like 
$\delta(\Dx-\Dx_0)$).

\item[Observation (III):]
If we define $\va$ to be the mean and $S$ to be the covariance matrix
of the probability distribution $G$, then 
$$\Gh(\vk) = 1 - ia_m k_m - {1\over 2}(a_m a_n+S_{mn})k_m k_n +
O(|\vk|^3),$$ 
so 
\beq{TaylorExpansionEq}
|\Gh(\vk)|^2 = 1 - S_{mn}k_m k_n + O(|\vk|^3).
\eeq
(Repeated indices are to be summed over, from $1$ to $3$.)

\end{description}
Observation (I) tells us that Thomson scattering causes a
low-pass filtering of the Fourier modes, {\ie} leaves very long wavelengthes
unaffected. $G(\Dx)=\delta(\Dx-\Dx_0)$ corresponds to no reionization, in which
case there is no smudging at all and no modes are suppressed.

\noindent
Observation (II), which is known as Riemann-Lebesgue's Lemma,
tells us that unless there is a finite probability for no
scattering at all, very high frequency modes get almost entirely damped
out. 

\noindent
Observation (III) gives us a Taylor expansion of $\Gh(\vk)$ around the
origin, which will prove useful below.

All quantities of physical interest ($C_l$,
$C(\theta)$, \etc) can be computed from the initial power spectrum
at $z_{rec}$ once the transfer function $\Gh(\vk,\qh)$ is known,
and explicit formulas for this are given in most standard texts. 
A crude rule of thumb simply identifies the comoving length scale 
$\lambda$ with the angle that this distance subtends on the last scattering
surface, a spherical region whose radius is our roughly our horizon radius 
$a_0\tau_0\approx 6000h^{-1}\Mpc$. 
Another rule of thumb is to identify a
multipole moment $\l$ with the angular scale $\theta\approx 180^{\circ}/l$. 
Thus for small angles, we have the rough correspondence 
$$\lambda\propto\theta\propto{1\over l}\propto{1\over k},$$
where the proportionality constants are given by the correspondence
$$\theta \approx 1^{\circ}\quad\sim\quad l\approx 200 \quad\sim\quad
\lambda\approx 100h^{-1}\Mpc.$$
Let us define the {\it smudging scale}
$$\lambda_c\equiv (\det S)^{1/6},$$
{\ie} as the geometric mean of the three eigenvalues of $S^{1/2}$.
For realistic scenarios, the smudging tends to be fairly isotropic. Thus 
all three eigenvalues of the covariance matrix $S$ tend to
be of the same order of magnitude, and 
the Taylor expansion\eqnum{TaylorExpansionEq}
gives  
\beq{CrudeTaylorEq}
|\Gh(\vk)|^2 \approx 1 - (\lambda_c k)^2
\eeq
for $k\ll\lambda_c^{-1}$.
When we detect a CBR photon arriving from some direction $\qh$ in the sky and
ask where it was at $z=z_{rec}$,
the smudging scale $\lambda_c$ is roughly the standard deviation 
of the answer (see Figure~\ref{lightningplot}).

\subsubsection{Approximating the transfer function $\Gh$}

For fairly late (say $z<200$) reionization, there is a substantial probability
that a given CBR photon manages to dodge all the reionized electrons without
being scattered a single time. Writing this probability 
as $e^{-\taut}$, where the quantity $\taut$ is called the {\it optical depth}
and will recur in subsequent chapters, 
the probability that a photon is scattered $n$ times will be 
$e^{-\taut}{\taut^n\over n!}$, 
a Poisson distribution. ($\taut$ is not to be confused with the conformal time
$\tau$.) Let us expand the propagator $G$ in a type of Born expansion
\beq{BornEq}
G = \sum_{n=0}^{\infty} e^{-\taut}{\taut^n\over n!}G_n,
\eeq
where $G_n$ is the probability distribution corresponding to the case where the
photon scatters exactly $n$ times. 
Omitting the trivial dependence on $\qh$ and $\qh'$, we have simply 
$$G_0(\Delta\vx) = \delta(\Delta\vx - \Delta\vx_0),$$
where $\Delta\vx_0$ is some constant,
since if the photon never scattered, we know exactly where it came from.
$G_1$ can be readily calculated analytically, which we will do in the next
section, after which the higher order terms can be calculated by repeated
integrations if desired.
A nice feature is that for late reionization, $\taut$ is often so small that 
a good approximation is obtained by dropping all but a few of the first terms
in \eq{BornEq}. 
For instance, reionization around $z=50$ with $\Omega=1$
and $h\Ob=0.03$ gives 
$e^{-\taut}\approx 0.70$,  $e^{-\taut}\taut \approx 0.25$ and 
$e^{-\taut}{\taut^2\over 2} \approx 0.044$, so if we keep only the $G_0$
and $G_1$ terms, we will be off by only $5\%$, and if we include the $G_2$ term
as well, we are off by less than $1\%$. 
Physically, this simply means that if it is not that likely that a photon
scattered even a single time, then it is quite unlikely that it scattered more
than a few times, so that we can safely neglect that possibility. 
If one nonetheless wants to include all terms, it is readily done by
Monte Carlo simulation that follow individual photons as they
propagate upwards in Figure~\ref{lightconeplot}, occasionally
changing direction as they Thomson scatter off of free electrons.
Some examples of this are shown in 
Figures~\ref{jesusplot}
and~\ref{lightningplot}.
Let us write the infinite sum\eqnum{BornEq} as
$$G = e^{-\taut} G_0 + \left(1-e^{-\taut}\right)G_{1+},$$
where $G_{1+}$ contains the contributions from all terms with $n\geq1$. 
We Fourier transform this with as before:
$$\Gh(\vk,\qh) =  
e^{-\tc} + \left(1-e^{-\tc}\right)\Gh_{1+}(\vk,\qh)$$
Since $\Gh_{1+}$ contains no delta-functions, its
Fourier transform approaches zero as $|\vk|\to\infty$ according
to Observation I. Thus
$$\Gh(\vk,\qh) \to
e^{-\tc}\quad\hbox{as}\quad k\to\infty.$$
Combining this with the estimate\eqnum{TaylorExpansionEq}, 
we thus know how the
transfer function behaves in both limits:
the approximation
\beq{TransferApproxEq}
|\Gh(\vk,\qh)| \approx 
\left(1-e^{-\taut}\right) e^{-{1\over 2}\vk^TS\vk} + e^{-\taut}
\eeq
becomes exact both as $k\to 0$ and as $k\to\infty$.
Using \eq{CrudeTaylorEq} yields the isotropic approximation
\beq{CrudeTransferApproxEq}
|\Gh(\vk,\qh)| \approx \left(1-e^{-\taut}\right) e^{-(\lambda_c k)^2/2} +
e^{-\taut}.
\eeq
This function is plotted in 
Figure~\ref{suppressionplot} for a few reionization scenarios.
The values of $\lambda_c$ used in this plot where 
computed through Monte Carlo
simulation in conjunction with Wayne Hu.

\subsubsection{The backward light cone}

We will conclude our discussion of how reionization suppresses
fluctuations by computing the function $G_1$ analytically. 
Let us assume that
$\Omega=1$. To simplify the calculations that follow, we will use comoving
coordinates $\vx$ and conformal time $\t$, so that all light rays (null
geodesics) make 45 degree angles with the $\t$ axis in $(\t,\vx)$-space, just
as in  Penrose diagrams.
Let us choose the scale factor $a_0$ such that 
$\t=1$ corresponds to today, {\ie} so that 
$$\t = {1\over\sqrt{1+z}}.$$
The situation is illustrated in 
Figure~\ref{lightconeplot},
where we have
suppressed one of the three spatial coordinates for the sake of the plot. 
We are
at the  apex of the cone, our {\it backward light cone}, 
and in the absence  of
reionization, the CBR
photons that reach us today have been moving on the surface of this cone
ever since recombination.

\subsubsection{Where where the photons at recombination?}

If we detect a CBR photon arriving from a given direction in the sky,
where was is at a redshift $z$?
The presence of free elections
makes it possible for a photon to change direction a number of
times by inelastic Thomson scattering. We neglect
other ways by which photons can change direction, such as
gravitational lensing.

To avoid double use of the letter
$z$, we will write comoving coordinates as $\vx = (u,v,w)$.
Suppose a photon arrives from the $w$-direction to the point
$\vx=0$ at some conformal time $\t$. Where was it at the
earlier time $\t_1 = \t-\Dt$?
If it has not been scattered during the interval $\Dt$, it has
followed a null geodesic, and its
position at $\t_1$ is given by
\beq{xzeroEq}
\vxzero(\t_1) = (0,0,\Dt).
\eeq
If it has been scattered exactly once during the interval, say at the
conformal time $\t_1+\ts$, and previously came from the direction
$(\theta,\phi)$ in spherical coordinates, then 
\beq{xoneEq}
\vxone(\t_1) = (u,v,w) =
\left[\ts\sin\theta\cos\phi,\ts\sin\theta\sin\phi,
\ts\cos\theta+(\Dt-\ts)\right].
\eeq
For $|\vx|\le \Dt$, this is readily inverted, yielding
\beq{InverseEq}
\cases{
\ts& = ${u^2+v^2+(\Dt-w)^2\over 2(\Dt-w)}$\cr
\theta& = $2 \arctan{\Dt-w\over\rad}$\cr
\phi& = $\arctan{v\over u}$
}
\eeq

To answer the question ``where did the photon come from if it
scattered exactly once", let us take $\ts$\ $\theta$ and $\phi$
to be independent random variables with probability distributions 
$\ft$, $\fth$ and $\fph$, 
and calculate the probability distribution for
the random variable $\vx(\t_1)$ as defined by 
\eq{xoneEq}.
This probability distribution is
\beq{fxoneEq}
G_1(\vx) = \cases{
\ft\left[\ts(\vx)\right] \fth\left[\theta(\vx)\right]
\fph\left[\phi(\vx)\right] \vol
&if $|\vx|\le \Dt$,\cr
0&if $|\vx| > \Dt$,}
\eeq
where the volume element is
$$\vol\equiv 
\left|\matrix{
\dd{\ts}{u}&\dd{\ts}{v}&\dd{\ts}{w}&\crr
\dd{\theta}{u}&\dd{\theta}{v}&\dd{\theta}{w}&\crr
\dd{\phi}{u}&\dd{\phi}{v}&\dd{\phi}{w}
}\right|
= {1\over\rad (\Dt-w)}.$$
For Thomson scattering, the angular dependence is 
${1\over\sigma}{d\sigma\over d\Omega} = 
{3\over 16\pi}\left(1+\cos^2\theta\right)$, so 
$$\cases{
\fth(\theta)& = $\int_0^{2\pi}
{3\over 16\pi}\left(1+\cos^2\theta\right) {d\Omega\over d\theta d\phi}
d\phi = {3\over 8}\left(1+\cos^2\theta\right)\sin\theta,$\cr
\fph(\phi)& = $1\over 2\pi.$
}$$
The function $\ft$ depends on the ionization history of the universe, 
so we will leave it arbitrary for the time being.
Figure~\ref{lightconeplot} 
shows a situation where the photon was scattered at 
$z=10$: we then know that it must have arrived on the smaller backward light
cone emanating from the scattering event. 
This corresponds to the probability distribution  
$\ft(\t)$ being a delta function. 
%%%%%%%%%%%%%%%%%
If we know that the photon scattered sometime around $z\approx 10$, 
and exactly once, but do not
know exactly when, we obtain a probability distribution 
$G_1(\vx)$ that is smeared out around the perimeter of
this $z=10$ backward light cone. 
For more physically realistic cases, where scattering occurs 
predominantly at higher redshifts, the resulting probability 
distribution tends to be unimodal.
%%%%%%%%%%%%%%%%%%%%

%%%%%%%%%%%%%%%%%%%%%%%%%END OF CBR SECTION %%%%%%%%%%%%%%%%%%%%%%%%%%%%

\newpage
\section{Cosmological Chemistry}

Almost all of the following chapters will involve the chemistry of the IGM, the
intergalactic medium, and frequent references will be made to constraints from
nucleosynthesis. 
The purpose of this section is to convince the reader that it
is quite straightforward to understand chemical equations,  even if one has
never been anywhere near a test tube.

\subsection{Evolution of the IGM}

As an example, let us take the evolution of the ionization state of the IGM in
the simplest case, neglecting helium and ionizing photons.
Let $x$ denote the ionization fraction and $n$ the total number density of
protons, free and bound. Then the densities of hydrogen atoms, free electrons
and free protons are $n_H = (1-x)n$, $n_e = xn$ and $n_p = xn$, respectively. 
Thermal ionization occurs through the reaction 
$$e^- + H \mapsto e^- + e^- + p^+,$$
{\ie} the electron is a catalyst that collides with the hydrogen atom and
uses its kinetic energy to ionize it. 
Thus the rate per unit volume for this reaction is proportional to 
$n_e n_H =n^2 x(1-x)$. Similarly, the rate for the recombination reaction 
$$e^- + p^+ \mapsto H + {\rm photon}$$
is proportional to $n_e n_p = n^2 x^2$.
Thus the time-evolution of $x$ is 
\beq{xdotEq}
\dot x = n[Ax(1-x) - Bx^2],
\eeq
where the 
proportionality constants $A$ and $B$ are functions of temperature
(since the particles come near each other more often if they are moving faster,
and since the reaction cross sections depend on the kinetic energies
of the particles).
The cosmologist generally takes such functions from the atomic 
physics literature without worrying to much about how they were calculated.
In addition to the equation specifying $\dot x$, one also needs an equation 
for $\dot T$, to be able to compute the evolution of temperature.
Such an equation is quite straightforward to interpret, the various
terms simply corresponding to the various cooling and heating
effects that one wishes to take into account.

Before integrating the above-mentioned equations numerically, it is  
often convenient to change the independent variable from time
to redshift. Then the IGM density is simply given by
$n = n_0(1+z)^3$, $n_0$ denoting the density today.
Using \eq{dzdtEq} for $dz/dt$, \eq{xdotEq} becomes
$$-{dx\over dz} = H_0 n_0{1+z\over\sqrt{1+\Oz z}}
[Ax(1-x) - Bx^2].$$

\subsection{Nucleosynthesis}

For a more complicated problem like nucleosynthesis
\footnote{This calculation would usually be referred to as nuclear physics
rather than chemistry, but the resulting equations are of the same
type --- the only difference being that nucleons (protons and neutrons) rather
than atoms are the fundamental entities.},  the formation of light elements
during the first three minutes after the Big Bang, the mathematical structure of
the problem is the same, merely more complicated. Instead of one single fraction
$x$ to keep track of, there will be a whole bunch of fractions $x_i$,
$i=1,2,...,n$: the fraction of protons in hydrogen, the fraction of protons in
helium-3, the fraction of protons in helium-4, the fraction of protons in
lithium, {\it etc}. The time derivative of each one of them is given by an
equation like the one above, where the left hand side is $\dot x_i$ and the
right hand side contains one term for each process that destroys or creates
species $i$. The resulting system of $n$ coupled first-order ordinary
differential equations is fed into a computer and solved numerically. The
numerical part is somewhat complicated by the fact that chemical equations tend
to be stiff, but there is a plethora of standard packages that handle this sort
of problems, for instance in the NAG library. 

When this calculation is carried out --- see 
Malaney \& Mathews (1993) for a recent review, 
results are obtained that are in fairly good agreement with the
primordial abundances of helium-4, helium-3, deuterium and lithium 
that are inferred from observational data. 
Pioneered by George Gamow and others
in the forties (Alpher, Bethe \& Gamow 1948\footnote{Gamow
added
his friend Hans Bethe to the author list of this so called 
$\alpha\beta\gamma$ paper as a joke...}), 
nucleosynthesis predictions are one of the most striking
successes of the Big Bang model.
Apart from being a powerful weapon against rival cosmological
theories, nucleosynthesis calculations also place an 
important constraint on the standard cosmological model,
since the
results turn out to depend strongly on the ratio between the densities of 
baryons and microwave background photons. This translates into a strong
dependence on $h^2\Ob$.
The most recent calculations to date (Smith {\etal} 1993; Walker {\etal} 1991)
give the 95\% confidence interval 
$$0.010 < h^2\Ob <0.015.$$
(Levels of confidence must be taken with a grain of salt in cosmology, where 
systematic errors often dominate.)
This interval keeps shifting around and getting narrower
as observational abundance data gets better and the the knowledge of 
the nuclear cross sections that go into the theoretical calculation improves.

\section{Astronomy Jargon}

Since the scales involved in cosmology are many orders of magnitude larger than
those usually encountered in physics, cosmologist tend to use larger units of
measurement.
A convenient unit of time is the Giga-year:
$$1\Gyr = 10^9\hbox{ years} \aet{3.15}{16}\s \approx 10^{16}\pi\>\s.$$

\noindent
An {\it astronomical unit}, or AU for
short, is the average distance between the earth and the sun. 
$$1\AU\aet{1.4960}{13}\cm.$$
For historical reasons, related to the parallax triangulation method by which the
distance to nearby stars is calculated, the standard astronomical distance unit
is the {\it parsec}. This is the distance at which an AU makes an angle of one
arcsecond (1/3600 of a degree) in the sky.  
Thus
$$1\pc = {360\times 3600\over 2\pi}\AU \aet{3.086}{18}\cm
\approx 3.2615\hbox{ light-years}.$$
This is still a tiny distance by cosmology standards.
The standard length unit in cosmology is the {\it megaparsec}:
$$1\Mpc \aet{3.086}{24}\cm.$$
The standard astrophysical unit of mass is {\it solar mass}, denoted 
$\Ms$. The mass of the sun is 
$$M_{\odot} \aet{1.989}{33}\g.$$

Apart from involving many powers of ten, some cosmological parameters 
are quite uncertain. It is convenient to write such parameters as
the product of a large round number that is in the right ballpark and a 
small dimensionless ``fudge factor".
For instance, the dimensionless quantity $h$ is ubiquitous in 
the cosmology literature. This is because 
the Hubble constant is factored as
$$H = \left({100\kms\over\Mpc}\right) h.$$
Whereas Hubble originally estimated that $h\approx 5$, most cosmologists
today assume that $0.5<h<0.8$. The jury is still out.
If some form of matter $x$ has a uniform density $\rho_x$, then this density is usually factored as 
$$\rho_x = \rho_{crit}\Omega_x,$$
where 
$$\rho_{crit}\equiv {3H^2\over 8\pi G} 
\aet{1.9}{-29}h^2{\g\over\cm^3}$$
is the critical density mentioned above.
Thus $\Ob$ is a measure of the baryon density, $\Omega_{igm}$ is a measure
of the IGM density, {\it etc}.

The only other unexplained astronomy jargon that I have been
able to spot in the subsequent chapters is some reference to different types of
stars. For historical reasons (ever heard that excuse before...?), 
stars are 
classified in categories called O, B, A, F, G, K, and M, where O-stars are
the hottest and M-stars the coolest. Within each category, there are
ten sub-categories specified by numbers, where for instance an O3 star is hotter
than an O9 star.

% While mentioning standard candles,  ref Spinrad:
% 28. CONFERENCE PAPER
%     Spinrad, H.
%      Spectroscopy and photometry of faint galaxies: hints at their evolution.
%    IN:  Objects of High Redshift. IAU Symposium No.92. (Objects of High
%    Redshift. IAU Symposium No.92, Los Angeles, CA, USA, 28-31 Aug. 1979).
%     Edited by: Abell, G.O.; Peebles, P.J.E. Dordrecht, Netherlands: Reidel,
%     1980. p. 39-48.
%       Pub type:  General or Review.
% Abstract: Discusses four optical methods to locate standard candles (giant E
%      galaxies), which should eventually lead to reasonably large samples at
%      high z. Redshift criteria and determinations are briefly discussed. The
%      observed faint galaxy colors fit well into a simple model of stellar
%      (galactic) evolution for systems with star-formation confined to a short
%      time interval. Finally a summary of the statistics of blue galaxies in
%      Coma-like clusters (Butcher-Oemler effect) is presented and interpreted.

\newpage
\section{An Abstract for Non-Cosmologists}

So in plain English, what is this thesis all about?
\medskip

It is now widely accepted that something heated and ionized the hydrogen that
fills intergalactic space, but we don't know quite when this happened.

If this is to have had interesting effects on the microwave background
fluctuations, then it must have occurred quite long ago, at redshifts
greater than 50 or so.
This is the topic of Chapters 3 and 4. 

Some theories predict that the hydrogen has indeed been 
ionized pretty much all along. In Chapter 5 we show to what extent such
scenarios are ruled out by recent measurements of the microwave background
spectrum.

To affect the microwave background, $100\%$ ionization is not required --- some
reasonable fraction like $30\%$ is quite sufficient, providing that
the ionization happens early enough.
By studying the spectra from quasars, however,
one gets more extreme constraints: 
in ``recent" times, for redshifts
less than about five, it appears that at least $99.999\%$ of the hydrogen atoms
must have been ionized.
It is still not entirely clear
whether photoionization alone is able to produce the high temperatures required to
achieve this. In Chapter 6 we provide an alternative model for how this may
have happened. 

Finally, there is increasing evidence that the primordial power spectrum 
may be a more complicated function than a simple power law, an assumption that
has been used in much of the literature, including Chapters 3, 4 and
6. Motivated by this evidence, the final chapter is devoted to 
developing a formalism that allows cosmological constraints to be placed
that are completely independent of the shape of power spectrum. 
Some supplementary material for Chapters 3 and 7 is given in the appendices.

\newpage

\def\fheight{14.5cm} \def\fwidth{14.5cm} % Light cone plot must remain square
\bfig
\vskip-1.5truecm 
\psfig{figure=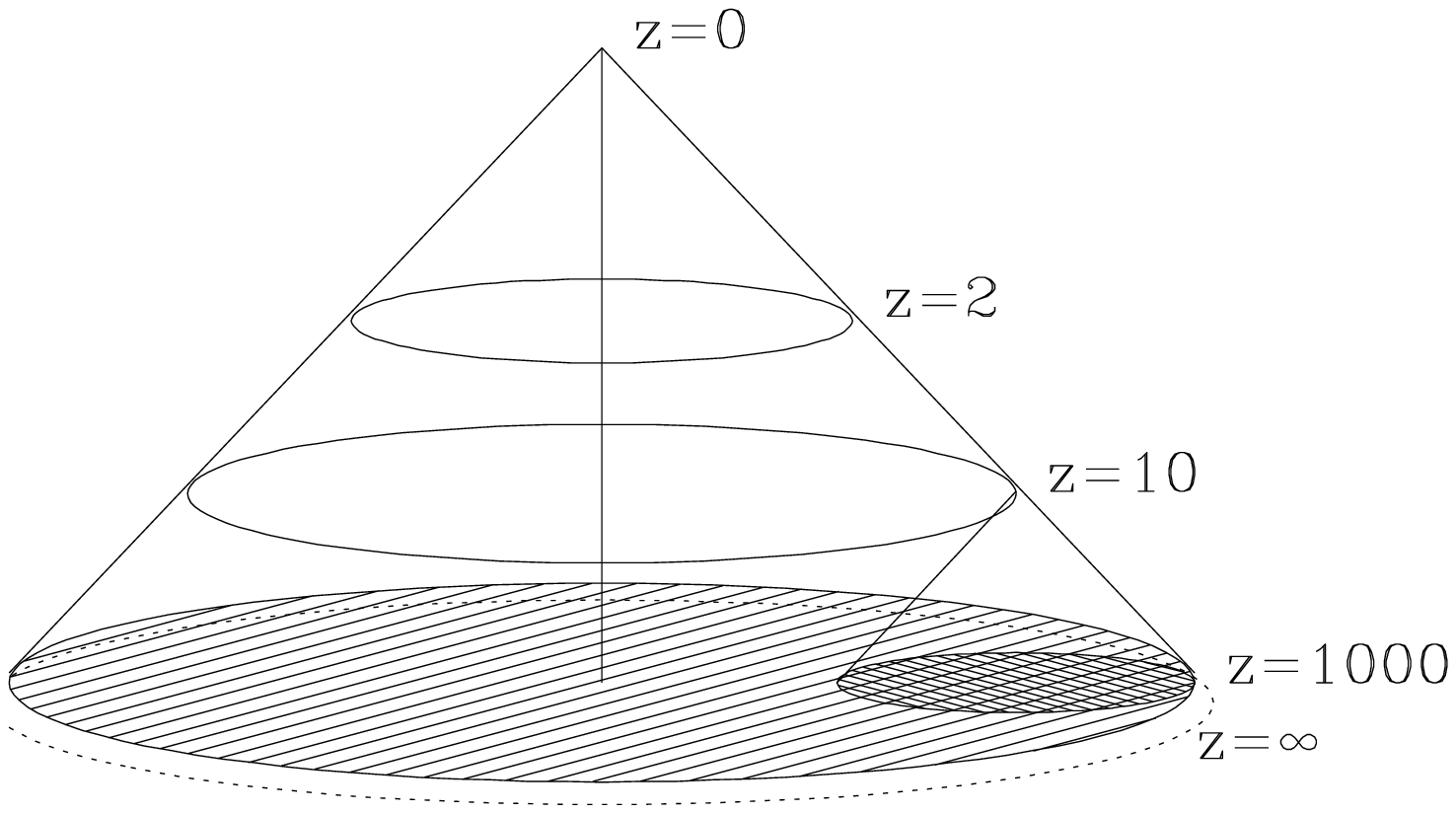,width=\fwidth,height=\fheight}
\vskip-1.5truecm 
\caption{Our backward light cone.}
\label{lightconeplot}

\mycaption{Our backward light cone in a flat $\Omega=1$ universe
is shown in comoving spatial coordinates,
with conformal time
on the vertical axis (one of the three spatial 
dimensions has been suppressed).
As can be seen, in these coordinates, light rays propagate in 
$45^{\circ}$ lines just as in Euclidean space. 
The horizontal circles are labeled with their corresponding redshifts,
the dotted circle corresponding to the Big Bang. 
If we (at the apex of the cone) detect a CBR photon arriving from the right
in the picture, it could have 
been anywhere in the shaded region at 
$z=10^3$, the recombination epoch. If we know that it has not been scattered
since $z=10$, it 
must have come from the double-hatched region.
}

\efig

\def\fheight{10.3cm} \def\fwidth{14.5cm}

\bfig
\psfig{figure=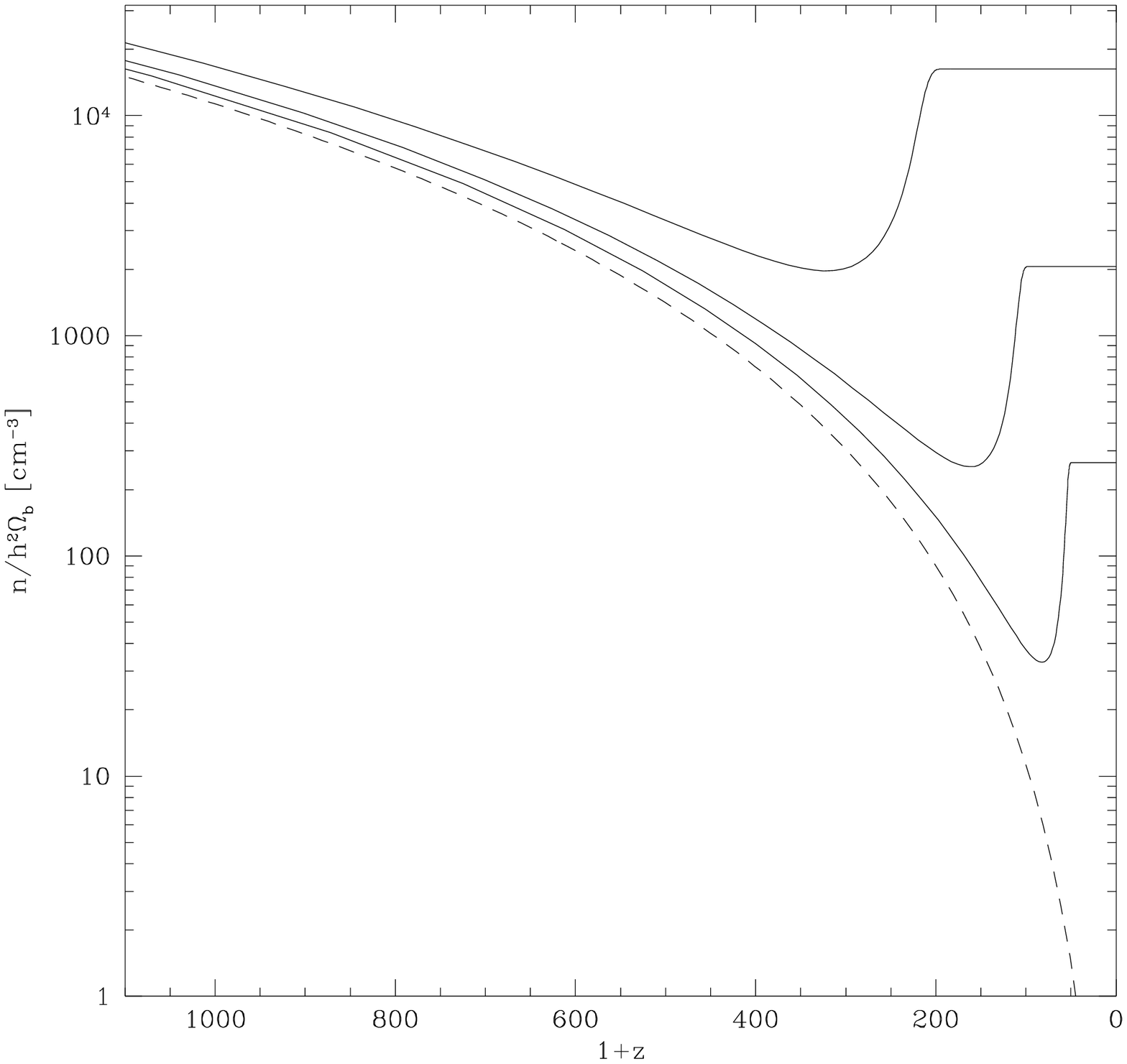,width=\fwidth,height=\fheight}
\nobreak
\caption{Evolution of a top hat overdensity.}
\label{znplot}
 
\mycaption{The time-evolution of the density in a spherical 
(``top hat") overdense region in a flat universe is 
plotted as a function of redshift $z$.
From top to bottom, the three solid curves correspond to virialization
redshifts $z_{vir}$ of 200, 100 and 50, respectively. 
The dashed curve gives the density of the IGM.
The number density is given in nucleons per unit volume, {\ie} $n=\rho/m_p$.
}
\efig

\bfig
\psfig{figure=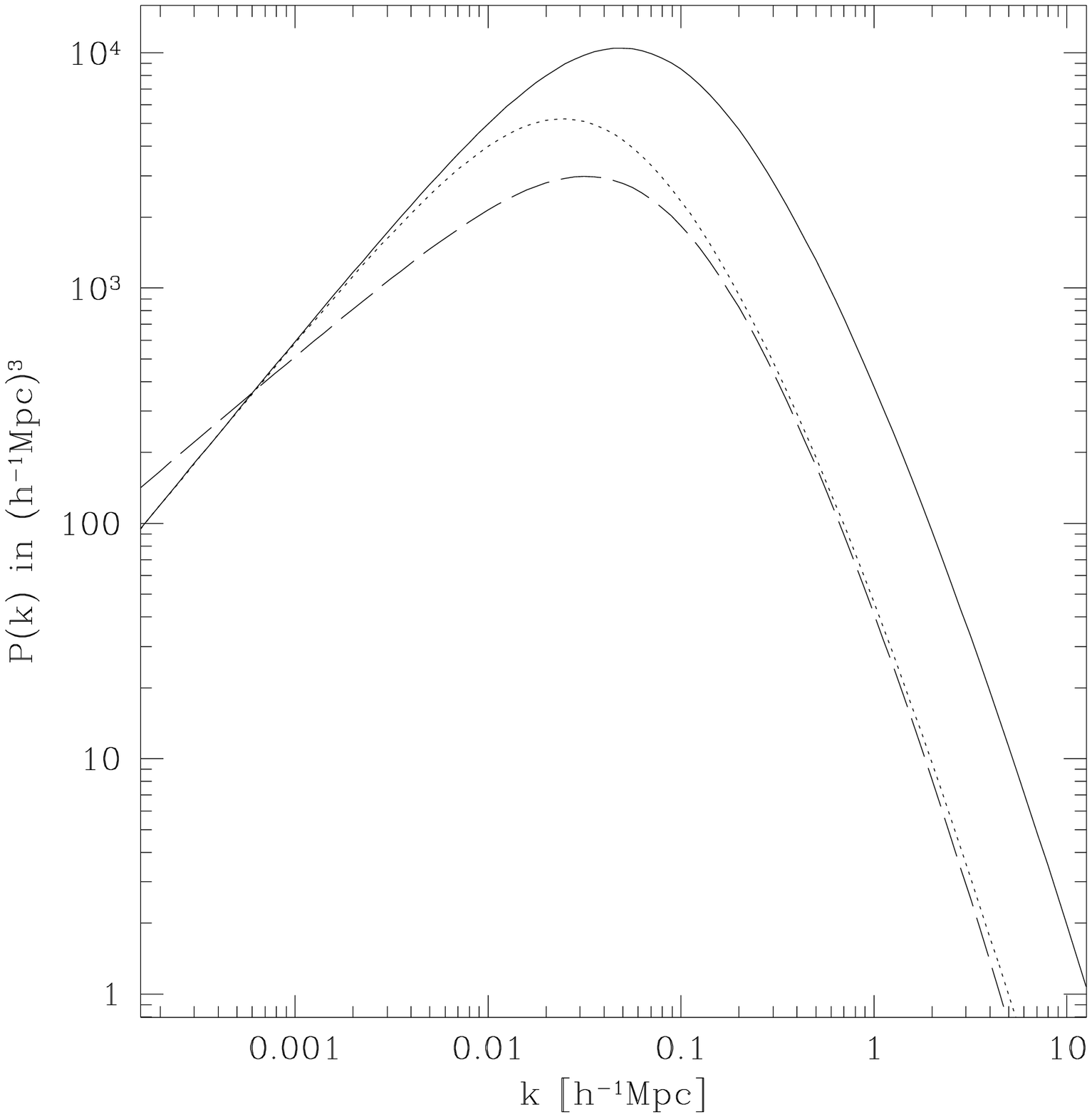,width=\fwidth,height=\fheight}
\nobreak
\caption{Assorted power spectra.}
\label{powerplot}
 
\mycaption{The power spectrum $P(k)$ is plotted
for a CDM model and two variations on the theme.
The solid curve is a standard CDM power spectrum with 
$h=0.5$, $\Oz=1$, and $n=1$ (Bond \& Efstathiou 1984). 
The dotted
curve has a lower ``shape parameter" $h\Oz=0.25$, which can be attained
in for instance the MDM and $\tau$CDM scenarios discussed in 
subsequent chapters, and corresponds to sliding the  
CDM curve to the left.
The dashed curve has a lower spectral index $n=0.7$, the so called
tilted model, which approximately corresponds to rotating the 
CDM curve clockwise.
All three power spectra have been normalized 
to COBE.
}
\efig

\bfig
\psfig{figure=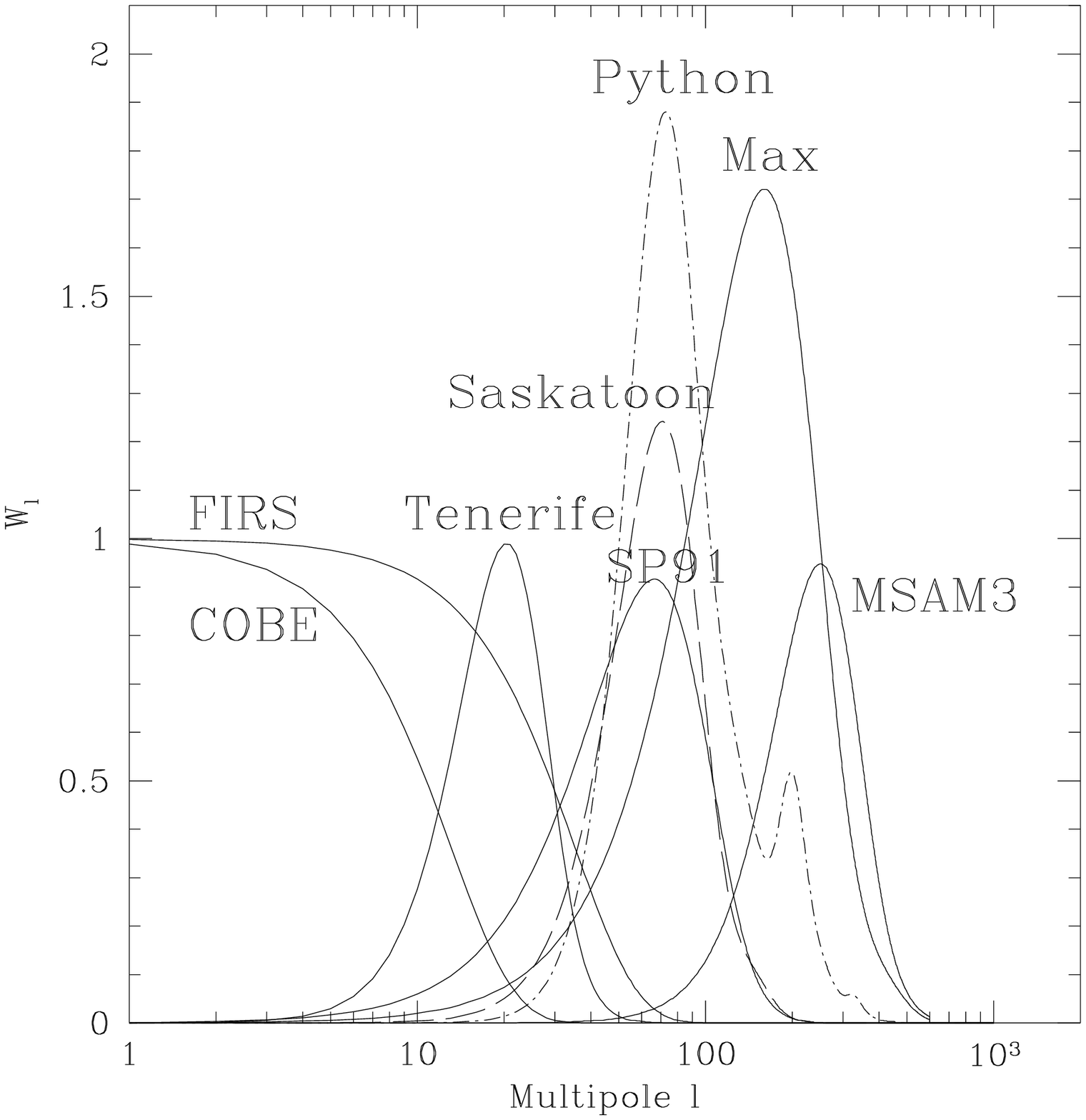,width=\fwidth,height=\fheight}
\nobreak
\caption{Assorted angular windows functions.}
\label{wlplot}
 
\mycaption{The angular window functions $W_l$
are plotted for a number of recent CBR experiments.
The data is courtesy of White, Scott \& Silk (1994).
}
\efig

\bfig
\psfig{figure=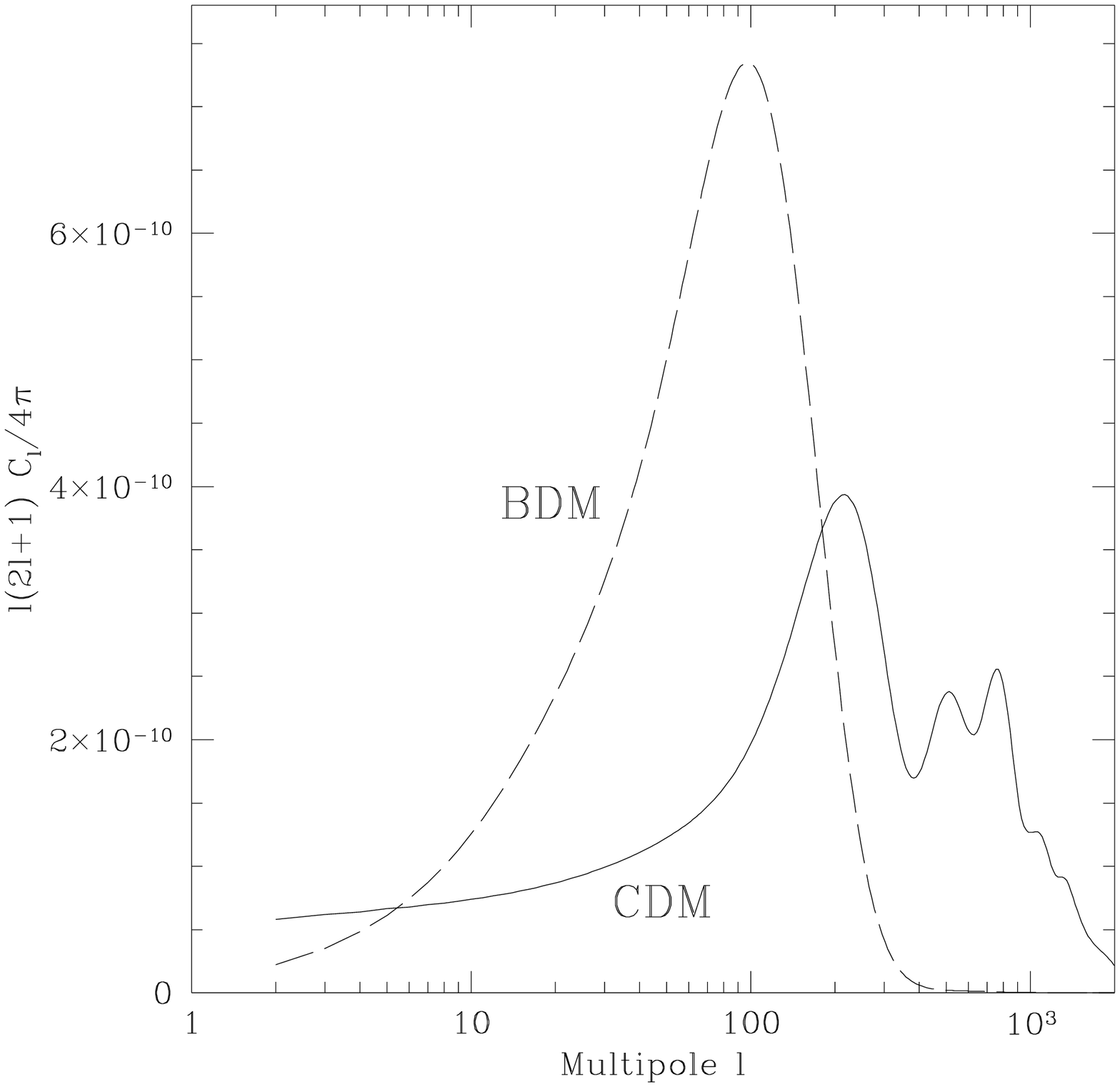,width=\fwidth,height=\fheight}
\nobreak
\caption{Assorted angular power spectra.}
\label{clplot}
 
\mycaption{The angular power spectrum $C_l$ is plotted
for a CDM and a BDM models.
The CDM model has $\Omega_0=1$, $\Ob=0.06$, $h=0.5$ and $n=1$.
The BDM model has $\Omega_0=\Ob=0.15$, $h=0.8$ and $n=-0.5$.
Both are normalized to COBE, {\ie} give the same value when integrated
against the window function for the COBE DMR $10^{\circ}$ pixel variance. 
The models are courtesy of Naoshi Sugiyama and Wayne Hu. 
}
\efig

%%%%%%%%%%%%%%%%%%%%%%%%%%%%%%%%%%%%%%%%%%%%%%%%%%%%%%%%%%%%%
% CBR FLUCTUATION SUPPRESSION:
%%%%%%%%%%%%%%%%%%%%%%%%%%%%%%%%%%%%%%%%%%%%%%%%%%%%%%%%%%%%%

\def\fheight{12.8cm} \def\fwidth{9.2cm}
\bfig
\psfig{figure=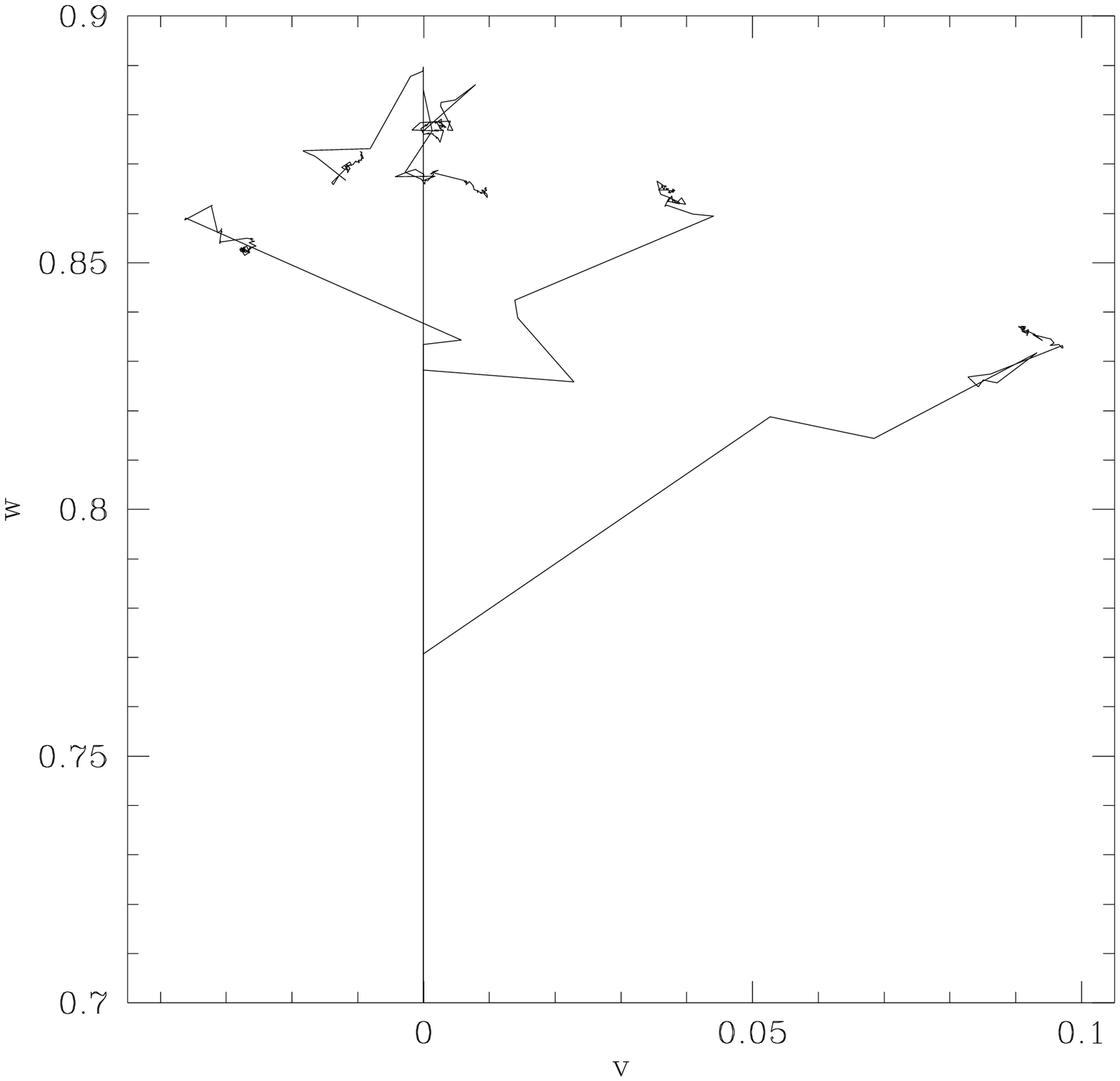,width=\fwidth,height=\fheight}
\nobreak
\caption{Monte Carlo photons, top view.}
\label{jesusplot}
 
\mycaption{The trajectories of seven photons in a Monte Carlo
simulation of a fully ionized universe
are shown in comoving coordinates $(u,v,w)$, with the
$u$-coordinate perpendicular to the page. 
(In terms of Figure~\ref{lightconeplot}, this is a view from ``above".)
The question being asked is where photons arriving from the w-direction
were in the past, so one can just as well model the photons as 
emanating from the apex of the light cone and traveling backwards in time.
With this terminology, they all enter this figure from below, 
and part ways as they Thomson scatter. Notice how their mean free
path gets shorter and shorter, which is due to the 
increase in electron density at higher redshifts.
}
\efig
\def\fheight{10.3cm} \def\fwidth{14.5cm}

\bfig
\psfig{figure=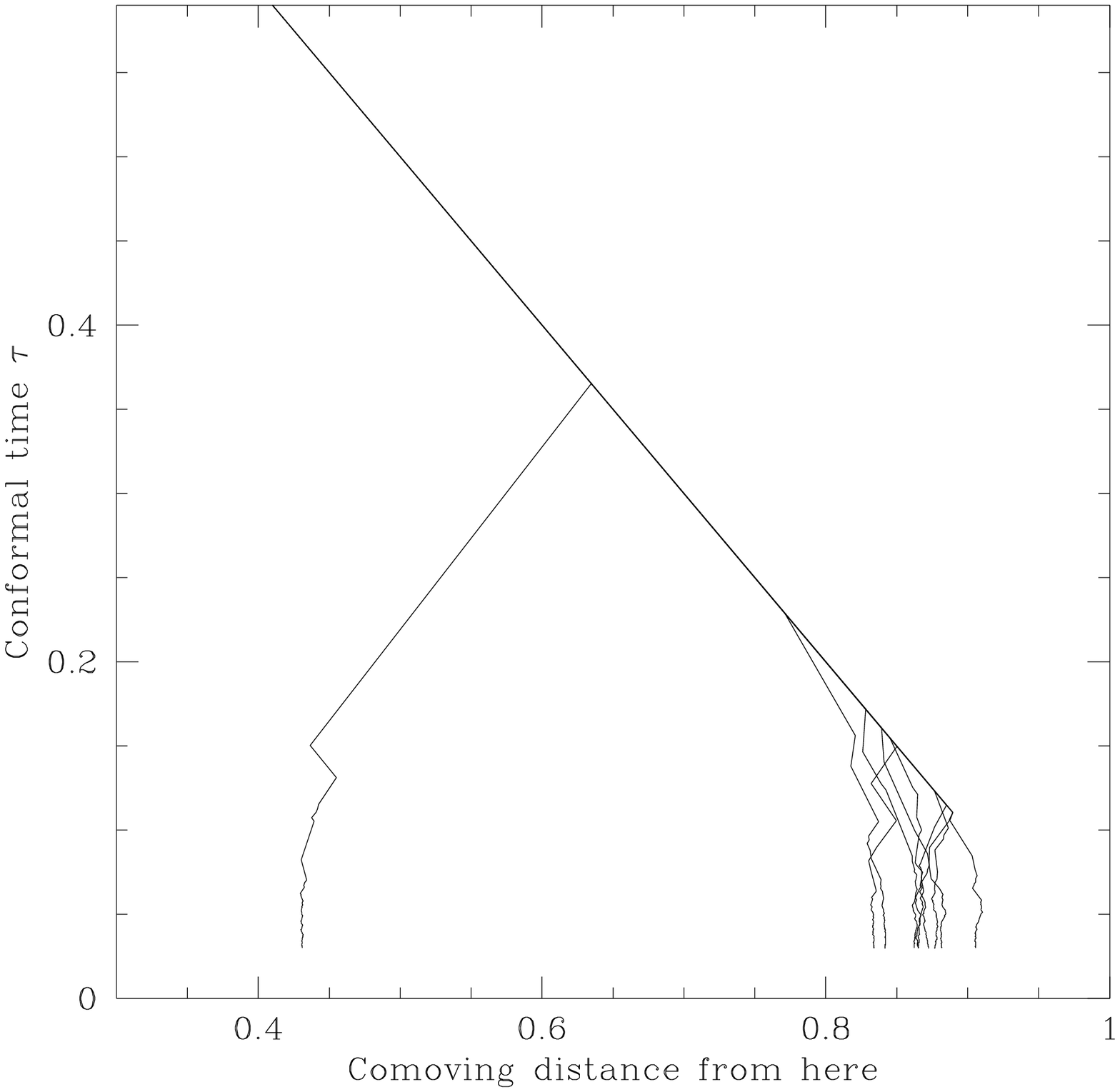,width=\fwidth,height=\fheight}
\caption{Monte Carlo photons, side view.}
\label{lightningplot}
 
\mycaption{The trajectories of ten photons in the Monte Carlo
simulation described in Figure~\ref{jesusplot} are plotted ``from 
the side" with reference to the light cone in
Figure~\ref{lightconeplot}.
The horizontal axis is $(u^2+v^2+w^2)^{1/2}$, the comoving
distance from 
``here" (the apex of the light cone).
The curves all end at $\tau\approx 0.3$, corresponding to 
$z\approx 10^3$, {\ie} standard recombination. 
The horizontal spread of these endpoints 
roughly corresponds to the smudging scale $\lambda_c$.
}

\efig

\bfig
\psfig{figure=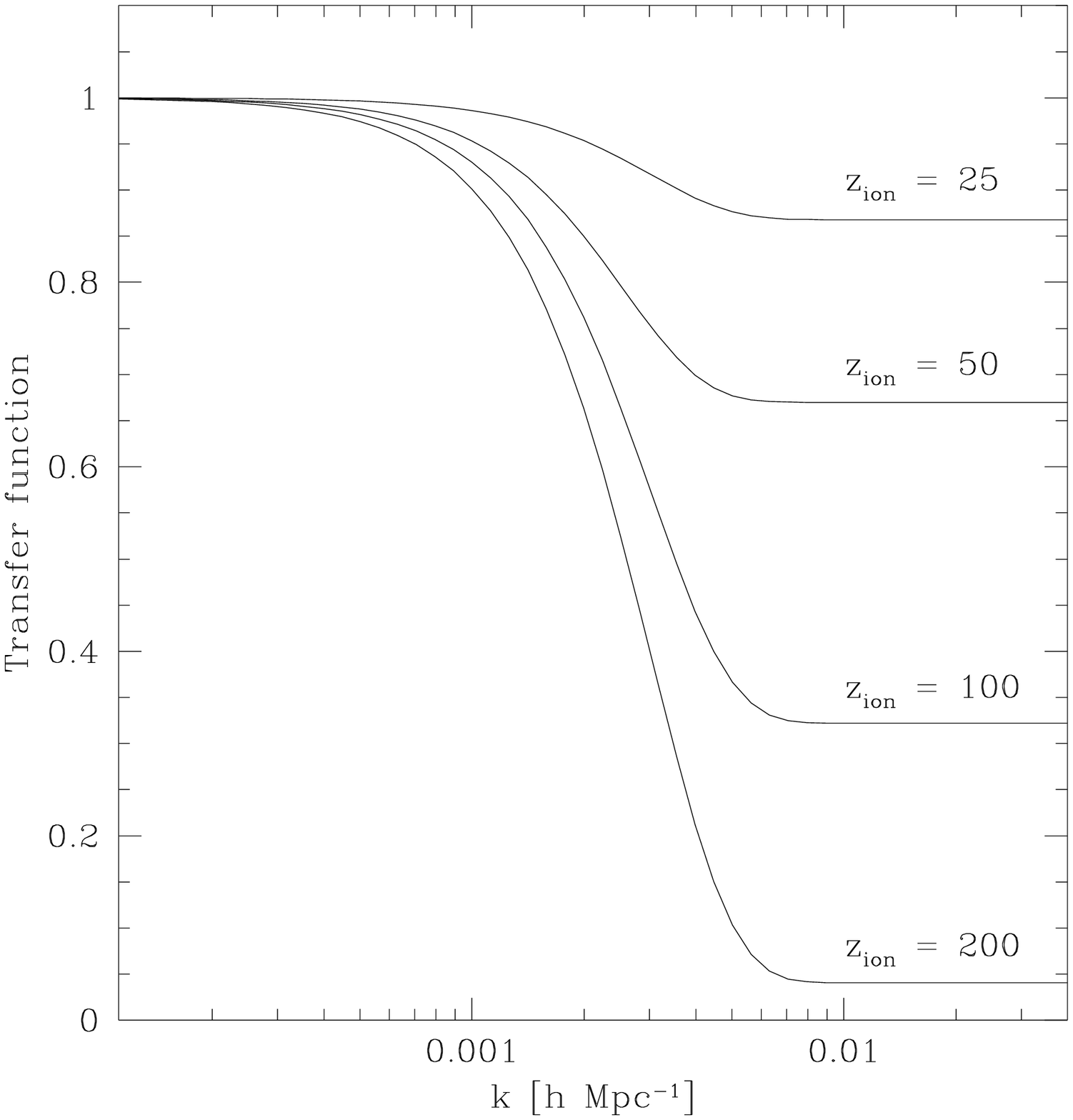,width=\fwidth,height=\fheight}
\nobreak
\caption{The transfer function $|\Gh|$.}
\label{suppressionplot}
 
\mycaption{The transfer function $|\Gh|$, {\ie} the factor by which 
the r.m.s. CBR fluctuations are suppressed by reionization, 
is plotted as a function of comoving wavenumber for scenarios 
where the universe becomes reionized at some redshift $z_{ion}$.
As $k\to\infty$, the suppression factor simply approaches $e^{-\taut}$.
For these scenarios, $h\Omega_{igm} = 0.03$, so 
$\taut\approx (z_{ion}/92)^{3/2}$.
The cutoff scales $\lambda_c$ where determined by Monte Carlo simulation,
and turn out to be depend only quite weakly on 
the ionization redshift $z_{ion}$. 
In all cases, degree-scale anisotropies
are seen to be suppressed whereas fluctuations on the scale probed by COBE are
almost unaffected. 
}
\efig

\cleardoublepage
\chapter{Early Reionization and CBR Fluctuations}
\label{reionchapter}

% On the Inevitability of Reionization:
% Implications for Cosmic Microwave Background Fluctuations

\def\Ob{\Omega_{igm}}
 
In this chapter, early photoionization of the 
intergalactic medium is discussed in a
nearly model-independent way, in order to 
investigate whether early structures
corresponding to rare Gaussian peaks in a CDM model can
photoionize the intergalactic medium sufficiently 
early to appreciably smooth out the microwave background fluctuations.
We conclude that this is indeed possible for a broad range
of CDM normalizations and is almost inevitable for unbiased CDM, 
provided that the
bulk of these early structures are quite small, no more massive than about
$10^8 M_{\odot}$. Typical parameter values predict that reionization
occurs around $z=50$, thereby suppressing fluctuations on degree scales
while leaving the larger angular scales probed by COBE 
relatively unaffected.
However, for non-standard CDM, incorporating mixed dark matter,
vacuum density or a tilted primordial power spectrum, early reionization
plays no significant role.

\section{Introduction}

The first quantitative predictions of cosmic microwave background
anisotropies in cold dark matter (CDM)-dominated cosmological models
recognized that reionization by rare, early-forming objects could play a
role in suppressing temperature fluctuations on small angular scales (Bond
\& Efstathiou 1984; Vittorio \& Silk 1984). Now that the COBE DMR
experiment has detected fluctuations on large angular scales
(Smoot {\etal} 1992) at a level
(within a factor of two) comparable to that predicted  by CDM models,
it is especially relevant to examine whether reionization can affect the
degree scale searches that are currently underway.

Cold dark matter models are generally characterized by a late epoch of
galaxy formation. However, the smallest and oldest objects first go
nonlinear at relatively large redshift. In this chapter we investigate,
for a wide range of CDM normalizations, power spectra and
efficiency parameters, whether reionization associated with energy
injection by early forming dwarf galaxies can reionize the universe
sufficiently early to smooth out primordial CBR temperature
fluctuations.  

Although we go into some detail in Appendix~\ref{fionappendix}
to make estimates of a certain efficiency parameter, our overall
treatment is fairly model-independent, and can be used as a
framework within which to compare various photoionization scenarios.
Our basic picture is roughly the following: 
An ever larger
fraction of the baryons in the universe falls into nonlinear structures
and forms galaxies. A certain fraction of these baryons form
stars or quasars which emit ultraviolet radiation, 
and some of this radiation escapes into the ambient 
intergalactic medium (IGM) and ends up
photoionizing and heating it. Due to cooling losses and
recombinations, the net number of ionizations per UV
photon is generally less than unity. 

Apart from photoionization, early galaxies can also ionize 
the IGM through
supernova driven winds, an ionization mechanism that will not be treated
in this chapter. Although such winds can ionize the IGM by $z=5$, early
enough to explain the absence of a Gunn-Peterson effect (Tegmark {\etal}
1993), the relatively low velocities of such winds makes them unable to
distribute the released energy throughout space at redshifts early enough
(by $z\approx 50$) to measurably affect the CBR.

Our approach will be to first write the ionization fraction of the
IGM as a product of a number of factors, and then discuss
the value of each of these factors in more detail. Let us
write
\beq{FirstFactorEq}
\v = \fs\fupp\fion,
\eeq
where
$$\cases{
\v&= fraction of IGM that is ionized,\cr
\fs&= fraction of baryons in nonlinear structures,\cr
\fupp&= UV photons emitted into IGM per proton in nonlinear structures,\cr
\fion&= net ionizations per emitted UV photon.
}$$
Let us first consider the case where the UV photons are produced by
stars,
and return to the quasar case later.
Using the fact that a
fraction $0.0073$ of the rest mass is released in stellar burning of
hydrogen to helium, we obtain
\beq{fuppEq}
\fupp \approx 0.0073\left({m_pc^2\over 13.6\eV}\right)
\fH\fmet\fuv\fesc,
\eeq
where
$$\cases{
\fH&= mass fraction hydrogen in IGM,\cr
\fmet&= mass fraction of hydrogen burnt,\cr
\fuv&= fraction of energy released as UV photons,\cr
\fesc&= fraction of UV photons that escape from galaxy.
}$$
We will take the primordial mass fraction of helium to be $24\%$, 
{\ie} $\fH=76\%$.
Now define the {\it net efficiency} 
$$\fnet = \fmet\>\fuv\>\fesc\>\fion,$$
and \eq{FirstFactorEq} 
becomes
\beq{SecondFactorEq}
\v\aet{3.8}{5}\>\fnet\>\fs.
\eeq
The key feature to note about this expression is that
since $3.8\times 10^5$ is such a large number, quite modest 
efficiencies $\fnet$ still allow $\v$ to become of order
unity as soon as a very small fraction of the
baryons are in galaxies. 
As will be seen in the next section, this means
that reionization is possible even at 
redshifts far out in the Gaussian tail of the
distribution of formation redshifts, at epochs long before
those when the bulk of the baryons go nonlinear. This appears to have
been first pointed out by Couchman and Rees (1986).

\section{The Mass Fraction in Galaxies}

In this section, we will discuss the parameter $\fs$.
Assuming the standard PS theory of structure
formation (Press \& Schechter 1974), the fraction of all mass that has
formed gravitationally bound objects of total (baryonic
and non-baryonic) mass greater than $M$ at redshift $z$
is the integral of the Gaussian tail, 
\beq{3fgEq}
\fs = \erfc\left[{\delta_c\over\sqrt{2}\sigma(M,z)}\right],
\eeq
where the complementary error function  $\erfc(x)\equiv
2\pi^{-1/2}\int_x^{\infty} e^{-u^2}du$ and $\sigma(M,z)$ is
the r.m.s. mass fluctuation in a sphere containing an
expected mass $M$ at redshift $z$.
$\sigma^2$ is given by top-hat filtering of the power
spectrum as  
\beq{3FilterEq}
\sigma(M,z)^2 \propto
\izi P(k) 
\left[{\sin kr_0\over(kr_0)^3} - {\cos
kr_0\over(kr_0)^2}\right]^2 dk,
\eeq
where $P(k)$ is the power spectrum at redshift $z$ and 
$r_0$ is
given by ${4\over 3}\pi r_0^3\rho = M$, 
$\rho = {3H^2\Omega\over 8\pi G}$ being
the density of the universe at redshift $z$. Although this approach
has been criticized as too simplistic, numerical
simulations (Efstathiou {\etal} 1988; Efstathiou \&
Rees 1988; Carlberg \& Couchman 1989) have shown that it
describes the mass distribution of newly formed structures
remarkably well. Making the standard assumption of a
Gaussian density field, Blanchard {\it et al.} (1992) have
argued that it is an accurate description at least in the
low mass limit. Since we are mainly interested in
extremely low masses such as $10^6\Ms$, it appears to
suffice for our purposes.

For our middle-of-the road estimate, we choose $\delta_c = 1.69$,
which is the linearly extrapolated overdensity at which a spherically
symmetric perturbation has collapsed into a virialized
object (Gott \& Rees 1975; Efstathiou {\etal} 1988; 
Brainerd \& Villumsen 1992).
We take
$\delta_c = 1.44$ (Carlberg \& Couchman 1989) for the optimistic
estimate, although the even lower value
$\delta_c = 1.33$ has been discussed (Efstathiou \& Rees 1988),
and $\delta_c = 2.00$ (Gelb \&
Bertschinger 1992) for the pessimistic estimate.
(Here and
throughout this chapter, parameter choices are referred to as optimistic
if they permit earlier reionization.)

The fact that $\sigma(M,z)\to\infty$ as $M\to 0$ implies
that if we consider arbitrarily small scales, then all
dark matter is in non-linear structures. Thus if
no forces other than gravity were at work, so that
the baryons always followed the dark matter, we would
simply have $\fs=1$ at all $z$. 
However, it is
commonly believed that galaxies correspond only to
objects that are able to cool (and fragment into stars) in
a dynamical time or a Hubble time (Binney 1977; Rees \&
Ostriker 1977; Silk 1977; White \& Rees 1978).
The former applies to ellipticals and bulges, the latter to disks. 
Let us define the {\it virialization redshift} $(1+\zvir)\equiv
(\sqrt{2}/\delta_c)\szmc$,
where $M_c$ is some characteristic cutoff mass which is
the total mass (baryonic and dark) of the first galaxies
to form. $\zvir$ is roughly the
redshift at which the bulk of all baryons 
goes non-linear. 
Using \eq{3fgEq} and the fact that 
$\sigma(M,z) = \sigma(M,0)/(1+z)$ in the linear regime of CDM, 
we thus have 
\beq{fsEq}
\fs = \erfc\left[{1+z\over 1+\zvir}\right].
\eeq
A common assumption is that $M_c\approx
10^6\Ms$, roughly the Jeans mass at recombination. 
Blanchard {\etal} (1992) examine the interplay
between cooling and gravitational collapse in considerable
detail, and conclude that the first galaxies to form have
masses in the range $10^7\Ms$ to  $10^8\Ms$, their redshift
distribution still being given by \eq{fsEq},
whereas Couchman \& Rees (1986) argue that the first
galaxies to form may have had masses as low as $10^5\Ms$. 

As our CDM power spectrum today, we will use that given by BBKS
(Bardeen {\etal} 1986) 
and an $n=1$ Harrison-Zel'dovich primordial spectrum:
$$P(k) \propto 
\left({q^{-1} \ln(1+2.34q)\over
\left[1+3.89q + (16.1q)^2 + (5.46q)^3
+ (6.71q)^4\right]^{1/4}}\right)^2 q,$$
where $q\equiv k/[h^2\Omega_0\Mpc^{-1}]$. Throughout this chapter, we
will take $\Omega_0 = 1$. 

Evaluating the $\sigma^2$-integral in
\eq{3FilterEq} numerically yields 
$$\sigma(10^5\Ms,0)\approx 33.7b^{-1}$$
for $h=0.8$ and 
$$\sigma(10^8\Ms,0)\approx 13.6b^{-1}$$
 for $h=0.5$, where the so
called bias factor $b\equiv\sigma(8h^{-1}\Mpc,0)^{-1}$ has
been estimated to lie between 0.8 (Smoot {\etal} 1992) and 
2.5 (Bardeen {\etal} 1986).
Our pessimistic, middle-of-the-road and optimistic CDM
estimates of $\zvir$ are given in 
Table~\ref{reiontable1}, 
and the
dependence of $\zvir$ on $M_c$ is plotted in 
Figure~\ref{reionfig1}.
This figure  
also shows three alternative models of structure
formation:  CDM with cosmological constant (Efstathiou {\etal} 1992);
tilted CDM (Cen {\etal} 1992) and 
MDM, mixed hot and dark matter
(Shafi
\& Stecker 1984; Schaefer \& Shafi 1992; Davis {\it et al.} 1992;
Klypin {\etal} 1993). For the model with cosmological constant,
we have taken a flat universe with $h=0.5$, $\Omega_0=0.4$ and
$\lambda_0=0.6$. For the tilted model, the power spectrum $P(k)$ is simply
multiplied by a factor $k^{n-1}$, where we have taken $n=0.7$.
For the tilted case, 
\eq{fsEq} still applies. 
For the MDM case,
however, perturbations in the cold component
grow slower than linearly with the scale factor
$(1+z)^{-1}$ and \eq{fsEq} is not valid. 
For the low masses we are considering, we have (Bond \& Szalay 1983) 
$$\sigma_{MDM}(M_c,z) \approx {\sigma_{MDM}(M_c,0)\over (1+z)^{\alpha}},
\quad\hbox{where}$$ 
$$\alpha \equiv {1\over 4}\left[\sqrt{25-24\Omega_{HDM}} - 1\right].$$
Using the parameters from Davis {\etal} (1992), who take
$\Omega=1$ and $\Omega_{HDM}=0.3$, the MDM version of 
\eq{fsEq} becomes 
$$\fs = 
\erfc\left[\left({1+z\over 1+\zvir}\right)^{\alpha}\right]$$
where $\alpha\approx 0.8$ and we redefine $\zvir$ by 
$$(1+\zvir) \approx \left[{\sqrt{2}\,\szmc\over 5.6 \delta_c}\right]^{1/\alpha},$$
with $\szmc$ referring to a pure CDM power spectrum.

For the $\Lambda$ case, perturbations grow approximately linearly until
the universe becomes vacuum dominated at $z \approx \Omega_0^{-1} - 1
= 1.5$, after which their growth slowly grinds to a halt. A numerical
integration of the Friedmann equation and the equation for perturbation 
growth using $h=0.5$, $\Omega_0 = 0.4$ and $\lambda_0 = 0.6$ gives 
$$\sigma_{\Lambda}(M_c,z) \approx 
1.2 {\sigma(M_c,0)\over (1+z)}$$
for $z\gg 3$.

\begin{table}
$$
\begin{tabular}{|l|cccccc|}
\hline
                   & Mixed & Tilted & Lambda & Pess. & Mid. & Opt.\\
\hline
$M_c$&$10^6\Ms$&$10^6\Ms$&$10^6\Ms$&$10^8\Ms$&$10^6\Ms$&$10^5\Ms$\\
Model&MDM&Tilted&Lambda&CDM&CDM&CDM\\
$h$&0.5&0.5&0.5&0.5&0.5&0.8\\
$b$&1&1&1&2&1&0.8\\
$\delta_c$&1.69&1.69&1.69&2.00&1.69&1.44\\
\hline
$\zvir$&2.9&10.1&8.4&4.8&17.2&41.4\\
\hline
\end{tabular}
$$
\vskip-0.3cm
\caption{Galaxy formation assumptions}
\vskip0.3cm
\label{reiontable1}
\end{table}

Since our $\Lambda$-model yields a value of 
$z_{vir}$ very similar to our tilted model, 
we will omit the former from future plots.

\section{Efficiency Parameters}
\label{reionsec3}

In this section, we will discuss the various parameters that give
$\fnet$ when multiplied together. The conclusions are summarized in 
Table~\ref{reiontable2}. 

\begin{table}
$$
\begin{tabular}{|l|rrr|}
\hline
                   & Pess. & Mid. & Opt.\\
\hline
$\fmet$&0.2\%&1\%&25\%\\
$\fesc$&10\%&20\%&50\%\\
$\fuv$&5\%&25\%&50\%\\
$\fion$&10\%&40\%&95\%\\
\hline
$\fnet$&$1\times 10^{-6}$&$2\times 10^{-4}$&$6\times 10^{-2}$\\
\hline
$\fupp$&4&190&24,000\\
\hline
\end{tabular}
$$
\vskip-0.3cm
\caption{Efficiency parameters used}
\vskip0.2cm
\label{reiontable2}
\end{table}

$\fmet$, the fraction of galactic hydrogen that is
burnt into helium during the early life of the galaxy
(within a small fraction of a Hubble after formation),
is essentially the galactic metallicity after the first wave of star
formation. 
Thus it is the product of the fraction of the hydrogen that forms
stars and the average metallicity per star (weighted by mass).
This depends on the stellar mass
function, the galactic star formation rate and the final metallicities
of the high-mass stars.  For our middle-of-the-road estimate, we follow
Miralda-Escud\'e \& Ostriker (1990) in taking $\fmet = 1\%$, half the
solar value.
An upper limit to $\fmet$ is obtained from the extreme scenario where
all the baryons in the galaxy form very massive and short-lived stars
with $M\approx 30\Ms$, whose metallicity could get as high as 25\%
(Woosley \& Weaver 1986). Although perhaps unrealistic, this is not ruled
out by the apparent absence of stars with such metallicities today, since
stars that massive would be expected to collapse into black holes.

In estimating $f_{esc}$, the fraction of the UV photons that despite 
gas and dust manage to escape from the galaxy where they are
created, we follow Miralda-Escud\'e \& Ostriker (1990). 

For $f_{uv}$, the fraction of the released energy that is radiated above
the Lyman limit, we also follow Miralda-Escud\'e \& Ostriker (1990). The
upper limit refers to the extreme $30\Ms$ scenario mentioned above. For
reference, the values of $\fuv$ for stars with various spectra are given in
Table~\ref{reiontable3},
together with some other spectral parameters that will be 
defined and used in
Appendix~\ref{fionappendix}. 
All these parameters involve spectral integrals, and have
been computed numerically.

The parameter $\fion$ is estimated in 
Appendix~\ref{fionappendix}. 

\begin{table}
$$
\begin{tabular}{|llrrrr|}
\hline
UV source&Spectrum $P(\nu)$&$\fuv$&$\euv$&$\Tpi$&$\uvsigfid$\\
\hline
O3 star&$T=50,000$K Planck&0.57&17.3\,eV&28,300K&2.9\\
O6 star&$T=40,000$K Planck&0.41&16.6\,eV&23,400K&3.4\\
O9 star&$T=30,000$K Planck&0.21&15.9\,eV&18,000K&3.9\\ 
Pop. III star&$T=50,000$K Vacca&0.56&18.4\,eV&36,900K&2.2\\ 
Black hole, QSO&$\alpha=1$ power law&&18.4\,eV&37,400K&1.7\\
?&$\alpha=2$ power law&&17.2\,eV&27,800K&2.7\\
?&$\alpha=0$ power law&&20.9\,eV&56,300K&0.6\\
?&$T=100,000$K Planck&0.89&19.9\,eV&49,000K&1.6\\
\hline
\end{tabular}
$$
\caption{Spectral parameters}
\label{reiontable3}
\end{table}

An altogether different mechanism for converting the baryons in
nonlinear structures into ultraviolet photons is black
hole accretion. If this mechanism is the dominant one, 
\eq{fuppEq} should be replaced by 
$$\fupp \approx
\left({m_pc^2\over 13.6\eV}\right) \fbh\facc\fuv\fesc,$$
where
$$\cases{
\fbh&= mass fraction of nonlinear structures that end up as black
holes,\cr 
\facc&= fraction of rest energy radiated away during accretion
process,\cr  
\fuv&= fraction of energy released as UV photons,\cr
\fesc&= fraction of UV photons that escape from host galaxy.
}$$
There is obviously a huge uncertainty in the factor $\fbh$. 
However, the absence of the factor $0.0073\times 0.76$ compared to
\eq{fuppEq} means that the conversion of matter into 
radiation is so much more efficient 
that the  black hole contribution might be important even
if $\fbh$ is quite small. For instance, $\facc=10\%$ and 
$\fesc=100\%$ gives $\fupp\approx 10^8\fbh\fuv$, which could easily
exceed the optimistic value  $\fupp\approx 24,000$ for the stellar
burning mechanism in 
Table~\ref{reiontable2}. 

In 
Figure~\ref{reionfig2},
the ionization fraction 
$\v(z)$ is plotted for various parameter values using
equations\eqnum{FirstFactorEq} and\eqnum{fsEq}.
It is seen that the ionization grows quite abruptly, so that we may speak
of a fairly well-defined {\it ionization redshift}.
Let us define the ionization redshift $\zion$ as the redshift when $\v$
becomes 0.5, {\ie}
\beq{zionEq}
1+\zion = (1+\zvir) \erfc^{-1}\left({1\over 2\fupp\fion}\right).
\eeq
This dependence of $\zion$ of the efficiency is shown in 
Figure~\ref{reionfig3}
for our various galaxy formation scenarios.
It is seen that the ionization redshift is fairly insensitive to the net
efficiency, with the dependence being roughly logarithmic for
$\fnet>0.0001$.

\section{Scattering History}

For a given ionization history $\v(z)$, the Thomson 
opacity out to a redshift $z$, the probability that a CBR
photon is Thomson scattered at least once after $z$, is 
$$\ps(z) = 1 - e^{-\tau(z)},$$ 
where the
optical depth for Thomson scattering is given by
$$\cases{
\tau(z)& = $\tau^*\izz{1+z'\over\sqrt{1+\Omega_0 z'}}\v(z')dz',$\crr
\tau^*& = ${3\Ob\over 8\pi}
\left[1 - \left(1-{\v_{He}\over 4\v}\right)f_{He}\right] 
{H_0c\st\over m_p G} \approx 0.057 h\Ob,$
}$$
where we have taken the mass fraction of helium to be 
$f_{He}\approx 24\%$ and assumed $x_{He} \approx x$,
{\ie} that helium never becomes doubly ionized and that the fraction
that is singly ionized equals the fraction of hydrogen that is
ionized. The latter is a very crude approximation, but makes a
difference of only $6\%$. 
We assume that $\Omega_0=1$ throughout this
chapter. 
$\Ob$ denotes the density of the intergalactic medium divided 
by the critical density, and is usually assumed to equal
$\Omega_b$, the corresponding density of baryons. 
the probability that a CBR
photon is Thomson scattered at least once after the standard
recombination epoch at $z\approx 10^3$.

The profile of the last scattering surface is given by the so called
visibility function 
$$f_z(z)\equiv {d\ps\over
dz}(z),$$ which is the probability distribution for the redshift 
at which a photon last scattered.
An illuminating special case is that of complete ionization at all
times, {\ie} $\v(z) = 1$, which yields
\beq{PsEq}
\ps(z) =  1-\exp\left(-{2\over
3}\tau^*\left[(1+z)^{3/2} - 1\right]\right) 
\approx 
1-\exp\left[-\left(z\over 92\right)^{3/2}\right]
\eeq
for $z \gg 1$ and $h\Ob=0.03$. 
Hence we see that in order for any
significant fraction of the CBR to have been rescattered
by reionization, the reionization must have occurred quite
early. 
Figures~\ref{reionfig4a} and~\ref{reionfig4b} 
show the opacity and last-scattering
profile for three different choices of $h \Ob$. 
In the optimistic case
$h \Ob = 0.1$, it is seen that even as low an ionization redshift as
$\zion=30$ would give a total opacity $P_s \approx 50\%$.
In 
Figures~\ref{reionfig5a} and~\ref{reionfig5b}, 
we have
replaced $z$ by the angle subtended by the horizon
radius at that redshift, 
$$\theta(z) = 2\arctan\left[{1\over
2\left(\sqrt{1+z}-1\right)}\right],$$
which is the largest angular scale
on which Thomson scattering at $z$ would affect the microwave
background radiation. In 
Figure~\ref{reionfig5b},
we have plotted the angular visibility
function $dP_s/d(-\theta)$ instead of  $dP_s/dz$, so that the curves are
probability distributions over angle instead of redshift.

In the {\it sudden approximation}, the ionization history is a step 
function
$$\v(z) = \theta(\zion - z)$$
for some constant $\zion$, and as was discussed in 
Section~\ref{reionsec3}, 
this
models the actual ionization history fairly well.
In this approximation, the visibility functions are identical to those
in 
Figures~\ref{reionfig4b} and~\ref{reionfig5b} 
for $z<\zion$, but vanish between $\zion$ and
the recombination epoch at $z\approx 10^3$.
Figure~\ref{reionfig6},
which is in a sense the most important 
plot in this chapter, shows the 
total opacity $P_s(\zion)$ as a function of $\fnet$ for
a variety of parameter values, as obtained by substituting 
\eq{zionEq} into\eqnum{PsEq}.
As can be seen, the resulting opacity is relatively insensitive to the
poorly known parameter $\fnet$, and depends mainly on the structure
formation model ({\ie} $\zvir$) and the cosmological parameter
$h\Ob$.

Thomson scattering between CBR photons and free electrons affects not
only the spatial but also the spectral properties of the CBR.
It has long been known that hot ionized IGM
causes spectral distortions to the CBR, known as the 
Sunyaev-Zel'dovich
effect. A useful measure of this distortion is the  Comptonization
$y$-parameter (Kompan\'eets 1957; Zel'dovich \& Sunyaev 1969; 
Stebbins \& Silk 1986; Bartlett \& Stebbins 1991)
$$y_c = \int\ktmc n_e\st c\> dt,$$ 
where the integral is to be taken from the reionization epoch to today.  
Let us estimate this integral by making the approximation that the IGM 
is cold and neutral until a redshift $z_{ion}$, at which it suddenly 
becomes ionized, and after which it 
is remains completely ionized with a constant temperature $T$.
Then for $\Omega=1$, $\zion\gg 1$, we obtain 
$$y_c = \ktmc\left({n_{e0}\st c\over H_0}\right)
\int_0^{\zion}\sqrt{1+z}dz \aet{6.4}{-8} h\Ob T_4\> z_{ion}^{3/2},$$
where $T_4\equiv T/10^4\K$ and $n_{e0}$, the electron density today, 
has been computed as before assuming a helium mass fraction of $24\%$
that is singly ionized. 
Substituting the most recent observational constraint 
from the COBE FIRAS experiment,
$y_c < 2.5\times 10^{-5}$ (Mather {\etal} 1994),
into this expression yields
$$\zion < 554T_4^{-2/3} \left({h\Ob\over 0.03}\right)^{-2/3},$$
so all our scenarios are consistent with this spectral constraint.

\section{Discussion}

A detailed discussion of how reionization affects the
microwave background anisotropies would be beyond the scope of this
chapter, so we will merely review the main features.
If the microwave background photons are rescattered at a redshift $z$,
then the fluctuations we observe today will be suppressed on angular
scales smaller than the angle subtended by the horizon at that redshift. 
This effect is seen in numerical integrations of the linearized Boltzmann
equation ({\eg} Bond \& Efstathiou 1984; Vittorio \& Silk 1984), and
can be simply understood in purely geometrical terms. 
Suppose we detect a microwave photon arriving from some direction in
space. Where was it just after recombination? In the absence of
reionization, it would have been precisely where it appears to be coming
from, say $3000$ Mpc away. If the IGM was reionized, however, the photon
might have originated somewhere else, scattered off of a free electron and
then started propagating towards us, so at recombination it might even
have been right here. Thus to obtain the observed anisotropy, we have to
convolve the anisotropies at last scattering with a window function
that incorporates this smoothing effect. Typical widths for the window
function appropriate to the last scattering surface range from a few
arc-minutes with standard recombination to the value of a few degrees
that we have derived here for early reionization models. 

In addition to
this suppression on sub-horizon scales, new fluctuations will be
generated by the first order Doppler effect and by the Vishniac
effect. The latter dominates on small
angular scales and is not included in the linearized Boltzmann
treatment because it is a second order effect.
The current upper limit on CBR fluctuations on the 1 arcminute scale
of $\Delta T/T < 9\tt{-6}$ (Subrahmanyan {\etal} 1993) provides
an interesting constraint on reionization histories through the
Vishniac effect.
In fact, according to the original calculations (Vishniac 1987), 
this would rule out most of the reionization histories in this chapter.
However, a more careful treatment (Hu {\etal} 1994) predicts a Vishniac
effect a factor of five smaller on this angular scale, so all
reionization histories in this chapter are still permitted.

The COBE DMR detection of $\Delta T/T$ has provided a normalization for
predicting CBR anisotropies on degree scales. Several experiments are
underway to measure such anisotropies, and early results that report
possible detections have recently become available from experiments at
the South Pole (Meinhold \& Lubin 1991; Shuster {\etal} 1993) and at
balloon altitudes (Devlin {\etal} 1992; Meinhold {\etal} 1993;
Shuster {\etal} 1993). There is some reason to believe that these detected
signals are contaminated by galactic emission. Were this the case, the
inferred CBR upper limits to fluctuations on degree scales might be
lower than those predicted from COBE extrapolations that adopt the
scale-invariant power spectrum that is consistent with the DMR result and
is generally believed to be the most appropriate choice on large scales
from theoretical considerations 
({\eg} Gorski {\etal} 1993; Kashlinsky 1992). 
In the absence of such contaminations, the
detected fluctuations in at least some degree-scale experiments are,
however, consistent with the COBE extrapolation ({\eg} Jubas \&
Dodelson 1993). The variation from field to field, repeated on
degree scales, also may argue either for galactic contamination or else for
unknown experimental systematics, or even non-Gaussian fluctuations.
The results of other recent experiments 
such as ARGO (de Bernardis {\etal} 1993),
PYTHON (Dragovan {\etal} 1993) and MSAM (Cheng {\etal} 1993) have reinforced the 
impression that the experimental data is not entirely self-consistent, and 
that some form of systematic errors may be important.

The controversy over the interpretation of the degree-scale CBR
fluctuations makes our reanalysis of the last scattering surface
particularly timely. We have found that canonical dark matter, tailored
to provide the 10 degree CBR fluctuations detected by the COBE DMR
experiment,
results in sufficiently early reionization (before $z \approx 50$)
over a fairly wide range of parameter space, to smooth out
primordial degree-scale  fluctuations. Our middle-of-the-road model
produces suppression by roughly a factor of two; it is difficult,
although not impossible, to obtain a much larger suppression.
This smoothing, because it is of order unity in scattering optical
depth, is necessarily inhomogeneous. We predict the presence of
regions with large fluctuations and many ``hot spots" and ``cold spots", 
corresponding to ``holes" in the last-scattering
surface, as well as regions with little small-scale power where the last 
scattering is more efficient. The
detailed structure of the CBR sky in models with reionization
will be left for future studies.
Here we
simply conclude by emphasizing that anomalously low values of $\Delta
T/T$ over degree scales are a natural corollary of reionization at
high redshift.

\def\fheight{10.3cm} \def\fwidth{14.5cm}

\newpage

\bfig
\psfig{figure=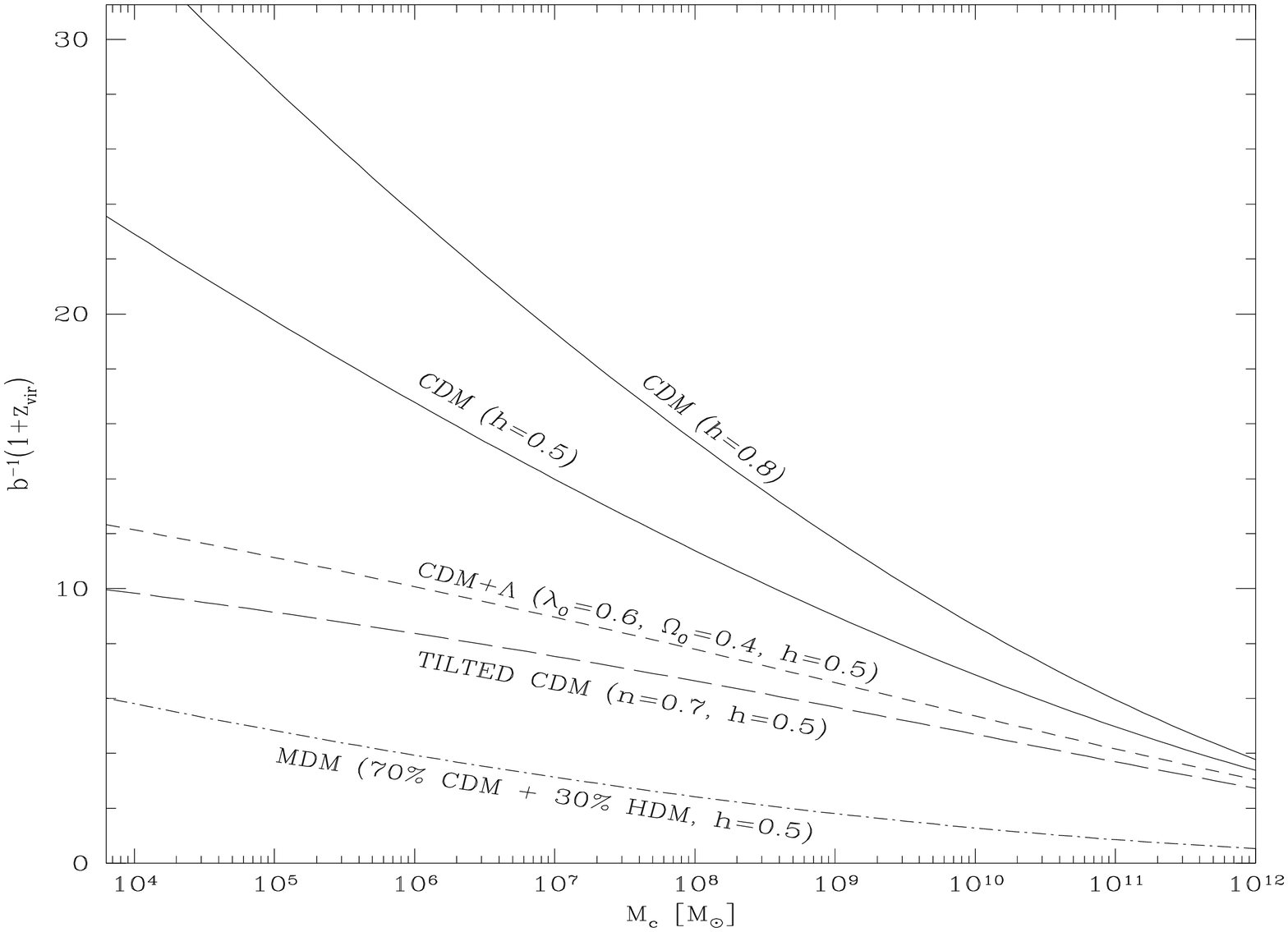,width=\fwidth,height=\fheight}
\caption{Virialization redshifts for objects of various masses.}
\label{reionfig1}
 
\mycaption{The virialization redshift, 
the redshift at which the bulk of the 
objects of mass $M_c$ form, is plotted for a number of cosmological 
models. In all cases shown, $\Omega_0+\lambda_0=1$ and $\delta_c = 1.69$.
}
\efig

\bfig
\psfig{figure=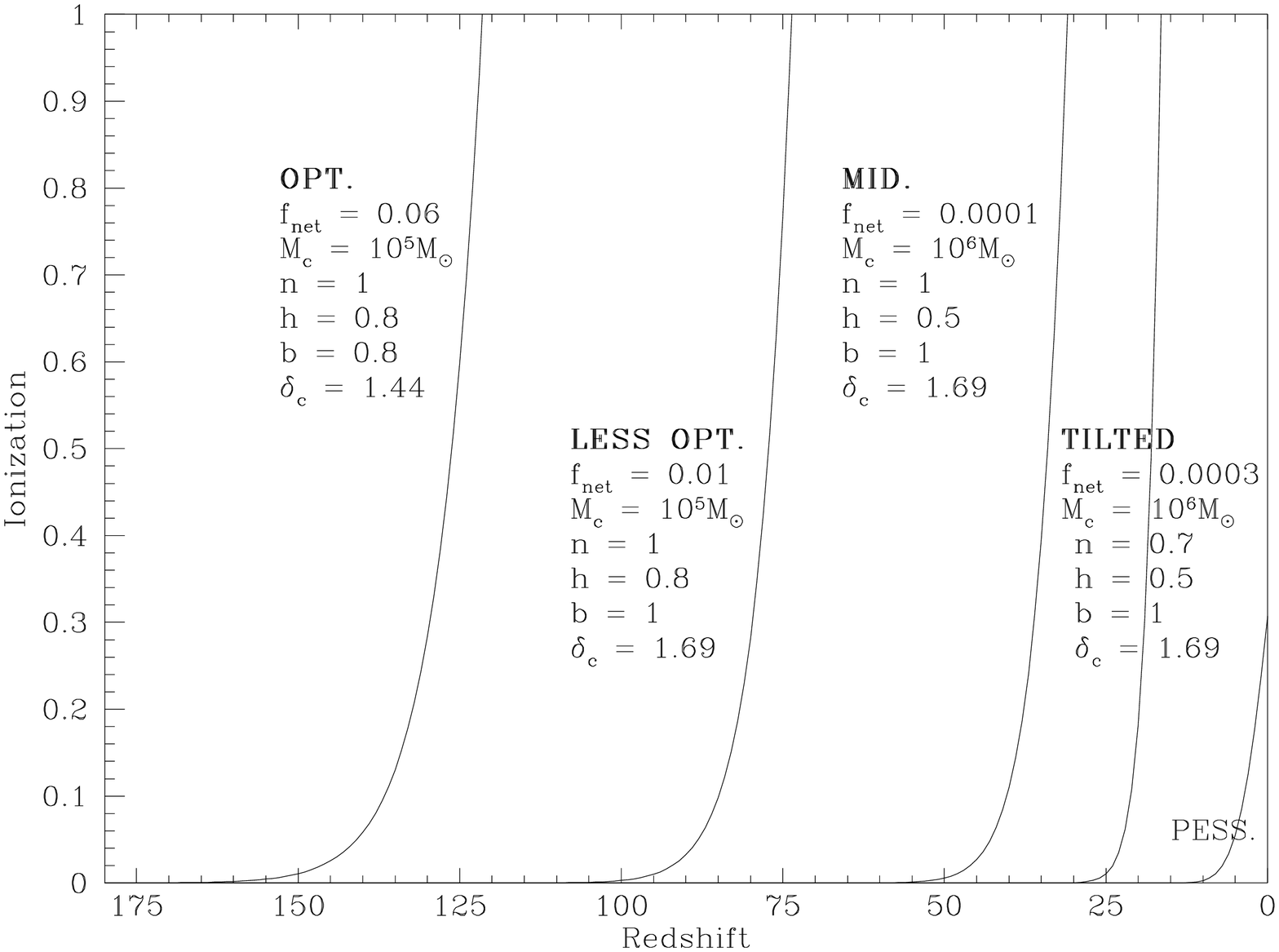,width=\fwidth,height=\fheight}
\nobreak
\caption{Volume fraction ionized for various scenarios.}
\label{reionfig2}
 
\mycaption{The volume fraction 
of the universe that is in ionized Str\"omgren bubbles
is plotted as a function of redshift for various parameter choices, 
corresponding to $n=1$ CDM (optimistic, less optimistic, middle-of-the-road
and pessimistic cases) and the tilted power spectrum (n=0.7) variant of CDM.
}
\efig

\bfig
\psfig{figure=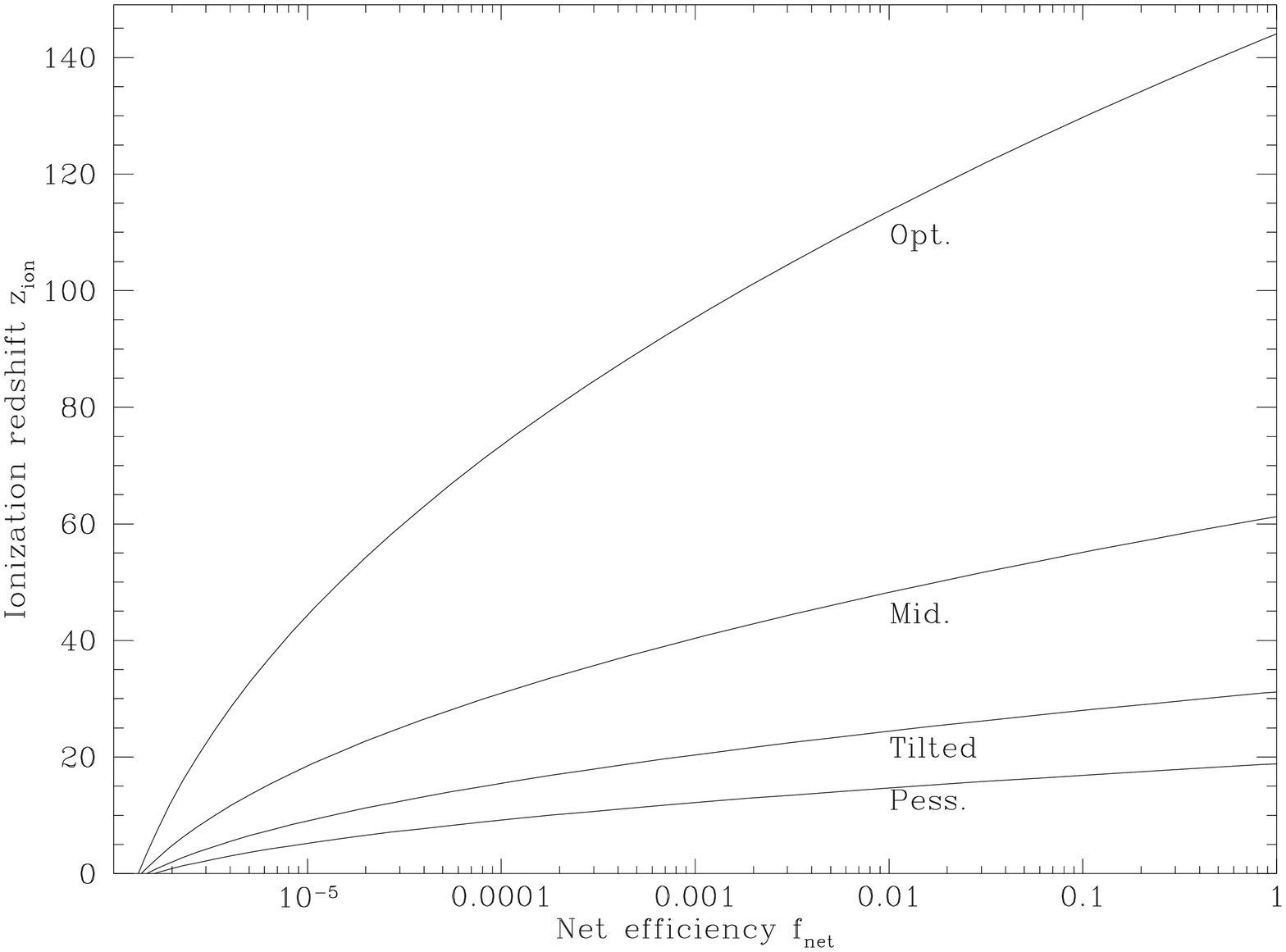,width=\fwidth,height=\fheight} 
\nobreak
\caption{Ionization redshift for various scenarios.}
\label{reionfig3}
 
\mycaption{The redshift at which $x=0.5$ plotted as a function of the net 
efficiency. The four curves correspond to four of the choices of $z_{vir}$
in Table~\ref{reiontable1}: 
$41.4$, $17.2$, $8.4$ and $4.8$ from top to bottom.
}
\efig

\bfig
\psfig{figure=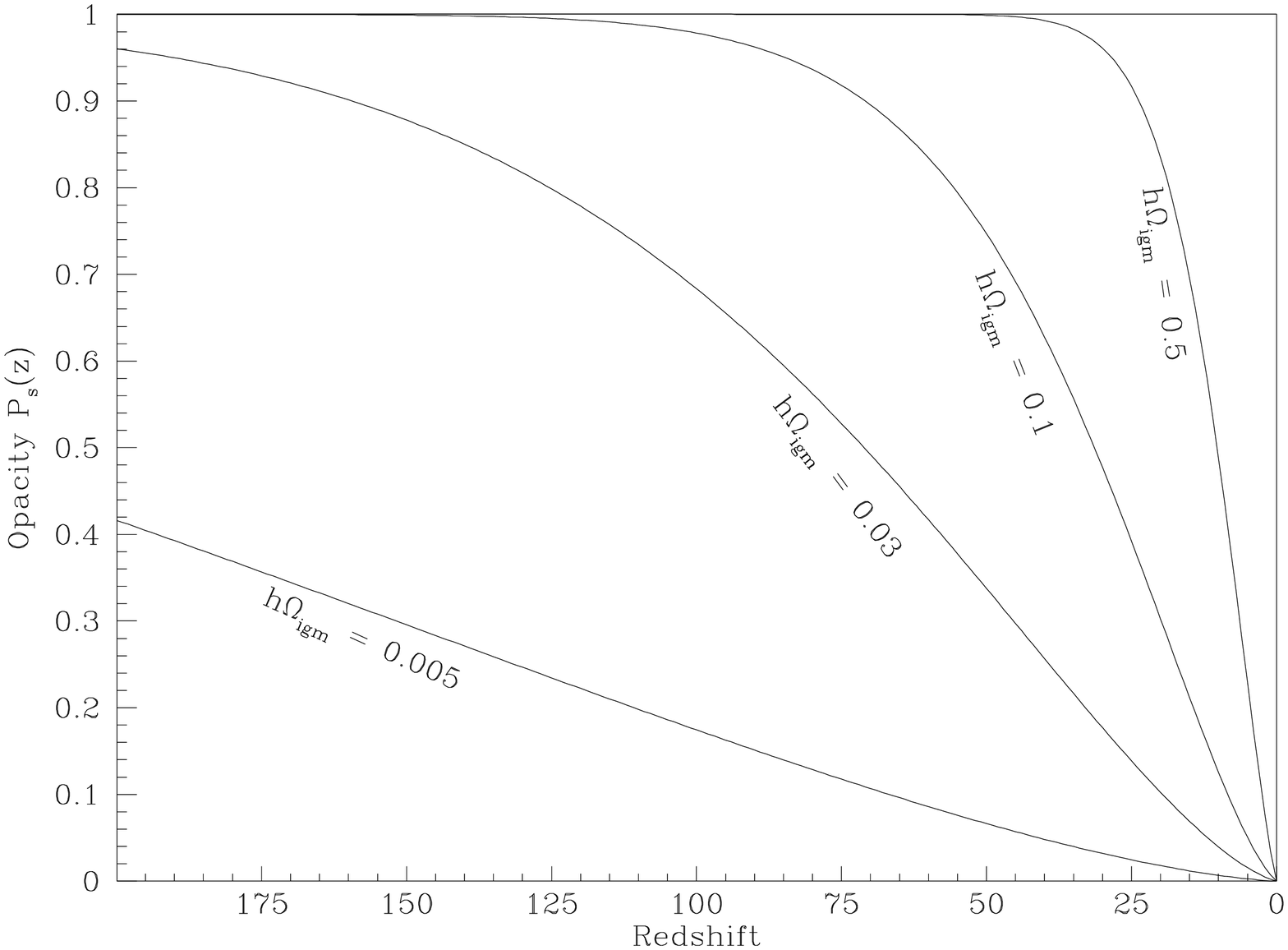,width=\fwidth,height=\fheight}
\nobreak
\caption{Opacity for completely ionized IGM.}
\label{reionfig4a}
 
\mycaption{The Thomson opacity $P_s(z)$, the probability that 
a CBR photon has been scattered at least once after the redshift $z$,
is plotted for four different choices of $h\Ob$ for the case where the
IGM is completely ionized at all times. For more realistic scenarios where
ionization occurs around some redshift $z_{ion}$, the opacity curves
simply 
level out and stay constant for $z\gg z_{ion}$.
}
\efig

\bfig
\psfig{figure=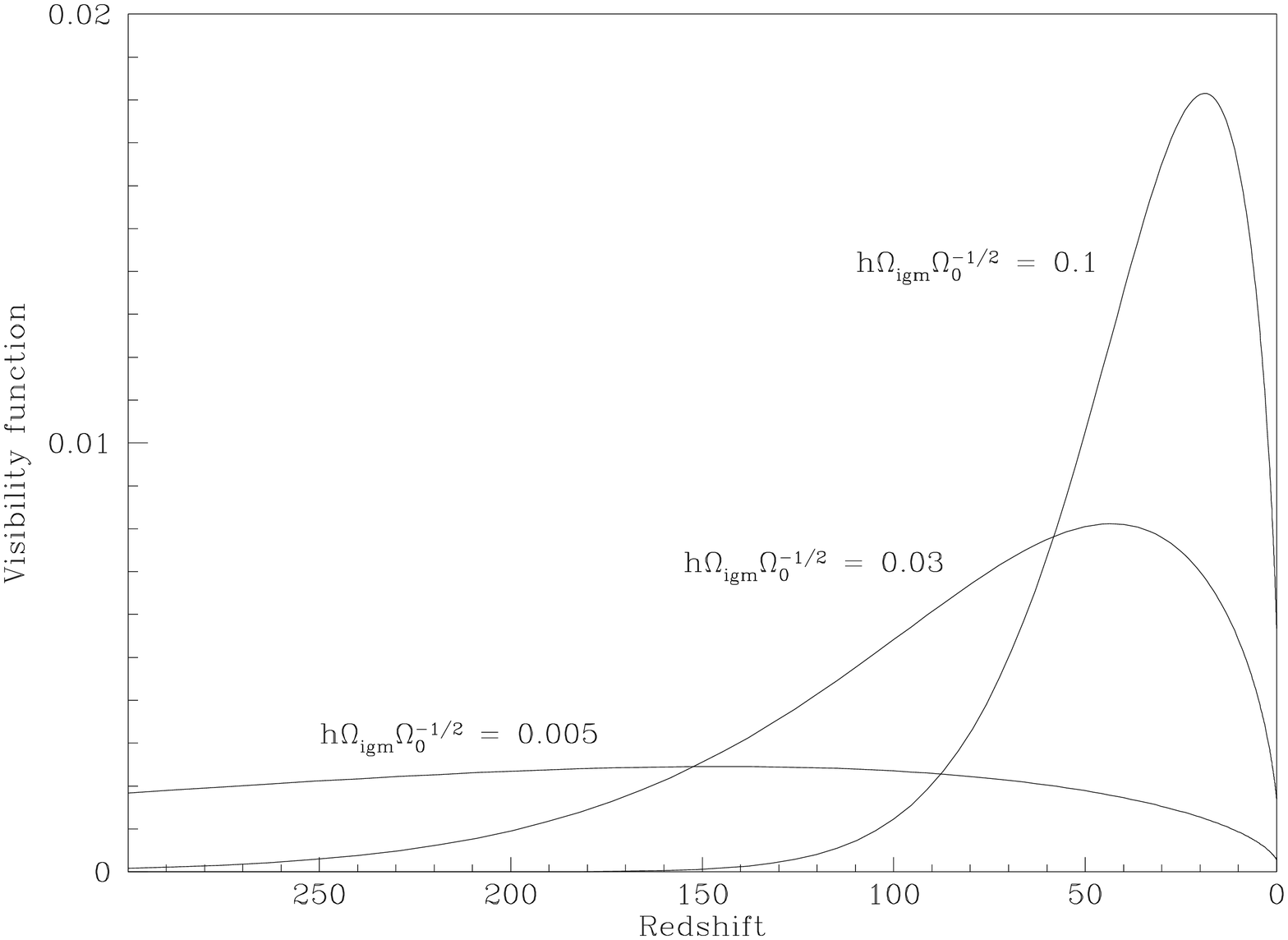,width=\fwidth,height=\fheight}
\nobreak
\caption{Last-scattering surface for completely ionized IGM.}
\label{reionfig4b}
 
\mycaption{The probability distribution for the redshift at which
a CBR photon was last scattered, the so called visibility 
function,  
is plotted for four different choices of $h\Ob$ for the case where the
IGM is completely ionized at all times. For more realistic scenarios where
ionization occurs around some redshift $z_{ion}$, the curves 
are unaffected for $z\ll z_{ion}$, vanish for 
$z_{ion}\ll z\ll 10^3$ and have a second bump around $z\approx 10^3$.
}
\efig

\bfig
\psfig{figure=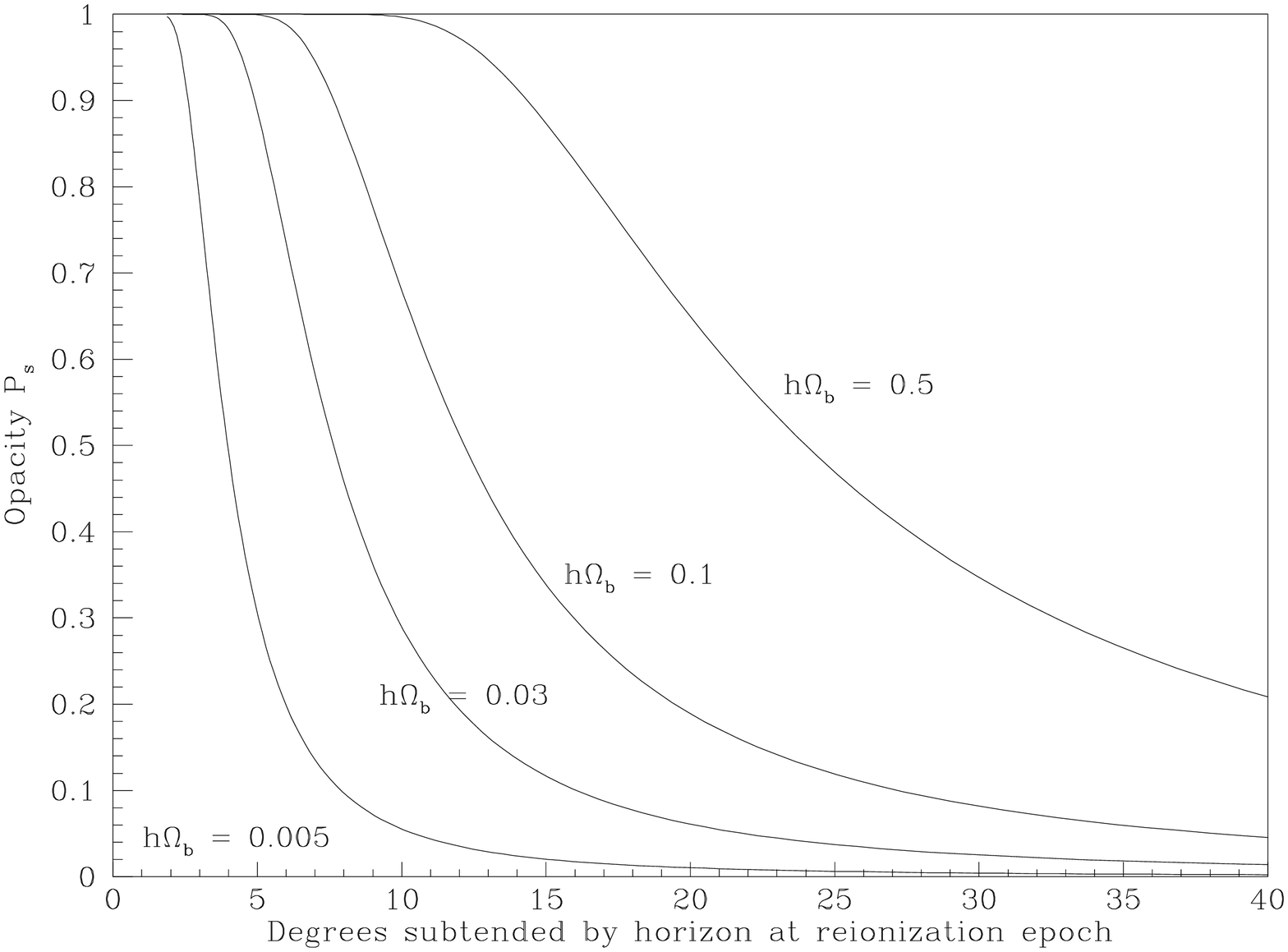,width=\fwidth,height=\fheight}
\nobreak
\caption{Opacity for completely ionized IGM as function of angle.}
\label{reionfig5a}
 
\mycaption{The total Thomson opacity $P_s$, the probability that 
a CBR photon has been scattered at least once since the recombination epoch, 
is plotted as a function of the angle in the sky that the horizon subtended at
the reionization epoch.
This is the largest angular scale on which fluctuations can be suppressed.}
\efig

\bfig
\psfig{figure=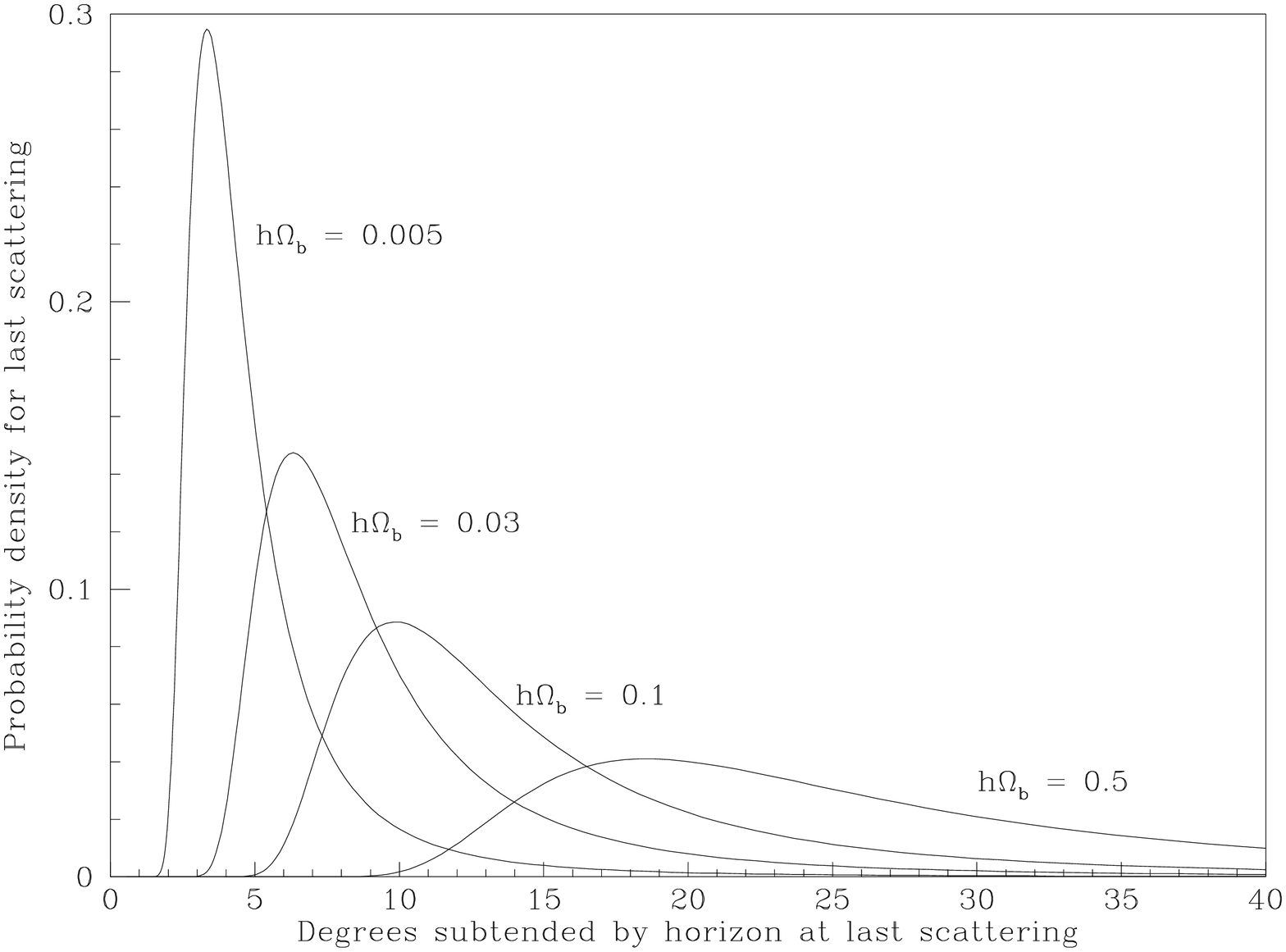,width=\fwidth,height=\fheight}
\nobreak
\caption{Last-scattering surface for completely ionized IGM as function of angle.}
\label{reionfig5b}
 
\mycaption{The probability distribution 
for the angle subtended by the horizon when
a CBR photon was last scattered, the so angular	 visibility 
function,  
is plotted for four different choices of $h\Ob$ for the case where the
IGM is completely ionized at all times. For more realistic scenarios where
ionization occurs around some redshift $z_{ion}$, corresponding to
an angle $\theta_{ion}$, 
the curves 
are unaffected for $\theta\gg \theta_{ion}$, vanish for 
$2^{\circ}\ll \theta\ll \theta_{ion}$ and have a second bump around 
$\theta\approx 2^{\circ}$, the horizon angle at recombination.
}
\efig

\bfig
\psfig{figure=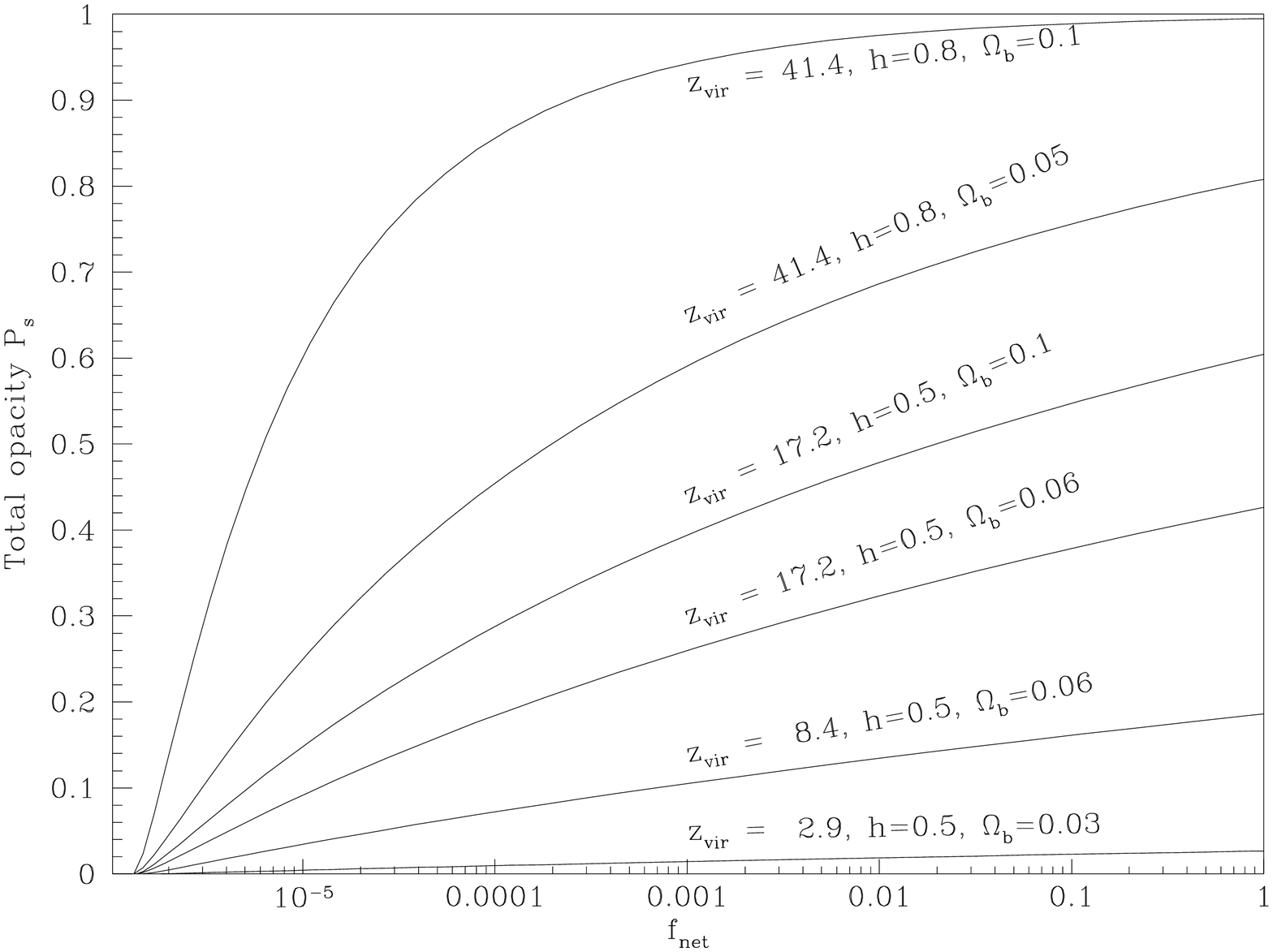,width=\fwidth,height=\fheight}
\nobreak
\caption{Total opacity for various models.}
\label{reionfig6}
 
\mycaption{The total opacity, the probability that a CBR photon has 
been scattered at least once since the recombination epoch, is plotted for
a variety of models.
}
\efig

\bfig
\psfig{figure=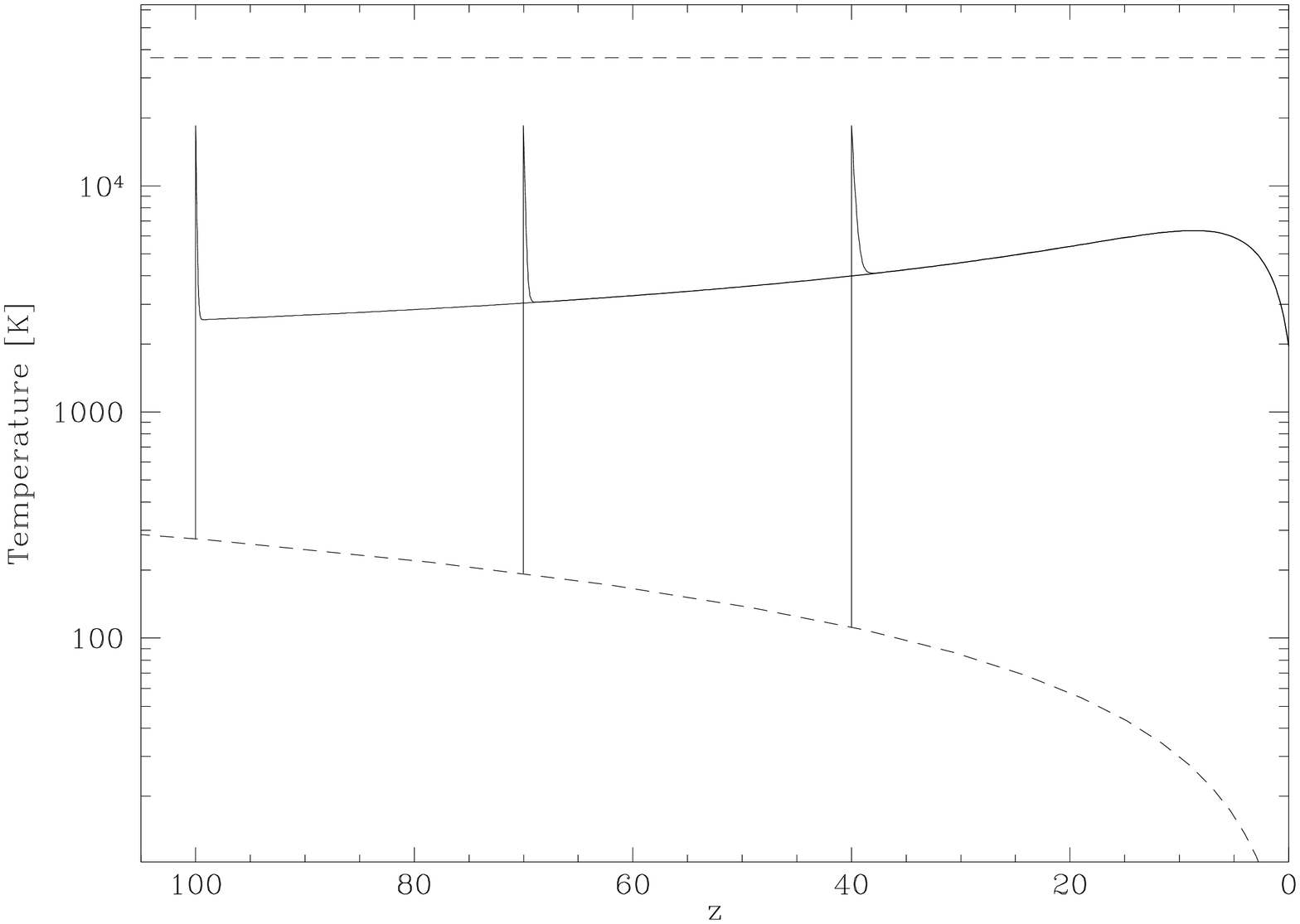,width=\fwidth,height=\fheight}
\nobreak
\caption{Temperature evolution in intergalactic Str\"omgren
\label{reionfig7}
bubbles.}
 
\mycaption{The temperature evolution is plotted 
for IGM exposed to a UV flux 
strong enough to keep it completely photoionized. In this example,
$h=0.5$, $\Ob=0.06$, and $T^* = 36,900$.  
The upper dashed line is $T^*$, the temperature corresponding to 
the average energy of the released photoelectrons,
towards which the
plasma is driven by recombinations followed by new photoionizations. 
The lower dashed line is the temperature of 
the CBR photons, towards which the plasma is driven Compton cooling. 
The three solid curves from left to right correspond to three different
redshifts for becoming part of a Str\"omgren bubble.
The first time the hydrogen becomes ionized, 
its temperature rises impulsively
to $T^*/2$. After this, Compton cooling 
rapidly pushes the temperature down to
a quasi-equilibrium level, where the Compton cooling rate equals the
recombination heating rate.
}
\efig

\bfig
\psfig{figure=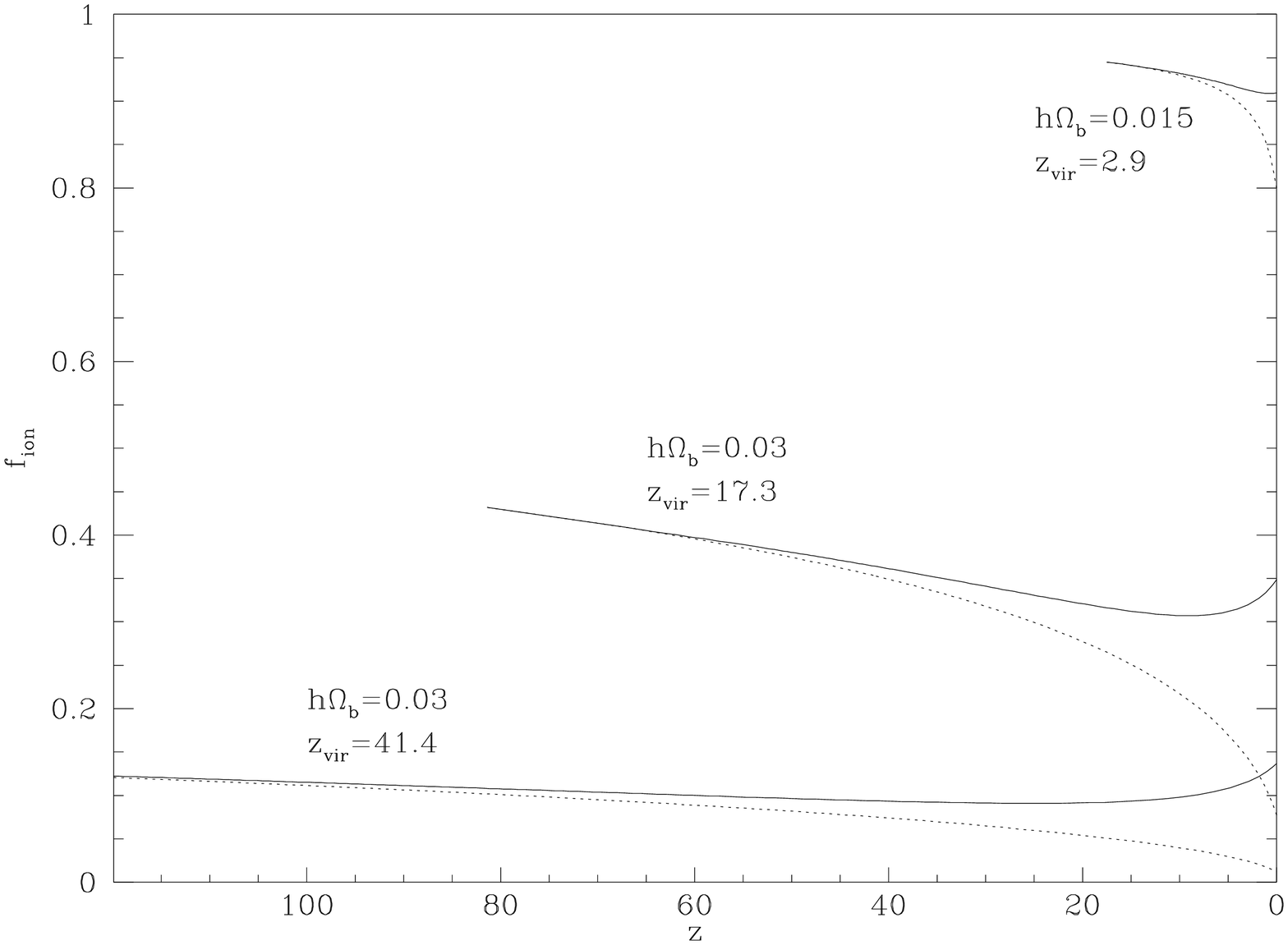,width=\fwidth,height=\fheight}
\nobreak
\caption{Ionization efficiencies for various scenarios}
\label{reionfig8}
 
\mycaption{The ionization efficiency, the 
fraction of the UV photons that produce a net 
ionization, is plotted for three different parameter combinations.
In all cases, $T^*=36,900\K$, the value appropriate for the radiation 
from the population 3 star in Table~\ref{reiontable3}.
The solid lines are the exact results from numerical integration of
\eq{ExpansionEq}. The dotted lines are the analytic fits, which are seen
to agree well in the redshift range of interest, which is typically
$z$ twice or three times $z_{vir}$.
}
\efig

\cleardoublepage

\chapter{Reionization in an Open Universe}
\label{openreionchapter}

% REIONIZATION IN AN OPEN CDM UNIVERSE:
% IMPLICATIONS FOR COSMIC MICROWAVE BACKGROUND FLUCTUATIONS

% \countdef\thetaEq=201
% \countdef\tauEq=202
% \countdef\thetaApproxEq=203
% \countdef\thetaEqThree=204

% \def\Ob{\Omega_{igm}}
\def\taut{\tau}

In this chapter, we
generalize the results of the previous chapter 
to CDM models with
$\Omega<1$. Such models have received recent interest because  
the excess power in the large--scale galaxy distribution 
is phenomenologically fit if the ``shape parameter"
$\Gamma = h\Omega_0\approx 0.25.$
It has been argued that essentially all CDM models may  require early
reionization to suppress degree-scale anisotropies in order to be consistent
with  experimental data, if the lowest degree-scale measurements are indeed
characteristic of the primordial temperature anisotropies. 
It is found that if
the cosmological constant  $\lambda = 0$, the extent of this suppression is
quite insensitive to $\Omega_0$,
as opposing effects partially cancel.
Given a $\sigma_8$-normalization today, 
the loss of small--scale power associated
with a lower $\Omega_0$ is partially canceled by higher optical depth from
longer lookback times and by structures forming at higher
redshifts as the universe becomes curvature--dominated at $z\approx
\Omega_0^{-1}$.
The maximum angular scale on which fluctuations are
suppressed decreases when $\Omega_0$ is lowered,
but this effect is also rather weak and unlikely to be
measurable in the near future.
For flat models, on the other hand, where
$\lambda_0 = 1 - \Omega_0$,  the negative effects of lowering $\Omega_0$
dominate,  and early reionization is not likely to play a
significant role if $\Omega_0\ll 1$. The same applies to 
for CDM models 
where the $\Gamma$ is lowered by increasing the number of
relativistic particle species.

\section{Introduction}

The inference from some, but not all, cosmic microwave background 
anisotropy experiments, is that many models, including the standard cold dark
matter (CDM) model, may produce excessive temperature fluctuations on degree and
sub--degree angular scales (Gorski {\etal} 1993; Vittorio \&
Silk 1992). Early reionization at redshift $z\simgt 30$  
% ONLY 20 percent tau is OK!!!! This corresponds to z\simeq 30.
produces an optical depth to scattering $\tau\simgt 20\%,$ and
suffices, with the known baryon density from primordial nucleosynthesis
constraints, to reconcile CDM with all observational limits 
(Sugiyama {\etal} 1993).
In 
Chapter~\ref{reionchapter}, 
early
photoionization of the intergalactic medium was discussed in a fairly
model--independent way, in order to  investigate whether early structures
corresponding to rare Gaussian peaks in a CDM model indeed could photoionize the
intergalactic medium sufficiently  early to appreciably smooth out the microwave
background fluctuations. In this chapter, the results of 
Chapter~\ref{reionchapter} 
will be
generalized to $\Omega<1$ models with non--zero cosmological constant
$\Lambda$.  Essentially all the notation used in this chapter was defined in
Chapter~\ref{reionchapter}
and, in the interest of brevity, some of the definitions will not be
repeated here.

Just as in 
Chapter~\ref{reionchapter},
our basic picture is the following: 
An ever larger fraction $\fs$ of the baryons in the universe 
falls into nonlinear
structures and forms galaxies. A certain fraction of these baryons form
stars or quasars which emit ultraviolet radiation. 
Some of this radiation escapes into the ambient 
intergalactic medium (IGM), which is consequently  
photoionized and heated. Due to cooling losses and
recombinations, the net number of ionizations per UV
photon, $\fion$, is generally less than unity. 

The results that we present here generalize the previous 
work to the case of an
open universe.
Lowering $\Oz$ has four distinct effects: 

\begin{enumerate}

\item
Density fluctuations gradually stop growing once 
$z\simlt\Oz^{-1}$. 
Thus given the observed power spectrum today, a lower $\Oz$ implies that 
the first structures formed earlier.

\item
Matter-radiation equality occurs later, which shifts the
turning-point of the CDM power spectrum toward larger scales.
This means less power on very small scales (such as $\sim 10^6\Ms$) relative to the
scales at which we normalize the power spectrum (namely galaxy cluster scales, $\sim 8h^{-1}\Mpc$ or even  
the much larger COBE scale). One consequence is that the first structures form later.

\item
The lookback time to a given ionization redshift becomes larger,
resulting in a higher optical depth.

\item
The horizon at a given ionization redshift subtends a
smaller angle on the sky, thus lowering the angular scale 
below which CBR fluctuations are suppressed.

\end{enumerate}

\noindent
Thus in terms of the virialization redshift $\zvir$ defined in 
Chapter~\ref{reionchapter}, 
the
redshift at which typical structures go nonlinear, 
effect 1 increases $\zvir$ whereas effect 2 decreases $\zvir$.
So these two effects influence $\fs$, the fraction of baryons in nonlinear
structures, in opposite directions.
As to effect 2, it should be noted that this applies not only to CDM, but to
any spectrum that ``turns over" somewhere between the very smallest nonlinear
scales  ($\sim 10^6\Ms$) and the very largest  
($\sim 10^{21}\Ms$)
scales at which we COBE-normalize. 
Yet another effect of lowering $\Oz$ is that 
the ionization efficiency $\fion$ drops slightly, at most by
a factor $\Oz^{1/2}$. This is completely negligible compared to the
above-mentioned effects, as $\zion$ depends only logarithmically on the
efficiency. 

In the following sections, we will 
discuss each of these four effects in greater detail,  
and then compute their
combined modification of the ionization history in a few scenarios.

\section{The Boost Factor}

When curvature and vacuum density are negligible, 
sub-horizon-sized density fluctuations simply grow as the
scale factor $a\propto (1+z)^{-1}$. 
Thus at early times $z\gg\Oz^{-1}$, we can write 
$$\delta =  {\boost(\Oz,\Lz)\over 1+z}\delta_0$$
for some function $\boost$ independent of $z$ that we will refer to as the
{\it boost factor}. Clearly $\boost(1,0) = 1$.
Thus if certain structures are assumed to form when $\delta$ equals some fixed
value, then given the observed power spectrum today, the boost factor tells us
how much earlier these structures would form than they would in a standard flat
universe. 
The boost factor is simply the inverse of the so called
growth factor, and can be computed analytically for 
a number of special cases (see {\eg} Peebles 1980).
For the most general case, the fit 
$$
B(\Oz,\Lz) \approx
{2\over 5\Oz}\left[\Oz^{4/7} - \Lz + 
\left(1+{\Oz\over 2}\right)\left(1+{\Lz\over 70}\right)\right]
$$ 
is accurate to within a few percent for all parameter values of cosmological
interest
(Carroll {\etal} 1992).
The exact results are plotted in 
Figure~\ref{openreionfig1} 
for the case $\Lz=0$
and the flat case $\Lz = 1-\Omega_0$.
Since we will limit ourselves to these two cases,
the simple power-law fits
$$\cases{
\boost(\Oz,0) & $\approx\Oz^{-0.63},$\crr
\boost(\Oz,1-\Oz) & $\approx\Oz^{-0.21},$
}$$
which are accurate to within $1\%$ for $0.2\leq\Oz\leq 1$,
will suffice for our purposes. 
Note that the standard rule of thumb that
perturbations stop growing at $1+z\approx\Oz^{-1}$, indicating 
$B\propto\Oz^{-1}$, is not particularly accurate in this context.

\section{The Power Spectrum Shift}

As mentioned above, lowering $h\Oz$ causes the first structures go 
nonlinear at a later redshift. This is quantified in the present section.

The standard 
CDM model with power-law initial fluctuations 
proportional to
$k^n$ predicts a power
spectrum that is well fitted by (Bond \& Efstathiou 1984;
Efstathiou {\etal} 1992)
$$P(k) \propto 
{q^n\over
\left(1+\left[aq+(bq)^{1.5} +
(cq)^2\right]^{1.13}\right)^{2/1.13}},$$
where $a\equiv 6.4$, $b\equiv 3.0$, $c\equiv 1.7$, 
$q\equiv (1h^{-1}\Mpc) k/\Gamma$ and the ``shape parameter" 
$\Gamma$
will be discussed further on.
Although this fit breaks down for scales comparable to the 
curvature scale  $r_{curv} = k_{curv}^{-1} = H_0^{-1}|1-\Oz|^{-1/2}$,
it is quite accurate for the much smaller scales that will be considered in
the present chapter. Rather, its main limitation is that it breaks down if 
$\Oz$ is so low that the baryon density becomes comparable to the 
density of cold dark matter. Thus for $\Ob\approx 0.05$,
the results cannot be taken too seriously for $\Oz<0.2$.
We will limit ourselves to the standard $n=1$ model here, 
 as the tilted ($n<1$) case was treated in 
Chapter~\ref{reionchapter}
and was seen
to be be essentially unable 
to reionize the universe early enough 
to be relevant to CBR anisotropies.  
The same applies to models with mixed hot and cold dark matter.

Let us define the {\it amplitude ratio} 
$$R(\Gamma,r_1,r_2) \equiv {\sigma(r_1)\over\sigma(r_2)},$$
where $\sigma(r_1)$ and $\sigma(r_2)$ are the {r.m.s.} 
mass fluctuation amplitudes in
spheres of radii $r_1$ and $r_2$, {\ie}
$$\sigma(r)^2 \propto
\izi P(k) 
\left[{\sin kr\over(kr)^3} - {\cos
kr\over(kr)^2}\right]^2 dk.$$ 
As in 
Chapter~\ref{reionchapter}, 
we normalize the power spectrum so that
$\sigma(8 h^{-1}\Mpc)$ equals some constant denoted $\sigma_8$,
and 
$M_c$ will denote the characteristic 
mass of the first galaxies
to form. The corresponding comoving length scale $r_c$ is 
given by $M_c = {4\over 3}\pi r_c^3\rho$, where $\rho$ is the mean density of
the universe. 
Thus given $\sigma_8$, 
what is relevant for
determining when the first galaxies form is the amplitude ratio
$$R(\Gamma, r_c, 8 h^{-1}\Mpc).$$
This ratio is computed numerically, and the results are 
plotted as a function of $\Gamma$ in 
Figure~\ref{openreionfig2} 
for a few different 
values of the cutoff mass $M_c$. 
It is easy to see why the amplitude ratio increases with $\Gamma$, since
on a logarithmic scale, a decrease in $\Gamma$ simply shifts the entire 
power spectrum towards lower $k$, thus decreasing the amount of power on very
small scales relative to that on large scales. The fit 
$$R(\Gamma, r_c, 8 h^{-1}\Mpc)
\approx 3 + 7.1\ln(1 h^{-1}\Mpc/r_c)\Gamma$$
is accurate to within $10\%$ for $0.05<\Gamma<2$ and
$100\pc<r_c<100\kpc$.

\section{The Optical Depth}

Since a lower $\Oz$ implies a larger $|dt/dz|$ and an older universe, 
the optical depth out to a given ionization redshift $\zion$
is greater for small $\Oz$.
For a given ionization history $\v(z)$, the 
optical depth for Thomson scattering is given by
$$\cases{
\taut(z)& = $\taut^*\izz{(1+z')^2\over
\sqrt{\Lz + (1+z')^2(1-\Lz+\Oz z')}}\v(z')dz',$\crr
\taut^*& = ${3\Ob\over 8\pi}
\left[1 - \left(1-{\v_{He}\over 4\v}\right)f_{He}\right] 
{H_0c\st\over m_p G} \approx 0.057 h\Ob,$
}$$
where we have taken the mass fraction of helium to be 
$f_{He}\approx 24\%$ and assumed $\v_{He} \approx\v$,
{\ie} that helium never becomes doubly ionized and that the fraction
that is singly ionized equals the fraction of hydrogen that is
ionized. The latter is a very crude approximation, but has the advantage that
the error can never exceed $6\%$. 
If the universe is fully ionized for all redshifts below $z$, the 
integral can be done analytically for $\Lz=0$:
\beq{tauEq}
\taut(z) = 
{2\taut^*\over 3\Oz^2}
\left[2-3\Oz + (\Oz z+3 \Oz-2)\sqrt{1+\Oz z}\,\right]
\approx 0.038{h\Ob z^{3/2}\over\Oz^{1/2}}
\eeq
for $z\gg\Oz^{-1}$. 
As is evident from the asymptotic behavior of the integrand, 
$\taut$ is independent of $\Lz$ in the high redshift limit. 
Thus optical depth of unity is attained if reionization occurs at 
$$z\approx 92 \left(h\Ob\over 0.03\right)^{-2/3}\Oz^{1/3}.$$

\section{The Angular Scale}

It is well known that reionization suppresses CBR fluctuations only on angular
scales below the horizon scale at last scattering. 
Combining the standard expressions for 
horizon radius ({\eg} Kolb \& Turner 1990) and angular size,
this angle is given by
\beq{thetaEq}
\theta = 2\tan^{-1}\left[
{\sqrt{1+z}\Oz^{3/2}/2\over
\Oz z-(2-\Oz)(\sqrt{1+\Oz z}-1)}
\right].
\eeq
For $z\gg \Oz^{-1}$, this reduces to
\beq{thetaApproxEq}
\theta\approx{\sqrt{\Oz\over z}},
\eeq
but as is evident from 
Figure~\ref{openreionfig3}, 
this is quite a bad
approximation
except for $z\gg 100$. If we substitute it into  
\eq{tauEq} nonetheless, to get a rough estimate, we
conclude that optical depth unity is obtained at an epoch 
whose horizon scale subtends the angle 
\beq{thetaEqThree}
\theta \approx 12^{\circ} \left({h\Ob\Oz\over 0.03}\right)^{1/3},
\eeq
{\ie}, the dependence on all three of these cosmological parameters is
relatively weak.
As discussed in 
Chapter~\ref{reionchapter}, 
fluctuations on angular scales much smaller than this
are suppressed by a factor 
$$P(z)\equiv 1-e^{-\tau(z)},$$
the {\it opacity}, which is
the probability that a photon was Thompson scattered after redshift $z$.
Its derivative, the {\it visibility function}
$f_z = dP/dz$, is the probability distribution for the redshift at which last
scattering occurred, the profile of the last scattering surface.  
The {\it angular visibility function}
$$f_{\theta}(\theta) = {dP_s\over d\theta} = 
\left({d\theta\over dz}\right)^{-1}{dP_s\over dz}$$ 
is plotted in 
Figure~\ref{openreionfig4} 
for the case where the universe never recombines 
(the curves for the more general case with reionization at
some redshift $\zion$ can be read off from 
Figure~\ref{openreionfig4} 
as described in the previous chapter).  
These functions give a good idea of the
range of angular scales on which suppression starts to become important.
In plotting these curves, the exact expression\eqnum{thetaEq}
has been used, rather
than the approximation\eqnum{thetaApproxEq}.
It is seen that the qualitative behavior indicated by 
\eq{thetaEqThree} is
correct: as $\Oz$ is lowered, the peak shifts down toward smaller
angular scales,
but the $\Oz$-dependence is quite weak.

\section{Cosmological Consequences}

We will now compare the effect of lowering $\Gamma$ in three 
cosmological models. The first model, which will be referred to as 
``open CDM" for short, has $\lambda_0 = 0$.
The second model, referred to as ``$\Lambda$CDM", has 
$\lambda_0 = 1-\Oz$. 
The {\it shape parameter} essentially tells us how early the
epoch of matter-radiation equality occurred, and 
is given by
$$\Gamma = h\Oz\left({g_*\over 3.36}\right)^{-1/2},$$
where $g_*	= 3.36$ corresponds to the standard model with
no other relativistic degrees of freedom than photons and 
three massless neutrino species. 
In open CDM and  $\Lambda$CDM, we have 
$\gamma_* = 3.36$, so that $\Omega_0 = \Gamma/h$. 
The third model, referred to as $\tau$CDM
(Dodelson, Gyuk \& Turner 1994b), has 
$\lambda_0=0$ and $\Omega = 1$, and achieves a lower value of $\Gamma$ 
by increasing $g_*$ instead. 

Including the effect of the boost factor, 
\eq{zionEq} in 
Chapter~\ref{reionchapter}
becomes
$$1+\zion = {\sqrt{2}\sigma_8\over\delta_c} 
\>R(h\Oz,r_c,8 h^{-1}\Mpc)
\>B(\Oz,\Lz)\>
\erfc^{-1}\left[{1\over 2\fupp\fion}\right].$$
The ionization redshift $\zion$ is plotted as a function of the shape parameter
in 
Figure~\ref{openreionfig5} 
for the various scenarios specified in 
Table~\ref{openreiontable1}. 
It is seen that for the open model,
the dependence 
on $\Oz=\Gamma/h$ is
typically much weaker than the dependence on other parameters. One reason for
this is that changes in the boost factor and the amplitude 
ratio partially cancel
each other. For $\Lambda$CDM, the $\Oz$-dependence is
stronger, since the boost factor is weaker.
In the  
$\tau$CDM model,
the dependence on $\Gamma/h$
is even stronger, as there is no boost factor whatsoever 
to offset the change in the amplitude ratio. 

\begin{table}
$$
\begin{tabular}{|l|cccc|} 
\hline
                   & Pess.& Mid. & Opt. & Very opt.\\
\hline
$\sigma_8$         & 0.5    & 1    & 1.1 &1.2\\
$\delta_c$         & 2.00   &1.69&1.44&1.33\\
$h$                & 0.5    &0.5&0.8&0.8\\
$M_c [\Ms]$        & $10^8$ &$10^6$&$10^5$&$10^5$\\
$f=\fion\fupp$     & 1      &120&23,000&$10^6$\\
$\erfc^{-1}[1/2f]$ & 0.48   &2.03&3.00&3.55\\
$h^2\Ob$           & 0.010  &0.013&0.015&0.020\\
\hline
\end{tabular}
$$
\caption{Parameters used}
\label{openreiontable1}

\end{table}

The scenarios in Table~\ref{openreiontable1} 
are similar to those in
Chapter~\ref{reionchapter}. 
In the one labeled ``very optimistic", 
the high value for $\fupp$, 
the net number of produced UV photons per proton, 
is obtained by assuming that the main source of
ionizing radiation is black hole accretion rather than conventional
stars.  Note that this speculative assumption still only increases
$\zion$ by $3.55/3.00 - 1\approx 18\%$, the
efficiency dependence being merely logarithmic.

Figure~\ref{openreionfig6}, 
in a sense the most important plot in this chapter, shows the opacity
as a function of $\Gamma/h$ for the 
various scenarios.
Because of the increase in optical depth due to larger lookback times, the 
open model now gives slightly 
larger opacities for lower $\Oz = \Gamma/h$.
However, this dependence is seen to be quite week. 
For $\Lambda$CDM, where the
boost factor contributes less,
the net result is seen to be the opposite; a slight decrease in the opacity for
lower $\Oz = \Gamma/h$. For the $\tau$CDM model, 
where there is neither a boost factor nor an increase in the lookback time, 
this drop in opacity is seen to be much sharper.
Note that the dependence on other uncertain parameters, 
summarized by the four
scenarios in Table~\ref{openreiontable1}, is quite strong. Indeed,
this dependence is stronger than the effect 
of moderate changes in $\Gamma/h$, so in the 
near future, it appears unlikely
that opacity limits will be able to constrain 
the shape parameter except perhaps
in the $\tau$CDM model.

The $\tau$CDM situation is summarized 
in 
Figure~\ref{openreionfig7}. 
To attain at least $50\%$ opacity, 
$h\Ob$ must lie above the
heavy curve corresponding to the scenario in question. On the other hand,
nucleosynthesis (Malaney \& Mathews 1993)
places a strict upper bound on this quantity if we assume that
$h\geq 0.5$\footnote{
The $\tau$CDM model also alters the nucleosynthesis process 
(Dodelson, Gyuk \& Turner 1994a), but this can only marginally
relax the bounds unless the $\tau$ neutrino is in a 
mass range incompatible with $\tau$CDM (Gyuk \& Turner 1994).}
It is seen that a shape parameter as low as
$\Gamma\simeq 0.25$, which would match large-scale structure observations
(Peacock \& Dodds 1994), is quite
difficult to reconcile with these two constraints. 

\section{Discussion}

Lowering $\Gamma$ is an attractive resolution of the problem 
that arises in reconciling the observed structure in the 
universe on large scales with observations on megaparsec scales. 
The empirical power spectrum is well fit by 
$\Gamma\approx 0.25$ (Peacock \& Dodds 1994). Kamionkowski 
\& Spergel (1994) have found that primordial adiabatic 
fluctuations in an open universe with $\Omega \approx 0.3$ are 
reconcilable with large-scale CBR anisotropy. On degree scales 
Kamionkowski, Spergel \& Sugiyama (1994) require reionization with 
optical depth $\tau\sim 1$ in order to reconcile the low density 
open model with recent experimental limits, if  the lowest 
limits are adopted.
In a low $\Omega_0$ $\Lambda$CDM model, the 
situation is not so critical, but reionization is required 
if the lowest limits (SP91) are adopted on degree 
scales (Gaier {\etal} 1992); 
$\tau\sim 0.5$
suffices however. A similar but slightly more favorable situation 
occurs in a $\tau$CDM model, 
where 
$\Gamma = h\Omega_0 (g_*/3.36)^{-1/2}$ is reduced by increasing 
$g_*$ by a factor of $\sim 4$, 
but some reionization is still required to match SP91.

We have found that reionization giving $\tau$ in the range 0.5 to 1
is readily produced and even natural in open models. 
This is because of the early 
formation of structure in combination with the increased age of the 
universe, effects which compensate for the flattening of 
the power spectrum due to the delay in matter domination.
However, the $\Lambda$CDM and $\tau$CDM models with low $\Omega_0$ 
fare less well in this 
regard, since the loss of small-scale power is not balanced 
by significantly earlier structure formation.
With $\Omega_b$ in the range given by standard nucleosynthesis, 
a significant optical depth $\tau\simgt 0.5$ is 
difficult to attain in either the $\Lambda$CDM or 
$\tau$CDM scenarios.

\def\fheight{10.3cm} \def\fwidth{14.5cm}

\newpage

\bfig
\psfig{figure=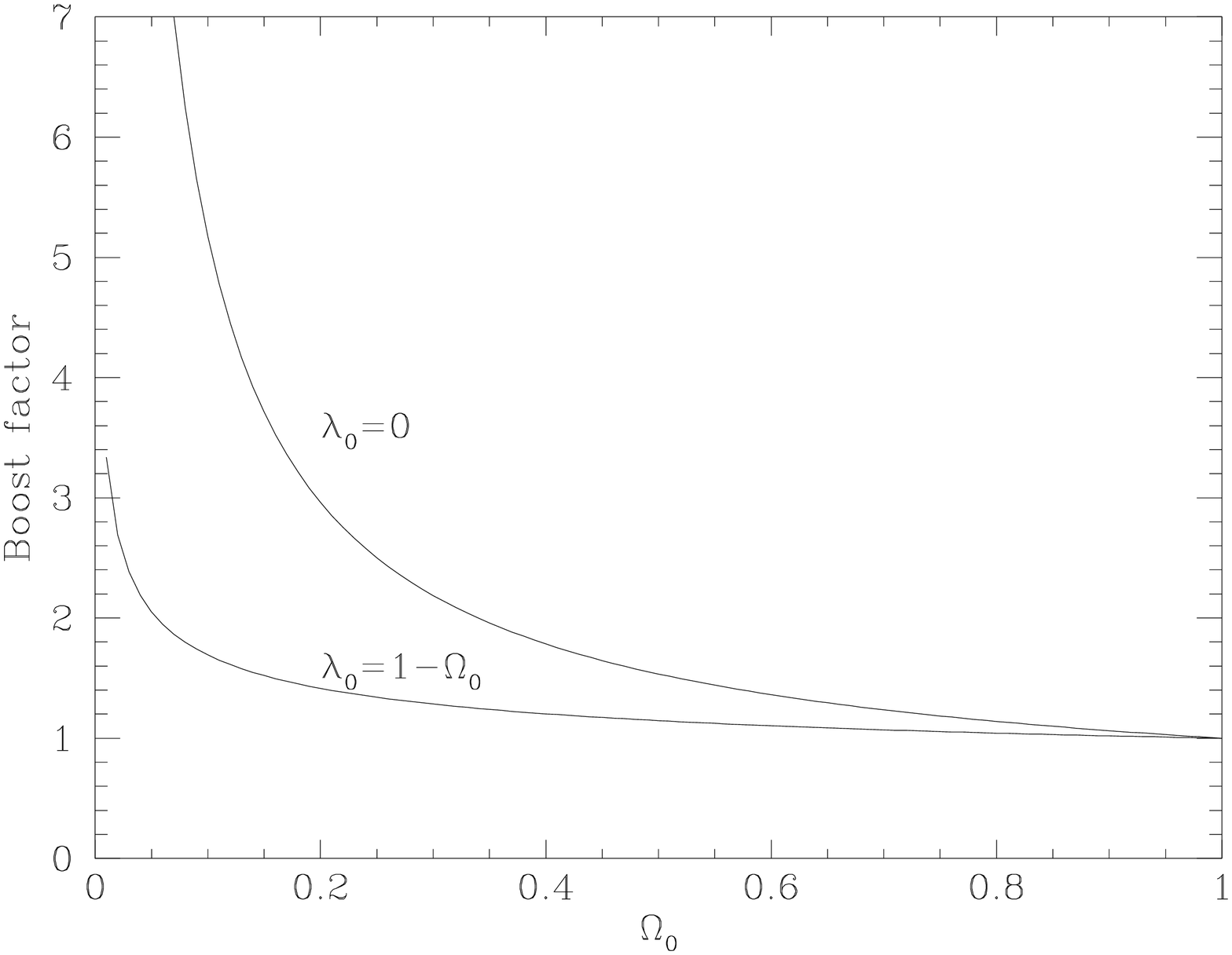,width=\fwidth,height=\fheight}
\nobreak
\caption{The boost factor.}
\label{openreionfig1}
 
\mycaption{The boost factor $B(\Omega_0,\lambda_0)$ is plotted as a function 
of $\Omega_0$ for two classes of cosmologies.
The upper curve corresponds to a standard open universe, 
{\it i.e.} $\lambda_0=0$, whereas the lower curve corresponds to flat
universes with 
$\lambda_0=1-\Omega_0$.
}
\efig

\bfig
\psfig{figure=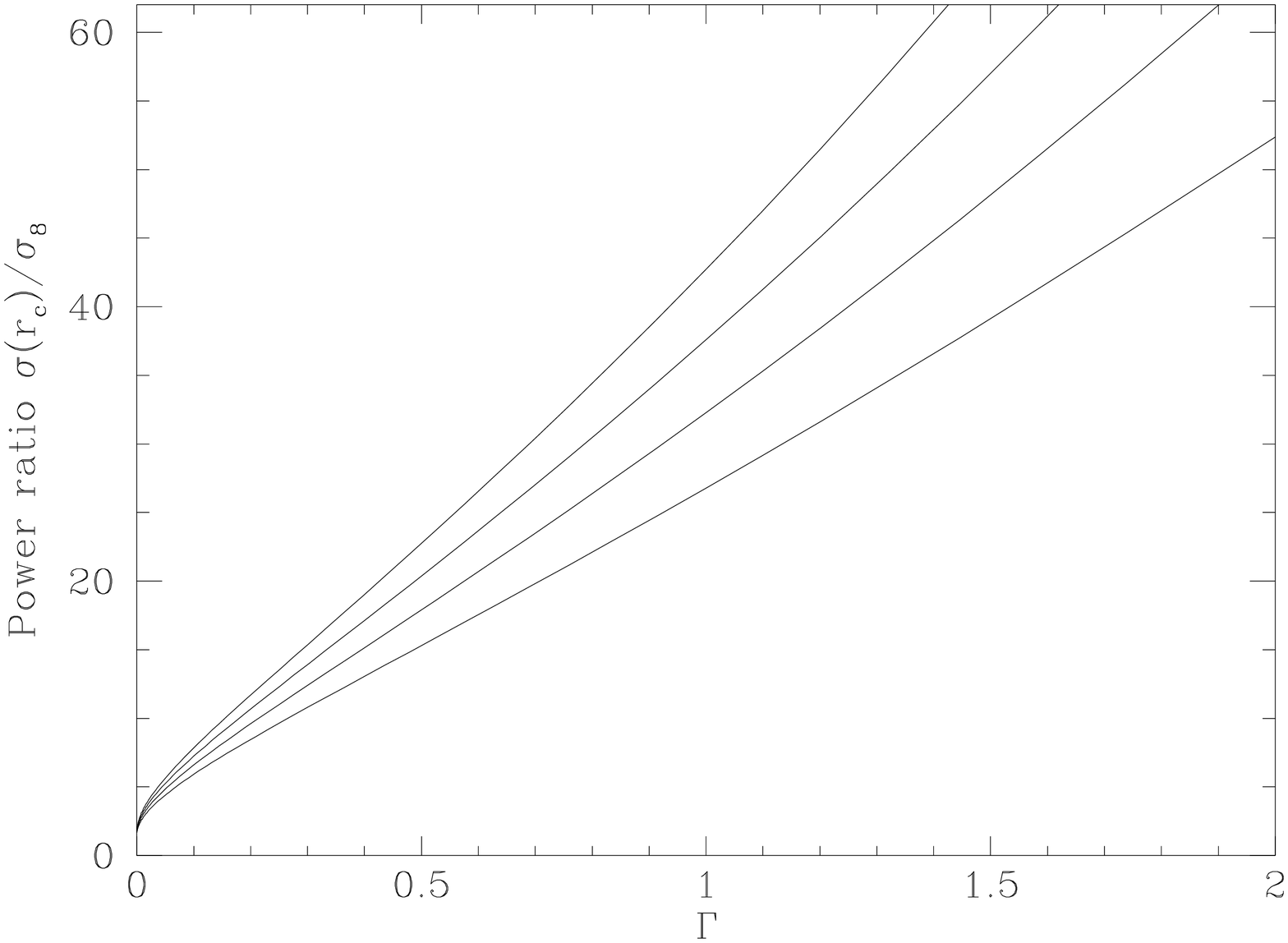,width=\fwidth,height=\fheight}
\nobreak
\caption{The amplitude ratio.}
\label{openreionfig2}
 
\mycaption{The ratio of the fluctuation 
amplitude on the small scale $r_c$ to that at 
$8h^{-1}$Mpc is plotted as a function of the shape parameter $\Gamma$.
From top to bottom, the four curves correspond to scales of 
$3.5 h^{-1}$kpc,
$7.5 h^{-1}$kpc,
$16 h^{-1}$kpc and
$35 h^{-1}$kpc, respectively.
For $h=0.5$ and $\Omega_0=1$, these four length scales correspond to
the masses 
$10^5M_{\odot}$,
$10^6M_{\odot}$,
$10^7M_{\odot}$ and
$10^8M_{\odot}$.
The weak additional dependence on $\Omega_0/h$ that would result from holding 
$M_c$ rather than $r_c$ fixed is clearly negligible, since as can be seen,
$M_c$ must vary by an entire order of magnitude to offset a mere
$20\%$ change in $\Gamma$. 
}
\efig

\bfig
\psfig{figure=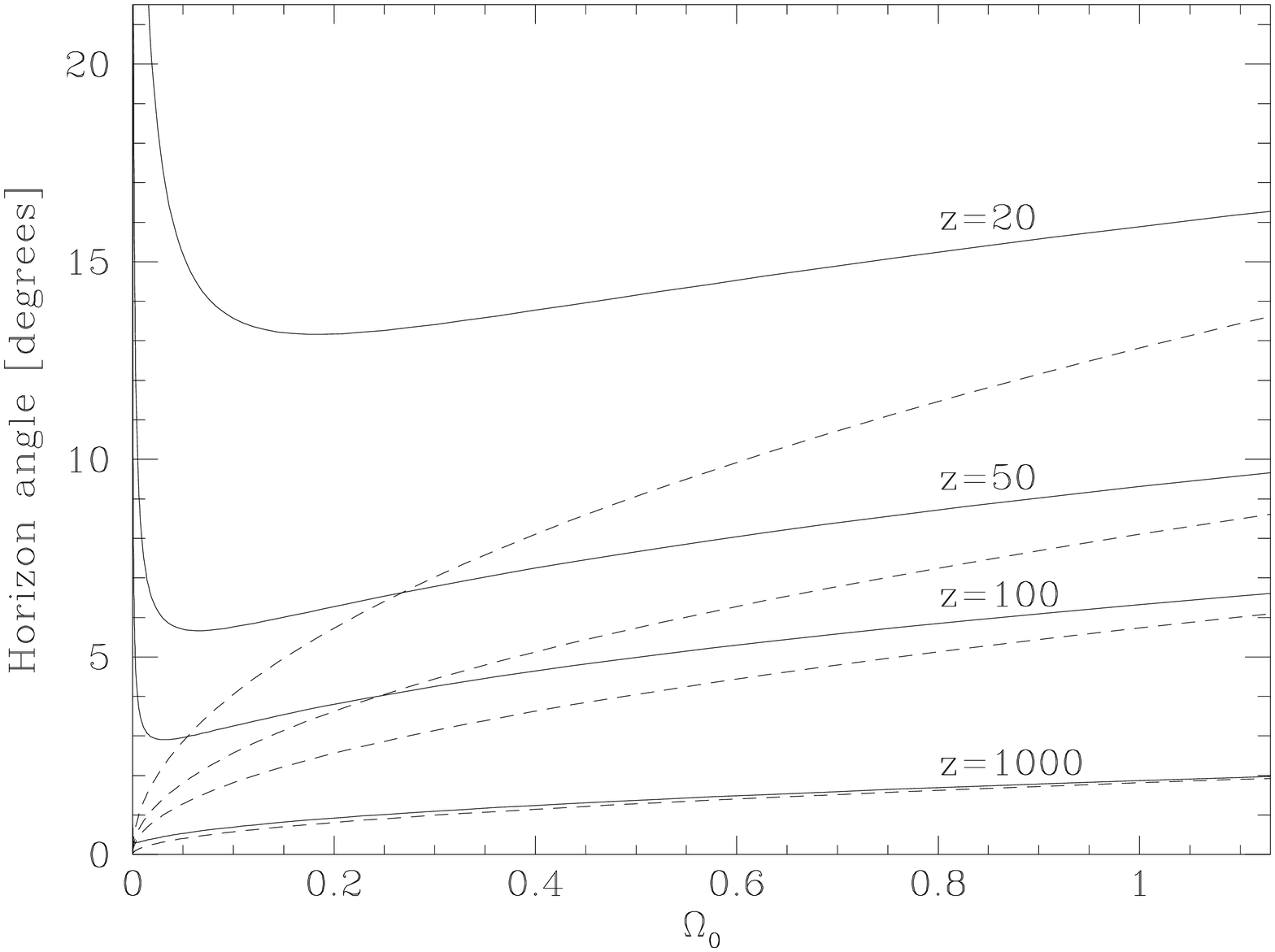,width=\fwidth,height=\fheight}
\nobreak
\caption{The horizon angle.}
\label{openreionfig3}
 
\mycaption{The angle in the sky subtended 
by a horizon volume at redshift $z$ is 
plotted as a function of $\Omega_0$ for the case with no cosmological 
constant.
The solid lines are the exact results for the four redshifts indicated,
and the dashed lines are the corresponding fits using the 
simplistic approximation $\theta\approx (\Omega_0/z)^{1/2}$.
}
\efig

\bfig
\psfig{figure=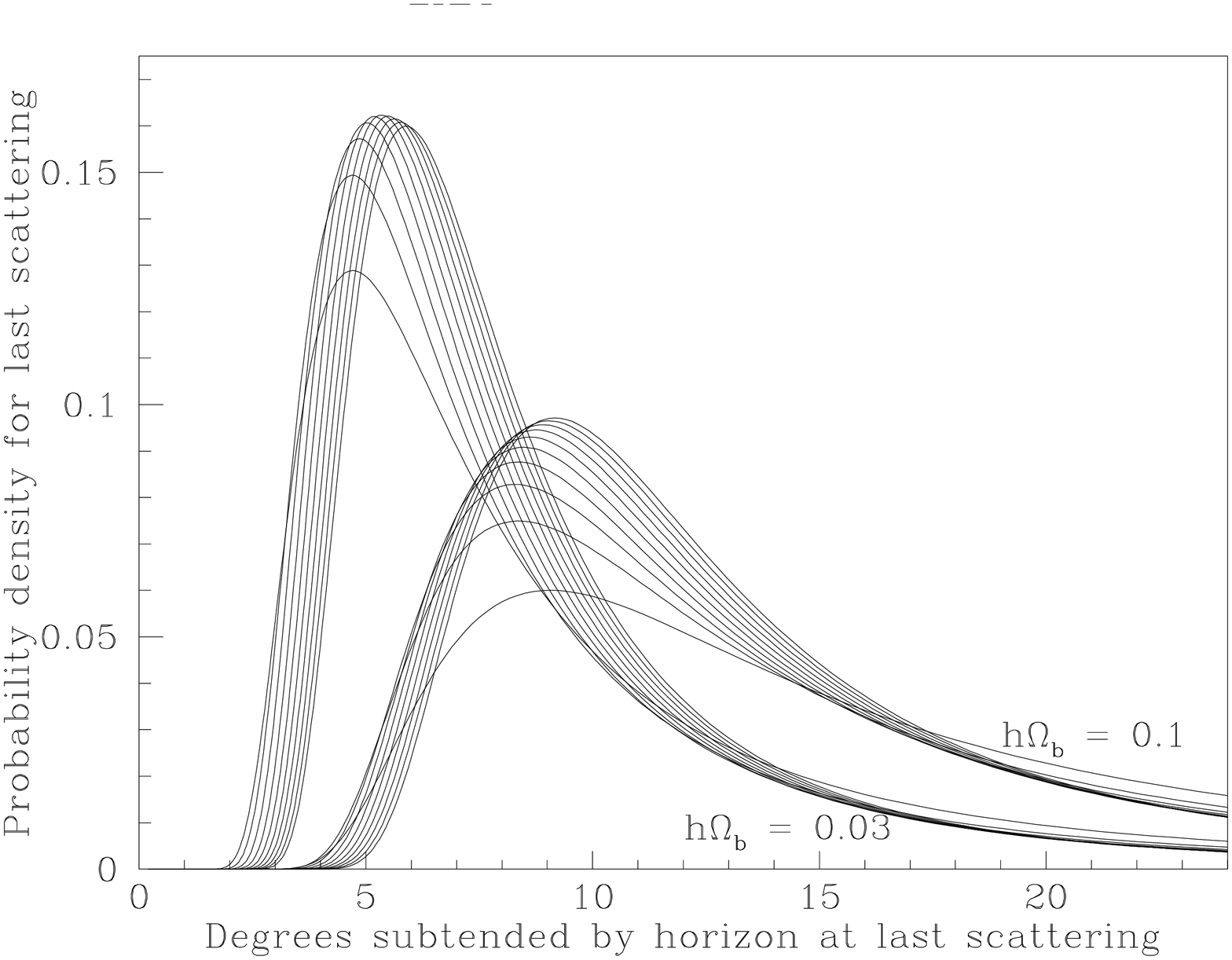,width=\fwidth,height=\fheight}
\nobreak
\caption{Visibility functions.}
\label{openreionfig4}
 
\mycaption{The angular visibility function for 
a fully ionized universe is 
plotted for different values of $\Omega_0$ and
diffuse baryon content $h\Ob$.
The left group of curves corresponds to $h\Ob = 0.03$ and
the right to $h\Ob = 0.1$.
Within each group, from left to right starting at the lowest
peak,
$\Omega_0 = $0.1, 0.2, 0.3, 0.4, 0.5, 0.6, 0.7, 0.8, 0.9 and 1.0.
}
\efig

\bfig
\psfig{figure=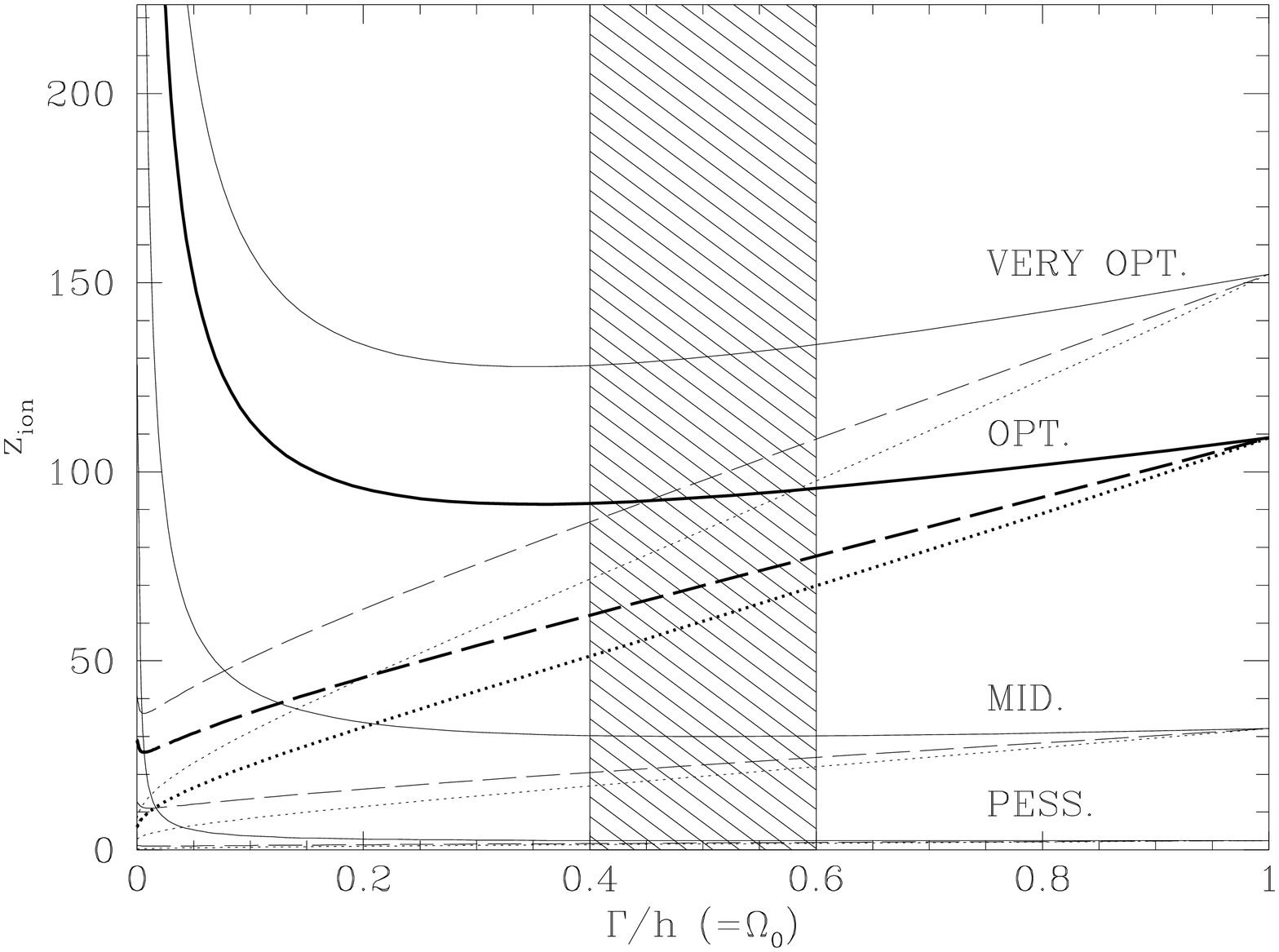,width=\fwidth,height=\fheight}
\nobreak
\caption{The ionization redshift.}
\label{openreionfig5}
 
\mycaption{The ionization redshift is plotted for 
the four scenarios described in 
Table~\ref{openreiontable1}.
The solid lines correspond to the open case where $\lambda_0 = 0$.
The dashed lines correspond to the flat case where 
$\lambda_0 = 1-\Omega_0$.
The dotted lines correspond to the $\tau$CDM model, where
$\lambda_0 = 0$ and $\Omega_0 = 1$. 
Note that the combination $\Gamma/h$ is really an $h$-independent 
quantity: for the open and flat cases, it is simply equal to 
$\Omega_0$, and for the $\tau$CDM case it depends only on 
$g_*$, the number of 
relativistic particle species.
}
\efig

\bfig
\psfig{figure=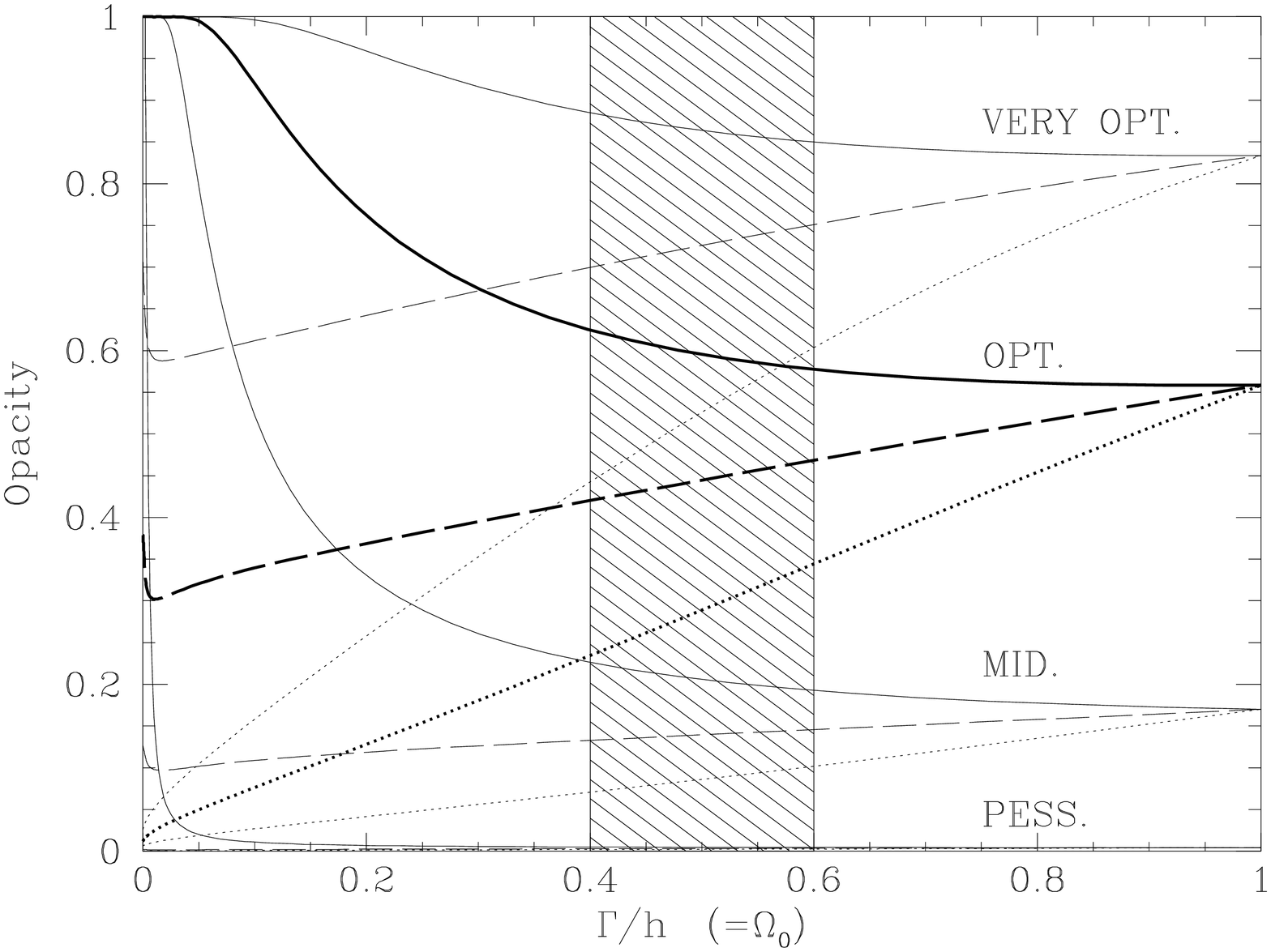,width=\fwidth,height=\fheight}
\nobreak
\caption{The opacity.}
\label{openreionfig6}
 
\mycaption{The opacity, the probability that a 
CBR photon is Thomson scattered
at least once since since the standard recombination epoch, is
plotted for the four scenarios described in 
Table~\ref{openreiontable1}.
The solid lines correspond to the open case where $\lambda_0 = 0$.
The dashed lines correspond to the flat case where 
$\lambda_0 = 1-\Omega_0$.
The dotted lines correspond to the $\tau$CDM model, where
$\lambda_0 = 0$ and $\Omega_0 = 1$. 
Just as in Figure~\ref{openreionfig5},
note that the combination $\Gamma/h$ is really an $h$-independent 
quantity: for the open and flat cases, it is simply equal to 
$\Omega_0$, and for the $\tau$CDM case it depends only on 
$g_*$, the number of 
relativistic particle species.
}
\efig

\bfig
\psfig{figure=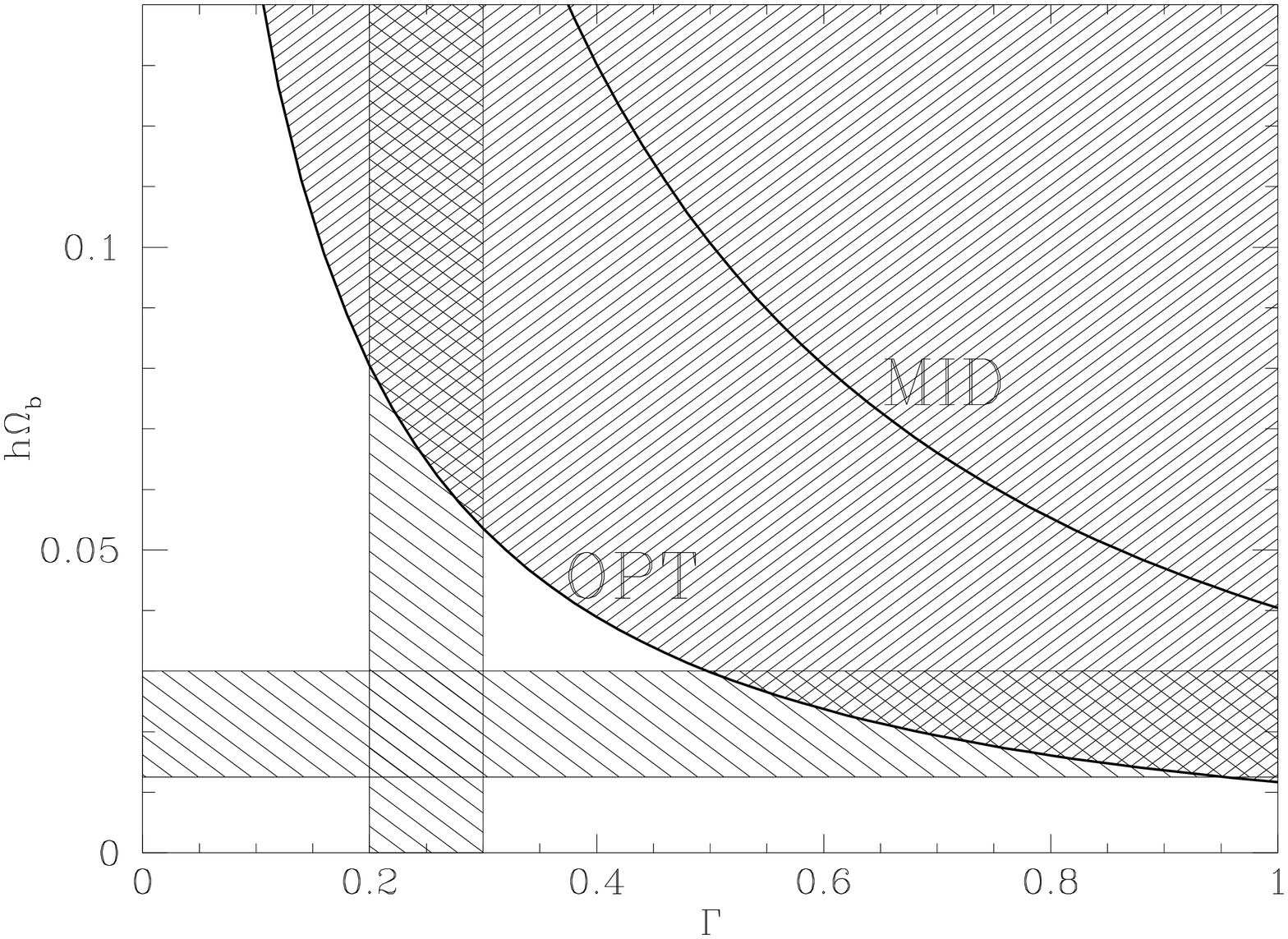,width=\fwidth,height=\fheight}
\nobreak
\caption{Reionization in $\tau$CDM.}
\label{openreionfig7}
 
\mycaption{The two 
curves show the baryon density required for $50\%$ opacity in
$\tau$CDM,
{\it i.e.} for reionization
to rescatter $50\%$ of the CBR photons.
The upper and lower heavy curves correspond to the
middle-of-the road and optimistic
scenarios, respectively. Thus even in the optimistic scenario,
$50\%$ opacity cannot be obtained outside of the fine-hatched region.
The horizontal shaded region corresponds to the values of
$h\Omega_b$ allowed by standard nucleosynthesis
($0.01 < h^2\Omega_b < 0.015$) in conjunction with the constraint
$0.5<h<0.8$.
The vertical shaded region corresponds to values of the ``shape
parameter" $\Gamma$ preferred by power spectrum measurements.
Note that these three regions do not intersect.
}
\efig

\cleardoublepage

\chapter{Did the Universe Recombine?}
\label{ychapter}
\def\i{x}

% DID THE UNIVERSE RECOMBINE? NEW SPECTRAL CONSTRAINTS ON REHEATING

In this chapter we will show that  
one still cannot conclusively assert that the
universe underwent a neutral phase, despite the new COBE FIRAS
limit $y <2.5\times 10^{-5}$ on
Compton $y$-distortions of the cosmic microwave background.
Although scenarios where the very early ($z\sim 1000$) ionization is
thermal (caused by IGM temperatures 
exceeding $10^4$K) are clearly
ruled out, there is a significant loophole for cosmologies with
typical CDM parameters if the dominant ionization mechanism is
photoionization.
If the ionizing radiation has a typical quasar spectrum, 
then the $y$-constraint implies roughly 
$h^{4/3}\Ob \Omega_0^{-0.28}<0.06$ for fully ionized models.
This means that BDM models with $\Omega_0\approx 0.15$ and reionization
at  $z\approx 1000$ are strongly constrained even in this very
conservative case, and can survive the $y$ test only if
most of the baryons form BDM around the reionization epoch. 

\section{Introduction}

Recombination of the primeval plasma is 
commonly assumed but was by no means inevitable. 
Theories exist that predict early reionization are as diverse as 
those invoking primordial seed fluctuations that underwent early 
collapse and generated sources of ionizing 
radiation, and models involving decaying or annihilating particles. 
The former class includes cosmic strings and 
textures, as well as primordial isocurvature baryon 
fluctuations. The latter category includes baryon 
symmetric cosmologies as well as decaying 
exotic particles or neutrinos.

The Compton $y$-distortion of the cosmic microwave 
background (CBR) provides a unique constraint on 
the epoch of reionization. In view of the extremely 
sensitive recent FIRAS limit of 
$y < 2.5\times 10^{-5}$, we have reinvestigated 
constraints on the early ionization history of the 
intergalactic medium (IGM), and have chosen to 
focus on what we regard as the most important of the 
non-standard recombination history models, namely 
the primordial isocurvature baryon scenario  involving a universe dominated by
baryonic dark matter (BDM), 
as advocated by 
Peebles (1987); Gnedin \&
Ostriker (1992) (hereafter ``GO");  Cen, Ostriker \&
Peebles (1993) and
others.  This class of models takes the simplest matter content 
for the universe, namely baryons, to constitute dark 
matter in an amount that is directly observed and is 
even within the bounds of primordial nucleosynthesis, 
if interpreted liberally, and can reconstruct 
essentially all of the observed phenomena that constrain 
large-scale structure. The BDM model is a non-starter 
unless the IGM underwent very early reionization, in 
order to avoid producing excessive CBR fluctuations 
on degree scales. Fortunately, early nonlinearity is 
inevitable with BDM initial conditions, 
$\delta\rho/\rho\propto M^{-5/12}$, corresponding to 
a power-spectrum $\expec{\delta_k^2}\propto k^{-1/2}$
for the observationally preferred choice of spectral index (Cen,
Ostriker \& Peebles 1993).

Is it possible that the
IGM has been highly ionized since close to the  
standard recombination epoch at $z\approx
1100$? Perhaps the most carefully studied BDM scenario in 
which this happens is that by GO. In their scenario, 
$\Oz=\Omega_{b0}\approx 0.15$. Shortly after recombination, a
large fraction of the mass condenses into faint stars or massive black
holes, releasing energy that reionizes the universe and heats it to
$T>10,000\K$ by $z=800$, so Compton scattering off of hot electrons
causes strong spectral distortions in the cosmic microwave background.
The models in GO give a Compton $y$-parameter between 
$0.96\times 10^{-4}$ and $3.1\times 10^{-4}$, and are thus all ruled
out by the most recent observational constraint from the COBE FIRAS
experiment, $y<2.5\times 10^{-5}$ (Mather {\etal} 1994).

There are essentially four mechanisms that can heat
the IGM sufficiently to produce Compton $y$-distortions:

\begin{itemize}

\item Photoionization heating from UV photons 
(Shapiro \& Giroux 1987; Donahue \& Shull 1991)

\item Compton heating from UV photons 

\item Mechanical heating from supernova-driven winds 
(Schwartz {\etal} 1975; Ikeuchi 1981; Ostriker \& Cowie 1981)

\item Cosmic ray heating 
(Ginzburg \& Ozernoi 1965)

\end{itemize}

\noindent
The second effect tends to drive the IGM temperature towards
two-thirds of the temperature of the ionizing radiation, whereas
the first effect tends to drive the temperature towards a lower value
$T^*$ that will be defined below. The third and fourth
effect can produce
much higher temperatures, often in the millions of degrees. 
The higher the
temperature, the greater the $y$-distortion.

In the GO models, the second effect dominates, which is why they fail so
badly. 
In this chapter, we wish to place limits that are
virtually impossible to evade. 
Thus we will use the most cautions assumptions possible, and 
assume that the
latter three heating mechanisms are negligible.

\section{The Compton $y$-Parameter}

Thomson scattering between CBR photons and hot electrons affects the
spectrum of the CBR. It has long been known that hot ionized IGM
causes spectral distortions to the CBR, known as the Sunyaev-Zel'dovich
effect. A useful measure of this distortion is the  Comptonization
$y$-parameter (Kompan\'eets 1957;  
Zel'dovich \& Sunyaev 1969;
Stebbins \& Silk 1986; Bartlett \& Stebbins 1991)
\beq{yDefEq}
y = \int\left({k\Te-k\Tp\over m_e c^2}\right) n_e\st c\> dt
= y^*\int{(1+z)\over\sqrt{1+\Oz z}}\DT_4(z)x(z)dz,
\eeq
where
$$y^* \equiv
\left[1 - \left(1-{x_{He}\over 4x}\right)Y\right] 
\left(k\times10^4\K\over m_e c^2\right)\left({3H_0\Ob\st c\over 8\pi G
m_p}\right) \aet{9.58}{8} h\Ob.$$
Here $\Te$ is the electron temperature, 
$\Tp$ is the CBR temperature,
$\DT_4\equiv(\Te-\Tp)/10^4\K$, $\Omega_{igm}$ is the fraction of 
critical density in 
intergalactic medium,
and $\i(z)$ is the fraction of the hydrogen that
is ionized at redshift $z$. 
Note that we may have
$\Ob\ll\Omega_b$,
{\ie} all baryons may not be in diffuse form.
The integral is to be taken from the reionization epoch to today.
In estimating the electron density $n_e$, 
we have taken the mass fraction of
helium to be  $Y\approx 24\%$ and assumed $x_{He} \approx x$,
{\ie} that helium never becomes doubly ionized and that the fraction
that is singly ionized equals the fraction of hydrogen that is
ionized. The latter is a very crude approximation, but makes a
difference of only $6\%$. 

Let us estimate this integral by making the approximation that the IGM 
is cold and neutral until a redshift $z_{ion}$, at which it suddenly 
becomes ionized, and after which it 
remains completely ionized with a constant temperature $\Te$.
Then for $\zion\gg 1$ and $\Te\gg  z_{ion}\times 2.7\K$ we obtain 
$$y \aet{6.4}{-8} h\Ob\Oz^{-1/2} T_4\> z_{ion}^{3/2},$$
where $T_4\equiv \Te/10^4\K$.
Substituting the most recent observational constraint 
from the COBE FIRAS experiment,
$y < 2.5\times 10^{-5}$ (Mather {\etal} 1994),
into this expression yields
\beq{zionLimitEq}
\zion < 554T_4^{-2/3}\Oz^{1/3}\left({h\Ob\over 0.03}\right)^{-2/3}.
\eeq
Thus the only way to have $\zion$ as high as $1100$ is to have temperatures
considerably below $10^4\K$. In the following
section, we will see to what extent this is possible.

\section{IGM Evolution in the Strong UV Flux Limit}

In this section, we will calculate the thermal evolution of 
IGM for which

\begin{itemize}

\item the IGM remains almost completely ionized at all times,

\item the Compton $y$-distortion is minimized given this constraint.

\end{itemize}

\subsection{The ionization fraction}

In a homogeneous IGM at temperature $T$ exposed to a
density of $\upp$ UV photons of energy {$h\nu>13.6\,\eV$} per proton,
the ionization fraction $\i$ evolves as follows: 
\beq{5IonizationEq}
 {d\i\over d(-z)} = 
{1+z\over\soz}
\left[\lpi(1-\i) + \lci\i(1-\i) - \lrec\i^2\right],
\eeq
where $H_0^{-1}(1+z)^{-3}$ times the rates per baryon for
photoionization, collisional ionization and recombination are given by
\beq{5RateEq}
\cases{
\lpi \aet {1.04}{12} \left[h\Ob\uvsigfid\right]\upp,&\crr
\lci \aet{2.03}{4} h\Ob T_4^{1/2} e^{-15.8/T_4,}&\crr
\lrec \approx 0.717 h\Ob 
T_4^{-1/2}\left[1.808-0.5\ln T_4 + 0.187 T_4^{1/3}\right],&
}
\eeq
and $T_4\equiv \Te/10^4\K$. 
Here $\uvsigfid$ is the spectrally-averaged 
photoionization cross section in units of $10^{-18}\cm^2$.
The differential cross section is given by
(Osterbrock 1974)
\beq{SigmaEq}
{d\uvsigfid\over d\nu}(\nu) \approx 
\cases{
0
&if $\nu < 13.6\,\eV$,\cr
6.30{e^{4-4\arctan(\epsilon)/\epsilon}\over
\nu^4\left(1-e^{-2\pi/\epsilon}\right)}
&if $\nu\ge 13.6\,\eV$,
}
\eeq
where 
$$\epsilon\equiv\sqrt{{h\nu\over 13.6\,\eV}-1}.$$
The recombination rate is the total to all hydrogenic levels
(Seaton 1959; Spitzer 1968). 
Recombinations directly to the ground
state should be included here, since as will become evident below,
the resulting UV photons are outnumbered by the UV photons 
that keep the
IGM photoionized in the first place, and thus can be neglected when
determining the equilibrium temperature.

At high redshifts, the ionization and recombination rates greatly exceed the
expansion rate of the universe, and the ionization level quickly
adjusts to a quasi-static equilibrium value for which the expression in
square brackets in \eq{5IonizationEq} 
vanishes. In the
absence of photoionization, an ionization fraction $x$ close to
unity requires $\Te>15,000\K$. Substituting this into
\eq{zionLimitEq} gives consistency with
$z_{ion}>1000$ only if $h\Ob < 0.008$, a value clearly inconsistent with
the standard nucleosynthesis constraints (Smith {\etal}
1993). Thus any reheating scenario that relies on collisional
ionization to keep the IGM ionized at all times may be considered
ruled out by the COBE FIRAS data.  

However, this does not rule out all ionized universe scenarios,
since photoionization can achieve the same
ionization history while causing a much smaller $y$-distortion.
The lowest temperatures (and hence the smallest
$y$-distortions) compatible with high ionization will be
obtained when the ionizing flux is so strong that 
$\lpi\gg\lci$.
In this limit, to a good
approximation, \eq{5IonizationEq} can be replaced by the
following simple model for the IGM:
 
\begin{itemize}

\item It is completely ionized ($\i=1$).

\item When a neutral hydrogen atom is formed
through recombination, it is instantly photoionized again.

\end{itemize}

\noindent
Thus the only unknown parameter is the IGM temperature $\Te$, which
determines the recombination rate, which in turn equals the
photoionization rate and thus determines the rate of heating.

\subsection{The spectral parameter $\Tpi$}

The net effect of a recombination and subsequent photoionization is to
remove the kinetic energy ${3\over 2}kT$ from the plasma and replace it with
the kinetic energy ${3\over 2}kT^*$, where $\Tpi$ is defined by
${3\over 2}k\Tpi\equiv\euv - 13.6\,\eV$ and $\euv$ is the average energy
of the ionizing photons.
Thus the higher the recombination rate, the faster this effect will tend to
push the temperature towards $\Tpi$. 

The average energy of
the ionizing photons is given by the spectrum $P(\nu)$ as
$\euv = h\expec{\nu}$,
where  $$\expec{\nu} = 
{\izi P(\nu)\sigma(\nu) d\nu\over
 \izi \nu^{-1}P(\nu)\sigma(\nu) d\nu}.$$
Here $\sigma$ is given by \eq{SigmaEq}.
Note that, in contrast to certain nebula calculations where all photons get
absorbed sooner or later, the spectrum should be weighted by the
photoionization cross section. This is because most photons never get
absorbed, and all that is relevant is the energy distribution of those
photons that do. Also note that $P(\nu)$ is the energy distribution
($W/Hz$), not the number distribution which is proportional to 
$P(\nu)/\nu$.

\begin{table}
$$
\begin{tabular}{|llrr|}
\hline
UV source&Spectrum $P(\nu)$&$\euv$&$\Tpi$\\
\hline
O3 star&$T=50,000$K Planck&17.3\,eV&28,300K\\
O6 star&$T=40,000$K Planck&16.6\,eV&23,400K\\
O9 star&$T=30,000$K Planck&15.9\,eV&18,000K\\
Pop. III star&$T=50,000$K Vacca&18.4\,eV&36,900K\\
Black hole, QSO&$\alpha=1$ power law&18.4\,eV&37,400K\\
?&$\alpha=2$ power law&17.2\,eV&27,800K\\
?&$\alpha=0$ power law&20.9\,eV&56,300K\\
?&$T=100,000$K Planck&19.9\,eV&49,000K\\
\hline
\end{tabular}
$$
\caption{Spectral parameters}
\label{ytable1}
\end{table}

The spectral parameters $\euv$ and $\Tpi$ are given in 
Table~\ref{ytable1}
for some selected spectra (this is merely a subset of 
Table~\ref{reiontable3}).
A power law spectrum $P(\nu)\propto
\nu^{-\alpha}$ with $\alpha=1$ fits observed QSO spectra rather well
in the vicinity of the Lyman limit (Cheney \& Rowan-Robinson 1981;
O'Brien {\etal} 1988), and is also consistent with the standard model
for black hole accretion. A Planck spectrum 
$P(\nu)\propto\nu^3/\left(e^{h\nu/kT}-1\right)$
gives a decent prediction of $T^*$ for stars with surface temperatures
below $30,000\K$. For very hot stars, more realistic spectra (Vacca 1993)
fall off much slower above the Lyman limit, thus giving higher values of 
$T^*$. As seen in 
Table~\ref{ytable1},
an extremely metal poor star of surface
temperature $50,000\K$ gives roughly the same $T^*$ as QSO radiation.
The only stars that are likely to be relevant to early
photoionization scenarios are extremely hot and short-lived ones, since 
the universe is less than a million years old at $z = 1000$, and fainter
stars would be unable to inject enough energy in so short a time.
Conceivably, less massive stars could play a the dominant role later on,
thus lowering $T^*$. However, since they radiate such a small fraction of
their energy above the Lyman limit, very large numbers would be needed,
which could be difficult to reconcile with the absence of observations of
Population~III stars today.

\subsection{The thermal evolution}

At the low temperatures involved, the two dominant cooling 
effects\footnote
{
Another cooling mechanism is collisional excitation of atomic hydrogen
followed by radiative de-excitation, which cools the IGM at a rate of
(Dalgarno \& McCray 1972) $$\hce \approx
7.5\tt{-19}  e^{-11.8/T_4} n^2 (1-x)x  \>\erg\, \cm^{-3} \s^{-1}.$$
The ratio of this cooling rate to the Compton cooling rate 
is 
$${\hce\over\hcomp}\approx
{\exp\left[9.2-11.8/T_4\right] (1-x)\over(1+z)T_4} h^2\Ob,$$
a quantity which is much smaller than unity for any reasonable
parameter values when $T < 10^4\K.$  
As will be seen, the temperatures
at $z\approx 1000$ are typically a few thousand K,
which with
$h^2\Ob < 0.1$ and $x>0.9$ renders collisional excitation cooling
more than nine orders of magnitude weaker than Compton cooling. 
Hence
we can safely neglect collisional excitations when computing the IGM
temperature, the reason essentially
being that the temperatures are so low that this process is suppressed
by a huge Boltzman factor.
}
are
Compton drag against the microwave background photons and cooling due to the
adiabatic expansion of the universe. Combining these effects, we obtain the
following equation for the thermal evolution of the IGM:
\beq{5Teq}
{dT\over d(-z)} = -{2\over 1+z}T + 
{1+z\over\soz}\left[\lcomp(\Tp-T) +
{1\over 2}\lrec(T)(\Tpi-T)\right],
\eeq
where
$$\lcomp = {4\pi^2\over 45}
\left({k\Tp\over\hbar c}\right)^4
{\hbar\st\over H_0 m_e}(1+z)^{-3}
\approx 0.00417 h^{-1}(1+z)$$
is $(1+z)^{-3}$ times the Compton cooling rate per Hubble time
and $\Tp = {\Tp}_0(1+z)$.
The factor of ${1\over 2}$ in front of $\lambda_{rec}$ is due to the
fact that the photoelectrons end up sharing their energy with the
protons. We have taken ${\Tp}_0\approx 2.726\K$ (Mather
{\etal} 1994). Numerical solutions to this equation are shown in
Figure~\ref{yfig1},
and the resulting $y$-parameters are given in
Table~\ref{ytable2}.

\begin{table}
$$
\begin{tabular}{|l|rrrrrcl|}
\hline
Model&$\Oz$&$\Ob$&h&$\Tpi$&$z_{ion}$&$y/0.000025$&Verdict\\
\hline
QSO BDM I&0.15&0.15&0.8&37,400K&1100&6.30&Ruled  out\\
QSO BDM II&0.15&0.15&0.8&37,400K&200&1.43&Ruled  out\\
O9 BDM&0.15&0.15&0.8&18,000K&800&2.91&Ruled out\\
QSO BDM III&0.15&0.04&0.8&37,400K&1100&0.67&OK\\
QSO CDM I&1&0.06&0.5&37,400K&1100&0.17&OK\\
QSO CDM II&1&0.03&0.8&37,400K&1100&0.16&OK\\
\hline
\end{tabular}
$$
\caption{Compton $y$-parameters for various scenarios}
\label{ytable2}
\end{table}

The temperature
evolution separates into two distinct phases. In the first phase,  
which is almost instantaneous
due to the high recombination rates at low temperatures, 
$T$ rises very rapidly, up to a quasi-equilibrium temperature slightly
above the temperature of the microwave background photons.
After this, in the second phase, $T$ changes only slowly,
and is approximately given by setting the expression
in square brackets in \eq{5Teq}  
equal to zero.
This quasi-equilibrium temperature is typically much lower
than $T^*$, since Compton cooling is so efficient at the high 
redshifts involved, and is given by 
\beq{DTeq}
\DT\equiv\Te-\Tp
\approx {\lrec\over 2\lcomp}(\Tpi-\Te)
\propto  
{1\over 1+z} g(\Te)h^2 \Ob (\Tpi-T),
\eeq
independent of $\Oz$,  
where $g(\Te)\simpropto T^{-0.7}$ encompasses the
temperature dependence of $\lrec$. 
We typically have $T\ll\Tpi$.     
Using this, making the crude approximation 
of neglecting the temperature dependence of $\lrec$,
and substituting \eq{DTeq} into 
\eq{yDefEq}
indicates that    
$$y\simpropto h^3\Ob^2\Omega_0^{-1/2} T_4^* z_{ion}^{1/2}.$$
Numerically selecting the best power-law fit, we find that this is
indeed not too far off: 
the approximation
\beq{yApproxEq}
y\approx 0.0012 h^{2.4}\Ob^{1.8}\Oz^{-1/2}(T^*_4)^{0.8}
(z_{ion}/1100)^{0.9}
\eeq
is accurate to about $10\%$ within the parameter range of cosmological
interest.
We have used \eq{yApproxEq} in
Figure~\ref{yfig3}
by setting  $y=2.5\times 10^{-5}$ and $z_{ion}=1100$. The shaded
region of parameter space is thus ruled out by the COBE FIRAS experiment
for fully ionized scenarios.

\section{Conclusions}

A reanalysis of the Compton $y$-distortion
arising from early reionization
shows that despite the radical sharpening of 
the  FIRAS limit on $y$, 
one still cannot conclusively assert that the universe underwent
a  neutral phase. 
Non-recombining scenarios where the ionization is
thermal, caused by IGM temperatures exceeding $10^4\K$, are clearly
ruled out. Rather, the loophole 
is for the dominant ionization mechanism to be photoionization.
We have shown that for spectra characteristic of both QSO radiation
and massive metal-poor stars, the resulting IGM
temperatures are so low that typical CDM models with no recombination
can still survive the FIRAS test by a factor of six. 
This conclusion is valid if the flux of ionizing radiation
is not so extreme that Compton heating becomes important. This
is not difficult to arrange, as 
the cross section for Thomson scattering is some
six orders of magnitude smaller than that for photoionization.

For BDM models, the constraints are sharper. 
Non-recombining ``classical" BDM models with
$\Omega_{igm}=\Omega_0\approx 0.15$ are ruled out even with the
extremely cautious reheating assumptions used in this chapter, the
earliest ionization redshift allowed being  $z \approx 130$.
Such models involving early non-linear 
seeds that on energetic grounds can 
very plausibly provide a photoionization 
source capable of reionizing the universe 
soon after the period of first recombination 
inevitably generate Compton distortions 
of order $10^{-4}$. These include 
texture as well as BDM models, both
of which postulate, and indeed require, 
early reionization ($z > 100$) to 
avoid the generation of excessive 
anisotropy in the cosmic microwave background 
on degree angular scales.

Thus BDM models with reionization at $z\approx 1000$ 
can survive the
$y$ test only if most of the baryons form BDM when reionization
occurs, and are thereby removed as a source of 
$y$-distortion, at least in the diffuse phase.
This may be difficult to arrange
at $z>100,$ since once the matter is reionized  
at this high a redshift, Compton drag is extremely 
effective in inhibiting any further gas collapse until 
$z<100.$ Since it takes only a small fraction 
of the baryons in the universe to provide a source 
of photons sufficient to maintain a fully 
ionized IGM even at $z\sim 1000,$ we suspect 
that most of the baryons remain diffuse 
until Compton drag eventually becomes ineffective. 
Moreover, the possibility that the IGM is only 
partially reionized at $z\sim 1000$ ({\it e.g.} GO), 
a situation which allows a lower value of the  
$y$-parameter, seems to us to be implausible 
as a delicate adjustment of ionization and 
recombination time-scales over a considerable 
range in $z$ would be required.  A 
complementary argument that greatly restricts 
the parameter space allowable for fully 
ionized BDM models appeals to temperature 
fluctuations induced on the secondary last 
scattering surface, both by first order 
Doppler terms on degree scales and by 
second order terms on subarcminute 
scales (Hu {\etal} 1994).
Thus, BDM models would seem to be in 
some difficulty because of the low 
limit on a possible $y$-distortion.

\def\fheight{10.3cm} \def\fwidth{14.5cm}

\newpage

\bfig
\psfig{figure=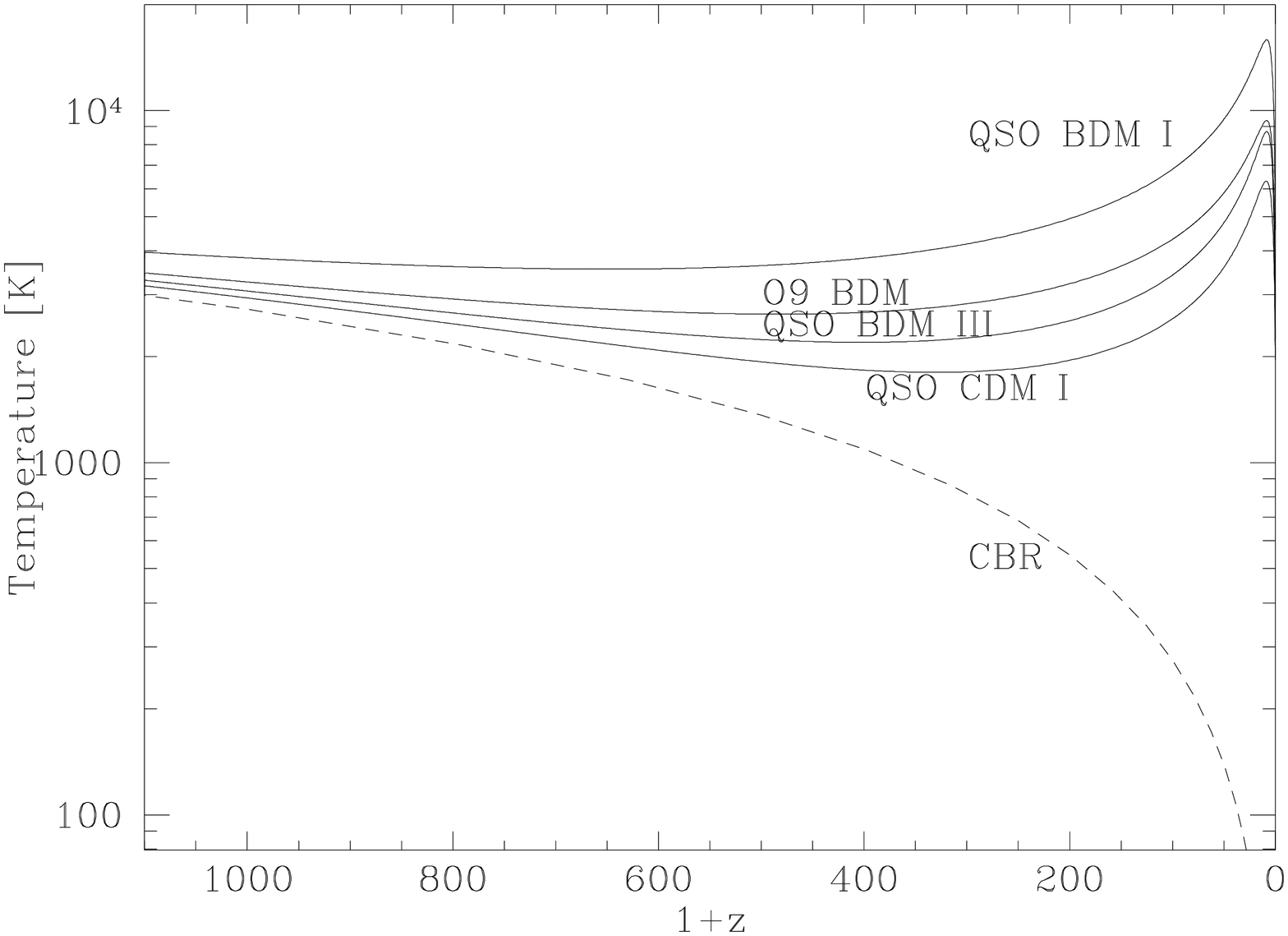,width=\fwidth,height=\fheight}
\nobreak
\caption{Thermal histories for various models.}
\label{yfig1}
 
\mycaption{The temperature of the 
photoionized IGM is plotted for four of the
cosmological models and spectra of ionizing radiation listed in
Table~\ref{ytable2}. 
The lowermost curve gives the temperature of the CBR photons.
}
\efig

\bfig
\psfig{figure=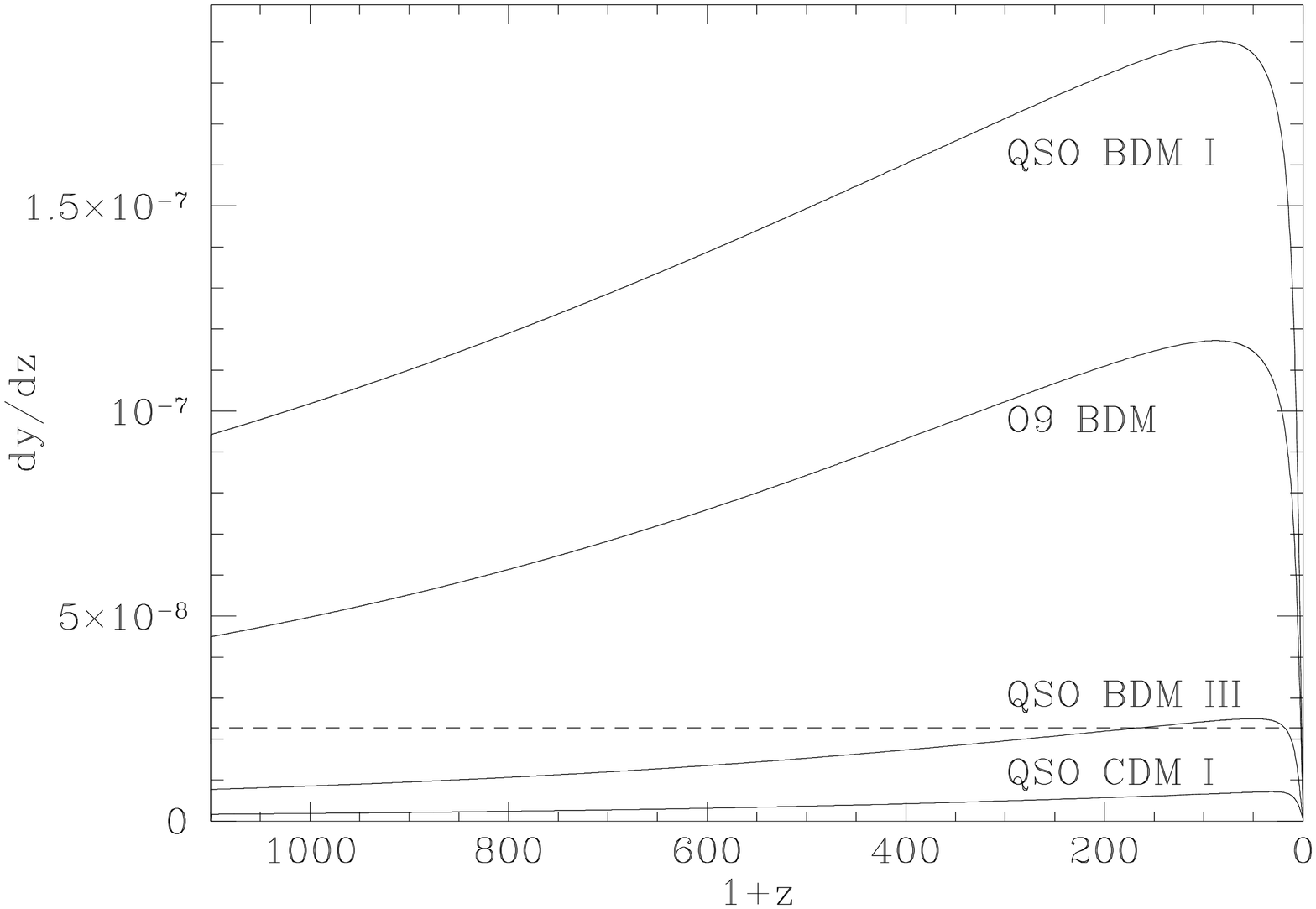,width=\fwidth,height=\fheight}
\nobreak
\caption{$dy/dz$ for various models}
\label{yfig2}
 
\mycaption{The contribution to the 
y-parameter from different redshifts is
plotted four of the cosmological models and spectra of ionizing
radiation listed in 
Table~\ref{ytable2}. 
Thus for each model, the area under
the curve is the predicted y-parameter. The area under the 
horizontal dashed line is $2.5\times 10^{-5}$, {\it i.e.} the COBE
FIRAS limit.} 
\efig

\bfig
\psfig{figure=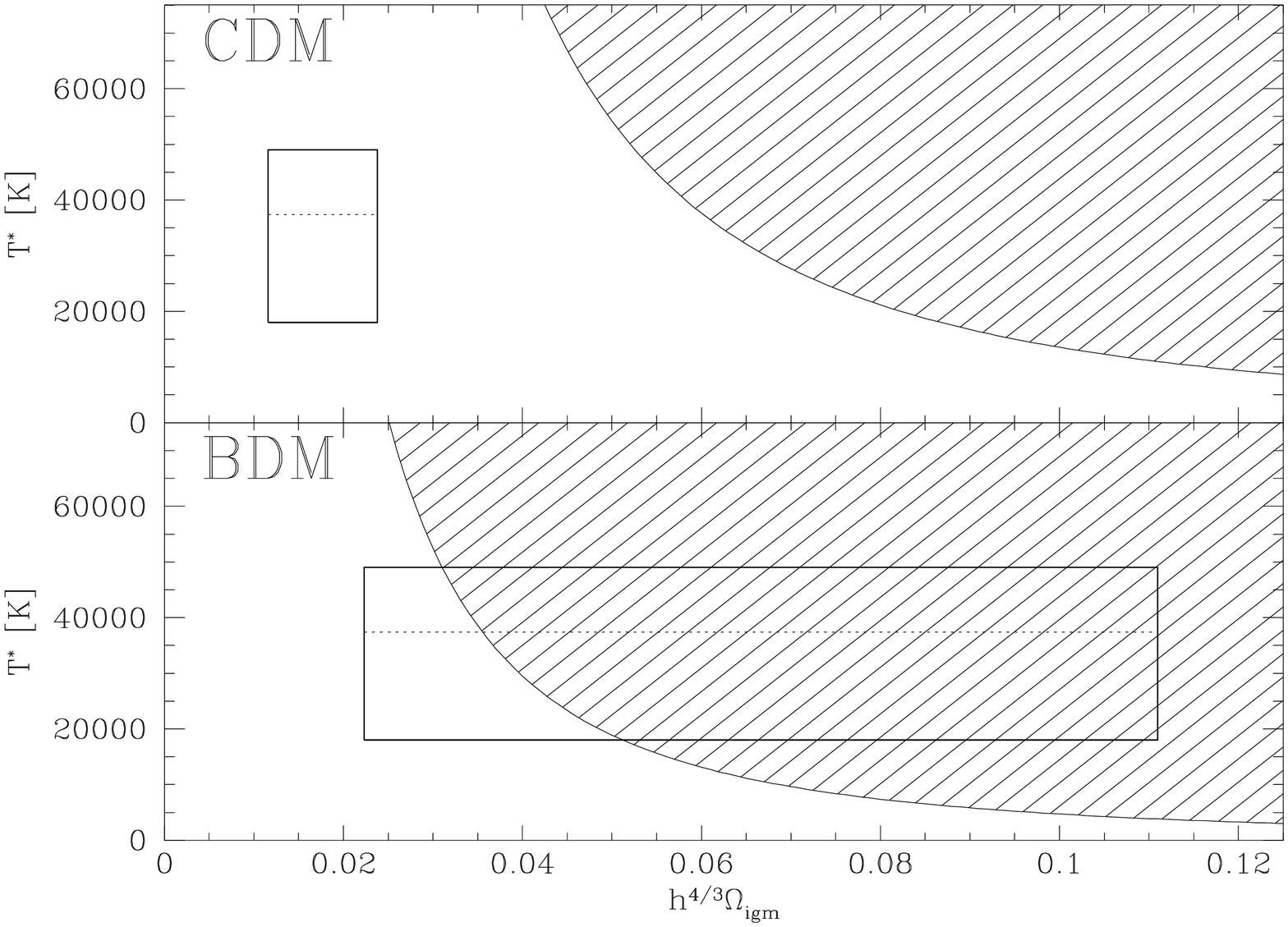,width=\fwidth,height=\fheight}
\nobreak
\caption{Predicted and ruled out regions of parameter space.}
\label{yfig3}
 
\mycaption{The hatched regions of parameter 
space are ruled out by the
the COBE FIRAS limit {$y<2.5\times
10^{-5}$} for $z_{ion}=1100$. $\Omega_0 = 1$ in the CDM plot and
$\Omega_0 = 0.15$ in the BDM plot.
The rectangular regions are the assumed parameter values
for the CDM and BDM models, respectively.
For CDM, the range $0.012 <  h^{4/3}\Omega_{igm} < 0.024$ is given by
the nucleosynthesis constraint $0.010 < h^2\Omega_{b} < 0.015$ and the
assumption that $0.5 < h < 0.8$. (If $\Omega_{igm} < \Omega_b$, the
rectangle shifts to the left.)
For the BDM models, $h=0.8$ and $0.03\leq\Omega_{igm}\leq\Omega_0$.
The vertical range corresponds to
feasible values of the spectral parameter
$T^*$.
The upper limit corresponds to highly speculative star with
surface temperature $100,000\K$ and $T^*=49,000\K$. The lower line
corresponds to an O9 star.
The dotted horizontal line
corresponds to the spectrum expected from quasars/accreting black
holes.
} 
\efig

\cleardoublepage
\chapter{Late Reionization by Supernova-Driven Winds}
\label{gpchapter}

% Late Reionization by Supernova Driven Winds

% \countdef\RadialEq=221
% \countdef\EqOfState=222
% \countdef\EconsEq=223
% \countdef\pdotEq=224
% \countdef\MainEq=225
% \countdef\SmallTimeSolEq=226
% \countdef\EtEq=227
% \countdef\ffEq=228
% \countdef\ffEqTwo=229
% \countdef\dTdzEq=230
% \countdef\LcompEq=231
% \countdef\IGMtempEq=232
% \countdef\EquilibrionizationEq=233
% \countdef\fgEq=234
% \countdef\FilterEq=235

\def\Ob{\Omega_b}
\pagestyle{myheadings}
\markboth{CHAPTER 6: LATE REIONIZATION...}{6.1. INTRODUCTION}

In this chapter, a model is presented in which supernova-driven winds
from early galaxies reionize the intergalactic medium by $z=5$.
This scenario can explain the observed absence of a
Gunn-Peterson trough in the spectra of high-redshift
quasars
providing that the bulk of these early galaxies are quite small,
no more massive than about $10^8 M_{\odot}$. It also
predicts that
most of the IGM was enriched to at least $10\%$ of current
metal content by $z=5$ and perhaps as early as $z=15$.
The existence of such
early mini-galaxies violates no spectral constraints
and is
consistent with a pure CDM model with $b\leq 2$. Since the final
radius of a typical ionized bubble is only around 100 kpc, the
induced modification of the galaxy autocorrelation function is
negligible, as is the induced angular smoothing of the CBR.
Some of the gas swept up by shells may be observable as
pressure-supported Lyman-alpha forest clouds.

\markboth{CHAPTER 6: LATE REIONIZATION...}{6.1. INTRODUCTION}
\section{Introduction}
\label{gpsec1}

The absence of a Gunn-Peterson trough in the spectra of
high-redshift quasars has provided strong
evidence for the intergalactic medium (IGM) being highly ionized as
early as $z=4$ (Gunn \& Peterson 1965; Steidel and Sargent
1987; Webb {\etal} 1992).  The hypothesis that
photoionization of the IGM by quasars could
account for this ionization (Arons \& McCray 1969; Bergeron \&
Salpeter 1970; Sherman 1980) has been challenged
(Shapiro 1986; Shapiro \& Giroux 1987;  Miralda-Escude \& Ostriker
1990). 
Other studies have
maintained that photoionization by quasars (Donahue \& Shull 1987)
or active galactic nuclei (Teresawa 1992) may nonetheless be
sufficient. However, in view of the large uncertainties in crucial
parameters such as ionizing fluxes, the issue of what reionized the
IGM must still be considered open.

In comparing the Gunn-Peterson constraints with our work in Chapters 3 and 4,
the crucial difference is the degree of ionization required. 
To affect the CBR, it does not really matter whether
the ionization fraction $x$ is $90\%$ or $99.999\%$, as this 
makes only a $10\%$ difference in the optical depth $\tau$.
The Gunn-Peterson limits constrain not $x$ but
$(1-x)$, the neutral fraction. Thus in this context, the difference between
$90\%$ and $99.999\%$ ionization is four orders of magnitude.
In Chapters 3 and 4, we found that photoionization
by early galaxies could easily reionize the IGM by a redshift 
$z=5$, but the issue here is whether photoionization alone can
provide the extremely high ionization fraction required to pass the Gunn-
Peterson test. 

In this chapter, we investigate an alternative  
reionization scenario, which produces considerably higher IGM
temperatures than those attained by the photoionization models in 
previous chapters. 
Supernova driven winds from luminous galaxies have long been conjectured
to be an important ionization source for the IGM (Schwartz
{\etal} 1975; Ikeuchi \& Ostriker 1986; Carlberg \& Couchman 1989).
Cold dark matter (CDM)-based models of structure formation
(Blumenthal {\it et al.} 1984; Efstathiou {\it et al.} 1985)
predict the formation of gravitationally bound objects of mass as
small as $10^7\,\Ms$ in large numbers before $z=5$. Recent work
(Blanchard {\it et al.} 1992) indicates that such objects can cool
rapidly and presumably fragment into stars.  These early mini-galaxies
would be expected to release great amounts of kinetic energy into the
surrounding IGM, thereby creating
large, fairly spherical voids filled with thin, hot, ionized
plasma.
We analyze the effect of expanding bubbles
driven by supernova winds from early mini-galaxies, and show that this
mechanism of distributing energy can indeed provide the required
ionization without violating any of the current spectral constraints.

\noindent
In 
Section~\ref{gpsec2},
we will treat the expansion of a shell in a
uniform, cold and neutral IGM. As these bubbles become
larger and more numerous and fill most of space, this obviously
becomes a very poor model of the IGM. In 
Section~\ref{gpsec3}
we estimate bulk
properties of this new processed IGM such as temperature, density
and ionization.

\markboth{CHAPTER 6: LATE REIONIZATION...}{6.2. THE EXPLOSION MODEL}
\section{The Explosion Model}
\label{gpsec2}

Since the pioneering work on spherically symmetric explosions by
Sedov (1959), a profusion of analytic solutions have been given by
numerous authors for models of ever-increasing complexity (Cox
\& Smith 1974; McKee \&
Ostriker 1977; Weaver {\it et al.} 1977; McCray \& Snow 1979; Bruhweiler
{\it et al.} 1980; Tomisaka {\it et al.} 1980; McCray \&
Kafatos 1987; Ostriker \& McKee 1988). Most of these models pertain
to bubbles in the interstellar medium of a galaxy, where the expansion
of the universe can be ignored. Ostriker \& McKee have given
asymptotic self-similarity solutions that incorporate this latter
complication, but unfortunately they are not sufficiently accurate for
our needs. The reason is that since neither energy nor momentum is
conserved in the regime before the shell becomes self-similar, there
is no accurate way to normalize the self-similar solution using the
initial data. 

Let $\rb$ and $\rd$ denote the average densities of baryonic and
non-baryonic matter in the universe. We will assume that all baryons
are in diffuse for early on, so that $\rb$ is also the density of the 
IGM. 
We will write $\rb = \Ob\rc$ and 
$\rd = \Od\rc$, where the critical density $\rc\equiv 3H^2/8\pi G$.
\noindent
We will use a three-phase model for the expanding bubbles:

\begin{itemize}

\item a dense, fairly cool spherical shell of outer radius $R$ and
thickness $R\delta$, containing a fraction $(1-\fm)$
of the total
baryonic mass enclosed,
 
\item uniform neutral ambient intergalactic medium (IGM) of density
$\rb+\rd$ and zero pressure outside,
 
\item a hot, thin, isothermal plasma of
pressure $p$ and temperature $T$ inside the shell.

\end{itemize}

\noindent
The shell is driven outwards by the pressure of the hot interior
plasma but slowed by the IGM and by gravity. The plasma is heated by
kinetic energy from supernova explosions and collisions with IGM
and cooled by bremsstrahlung and Compton drag against the cosmic
background radiation.
 
\subsection{The expanding shell}

We assume that the expanding shell sweeps up almost all baryonic IGM
that it encounters and loses only a small fraction of it through
evaporation into the interior, so that its total mass is
given by $m(t) = {4\over 3}\pi R(t)^3 (1-\fm)\rb$, where the
constant $\fm\ll 1$. Since $\dot{\rb}/\rb = -3H$ for any cosmological
model, we get  $${\md\over m} = \left(R^3 \rb\right)^{-1}
{d\over dt} \left(R^3 \rb\right) = 
3\left(\rdr-H\right)\hbox{ if }\rdr > H\hbox{, zero otherwise.}$$
(The shell will acquire new mass when it is expanding faster than
the Hubble flow, and will never lose mass.)
It turns out that the Hubble flow catches up with the shell
only as $t\to\infty$, so we will always have $\Rd>HR$ and $\md>0$.

When new mass is swept up,
it must be accelerated from the velocity $HR$ to $\Rd$, so the shell
experiences a net braking force $(\Rd-HR)\md$. 
The interior
pressure $p$ drives the shell outward with a force 
$pA = 4\pi R^2p = 3mp/\rb R$ in the thin shell approximation 
$\delta, \fm \ll 1$.  Finally there is a gravitational braking force,
which in the thin-shell approximation (Ostriker \& McKee 1988)
gives the deceleration ${4 \over 3} \pi G R(\rd + {1\over 2}\rb)$.
Adding these three force
terms, the radial equation of motion becomes
\beq{RadialEq}
\Rdd = {8\pi pG \over \Ob H^2R} - 
{3 \over R}\left(\Rd-HR\right)^2 - 
\left(\Od+{1\over 2}\Ob\right){H^2 R \over 2}.
\eeq

\subsection{The interior plasma}

The equation of state for the plasma in the hot interior gives the
thermal energy 
\beq{EqOfState}
\Et = {3 \over 2} pV = 2\pi pR^3,
\eeq
and energy conservation for the interior yields
\beq{EconsEq}
\dot \Et = L - pdV/dt = L-4\pi p R^2\Rd,
\eeq
where the luminosity $L$ incorporates all sources of heating
and cooling of the plasma. We will consider five contributions to
$L$ and write  
$$L = \Lsn-\Lc-\Lb-\Li+\Lh,$$
where $\Lsn$ is the energy injection from supernova explosions, 
$\Lc$ the cooling by Compton drag against the CBR,
$\Lb$ the cooling by bremsstrahlung,
$\Li$ the cooling by ionization of neutral IGM and  
$\Lh$ the heating from collisions between the shell and the IGM.

In stellar burning from zero to solar
metallicity, the mass fraction $0.02 \times 0.007$ is released,
mostly as radiation. Due to low cross-sections, only a negligible
fraction of this radiation will contribute towards heating the gas,
so we will only be interested in the energy that
is released in kinetic form. From empirical observations of active
galactic winds (Heckman 1990) about $2\%$ of the total luminosity from
a galaxy is mechanical. Another empirical observation is that for a
solar neighborhood initial stellar mass function, one has 
roughly one supernova for every $150\Ms$ of baryons that form
stars, with a  typical kinetic energy output of
$10^{51}$ ergs per explosion. Both of these observations lead to the
same estimate $$\Lsn = {\fsn M_b c^2\over\tb} \theta(\tb-t) \approx
1.2\Ls{M_b\over\Ms}\theta(\tb-t),$$  where the efficiency
$\fsn\aet{4}{-6}$ and where we have assumed that the  energy is
released at a constant rate during a  period $\tb\aet{5}{7}$ years.

Now let us examine cooling. 
The interior baryon density is $\ri = \rb\fm/(1-\delta)^3$
whereas the shell density is $\rs =
\rb(1-\fm)/(1-(1-\delta)^3) \approx \rb/3\delta$ if $\fm, \delta
\ll 1$. 
Compton drag against the
microwave background radiation causes energy loss at a rate
(Kompaneets 1957) 
\beq{LcompEq}
\Lc = {4\pi^2\over
15}\left(\st c n_e\right) \left({kT_e\over m_ec^2}\right) 
\left({k\Tg\over\hbar c}\right)^4\hbar c V,
\eeq
where $\st$ is the Thomson cross section, $V={4\over 3}\pi R^3$, 
and $T_e = T$, the temperature of the interior plasma, which is given
by
$$E_t = \left({3\over 2}+{3\over 2}\right)kT{\fm\rb\over m_p}V.$$
(We will assume almost complete ionization and low metallicity, so
that  $n_e\approx \fm n_b$.) Using \eq{EqOfState}, we see that
Compton drag causes cooling on a timescale 
$${\Et\over\Lc} = 
 {45\over 4\pi^2}\left({\hbar c\over k\Tg}\right)^4{m_e\over\st\hbar}
\aet{2}{12}\,\hbox{years}\times(1+z)^{-4},$$
that is, it becomes important only at high redshifts. 
It turns out that $\Lb\ll\Lc$ in our regime of interest, so we will
simply make the approximation $\Lb\approx 0$.
Assuming that the ambient IGM is completely neutral, the power
required to ionize the hydrogen entering the interior is
simply 
$$\Li = \fm n_b\rydberg\times 4\pi R^2\left[\Rd-HR\right],$$
where $\rydberg \approx 13.6\,eV$.

The equation of motion\eqnum{RadialEq} assumes that the
collisions between the expanding shell and the ambient IGM are
perfectly inelastic. The kinetic energy dissipated has one of three
fates: It may 

\begin{description}

\item[(a)] radiate away in shock cooling processes,
 
\item[(b)] ionize the swept up IGM, or
 
\item[(c)] heat the shell and by conduction heat the
interior plasma.
 
\end{description}

\noindent
Let $\fcoll$ denote the fraction that is reinjected
into the interior plasma through processes (b) and (c). 
This is one of the major uncertainties of the model.
Now a straightforward kinematic calculation of the kinetic
energy loss per unit time gives 
$$\Lh = \fcoll {3m\over 2R}\left(\Rd-HR\right)^3.$$
Making accurate estimates of $\fcoll$ is difficult, so we simply
use the two extreme cases $\fcoll = 0$ and $\fcoll = 1$ in the
simulations. Perhaps surprisingly, the results will be seen to be
relatively independent of the choice of $\fcoll$.

\subsection{Solutions to the equations}

Combining\eqnum{EqOfState} and\eqnum{EconsEq} leaves us
with
\beq{pdotEq}
\pd = {L\over 2\pi R^3} - 5 {\Rd\over R}p.
\eeq
The system of equations\eqnum{RadialEq} and\eqnum{pdotEq}
reduces to that derived by Weaver {\it et al.} (1977) in the
special case where $L(t)$ is constant and the expansion of
the universe is ignored.

Let us define dimensionless variables as follows:

\smallskip
{
\raggedright
\noindent
\baselineskip20pt
\tabskip = 1em
\halign{\hglue1.3cm$#\equiv$&$#, $&$#_*\equiv$&$#$&$#$\hfill\cr
\t&t/t_*&t&{2\over 3}H_0^{-1}(1+z_*)^{-3/2}&\cr
\h&H/H_*&H&{2\over 3}t_*^{-1}&\cr
\l&L/L_*&L&\fsn M_b c^2\tb^{-1}&\aet{1.2}{5}\Ls\times M_5\cr 
\e&E/E_*&E&L_*\tb&\aet{7.2}{53}\,\ergs\times M_5\cr 
\r&R/R_*&R&L_*^{1/5}G^{1/5}t_*&\approx 0.13\,\Mpc \times
 h^{-1}(1+z_*)^{-3/2}M_5^{1/5}\cr 
\p&p/p_*&p&L_*^{2/5}G^{-3/5}t_*^{-2}&\aet{1.4}{-16}\,Pa
\times h^2(1+z_*)^{3}M_5^{2/5}\cr}
}
\smallskip\noindent
Here we have taken $h=0.5$ and defined $M_5 \equiv M_b/10^5\,\Ms$.
If $\Omega\equiv\Ob+\Od$=1, then $t_*$ is the age of the
universe at the redshift $z_*$ when the shell begins its expansion,
i.e. the Big Bang occurred at $\t=-1$ and the shell starts expanding
at $\t=0$. For this simple case, we have $\h =
(1+\t)^{-1}=(1+z_*)^{-3/2}(1+z)^{3/2}$.
\goodbreak

Equations\eqnum{RadialEq} and\eqnum{pdotEq} now become
\beq{MainEq}
\cases{
\r''(\t) =&${18\pi\over\Ob}\h(\t)^{-2}{\p(\t)\over \r(\t)} 
-3\left(1-{2\over 3}{\h(\t)\r(\t)\over\r'(\t)}\right)^2
{\r'(\t)^2\over \r(\t)}
-\left({2\over 9}\Od+{1\over 9}\Ob\right) \h(\t)^2\r(\t)$\crr 
\p'(\t) =&${\l(\t)\over 2\pi \r(\t)^3} - 
5{\r'(\t)\over \r(\t)}\p(\t)$
}
\eeq
Here $\l=\ls-\lc-\li+\lh$, where 
\smallskip
{
\baselineskip20pt
\tabskip = 1em
\halign{\hglue3cm$#$&$#$&$#$\hfill\cr
\ls&=&\theta(\tb-t_*\t),\cr
\lc&\approx&0.017 h^{-1}(1+z_*)^{-3/2}(1+z)^4\r^3\p,\cr
\li&\approx&2.2\fm\Ob M_5^{-2/5}\times\h^2\r^2\hubpar\hbox{, and}\cr 
\lh&=&{1\over
3}\fcoll\Ob\times
  \left(\h\r\right)^2\hubpar^3.\cr
}}
\smallskip\noindent
In computing $\lc$, we have taken $T_{\gamma 0} =
2.74K$. The interior temperature, the thermal energy and the kinetic
energy are given by  
$$T \aet {4.5}{5}\,K\times 
 {M_5^{2/5}\over\fm\Ob}{\p(\t)\over\h(\t)^2},$$
$$\et = {2\pi\over\taub}\r^3\p,$$
$$\ek = {m\Rd^2/2 \over L_*\tb} = {\Ob\over 9\taub}\h^2\r^3\r'^2.$$

The solution to the system\eqnum{MainEq} evolves through three
qualitatively different regimes: $\t\ll 1$, $\t\approx 1$ and $\t\gg
1$.

\begin{description}

\item[a)] 
In the limit of small times $\t\ll 1$, gravity and Hubble
flow are negligible and we obtain the asymptotic power
law solution 
$$\r(\t) = a\t^{3/5},\quad \p(\t) = b\t^{-4/5}$$
\beq{SmallTimeSolEq}
\hbox{where } 
a\equiv\left({375/\Ob\over 77-27\fcoll}\right)^{1/5}
\quad\hbox{and}\quad
b\equiv{7\Ob\over 150\pi}a^2,
\eeq
as may be verified by direct
substitution. This solution reduces to that found by Weaver {\it et al.}
in the special case $\fcoll = 0$. 
Since the total energy injected is
simply $\ein=\t/\taub$, this gives  $${\et\over\ein} = {35\over
77-27\fcoll}\quad\hbox{and}\quad
{\ek\over\ein} = {15\over 77-27\fcoll}$$ for small $\t$.
Hence even though $\et+\ek=\ein$ only for the most optimistic
case $\fcoll = 1$, we see that no more than 
$1-{35+15\over 77}\approx 25\%$ of the injected energy is lost as
radiation even in the worst case $\fcoll=0$.

\item[b)] 
The behavior in the intermediate regime is a complicated interplay
between several different effects:

\begin{enumerate}

\item After approximately $5\times10^7$ years, the supernova explosions
cease, which slows the expansion. In this pressure-driven snowplow
phase, we would asymptotically have $R\propto t^{2/7}$, $t^{4/13}$,
$t^{1/3}$, $t^{4/11}$ or $t^{2/5}$ if there were no gravity, no Hubble
flow and no cooling with $\fcoll=0$, ${1\over 8}$, ${1\over 3}$,
${5\over 8}$ or $1$, respectively.
 
\item Cooling (and $pdV$) work reduces the pressure and the
thermal energy to virtually zero, which slows the expansion.
With zero pressure, we would approach the momentum-conserving
snowplow solution $R\propto t^{1/4}$ if there were no gravity and no Hubble
flow.
 
\item The density of the IGM drops and the IGM already has an outward
Hubble velocity before it gets swept up, which
boosts the expansion and adds kinetic energy to the shell.
 
\item Gravity slows the expansion.
 
\item 
Dark
matter that has been accelerated outward by the shell catches up with
it again and speeds up the expansion.
(This last effect has been neglected in the equations above, since it
generally happens too late to be of importance for our purposes.)

\end{enumerate}
 
\item[c)]
As $t\to\infty$, the shell gets
frozen into the Hubble flow, {\ie} $R\propto t^{2/3}$ if $\Omega=1$. An
approximate analytic solution for $\t\gg 1$ is given by Ostriker\&
McKee (1988), but since neither energy nor momentum is conserved in the
intermediate regime,  there is no simple way to connect this solution
with the short-time solution above. 

\end{description}

\noindent
Numerical solutions for the comoving radius $(1+z)R$
are plotted in 
Figure~\ref{gpfig1} 
for different values of $z_*$ and $\fcoll$.
The asymptotic solution\eqnum{SmallTimeSolEq} has been used to
generate initial data at $\t=0.01$ for the numerical integration. 
In this Figure, 
we have truncated $R$ when the interior temperature
drops below 15,000K, after which newly swept up IGM fails to become
ionized.
Figures~\ref{gpfig2a}, \ref{gpfig2b} and~\ref{gpfig2c}  
show what becomes of the injected energy for
different parameter values.
Note that the relative fractions are approximately constant early on,
while the supernovae inject energy, in accordance with the
asymptotic solution\eqnum{SmallTimeSolEq}. The reason that the total energy
exceeds $100\%$ of the input is that the shell gobbles up
kinetic energy from swept-up IGM that already has an outward
Hubble velocity.

\markboth{CHAPTER 6: LATE REIONIZATION...}{6.3. COSMOLOGICAL CONSEQUENCES}
\section{Cosmological Consequences}
\label{gpsec3}

Once the expanding bubbles discussed in the previous section have
penetrated most of space, the IGM will presumably have a frothy
character on scales of a few 100 kpc, containing thick and fairly cool
shell fragments separated by large, hot, thin and ionized regions that
used to be bubble interiors.

In 
Section~\ref{gpsec3.1},
we calculate at what point the IGM becomes frothy,
more specifically what fraction of space is covered by expanding
shells at each $z$. 
In~\ref{gpsec3.2} 
we discuss the resulting enrichment of the
IGM with heavy elements. 
In~\ref{gpsec3.3} 
the thermal history of the IGM after
this epoch is treated.
Finally, in~\ref{gpsec3.4} 
the residual ionization is computed,
given this thermal history, and we discuss the circumstances under
which the Gunn-Peterson constraint is satisfied.

\subsection{IGM porosity}
\label{gpsec3.1}

Assuming the standard PS theory of structure
formation (Press \& Schechter 1974), the fraction of all mass that has
formed gravitationally bound objects of total (baryonic
and non-baryonic) mass greater than $M$ at redshift $z$ is
$$1-\erf\left[{\delta_c\over\sqrt{2}\sigma(M)}\right],$$  
where $\erf(x)\equiv 2\pi^{-1/2}\int_0^x e^{-u^2}du$ and 
$\sigma^2$ is the linearly extrapolated r.m.s. mass fluctuation in a
sphere of radius $r_0$. The latter is given by top-hat filtering of
the power spectrum as
\beq{FilterEq}
\sigma^2\equiv\left({\sigma_0\over 1+z}\right)^2 \propto
{1\over(1+z)^2}\izi
\left[{\sin kr_0\over(kr_0)^3} - {\cos kr_0\over(kr_0)^2}\right]^2
 P(k) dk,
\eeq
where $r_0$ is given by ${4\over 3}\pi r_0^3\rho = M$ and where
$P(k)$ is the power spectrum.
Although this approach has been criticized as too simplistic,
numerical simulations (Efstathiou {\it et al.} 1988; Efstathiou \&
Rees 1988; Carlberg \& Couchman 1989) have shown that it describes
the mass distribution of newly formed structures remarkably well.
Making the standard assumption of a Gaussian density field, Blanchard
{\it et al.} (1992) have argued that it is an accurate description at least in
the low mass limit. Since we are interested only in extremely low
masses such as $10^6\Ms$, it appears to suffice for our purposes.

We choose $\delta_c = 1.69$, which is the
linearly extrapolated overdensity at which a spherically symmetric
perturbation has collapsed into a virialized
object (Gott \& Rees 1975).
Letting $\fg$ denote the fraction of all baryons
in galaxies of mass greater than $M$ at $z$, this would imply that 
\beq{fgEq}
\fg \approx
1-\erf\left[{1.69(1+z)\over\sqrt{2}\sigma_0(M)}\right]
\eeq
if no other
forces than gravity were at work. However, it is commonly believed that
galaxies correspond only to such objects that are able to cool (and
fragment into stars) in a dynamical time or a Hubble time (Binney 1977;
Rees \& Ostriker 1977; Silk 1977; White \& Rees 1978). Hence the
above value of $\fg$ should be interpreted only as an upper limit.

A common assumption is that the first galaxies to form have a
total (baryonic and dark) mass $M_c\approx 10^6\Ms$, roughly the Jeans
mass at recombination.  Blanchard {\it et al.} (1992) examine the interplay
between cooling and gravitational collapse in considerable detail, and
conclude that the first galaxies to form have masses in the range
$10^7\Ms$ to  $10^8\Ms$, their redshift distribution still being given
by \eq{fgEq}. To keep things simple we will assume that all
early galaxies have the same mass $M_c$ and compare the results for
$M_c = 2\times 10^6\Ms, 10^8\Ms$ and $10^{11}\Ms$. 

Let $R(z;z_*)$ denote the radius of a shell at $z$ that was
created at $z_*$ by a galaxy of baryonic mass $M_b = \Ob M_c$ as in 
Section~\ref{gpsec2}.
Then the {\it naive filling factor}, the
total bubble volume per unit volume of the universe, is 
\beq{ffEq}
\ff(z) = \int_z^{\infty} 
{4\over 3}\pi R(z;z_*)^3 {\rb\over M_b}
{df_g(z_*)\over d(-z_*)} \,dz_*
= \phi_*(1+z)^3\int_z^{\infty} 
{r(z;z_*)^3\over (1+z_*)^{9/2}}
\,{df_g(z_*)\over d(-z_*)} \,dz_*,
\eeq
where
$$\phi_*\approx 1600h^{-1}M_5^{-2/5}(\Ob/0.06).$$
Clearly nothing prohibits $\ff$ from exceeding unity. This
means that nearby shells have encountered each other and that certain
volumes are being counted more than once. If the locations of the
bubbles are uncorrelated, then the fraction of the universe that will
be in a bubble, the {\it porosity}, is given by  
$$\P \equiv 1-e^{-\phi}.$$

If the early
galaxies are clustered rather than Poisson-distributed, this value
is an overestimate. 
For an extreme (and very unrealistic) example, if they would always
come in clusters of size $n$ and the clusters would be much smaller
than the typical bubble size of 100 kpc, then it is easy to see
that $\P \approx 1-e^{-\phi/n}.$ For more realistic cases, simple
analytic expressions for $P$ are
generally out of reach. Since we expect the clustering to be quite
weak, we will use the Poisson assumption for simplicity.

The uppermost panels of 
Figures~\ref{gpfig3a} and~\ref{gpfig3b}  
contain $\P(z)$ for various
parameter values, calculated numerically from \eq{ffEq} using
the numerical solutions for $\r(z;z_*)$. 
It is seen that the lower mass in 
Figure~\ref{gpfig3a} 
($2\times 10^6\Ms$ 
versus $10^8\Ms$) gives higher filling factors, so that the expanding
shells fill almost $100\%$ of space by $z=5$ for three of the four 
choices of $f_g(5)$. In~\ref{gpfig3b}, we see that almost $20\%$ of 
the baryons must be in galaxies by $z=5$ to achieve this.
The greater efficiency of small galaxies 
is to be expected, since $\phi_*\propto M^{-2/5}$. Although
some parameters still yield the desired $P\approx 100\%$ by $z=5$ 
in Figure~\ref{gpfig3b},
using present-day masses like $M_c = 10^{11}\Ms$ fails dismally
(not plotted) for all choices of the other parameters. Roughly, the
largest $M_c$ that works is $10^8\Ms$

As can be seen, the dependence on
$\fcoll$ (dashed versus solid lines) is rather weak.

In order to calculate $\sigma_0(M_c)$ from the fluctuations observed
on larger scales today, we need detailed knowledge of the power
spectrum down to very small scales, something which is fraught with
considerable uncertainty. 
For this reason, we have chosen to label the curves by the more physical
parameter $f_g(5)$, the fraction of all baryons that have formed galaxies
by $z=5$. The four sets of curves correspond to fractions of $50\%$,
$20\%$, $10\%$ and $1\%$.
These percentages should be compared with observational
estimates of metallicity, as will be discussed in 
Section~\ref{gpsec3.2}.

The second column of 
Table~\ref{gptable1}  
contains the values $\sigma_0(M_c)$
necessary to obtain various values of $f_g(5)$, calculated by
inverting the error function in \eq{fgEq}. 
The last four columns contain the bias factors necessary to yield this
value of $\sigma_0(M_c)$ for two choices of power spectra
(CDM and n=0.7 tilted CDM) and two choices of cutoff
mass ($M_c = 2\times 10^6\Ms$ and $M_c =10^8\Ms$).
Thus $b =\gamma/\sigma_0(M_c)$, where we define
$\gamma$ to be the ratio between $\sigma$ at $M_c$ and 
$\sigma$ at $8h^{-1}\,\Mpc \equiv b^{-1}$.
Performing the integral\eqnum{FilterEq} numerically with the 
CDM transfer function given by Bardeen {\etal} 1986 (BBKS), 
$h=0.5$, $\Omega=1$, $\Omega_b \ll 1$ and an $n=1$ Harrison-Zel'dovich
initial spectrum gives $\gamma\approx 19.0$ for $M_c=2\times 10^6\Ms$ and
$\gamma\approx 13.6$ for $M_c=10^8\Ms$. Using the CDM transfer function
of Bond and Efstathiou (1984) instead gives $\gamma\approx 18.1$ and
$\gamma\approx 13.7$, respectively. The BBKS transfer function is 
more applicable here since it includes the logarithmic dependence that
becomes important for very low masses. The BBKS transfer function with a
tilted (n=0.7) primordial spectrum yields the significantly lower values 
$\gamma\approx 9.71$ and
$\gamma\approx 7.93$, respectively.

\begin{table}
$$
\begin{tabular}{|rrrrrr|}
\hline
$\fg(5)$&$\sigma_0(M_c)$&$b_{cdm,6}$&$b_{cdm,8}$&$b_{tilted,6}$&
$b_{tilted,8}$\\
\hline
1\%&3.94&4.8&3.5&2.5&2.0\\
10\%&6.18&3.1&2.2&1.6&1.3\\
20\%&7.92&2.4&1.7&1.2&1.0\\
50\%&15.06&1.3&0.9&0.6&0.5\\
\hline
\end{tabular}
$$
\caption{Correspondence between various ways of 
normalizing the power spectrum}
\label{gptable1}
\end{table}
\noindent
Basically, 
Table~\ref{gptable1}  
shows that any of our values of $f_g(5)$ become
consistent with a feasible bias factor for some choice of
power spectrum and cutoff mass. 
\bigskip

\subsection{IGM enrichment}
\label{gpsec3.2}

These values of $f_g(5)$ should be compared with observational
estimates of metallicity, since if the
stars in these early mini-galaxies produce the same fractions of heavy
elements as do conventional stars, then these percentages are directly
linked to the fraction of currently observed metals that were made
before $z=5$. Some of the enriched shells may be observable as quasar 
absorption line
systems, as intracluster gas, and, indirectly, as in the
metallicities of old disk and halo stars.

Observations of
iron abundances in intracluster gas by HEAO-1,
Exosat and Ginga ({\eg} Mushotzky 1984; 
Hughes {\it et al.} 1988; Edge 1989;
Hatsukade 1989) have shown that most clusters have abundances
between $25\%$ and $50\%$ of the solar value. Einstein observations
have showed the presence of a large variety of other heavy elements in
the intracluster gas (Lea {\it et al.} 1982; Rothenflug {\it et al.} 1984).
Most of this gas and some of these metals are believed to be
``primordial",  since the gas mass in clusters is typically several
times greater than the observed stellar mass in the cluster galaxies
(Blumenthal {\it et al.} 1984; David {\it et al.} 1990; Arnaud {\it et al.}
1991).

There are indications that the most of these heavy elements
may have been produced as recently as around $z=2-3$, and that the
metallicity in the halo gas of some $z\approx 3$ galaxies inferred
from QSO absorption line studies are as low as $0.1\%$ of the
solar value (Steidel 1990).
However, this and other
observations of extremely metal-poor objects 
(Pettini {\etal} 1990) does not necessarily rule out our
scenario, since it is highly uncertain whether all the hydrogen in the
swept-up IGM would get thoroughly mixed with the metal-rich supernova
ejecta.

\subsection{IGM temperature}
\label{gpsec3.3}

Let $T(z;z_*)$ denote the temperature of the interior of a bubble at
$z$ that was created at $z_*$ as in 
Section~\ref{gpsec2}.
Then the volume-averaged temperature of the IGM is 
\beqa{IGMtempEq}
\Tigm(z) 
&\equiv&\int_z^{\infty} 
{4\over 3}\pi R(z;z_*)^3 
T(z;z_*) {\rb\over M_b}
{df_g(z_*)\over d(-z_*)} \,dz_*\nonumber\\
&=& \phi_*(1+z)^3\int_z^{\infty} 
{r(z;z_*)^3\over (1+z_*)^{9/2}}\,
T(z;z_*)
{df_g(z_*)\over d(-z_*)} \,dz_*.
\eeqa
When $\phi$ becomes of order unity, the IGM swept up by the expanding
shells is no longer cold, neutral and homogeneous, so the
treatment in 
Section~\ref{gpsec2}
breaks down. The resulting temperatures will
be underestimated, since less thermal energy needs to be expended on
heating and ionization.  As can be seen in 
Figures~\ref{gpfig3a} and~\ref{gpfig3b},  
the
transition from $\phi\ll 1$, where the treatment in 
Section~\ref{gpsec2}
is valid, to $\phi\gg 1$, where the IGM becomes fairly uniform, is quite
rapid. Since $\Tigm$
defined above is proportional to the thermal energy per unit volume,
energy conservation leads us to assume that $\Tigm$ remains fairly
constant during this transition and therefore is a good estimate of
the bulk IGM temperature immediately afterwards.
From this time on, we will approximate the IGM outside the
scattered dense and cold shell remnants by a uniform isothermal 
plasma.
Applying equations\eqnum{EqOfState},\eqnum{EconsEq} and\eqnum{LcompEq}
to the IGM yields the
following equation for its thermal evolution: 
\beq{dTdzEq}
-{d\over dz} T_5 = -\left[{2\over 1+z} + A(1+z)^{3/2}\right] T_5
+ \linj,
\eeq
where the first term encompasses cooling from adiabatic expansion 
and the second term Compton cooling. $T_5\equiv \Tigm/10^5\,K$, 
$A\equiv
1.5(t_0/t_{comp})\approx 0.0042h^{-1}$ and $\linj\equiv t_0
L_{inj}/k\times10^5K$, where  $ L_{inj}$ is the power injected into
the IGM per proton from all heat sources combined.
The Compton cooling term is seen to increase with
redshift, equaling the adiabatic term at $z\approx 17h^{0.4} - 1$.

In the most pessimistic case of no reheating whatsoever, {\ie} for
$\linj=0$, \eq{dTdzEq} has the solution 
$$T \propto (1+z)^2 e^{0.4A(1+z)^{5/2}}.$$
A more optimistic assumption is that some fraction $\finj$ of the
total energy released from stellar burning in newly formed galaxies
continues to heat the IGM, i.e. 
$$\linj = f_{inj}\left(0.02\times0.007 m_p c^2\over
k\times10^5K\right) {df_g\over d(-z)} \approx 
\left({2.1\times 10^4\over \sigma_0}\right)
f_{inj}\exp\left(-{1\over 2}
\left[{1.69(1+z)\over\sigma_0(M_c)}\right]^2\right),$$ 
which would also incorporate other modes of energy injection such
as radiation. 
The middle panels in 
Figures~\ref{gpfig3a} and~\ref{gpfig3b}  
show the temperature resulting from
$\finj=0, 0.001$ and $0.02$ and different initial values. For $\P<0.8$,
$\Tigm$ has been calculated numerically from 
\eq{IGMtempEq} by
using the numerical solutions for $r(z;z_*)$. Then the value at the
redshift for which $\P=0.8$ has been used as initial data for
\eq{dTdzEq}. For comparison, two horizontal lines have been
added showing what temperatures would be required to obtain neutral
fractions of $10^{-6}$ and $10^{-5}$ in {\it equilibrium}, using
\eq{EquilibrionizationEq}.
Since the plasma is in fact out of equilibrium, these highly ionized
states can be maintained at much lower temperatures, as is seen in the
bottom plots.

\subsection{IGM ionization and the Gunn-Peterson effect}
\label{gpsec3.4}

Assuming that any neutral hydrogen in the remains of the shells
will have insufficient time to diffuse far into the hot
ionized regions that used to be shell interiors, we can treat the
latter as an isolated mixture of gas and plasma where the ionization
fraction $\chi$ evolves as  $$\dot\chi =
n\chi\left[(1-\chi)\Lion(T)-\chi\Lrec(T)\right],$$ 
and where the rates
for collisional ionization and recombination are given by
(Stebbins \& Silk 1986)
$$\Lion = \left<\sigma_{ci}v\right>\approx
7.2\,a_0^2\left({kT\over m_e}\right)^{1/2} e^{-\rydberg/kT},$$
$$\Lrec = \left<\sigma_{rec}v\right>\approx
{64\pi\over 3\sqrt{3\pi}}\alpha^4a_0^2c
\left({kT\over\rydberg}\right)^{-2/3},$$
where $a_0$ is the Bohr radius and $\alpha$ is the fine structure
constant. Changing the independent variable to redshift, the $\Omega=1$
case leaves us with
$$-{d\over dz}\ii = {3\over 2}{\fm\Ob\over\omdc}(1+z)^{1/2}\ii
\bigl[(1-\ii)\lion(\ii)-\ii\lrec(T)\bigr],$$
$$\lion(T) \equiv
n_{c0}t_0\Lion(T)\approx \left[5.7\times 10^4 h\Ob\right]
T_5^{1/2}e^{1.58/T_5}$$  
$$\lrec(T) \equiv n_{c0}t_0\Lrec(T)
\approx\left[0.16 h\Ob\right] T_5^{-2/3}.$$
For large enough $z$, the ionization fraction will adjust rapidly
enough to remain in a quasistatic equilibrium and hence be given by
$\dot\ii=0$, {\ie}
\beq{EquilibrionizationEq}
\ii =
\left[1+{\Lrec(T)\over\Lion(T)}\right]^{-1}\approx
\left[1+2.8\times10^{-6}T_5^{-7/6}e^{1.58/T_5}\right]^{-1}.
\eeq

The observed absence of a Gunn-Peterson trough in the spectra of
high-redshift quasars strongly constrains the density of neutral
hydrogen in the IGM. The most thorough study to date, involving eight
quasars (Steidel \& Sargent 1987), concluded that
$$\Omega_{H_I}(z=2.64)<(1.2\pm 3.1)\times 10^{-8}h_{50}^{-1}$$
if $\Omega=1$.
In our model this corresponds to
$(1-\ii) < (1.2\pm 3.1)\times 10^{-8}/(\fm\Ob)\aet{2}{-6}$ 
for $\Ob=0.06$ and $f_m$=0.1. Thus we
are helped not only by the IGM being ionized, but also by it being
diffuse. 
In a recent study of a single quasar, Webb {\etal} (1992) find the data
consistent with either $\Omega_{H_I}(z=4.1)=0$ or
$\Omega_{H_I}(z=4.1)=1.5\times
10^{-8}h_{50}^{-1}$, depending on model assumptions. We will
use the latter value as an upper limit.
Finally, recent Hubble Space Telescope spectroscopy of 3C 273 has been
used to infer that $\Omega_{H_I}(z = 0.158)<1.4\times
10^{-7}h_{50}^{-1}$. The constraints from these three studies are
plotted in 
Figures~\ref{gpfig3a} and~\ref{gpfig3b}  
together with the ionization levels
predicted by our scenario.

To achieve $\ii = 10^{-4}$, $10^{-5}$ and $10^{-6}$ in equilibrium
would by \eq{EquilibrionizationEq} require $T > 5.5\times
10^4\,K$, $T > 1.1\times 10^5\,K$ and $T > 3.6\times 10^5\,K$,
respectively.
As can be seen from the numerical solutions in the bottom panels of
Figures~\ref{gpfig3a} and~\ref{gpfig3b},  
the recombination rate is generally too slow for equilibrium
to be established, and the IGM remains almost completely ionized even when
$T\ll 15,000 K$ and equilibrium would have yielded $\ii\approx 0$. 
In both 3a and 3b, a very moderate reheating ($f_{inj} =
0.001$, heavy lines) is seen to suffice to satisfy the three
observational constraints. In the absence of any reheating whatsoever,
the only models that satisfy the constraints are those with very low
density ($\Ob=0.01$ or $\fm=0.01$). 

In summary, the only parameters that are strongly
constrained by the Gunn-Peterson test are $M_c$ and $\sigma_0(M_c)$.

\subsection{Other spectral constraints}

Let us estimate to what extent Compton cooling of the hot plasma
will distort the cosmic microwave background radiation (CBR). Since
for $T_e\gg\Tg$ the Comptonization y-parameter (Stebbins \& Silk
1986) 
$$y_C\equiv\int_t^{t_0}
{kT_e\over m_ec^2}n_e\st c\,dt$$ 
is linear in the plasma
energy density  
$\left({3\over 2}+{3\over 2}\right)kT_en_e$ at each fixed
time $(T_e = \Tigm)$, all that counts is the spatially averaged thermal
energy density at each redshift. 
Since the former is simply $\et(z;z_*)$
times the density of injected energy $f_g\fsn\Ob\rho c^2$,
the calculation reduces to mere energetics and we obtain
$$y_C = y_*\int_z^{\infty} 
{df_g(z_*)\over d(-z_*)} 
  \int_0^{z_*} \sqrt{1+z}\> \et(z;z_*) dz\,dz_*,$$
where
$$y_* \equiv{1\over 8} \fsn \Ob^2 {\st cH\over m_pG}
\aet{9}{-7}h\Ob^2.$$
The current observational upper limit on $y$ is  
$2.5\times 10^{-5}$ (Mather {\etal} 1994), so even if we 
take $f_g(0)$ as high as
$100\%$ and make a gross overestimate of the integral by making all
our galaxies as early as at $z_*=30$ and by
replacing $\et(z;z_*)$ by 
its upper bound $60\%$ for all $z$, $z_*$, our $y$ is
below the observational limit by three orders of magnitude for
$\Omega=1$. \smallskip

Now let us estimate the optical depth of the IGM. It has long been
known that reionization can cause a
spatial smoothing of the microwave background as CBR photons Thomson
scatter off of free electrons. Since $n_e  = \chi_{IGM}
n_b$, the optical depth for Thomson scattering, i.e. the number of
mean free paths that a CBR photon has traversed when it reaches our
detectors, is  
$$\tau_t = \int_{t_{rec}}^{t_0}\st \chi_{IGM} n_bc\, dt
= \tau_t^* \int_0^{z_{rec}}\sqrt{1+z} \chi_{IGM}
dz,$$ 
where
$$\tau_t^*\equiv {3\over 8\pi}\fm\Ob{H_0c\st\over
m_pG}\approx 0.07\Ob h.$$
Let us evaluate the integral by making the approximation that
$\chi_{IGM}$ increases abruptly from $0$ to $1$ at some redshift
$z_{ion}$. Then even for $z_{ion}$ as high as 30, 
$\tau_t \approx 7.9 h\Ob\fm\approx 0.02 \ll 1$ for our fiducial
parameter values $h=0.5, \Ob=0.06$ and $\fm=0.1$, so the probability
that a given CBR photon is never scattered at all is 
$e^{-0.02}\approx 98\%$. Hence this scenario for late reionization
will have only a very marginal smoothing effect on the CBR.
If the shells are totally ionized as well, then the factor $\fm$
disappears from the expressions above which helps only slightly.
Then $z_{ion}=15$ would imply that $8\%$ of the CBR
would be spatially smoothed on scales of a few degrees.

\markboth{CHAPTER 6: LATE REIONIZATION...}{6.4. DISCUSSION}
\section{Discussion}

We have calculated the effects of supernova driven winds from
early galaxies assuming a Press-Schechter model of galaxy formation
and a CDM power spectrum. The calculations have shown that
reionization by such winds can indeed explain the observed
absence of a Gunn-Peterson effect if a number of conditions are
satisfied:

\begin{enumerate}

\item The masses of the first generation of galaxies must be very small,
not greater than about $10^8\Ms$.
 
\item There is enough power on these small scales to get at least $10\%$
of the baryons in galaxies by $z=5$.
 
\item Except for the case where $\Ob$ is as low as $0.01$, there must be
some reheating of the IGM after $z=5$ to prevent the
IGM from recombining beyond allowed levels.
 
\item The commonly used thin-shell approximation
for
expanding bubbles must remain valid over cosmological timescales, with
the mass fraction in the interior remaining much less than
unity.

\end{enumerate}
 
\smallskip

Whether 1) is satisfied or not depends crucially on the model for
structure formation. This scenario is consistent with a pure CDM model
and some low-bias tilted CDM models, but not with top-down models like
pure HDM.

\smallskip

Observations of nearly solar abundances of heavy elements in
intracluster gas have given some support for 2),
which is roughly  equivalent to requiring that at least $10\%$ of the
heavy elements in the universe be made before $z=5$ (or
whenever $\phi\gg 1$). As discussed in 
Section~\ref{gpsec3.2},
the observations of some extremely metal-poor objects in QSO absorption
line studies do not necessarily rule out our scenario, since it is
highly uncertain whether all the hydrogen in the swept-up IGM would get
thoroughly mixed with the metal-rich supernova ejecta. 
The fact that large numbers of mini-galaxies are not seen today need not
be a problem either. Possible explanations for this
range from mechanisms for physically destroying them 
(Dekel \& Silk 1986, for instance) to the fact that 
the faint end of the luminosity function is still so poorly known
that old dwarf galaxies in the field may be too faint to see
by the present epoch (Binggeli {\etal} 1988).
\smallskip

To violate 3), the actual reheating would have to be extremely small.
A current IGM temperature between $10^4K$ and $10^5K$ suffices,
depending on other parameter values, since the low density IGM never has
time to reach its equilibrium ionization.
\smallskip

The thin-shell approximation 4) is obviously a weak point in the
analysis, because of the simplistic treatment of the dense shell and
its interface with the interior bubble. For instance, could the shell
cool and fragment due to gravitational instability before it
collides with other shells?
An approximate analytic model for such instability has been
provided by Ostriker \& Cowie (1981). Their criterion is that
instability sets in when $\Xi>1$, where
$$\Xi\equiv {2G\rho_{shell}R^2\over \Rd v_s}$$
and the sound speed $v_s = \sqrt{5 kT/3 m_p}$. In terms of
our dimensionless variables, this becomes
$$\Xi\approx 0.011 \times M_5^{1/5} T_5^{-1/2} 
\left({\Ob\over 0.06}\right)\left({\delta\over 0.1}\right)^{-1}
\left({1+z\over 1+z_*}\right)^3{r^2\over r'},$$ 
which indicates that with our standard parameter values,
gravitational instability does not pose problems even with fairly low
shell temperatures.
The reason that the shell density is not limited to four times the 
ambient IGM density is that 
the jump condition is not adiabatic, due mainly to effective Compton
cooling at the high redshifts under consideration.

After the critical $z$ (typically between 20
and 5) at which the expanding shells have collided with
neighbors and occupied most of space, the IGM is ``frothy" on
scales around 100 kpc, with dense cool shell remnants scattered
in a hot thin and fairly uniform plasma. Since the dark matter
distribution is left almost unaffected by the expanding bubbles,
formation of larger structures such as the galaxies we observe today 
should remain fairly unaffected as far as concerns gravitational
instability. There is indirect influence, however: the
ubiquitous metals created by the early mini-galaxies would enhance the
ability of the IGM to cool, which as mentioned in 
Section~\ref{gpsec3.1}
is commonly believed to be crucial for galaxy formation. 

Blanchard {\it et al.} (1992) argue that if the IGM has a temperature
higher than the virial temperature of a dark halo, pressure support
will prevent it from falling into the potential well and thus stop
it from forming a luminous galaxy. The virial temperature they
estimate for an object of mass $M$ formed at a redshift $z$ is
approximately
$$T_{vir} \aet{5.7}{5}K\,\left({M\over 10^{12}\Ms}\right)^{2/3}(1+z)$$
for $h=0.5$, so requiring $T_{vir}>T_{IGM}$ for say $T_{IGM}=10^6K$ at
$z=5$ would give a minimum galaxy mass of about $10^{11}\Ms$.
Such arguments indicate that the IGM reheating of our scenario might
produce a ``mass desert" between the earliest mini-galaxies and the
galaxies we see today: 
The first generation of galaxies, mini-galaxies
with masses of perhaps $10^6$ or $10^8 \Ms$, would keep forming until
their expanding bubbles had occupied most of space and altered the
bulk properties of the IGM. After that, formation of galaxies much
smaller than than those of today would be suppressed, since the IGM
would be too hot. 
Eventually, as the IGM cools by adiabatic expansion, a
progressively larger fraction of the IGM can be accreted by dark
matter potential wells. 
Indeed, even with the volume averaged IGM temperature remaining hot due
to some form of reheating,  cooling flows in deep potential wells, in
particular galaxy clusters, would not be suppressed. Late formation of
galaxies is therefore possible.

Pressure balance between the shell and the interior
during the expansion would give the ration
$T_{shell}/T_{interior} = \rho_{b,interior}/\rho_{b,shell}\approx
3\delta\fm\approx 0.03$ for $\delta=\fm=0.1$, so the 
shell fragments would expected to contain non-negligible fractions of
neutral hydrogen and thus absorb some Lyman-alpha. 
A typical shell radius is about 100 kpc for $M_c=10^6\Ms$, 
a size comparable to that of the clouds of the Lyman-alpha forest.
As to the number density of Lyman-alpha
clouds, the observed velocity separations greatly exceed those we
would expect if all shell fragments were to be identified with Lyman
alpha clouds. Thus the majority of these fragments must have been
destroyed by some other process. There are a number of ways in which this
could occur, for instance through photoionization by UV flux from the
parent galaxy or by collapse to form
other dwarf galaxies. The resulting numbers resemble the abundance of
minihalos in an alternative explanation of the Lyman-alpha forest
(Rees 1986). Strong evolution, in the sense of an increasing
cloud abundance with decreasing redshift, is expected to 
occur as cooling becomes effective.

\def\fheight{10.3cm} \def\fwidth{14.5cm}

\newpage

\bfig
\psfig{figure=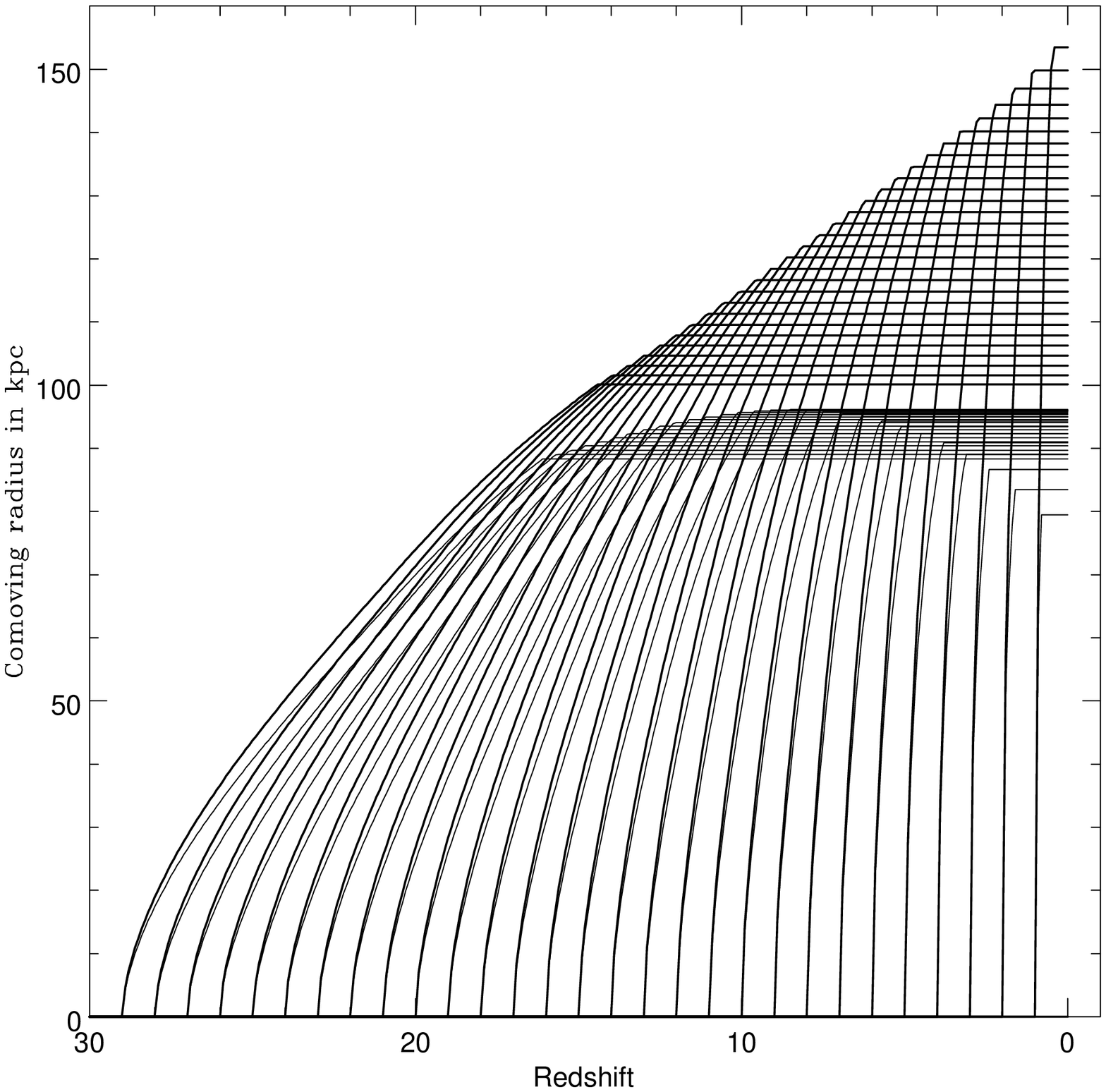,width=\fwidth,height=\fheight}
\nobreak
\caption{Comoving radius of expanding shell.}
\label{gpfig1}
 
\mycaption{The comoving shell radius $(1+z)R$ is plotted for galaxies
of total mass $2\times 10^6\Ms$, forming at integer
redshifts from 1 to 29.
Here $\Omega=1$, $\Ob=0.06$, $h=0.5$, and $\fm=0.1$.
$\fcoll=1$ for the upper set of lines and $\fcoll=0$ for the 
lower set. R has been truncated when $T$ drops below 15,000 K, 
after which
newly swept up IGM fails to become ionized.
}
\efig

\bfig
\psfig{figure=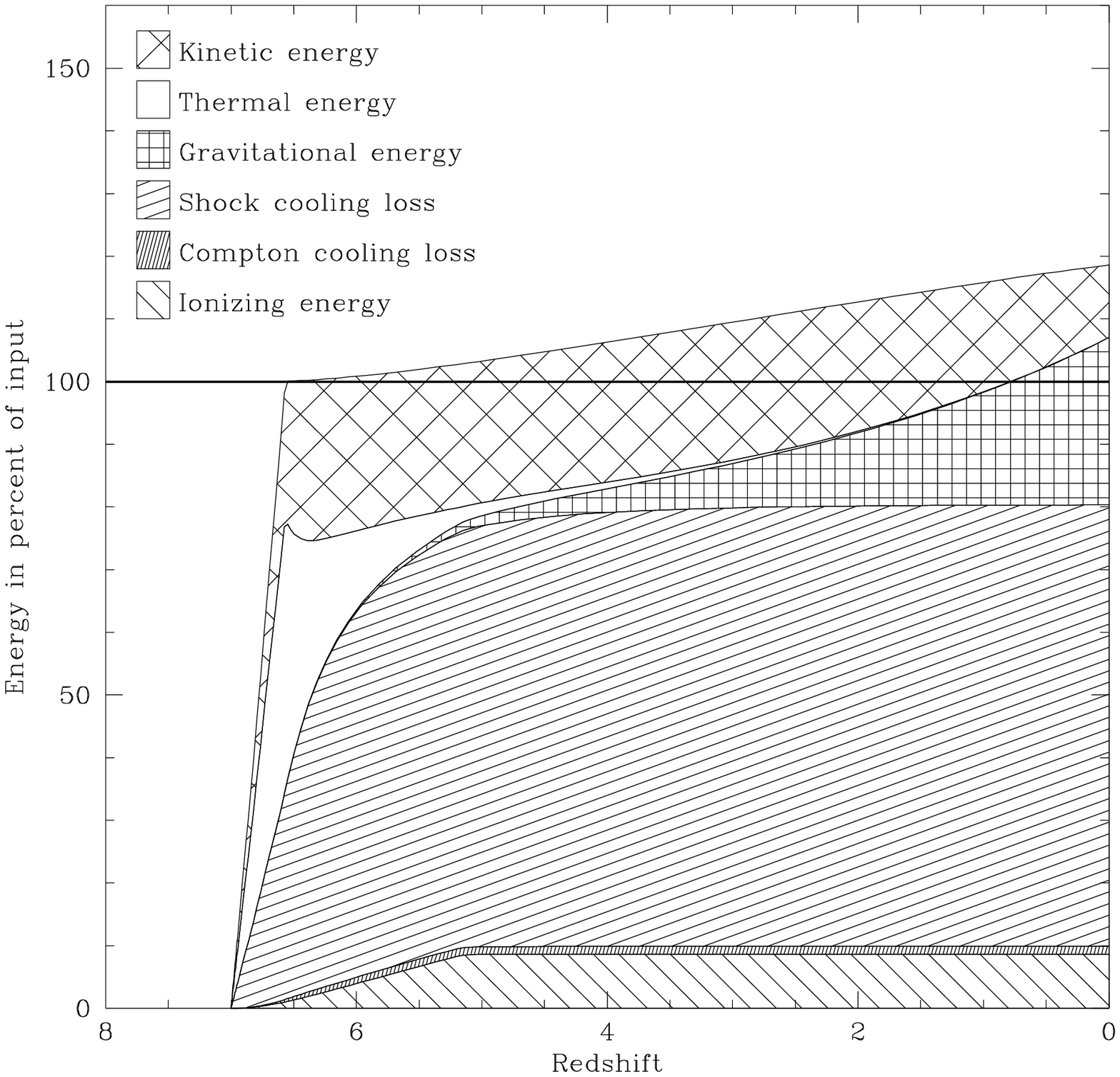,width=\fwidth,height=\fheight}
\caption{Energetics of expanding shell, example 1.}
\label{gpfig2a}

\mycaption{This and the two following 
figures show the
energy contents of an expanding bubble
as a function of redshift, for different choices of $\fcoll$
and $z_*$.
$\Omega=1$, $\Ob=0.06$, $h=0.5$ and $\fm=0.1$ for all three plots.
Figures~\ref{gpfig2a} and~\ref{gpfig2b}
illustrate the difference between $\fcoll=0$
and $\fcoll=1$ (there is no shock cooling loss in the second case).
Figure~\ref{gpfig2c}
has $\fcoll=0$ and illustrates that the Compton cooling
loss is larger at higher redshift.}
\efig

\bfig
\psfig{figure=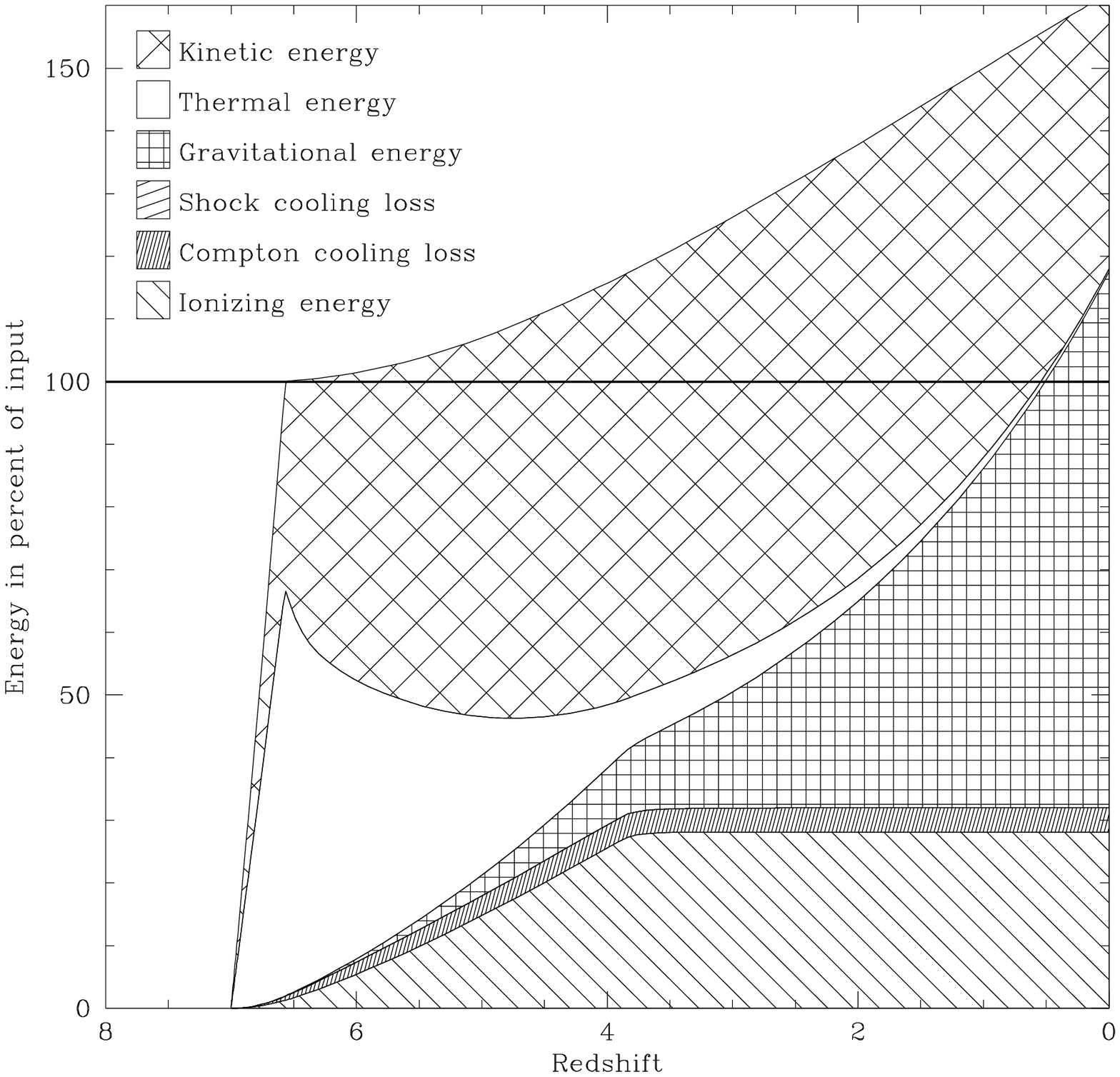,width=\fwidth,height=\fheight}
\caption{Energetics of expanding shell, example 2.}
\label{gpfig2b}
\efig

\bfig
\psfig{figure=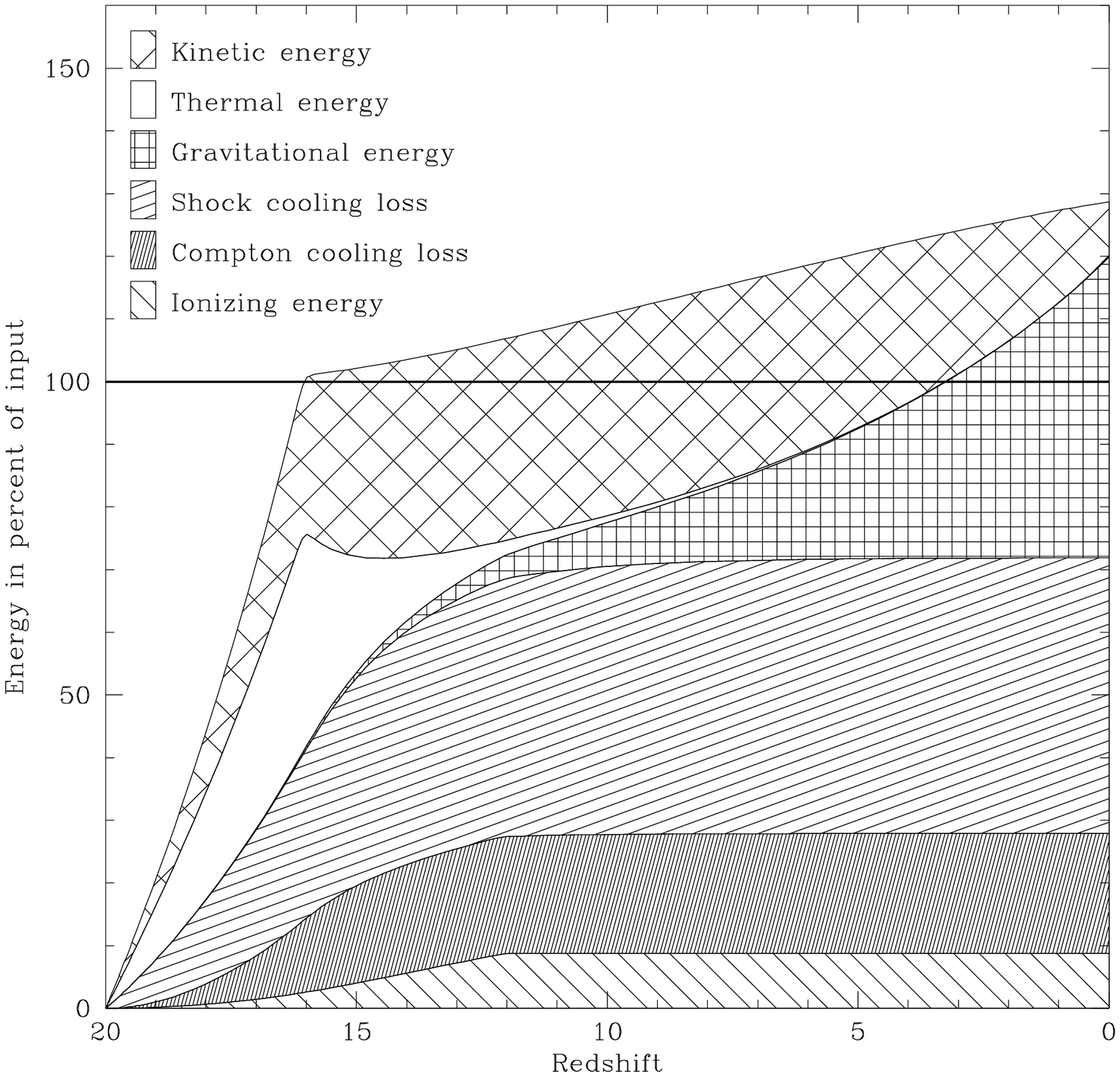,width=\fwidth,height=\fheight}
\caption{Energetics of expanding shell, example 3.}
\label{gpfig2c}
\efig

\bfig
\psfig{figure=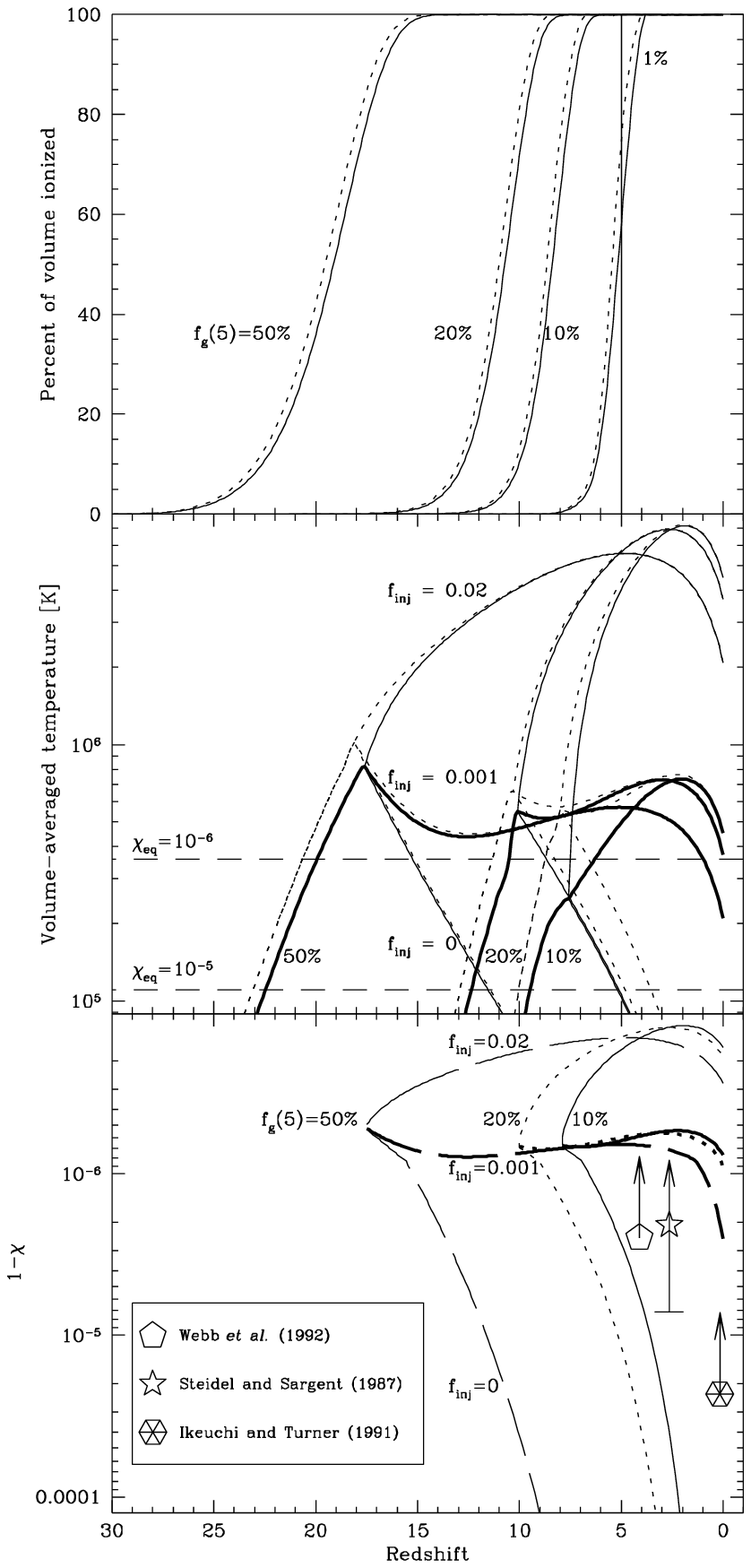,width=32cm,height=23cm}
\vskip8.5truecm
\caption{IGM evolution for $M_c=2\times 10^6\Ms$.}
\label{gpfig3a}
\efig

\bfig
\psfig{figure=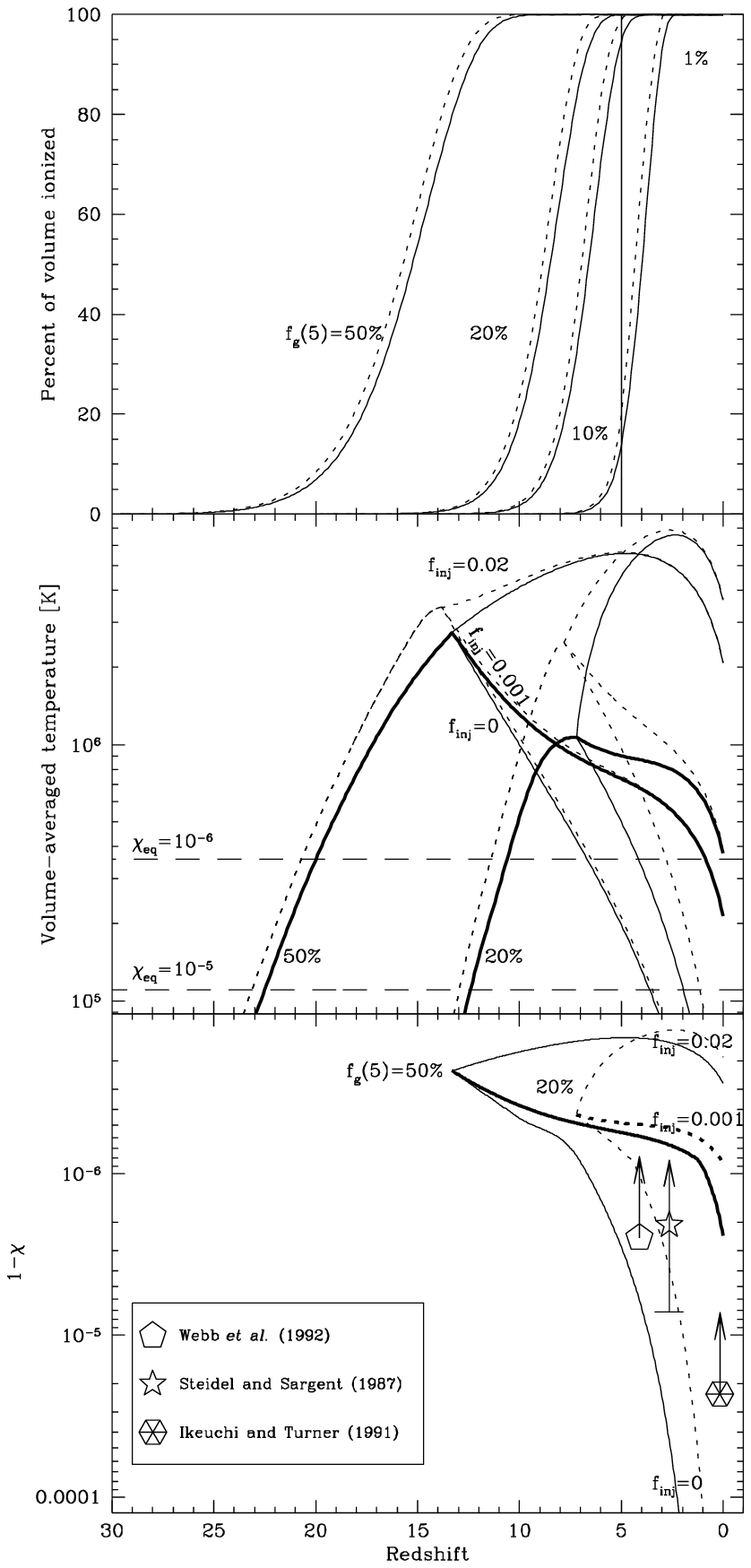,width=32cm,height=23cm}
\vskip8.5truecm
\caption{IGM evolution for $M_c=10^8\Ms$.}
\label{gpfig3b}
\efig

\begin{figure} % Since LATeX is stupid, and places this elsewhere otherwise.
\centerline{Figures~\ref{gpfig3a} and~\ref{gpfig3b}: IGM evolution.} 

\mycaption{
Three different properties of the IGM 
(filling factor, temperature and ionization) are plotted as a function
of redshift for different choices of $M_c$, $f_g(5)$ and $\fcoll$.
$\Omega=1$, $\Omega_b=0.06$, $h=0.5$ and $f_m=0.1$ in the two
previous figures,
\ref{gpfig3a} and \ref{gpfig3b}.
In all panels, the different families of curves correspond to different
values of $f_g(5)$; 50\%, 20\%, 10\% and 1\% from left to right, with
the rightmost cases being omitted where they fail dismally.
In the porosity and temperature plots (the upper
two panels of 
\ref{gpfig3a} and \ref{gpfig3b}),
dashed lines correspond to $\fcoll=1$ and
solid ones to $\fcoll=0$, whereas only the
pessimistic $\fcoll=0$ case is plotted in the ionization plots (the lower
third). 
In the temperature and ionization plots 
(the lower two panels), the three branches of each curve correspond
to the three reheating scenarios: $f_{inj} = 0$,
$f_{inj} = 0.001$ and $f_{inj} = 0.02$.
}
\efig

\cleardoublepage
\pagestyle{myheadings}
\chapter{Power Spectrum Independent Constraints}
\label{pindepchapter}

% Power Spectrum Independent Constraints
% on Cosmological Models

\def\x{\eta}
\def\N{N}
 
In this chapter, a formalism is presented that allows cosmological
experiments to be tested for consistency, and allows a simple frequentist
interpretation of the resulting significance levels.
As an example of an application, this formalism is used to place 
constraints on bulk flows of galaxies using the
results of the microwave background anisotropy experiments COBE and SP91,
and a few simplifying approximations about the experimental 
window functions.
It is found that if taken at face value, with the quoted errors, 
the recent detection by Lauer and Postman of a bulk
flow of 689 km/s on scales of 150$h^{-1}$Mpc
is inconsistent with SP91 at a 95\% confidence level 
within the framework of a Cold Dark Matter (CDM)
model.
The same consistency test is also used to place constraints that are
completely 
model-independent, in the sense that they hold for any 
power spectrum whatsoever
--- the only assumption being that the random fields are Gaussian.
It is shown that the resulting infinite-dimensional optimization 
problem reduces
to a set of coupled non-linear equations that can readily be solved
numerically.
Applying this technique to the above-mentioned example, we find that the
Lauer and Postman result is inconsistent with SP91
even if no
assumptions whatsoever are made about the power spectrum.

\markboth{CHAPTER 7: POWER SPECTRUM INDEP{.} CONSTRAINTS}{7.1. INTRODUCTION}
\section{Introduction}
\label{pindepsec1}

Together with the classical cosmological parameters $h$, $\Omega$, {\etc},  
the power spectrum $P(k)$ of cosmological density fluctuations is one of
the most sought-for quantities in modern cosmology, vital for
understanding both the formation of large-scale
structure and the fluctuations in the cosmic microwave background radiation
(CBR). 

The traditional approach has been to assume some functional
form for $P(k)$
(like that predicted by the cold dark matter (CDM) scenario, for instance),
and then investigate whether the predictions of the model are 
consistent with experimental data or not.
The large amounts of data currently being produced by 
new CBR experiments
and galaxy surveys, all probing different parts of the power spectrum,
allow a new and more attractive approach. We can now begin to probe exact
shape of the function $P(k)$, without making any prior assumptions about $P(k)$.
More specifically, we measure different weighted averages of the function, the
weights being the experimental window functions. 

This new approach is quite timely (Juszkiewicz 1993), as there are now many
indications that the primordial power spectrum may have been more 
complicated than an $n=1$ power law.
There are several sources of concern about the standard CDM
cosmology, with inflation leading to $\Omega\approx 1$ and a primordial
$n\approx 1$ Harrison-Zel'dovich power spectrum.
Compared to COBE-normalized CDM, observational data shows
unexpected
large-scale bulk flows  (Lauer \& Postman 1994), 
too weak density correlations on small scales (Maddox {\etal} 1990), 
a rather quiet local velocity
field (Schlegel {\it et al.} 1993) and a deficit of hot x-ray clusters 
(Oukbir \& Blanchard 1992).
The combined data from the COBE DMR (Smoot {\etal} 1992) and the Tenerife 
anisotropy experiment (Hancock {\etal} 1994) point to a spectral
index exceeding unity (Watson \& Guti\'errez de la Cruz 1993)
which, if correct, cannot be explained by any of the
standard inflationary models.
The recent possible detections of halo gravitational microlensing events (Alcock
{\etal} 1993) give increased credibility to the possibility that the dark matter
in our galactic halo may be baryonic. If this is indeed the case, models
with $\Omega<1$ and nothing but 
baryonic dark matter (BDM) (Peebles 1987; 
Gnedin \& Ostriker 1992,
Cen, Ostriker \& Peebles 1993) become rather appealing. 
However, in contrast to CDM with inflation, 
BDM models do not include a physical
mechanism that makes a unique prediction for what the primeval power spectrum
should be. Rather, the commonly assumed $P(k)\propto k^{-1/2}$ is 
chosen {\it ad hoc} to fit observational data. 
Moreover, for fluctuations near the curvature scale in open 
universes, where the
$\Omega=1$ Fourier modes are replaced by hyperspherical Bessel functions
with the curvature radius as a built-in length scale,
the whole notion of scale-invariance loses its meaning 
(Kamionkowski \& Spergel 1993). 

In summary, it may be advisable to avoid   
theoretical prejudice as to the shape of the primordial
power spectrum.
In this spirit, we will develop a consistency test that
requires no such assumptions whatsoever about the form of the
power spectrum. This approach was pioneered by 
Juszkiewicz; G\'orski and Silk (1987), who developed a formalism for
comparing two experiments in a power-spectrum independent manner. 
We generalize this method
to the case of more than two experiments, and then use the formalism to 
assess the consistency of three recent observational results:  the
CBR anisotropy measurements made by the COBE Differential Microwave Radiometer
(Smoot {\etal} 1992), the South Pole anisotropy experiment (SP91, Gaier {\etal}
1992), and the measurement of bulk velocity of Abell clusters in a 150 $h^{-1}
\,{\rm Mpc}$
sphere (Lauer \& Postman 1994, hereafter LP).

In 
Section~\ref{pindepsec2}, 
we develop a formalism for testing cosmological models for
consistency. 
In 
Section~\ref{pindepsec3}, 
we apply this formalism to the special case of cold
dark matter (CDM) and the LP, SP91 and COBE experiments. In 
Section~\ref{pindepsec4}, 
we solve
the variational problem that arises in consistency tests of models where we
allow arbitrary power-spectra, and apply these results to the LP,
SP91 and COBE experiments. 
Section~\ref{pindepsec5} 
contains a discussion of our results.
Finally, two different goodness-of-fit parameters are compared
in Appendix A,
and the relevant window functions are derived in
Appendix B.

\markboth{CHAPTER 7: POWER SPECTRUM INDEP{.} CONSTRAINTS}{7.2. CONSISTENCY
TEST FOR COSMOLOGICAL MODELS}
\section{Consistency Tests for Cosmological Models}
\label{pindepsec2}

In cosmology, a field where error bars tend
to be large, conclusions can depend crucially on the
probabilistic interpretation of 
confidence limits. 
Confusion has sometimes arisen
from the fact that large-scale measurements of microwave background
anisotropies and bulk flows are fraught with two quite distinct sources of
statistical uncertainty, usually termed experimental noise and
cosmic variance.
In this section, we present a detailed prescription for testing any
model for consistency with experiments, and discuss the 
appropriate probabilistic
interpretation of this test.
By {\it model} we will mean not merely a model for the  underlying
physics, which predicts the physical quantities that we wish to
measure, but also a model for the various experiments. 
Such a model is allowed to contain any number of free parameters.
In subsequent sections, we give examples of both a very narrow class of models
(standard CDM where the only free parameter
is the overall normalization of the power spectrum), 
and a wider class of models 
(gravitational instability with Gaussian adiabatic fluctuations
in a flat universe with the standard recombination history, 
the power spectrum being an arbitrary function).  

Suppose that we are interested in $\N$ physical quantities 
$\c_1, ...,\c_{\N}$, and have $\N$ experiments $E_1,...,E_{\N}$
devised such that the experiment $E_i$ measures the quantity $\c_i$.
Let $\s_i$ denote the number actually obtained by the
experiment $E_i$. Because of experimental noise, cosmic variance,
{\etc}, we do not expect $\s_i$ to exactly equal $\c_i$. Rather,
$\s_i$ is a random variable that will yield different values each
time the experiment is repeated. By repeating the experiment $M$
times on this planet and averaging the results, the uncertainty
due to experimental noise can be reduced by a factor $\sqrt M$.
However, if the same experiment were carried out in a number of
different horizon volumes throughout the universe (or, if we have
ergodicity, in an ensemble of universes with different realizations of
the underlying random field), the results would also be expected to
differ. This second source of uncertainty is known as cosmic variance.
We will treat both of these uncertainties together by simply
requiring the model to specify the probability distribution for the
random variables $\s_i$.

Let us assume that the random variables $s_i$ are all independent, so
that the joint probability distribution is 
simply the product of the individual probability distributions, which
we will denote $f_i(s)$.
This is an excellent approximation for the microwave
background and bulk flow experiments we will consider.
Finally, let $\sh_1, ...,\sh_{\N}$ denote the numbers actually obtained in one
realization of the experiments.

The general procedure for statistical testing will be as follows: 

\begin{itemize}

\item First, define a parameter $\x$ that is some sort of measure of
how well the observed data $\s_i$ agree with the probability
distributions $f_i$, with higher $\x$ corresponding to a better fit.

\item Then compute the probability distribution $f_{\x}(\x)$ of this
parameter, either analytically or by employing Monte-Carlo
techniques.

\item Compute the observed value of $\x$, which we will denote $\xh$.

\item Finally, compute the probability $P(\x<\xh)$, {\ie} the
probability of getting as bad agreement as we do or worse.

\end{itemize}

We will now discuss these four steps in more detail.

\subsection{Choosing a goodness-of-fit parameter}
\label{pindepsec2.1}

Obviously, the ability to reject models at a high level of
significance depends crucially on making a good choice of
goodness-of-fit parameter $\eta$.
In the literature, a common choice is the {\it likelihood product}, {\ie}
$$\x_l\propto\prod_{i=1}^{\N} f_i(\s_i).$$
In this chapter, we will instead use the {\it probability product}, 
{\ie} the product of the
probabilities $P_i$ that each of the experiments yield results at least as extreme as 
observed. 
Thus if the observed $\sh_i$ is smaller than the median of the 
distribution $f_i$, 
we have $P_i= 2P(\s_i < \sh_i)$, whereas  $\sh_i$ larger than the median
would give  $P_i= 2P(\s_i > \sh_i)$. The factor of two is present because
we want a two-sided test. Thus  $P_i = 1$ if $\sh_i$ equals the median,
$P_i = 2\%$ if $\sh_i$ is at the high 99th percentile, {\it etc.}

These two goodness-of-fit parameters are compared in Appendix A for a few
explicit examples, and the conclusions can be summarized as follows:

\begin{enumerate}

\item Rather unphysical probability distributions
can be concocted that ``fool" either one of these two parameters but not the
other. Thus neither the likelihood product nor the probability product can be
hailed as fundamentally better than the other.

\item For probability distributions encountered in the type of cosmology
applications discussed in this chapter, always smooth and unimodal 
functions, the likelihood product and the probability product yield 
very similar results. 

\item As described in the following sections, 
the probability distribution of the
probability product can always be computed analytically. The probability
distribution  for the likelihood product, however, depends on the distributions
of the underlying random variables, and except in a few fortuitous simple cases,
it must be computed numerically through either repeated convolutions or
Monte Carlo simulations.
Thus the probability product is considerably more convenient to use. 

\end{enumerate}

\noindent
Since 1) and 2) indicate that neither of the two goodness-of-fit
parameters is preferable over the other on scientific grounds,
the authors feel that 3) tips the balance in favor of the probability
product.

\subsection{Its probability distribution}

Apart from the simple interpretation of the 
probability product $\eta$, it has the advantage that
its probability distribution can be calculated 
analytically, and is completely independent of the
physics of the model --- in fact, it depends only 
on $\N$. We will now give the exact distributions.

By construction, $0\leq\eta\leq 1$.
For $\N=1$, $\x$ will simply have a uniform distribution, {\ie}
$$\fx(\x) = \cases{
1&if $0\leq\x\leq 1$,\cr
0&otherwise.
}$$
Thus in the general case, $\eta$ will be a product of $N$ independent
uniformly distributed random variables. 
The calculation of the probability distribution for $\x$ is straightforward,
and can be found in a number of standard texts. The result is
$$\fx(\x) = -f_z(-\ln\x){dz\over d\x} = 
\cases{
{1\over (\N-1)!}(-\ln\x)^{\N-1}&if $0\leq\x\leq 1$,\crr
0&otherwise.
}$$

\subsection{The consistency probability}

The probability $P(\x<\xh)$, the
probability of getting as bad agreement as we do or worse,
is simply the cumulative distribution function $\Fx(\xh)$,
and the integral can be carried out analytically for any $\N$:
\beq{FetaEq}
\Fx(\xh) \equiv P(\x<\xh) = \int_0^{\xh}\fx(u)du =
\xh\theta(\xh)\sum_{n=0}^{\N-1}{(-\ln\xh)^n\over n!},
\eeq
where $\theta$ is the Heaviside step function, and $\Fx(\xh)=1$ for
$\xh\geq 1$. Since the product of $\N$ numbers between zero and one tends
to zero as $\N\to\infty$, it is no surprise that 
$$\Fx(\xh)\to\theta(\xh)\xh e^{-\ln\xh} = \theta(\xh)$$
as $N\to\infty$, {\ie} that $\fx(\xh)\to\delta(\xh)$.
The function $\Fx(\xh)$ is plotted in 
Figure~\ref{pindepfig1}, 
and 
the values of $\xh$ for which $\Fx(\xh) = 0.05$, 0.01 and 0.001,
respectively, are given in 
Table~\ref{pindeptable1} 
for a few $N$-values.
For example, if three experimental results give a goodness-of-fit parameter
$\xh = 0.0002$ for some model, then this model is ruled out at a
confidence level of $99\%$.
Thus if the model where true and the experiments where repeated in very
many different horizon volumes of the universe, such a low
goodness-of-fit value would be obtained less than $1\%$ of the time.

\begin{table}
$$
\begin{tabular}{|l|llll|}
\hline
Confidence level&$\N=1$&$\N=2$&$\N=3$&$\N=4$\\
\hline
95\%&0.05&0.0087&0.0018&0.00043\\
99\%&0.01&0.0013&0.00022&0.000043\\
99.9\%&0.001&0.000098&0.000013&0.0000021\\
\hline
\end{tabular}
$$
\caption{Probability product limits}
\label{pindeptable1}
\end{table}

\subsection{Ruling out whole classes of models}

If we wish to use the above formalism to test a whole set of models,
then we need to solve an optimization problem to find the one model 
in the set for which the consistency probability is maximized. 
For instance, if the family of models under consideration is standard $n=1$,
$\Gamma=0.5$ CDM (see 
Section~\ref{pindepsec3}), 
then the only free parameter is the overall
normalization constant $A$. Thus we can write the consistency probability as
$p(A)$, and use some numerical method to find the normalization $A_*$ for which
$p(A)$ is maximized. 
After this, the
statistical interpretation is clear: if the experiments under consideration are
carried out in an ensemble of CDM universes, as extreme results as those
observed will only be obtained at most a fraction $p(A_*)$ of the time,
whatever the true normalization constant is. Precisely this case will be
treated in the next section. 
For the slightly wider class of models
consisting of CDM power spectra with arbitrary $A$, $n$  and $\Gamma$, the
resulting optimization problem would be a three-dimensional one, and 
the maximal consistency probability would necessarily satisfy
$$p(A_*,n_*,\Gamma_*) \ge p(A_*,1,0.5) = p(A_*).$$
An even more general class of models is the set of all models where the
random fields are Gaussian, {\ie} allowing completely arbitrary power spectra
$P$. In section 4, we will show that the resulting infinite-dimensional
optimization problem can in be reduced to a succession of two finite-dimensional
ones.

\markboth{CHAPTER 7: POWER SPECTRUM INDEP{.} CONSTRAINTS}{7.3. CDM
CONFRONTS SP91, COBE \& LAUER-POSTMAN}
\section{Cold Dark Matter Confronts SP91, COBE and Lauer-Postman}
\label{pindepsec3}

As an example of an application of the formalism presented in the
previous section, we will now test the standard cold dark matter (CDM)
model of structure formation for consistency with the SP91 CBR
experiment and the Lauer-Postman bulk flow experiment.

Let $E_1$ be the Lauer-Postman (LP for short) measurement of bulk flows of
galaxies in a $150h^{-1}\Mpc$ sphere (Lauer\& Postman, 1994).
Let $E_2$ be the 1991 South Pole CBR anisotropy experiment, SP91 for
short (Gaier {\etal} 1992). 
Let $E_3$ be the COBE DMR experiment (Smoot {\etal} 1992). 
All of these experiments probe scales that are well described by linear
perturbation theory, and so as long as the initial fluctuation are Gaussian,
the expected results of the experiments can
be expressed simply as integrals over the power spectrum of the matter
perturbation:
$$
\left<s_i\right>=\int W_i(k)P(k)dk.
$$
Here $s_{sp}$ and $s_{cobe}$ are the mean-square temperature fluctuations
measured by the experiments, and $s_{lp}\equiv (v/c)^2$ is the squared bulk 
flow.
The corresponding
window functions $\Wlp$, $\Wsp$ and $\Wc$ are derived in Appendix B, and
plotted in 
Figure~\ref{pindepfig2}.
These window functions assume  that the initial 
perturbations were adiabatic, that $\Omega=1$, and that recombination
happened in the standard way, 
{\i.e.} a last-scattering surface at $z\approx 1000$. The
SP91 window function is to be interpreted as a lower limit to the true window
function, as it includes contributions only from the Sachs-Wolfe effect, not
from Doppler motions or intrinsic density fluctuations of the
surface of last scattering.

Now let us turn to the probability distributions for the random variables
$\slp$, $\ssp$ and $\scobe$.
The standard 
CDM model with power-law initial fluctuations 
$\propto k^n$ predicts a power
spectrum that is well fitted by 
(Bond \& Efstathiou 1984) 
$$P(k) = 
{A q^n\over
\left(1+\left[aq+(bq)^{1.5} +
(cq)^2\right]^{1.13}\right)^{2/1.13}},$$
where $a=6.4$, $b=3.0$, $c=1.7$ and 
$q=(1h^{-1}\Mpc) k/\Gamma$.
For the simplest model, $\Gamma=h$, but certain additional complications
such as a non-zero cosmological constant $\Lambda$ and a non-zero fraction
$\Omega_{\nu}$ of hot dark matter can be fitted with reasonable accuracy
by other values of $\Gamma$ (Efstathiou, Bond \& White 1992).
Thus the model has three free parameters: $n$, $\Gamma$ and the overall
normalization A. Integrating the power spectrum against
the three window functions yields the values of $c_i$ given in
Table~\ref{pindeptable2}. 
The two rightmost columns contain the 
quotients $\clp/\csp$ and $\clp/\ccobe$,
respectively. 
\begin{table}
$$
\begin{tabular}{|cc|ccccc|}
\hline
$n$&$\Gamma$&LP&SP91&COBE&LP/SP91&LP/COBE\\
\hline
1&0.5&$9.2\tt{-7}$&$1.6\tt{-8}$&$2.0\tt{-8}$&56.7&45.1\\
0.7&0.5&$1.7\tt{-6}$&$2.9\tt{-8}$&$5.1\tt{-8}$&57.2&32.7\\
2&0.5&$1.4\tt{-7}$&$2.6\tt{-9}$&$1.2\tt{-9}$&53.9&112.2\\
1&0.1&$1.4\tt{-6}$&$2.3\tt{-8}$&$4.3\tt{-8}$&57.9&31.9\\
1&10&$2.3\tt{-7}$&$4.1\tt{-9}$&$4.6\tt{-9}$&55.9&49.6\\
\hline
\end{tabular}
$$
\caption{Expected r.m.s. signals for CDM power spectrum with 
$A=(1h^{-1}\Mpc)^3$}
\label{pindeptable2}
\end{table}
\noindent
As can be seen, the dependence on $\Gamma$ is quite weak, and the quotient
$\clp/\csp$ is quite insensitive to the spectral index $n$ as well.
Let us for definiteness assume the canonical values $n=1$ and $\Gamma=0.5$ 
in what
follows.

These values $\c_i$ would be the average values of the probability
distributions for $\ssp$ and $\slp$ if there where no experimental
noise. We will now model the full probability distributions of the three
experiments, including the contribution from experimental noise.

For a bulk flow experiment, the 
three components $v_x$, $v_y$ and $v_z$ of the velocity vector $\vv$ are
expected to be independent Gaussian random variables with zero mean, and  
$$\expec{|\vv|^2} = \clp.$$
However, this is not quite the random variable $\slp$ that we measure,
because of errors in distance estimation, {\it etc.} Denoting the
difference between the observed and true bulk velocity vectors by $\ve$,
let us assume that the three components of $\ve$ are identically
distributed and independent Gaussian random variables. This should be a
good approximation, since even if the errors for individual galaxies are
not, the errors in the average velocity $\ve$ will be approximately
Gaussian by the Central Limit Theorem. Thus the velocity vector that we
measure, $\vv+\ve$, is also Gaussian, being the sum of two Gaussians. 
The variable that we actually measure is $\slp = |\vv+\ve|^2$, so 
$$\slp = {1\over 3}\left(\clp + \vlp\right)\chi^2_3,$$
where $\chi^2_3$
has a chi-squared distribution with three degrees of freedom, 
and $\vlp$ is the variance due to experimental noise, {\ie} the
average variance that would be detected even if the true power spectrum
were $P(k)=0$. 
The fact that the expectation value of the 
detected signal $\slp$ (which is usually
referred to as the {\it uncorrected} signal in the literature) 
exceeds the true signal $\clp$ is usually referred to as 
{\it error bias} 
(LP; Strauss, Cen \& Ostriker 1993 --- hereafter SCO). 
Error bias is ubiquitous to all 
experiments of the type discussed in this chapter, including 
CBR experiments, since the 
measured quantity is positive definite and the noise errors
contribute squared. In the
literature, experimentally detected signals are
usually quoted after error bias has been corrected for, {\ie}
after the noise has been subtracted from the uncorrected signal in
For LP, the uncorrected signal is 807 km/s, whereas the 
signal quoted after error bias correction is 689 km/s.

For the special case of the LP experiment, detailed
probability distributions have been computed using Monte-Carlo
simulations 
(LP, SCO), 
which incorporate such experiment-specific complications as
sampling errors,  asymmetry in the error ellipsoid, {\it etc.}
To be used here, such simulations would need to be carried
out for each value of $\clp$ under consideration. 
Since the purpose of this section is merely to
give an example of the test formalism, the above-mentioned
$\chi^2$-approximation will be quite sufficient for our needs. 

% \bigskip
% (The following is not quite satisfactory - I want to do it properly when
% I find out how they ACTUALLY analyzed their numbers:)

For the SP91
nine-point scan, the nine true values $\Delta T_i/T$ 
are expected to be Gaussian random variables that to a good approximation
are independent. They have zero mean, and 
$$\expec{\left|{\Delta T_i/T}\right|^2} = \csp.$$ 
Denoting the difference
between the actual and observed values by $\delta_i$, we make the
standard assumption that these nine quantities are identically distributed
and independent Gaussian random variables. 
Thus the temperature fluctuation that we measure at
each point,  $\Delta T_i/T+\delta$, is
again Gaussian, being the sum of two Gaussians. 
The variable that we
actually measure is
$$\ssp = 
{1\over 9} \sum_1^9 \left({\Delta T_i\over T}+\delta_i\right)^2
= {1\over 9}\left(\csp + \vsp\right)\chi^2_9,$$
where $\chi^2_9$ has a chi-squared
distribution with nine degrees of freedom, and $\vsp$ is the variance
due to experimental noise, the error bias, 
{\ie} the average variance that would be
detected even if the true power spectrum were $P(k)=0$. 

We will use only the signal from highest of the four
frequency channels, which is the one likely to be the least affected
by galactic contamination. Again, although Monte-Carlo
simulations would be needed to obtain the exact probability
distributions, we will use the simple $\chi^2$-approximation here. 
In this case, the main experiment-specific complication is the
reported gradient removal, which is a non-linear operation and thus
does not simply lead to a $\chi^2$-distribution with fewer degrees of
freedom. 

The amplitude of the COBE signal can be characterized by the variance in
$\Delta T/T$ on an angular scale of $10^\circ$.  This number can be
estimated from the COBE data set as
$\scobe=\sigma^2_{10^\circ}= ((11.0\pm 1.8)\times
10^{-5})^2$ (Smoot {\etal} 1992).   
The uncertainty in this quantity is purely due to
instrument noise, and contains no allowance for cosmic variance.  We
must fold in the contribution due to cosmic variance in order to determine
the probability distribution for $\scobe$.
We determined this probability distribution by performing Monte-Carlo
simulations of the COBE experiment.  We made simulated COBE maps with a
variety of power spectra (including power laws with indices ranging
from $0$ to $3$, as well as delta-function power spectra of the sort described
in 
Section~\ref{pindepsec2}). 
We included instrumental noise in the maps, and excluded
all points within $20^\circ$ of the Galactic plane.  By estimating
$\scobe$ from each map, we were able to construct a probability 
distribution corresponding to each power spectrum. In all cases, the
first three moments of the distribution were well approximated by
\beq{COBEfitEq1}
\cases{
\mu_1&$\equiv\expec{\scobe}=\ccobe$,\crr
\mu_2&$\equiv\expec{\scobe^2}-
\expec{\scobe}^2\le 0.063 \ccobe^2+
1.44\times 10^{-21}$,\crr
\mu_3&$\equiv\expec{\scobe^3}=0.009 \ccobe^3.$\cr
}
\eeq
Furthermore, in all cases the probability distributions were well modeled
by chi-squared distributions with the number of degrees of freedom,
mean, and offset chosen to reproduce these three moments.  Note that
the magnitude of the cosmic variance depends on the shape of the power
spectrum as well as its amplitude.  The inequality in the above expression
for $\mu_2$ represents the largest cosmic variance of any of the power
spectra we tested.  Since we wish to set conservative limits on models,
we will henceforth assume that the cosmic variance is given by this
worst-case value.
Thus we are assuming that the random variable 
$(\scobe-s_0)/\Delta s$ has a chi-squared distribution with 
$\delta$ degrees of freedom, where 
\beq{COBEfitEq2}
\cases{
\s_0&$=\mu_1-2\mu_2^2/\mu_3$,\crr
\Delta s&$=\mu_3/4\mu_2$,\crr
\delta&$=8\mu_2^3/\mu_3^2$.\cr
}
\eeq

The results obtained using these three probability distributions are
summarized in 
Tables~\ref{pindeptable3a},~\ref{pindeptable3b} and~\ref{pindeptable3c}. 
In~\ref{pindeptable3a} and~\ref{pindeptable3b}, $\N=2$, and the question
asked is whether LP is consistent with COBE and SP91, respectively. In
Table~\ref{pindeptable3c},
$\N=3$, and we test all three experiments for consistency
simultaneously. In each case, the optimum normalization (proportional to
the entries labeled ``Signal") is different, chosen such that
the consistency probability for the experiments under consideration is
maximized.
As can be seen, the last two tests rule out CDM at a significance
level of  
$95\%$, {\ie}
predict that in an ensemble of universes, results as extreme 
as those we observe would be obtained less than $5\%$ of the time.
Note that 
using both COBE and SP91 to
constrain LP yields a rejection that is no stronger than that
obtained when ignoring COBE. In the latter case, the best fit is
indeed that with no cosmological power at all, which agrees well with
the observation of SCO
that sampling
variance would lead LP to detect a sizable bulk
flow (before correcting for error bias) even if there where none. 

\begin{table}[p]
$$
\begin{tabular}{|l|rrl|}
\hline
 &LP&COBE&Combined\\
\hline
Noise&$420\kms$&$9.8\mK$&\\
Signal&$169\kms$&$33.8\mK$&\\
Noise+Signal&$453\kms$&$35.2\mK$&\\
Detected&$807\kms$&$35.2\mK$&\\
\hline
$\xh$&0.046&1.00&0.046\\
$P(\x<\xh)$&0.046&1.00&0.19\\
\hline
\end{tabular}
$$
\vskip-0.3cm
\caption{Are LP and COBE consistent with CDM?}
\vskip0.2cm
\label{pindeptable3a}
\end{table}

\begin{table}[p]
$$
\begin{tabular}{|l|rrl|}
\hline
 &LP&SP91&Combined\\
\hline
Noise&$420\kms$&$26.4\mK$&\\
Signal&$0\kms$&$0\mK$&\\
Noise+Signal&$420\kms$&$26.4\mK$&\\
Detected&$807\kms$&$19.9\mK$&\\
\hline
$\xh$&0.023&0.35&0.0079\\
$P(\x<\xh)$&0.023&0.35&0.046\\
\hline
\end{tabular}
$$
\vskip-0.3cm
\caption{Are LP and SP91 consistent with CDM?}
\vskip0.2cm
\label{pindeptable3b}
\end{table}

\begin{table}[p]
$$
\begin{tabular}{|l|rrrl|}
\hline
 &LP&SP91&COBE&Combined\\
\hline
Noise&$420\kms$&$26.4\mK$&$9.8\mK$&\\
Signal&$168\kms$&$26.9\mK$&$33.8\mK$&\\
Noise+Signal&$452\kms$&$37.7\mK$&$35.1\mK$&\\
Detected&$807\kms$&$19.9\mK$&$35.2\mK$&\\
\hline
$\xh$&0.046&0.039&0.97&0.0017\\
$P(\x<\xh)$&0.046&0.039&0.97&0.046\\
\hline
\end{tabular}
$$
\vskip-0.3cm
\caption{Are LP, SP91 and COBE all consistent with CDM?}
\vskip0.2cm
\label{pindeptable3c}
\begin{quote}
Tables~\ref{pindeptable3a},~\ref{pindeptable3b} and~\ref{pindeptable3c}
show the consistency probability calculations.
The first line in each table gives the experimental noise, {\ie} 
the detection that would be expected in the absence of any cosmological 
signal.
The second line is the best-fit value for the cosmological signal $c$,
the value that maximizes the combined consistency probability
in the lower right corner of the table.
The third line contains 
the expected value of an experimental detection, and
is the sum in quadrature of the two preceding lines.
The fourth line gives the goodness-of-fit parameter for 
each of the experiments, i.e. the probability that they would yield 
results at least as extreme as they did. 
The rightmost number is the combined goodness-of-fit parameter, which is 
the product
of the others.
The last line contains the consistency probabilities, the probabilities
of obtaining goodness-of-fit parameters at least as low as
those on the preceding line.
\end{quote}
\vfill
\end{table}

\markboth{CHAPTER 7: POWER SPECTRUM INDEP{.} CONSTRAINTS}{7.4. ALLOWING
ARBITRARY POWER SPECTRA}
\section{Allowing Arbitrary Power Spectra}
\label{pindepsec4}

In this section, we will derive the mathematical formalism
for testing results from multiple experiments for consistency, without making
any assumptions whatsoever about the power spectrum. 
This approach was pioneered by Juszkiewicz
{\etal} (1987) for the case $\N=2$. Here we generalize the
results to the case of arbitrary $\N$.
Despite the fact that the original optimization problem is
infinite-dimensional, the necessary calculations will be seen to be of a
numerically straightforward type, the case of $\N$ 
independent constraints leading
to nothing more involved than numerically solving a system of $N$ coupled
non-linear equations. After showing this, we will discuss some inequalities
that provide  both a good approximation of the exact results and a 
useful qualitative understanding of them.

\subsection{The optimization problem}
\label{pindepsec4.1}

Let us consider $N = n+1$ experiments numbered 0, 1, ..., $n$ that
probe the cosmological power spectrum $P(k)$. We will think of each
experiment as measuring some weighted average of the power spectrum,
and characterize an experiment $E_i$ by its window function
$W_i(k)$ as before. 

Purely hypothetically, suppose we that we had repeated the same
experiments in many different locations in the universe, and for all
practical purposes knew the quantities $c_1,...,c_n$ exactly. Then for
which power spectrum $P(k)$ would $c_0$ be maximized, and what would
this maximum be? If we experimentally determined $c_0$ to be larger
than this maximum value, our results would be inconsistent, and we
would be forced to conclude that something was fundamentally wrong
either with our theory or with one of the experiments. In this section, we will
solve this hypothetical problem. After this, it will be seen that the real
problem, including cosmic variance and experimental noise, can be solved in
almost exactly the same way.

The extremal power spectrum we are looking for is the solution to the
following linear variational problem: 

Maximize
\beq{ContMinEq}
\intk P(k) W_0(k) dk
\eeq
subject to the constraints that
$$\cases{
\intk P(k) W_i(k) dk = c_i&for $=1,...,n$,\crr
P(k) \geq 0&for all $k\geq 0.$
}$$

This is the infinite-dimensional analogue of the so called
linear programming problem, and its solution is quite analogous to
the finite-dimensional case.
In geometrical terms, we think of each power spectrum as a point in the
infinite dimensional vector space of power spectra (tempered
distributions on the positive real line, to be precise), and limit
ourselves to the subset $\Omega$ of points where all the above
constraints are satisfied. We have a linear function on this space,
and we seek the point within the subset $\Omega$ where this function is
maximized.
We know that a differentiable functional on a bounded region takes its
maximum either at an interior point, at which its gradient will
vanish, or at a boundary point. 
In linear optimization problems like the one above, the
gradient (here the variation with respect to $P$, which is simply
the function $W_0$) is simply a constant, and will never vanish.
Thus any maximum will always be attained at a boundary point.
Moreover, from the theory of linear programming, we know that if
there are $n$ linear constraint equations, then the optimum point
will be a point where all but at most $n$ of the coordinates are zero.
It is straightforward to generalize this result to our
infinite-dimensional case, where each fixed $k$ specifies
a ``coordinate" $P(k)$, and the result is that the solution to the
variational problem is of the form
$$P(k) = \sumi p_i \delta(k-k_i).$$
This reduces the optimization problem from an infinite-dimensional
one to a $2n$-dimensional one, where only the constants $p_i$ and $k_i$
remain to be determined:

Maximize 
\beq{DiscMinEq}
\sumj p_j W_0(k_j)
\eeq
subject to the constraints that

$$\cases{
\sumj p_j W_i(k_j) = c_i&for $i=1,...,n$,\crr
p_i \geq 0&for $i=1,...,n$.
}$$
This problem is readily solved using the method of Lagrange multipliers:
defining the Lagrangian 
$$L = \sumj p_j W_0(k_j) - 
\sumi\lambda_i\left[\sumj p_j W_i(k_j) - c_i\right]$$
and requiring that all derivatives vanish leaves the following set of $3n$
equations to determine the $3n$ unknowns $p_i$, $k_i$ and
$\lambda_i$:
$$
\cases{
W_0(k_i) - \sumj \lambda_j W_j(k_i)&$= 0$,\crr
\left[W'_0(k_i) - \sumj \lambda_j W'_j(k_i)\right]p_i&$= 0$,\crr
c_i - \sumj p_j W_i(k_j)&$= 0$.
}
$$
Introducing matrix notation by defining the $k_i$-dependent quantities
$A_{ij}\equiv W_j(k_i)$, $B_{ij}\equiv W'_j(k_i)$, 
$a_i\equiv W_0(k_i)$ and $b_i\equiv W'_0(k_i)$ brings out the
structure of these equations more clearly:
If $p_i\neq 0$, then 
$$
\cases{
A\vl&$= \va$,\crr
B\vl&$= \vb$,\crr
A^T\vp &$= \vc$.
}
$$
If $A$ and $B$ are invertible, then eliminating $\vl$ from the
first two equations yields the following system of $n$
equations to be solved for the $n$
unknowns $k_1$, ..., $k_n$: 
\beq{nonlinEq}
A^{-1}\va = B^{-1}\vb.
\eeq
Although this system is typically coupled and non-linear and out of
reach of analytical solutions for realistic window functions, solving
it numerically is quite straightforward. 
A useful feature is that once this system is solved, $\va$, $\vb$, $A$ and
$B$ are mere constants, and the other unknowns are simply given by
matrix inversion: 
$$
\cases{
\vl&$= A^{-1}\va$\crr
\vp&$= \left(A^{-1}\right)^T\vc$\crr
}
$$
Since the non-linear system\eqnum{nonlinEq} may have more than
one solution, all solutions should be substituted back 
into\eqnum{DiscMinEq} to determine which one is the
global maximum.
Furthermore, to make statements about the solution to our original
optimization problem\eqnum{ContMinEq}, we need to consider also
the case where one or more of the $n$ variables $p_1$,...,$p_n$ vanish.
If exactly $m$ of them are non-vanishing, then without loss of
generality, we may assume that these are the first $m$ of the $n$
variables. Thus we need to solve the maximization 
problem\eqnum{DiscMinEq} separately for the cases where $P(k)$ is
composed of $n$ delta functions, $n-1$ delta functions, {\etc},
all the way down to the case where $P(k)$ is single delta function. 
These solutions should then be substituted back 
into\eqnum{ContMinEq} to determine which is the 
global maximum sought
in our original problem. 
Thus the solutions depend on the window functions $W_i$ and the
signals $c_i$ in the following way:

\begin{itemize}

\item From the window functions alone, we can determine a discrete
and usually finite number of candidate wavenumbers $k$ where delta
functions can be placed.

\item The actual signals $c_i$ enter only in determining the
coefficients of the delta functions in the sum, i.e. in determining
what amount of power should be hidden at the various candidate
wavenumbers.

\end{itemize}

\noindent
If we have found an the optimal solution, then a small change in the 
signal vector $\vc$ will typically result in a small change in $\vp$
and no change at all in the number of delta functions in $P(k)$ or
their location. If $\vc$ is changed by a large enough amount, the
delta functions may suddenly jump and/or change in number as a
different solution of\eqnum{DiscMinEq} takes over as global
optimum or one of the coefficients $p_i$ becomes negative, the latter
causing the local optimum to be rejected for constraint violation.
Thus within certain limits, we get the extremely simple result that
for the optimal power spectrum $P(k)$,
$$c_0 = \intk P(k) W_0(k) dk =
\left(A^{-1}\va\right)\cdot\vc.$$
Thus within these limits, 
$c_0$ depends linearly on the observed
signal strengths $c_i$. This is exactly analogous to what happens in
linear programming problems.

\subsection{A useful inequality}

Before proceeding further, we will attempt to provide a more
intuitive understanding of the results of the previous section, and show how to
determine how complicated a calculation is justified. 
For the special case of only
a single constraint, {\ie} $n=1$, we obtain simply $P(k) = p_1\delta(k-k_1)$,
where $k_1$ is given by   
$$W_0'(k_1)W_1(k_1) = W_1'(k_1)W_0(k_1).$$
For the case of $n$ constraints, let us define the functions 
$$\f_i\equiv  {W_0(k)\over W_1(k)} c_i.$$
Then we see that for $n=1$, $k_1$ is simply the wavenumber for which the
function $\f_1$ is maximized, and that the maximum signal possible is simply
$c_0 = \f_1(k_1)$.
Thus the maximum signal in experiment $0$ that is consistent with
the constraint from experiment $i$ is obtained when 
the power is concentrated where the function $\f_i$ is
large. In other words, if we want to explain a high signal $c_0$ in the face
of low signals in several constraining experiments, then the best
place to hide the necessary power from the $i^{th}$ experiment is where $f_i$
takes its maximum. 
These functions are plotted in 
Figure~\ref{pindepfig4} 
for the experiments discussed in
Section~\ref{pindepsec3}, 
the optimization problem being the search for 
the maximum LP signal that is consistent with the constraints from SP91 
and COBE. For illustrative purposes, we here assume that $\csp$ and $\ccobe$ are
known exactly, and given by the detected signals 
$\ssph^{1/2}\approx 19.9\mK$ and $\scobeh^{1/2}\approx 33.8\mK$
(we will give a
proper treatment of cosmic variance and noise in the following section).
Using the $n=1$ constraint for each constraining experiment separately, the
smallest of the functions thus sets an
upper limit to the allowed signal $c_0=\csp$.
Thus the limit is given by the highest point in the hatched region in
Figure~\ref{pindepfig4}, 
{\ie} 
$$c_0\leq c_{max}^{(1)}\equiv\sup_k\>\min_i f_i(k).$$
We see that using the SP91 constraint
alone, the LP signal would be maximized if all power were at $k\approx
(940\Mpc)^{-1}$. Since this flagrantly violates the COBE constraint, the best
place to hide the power is instead at $k\approx (100\Mpc)^{-1}$.

By using the above formalism to impose all the constraints at once, the
allowed signal obviously becomes lower. If the constraints are equalities rather
than inequalities, then this stronger limit can
never lie below the value at $(k_*\approx 250\Mpc)^{-1}$, where 
$\fsp(k_*)=\fcobe(k_*)$, since this is the signal that would result from a power
spectrum of the form $P(k)\propto\delta(k-k_*)$. Thus for the particular window
functions in our example, where the constraint from the $n=2$ calculation cannot 
be more than a factor $\fsp(80\Mpc)/\fsp(250\Mpc)\approx 1.05$ stronger than
the simple $n=1$ limits, the latter are so close to the true optimum that they 
are quite sufficient for our purposes. 
If the constraints are upper limits
rather than equalities, then the limit on $c_0$ is more relaxed, and is always
the uppermost point in the hatched region, {\ie} $c_{max}^{(1)}$.

\subsection{Including noise and cosmic variance}

To correctly handle cosmic variance and instrumental noise, we need to use the
formalism developed in 
Section~\ref{pindepsec2}. 
Thus given the probability distributions for the various experimental results
$s_i$, we wish to find the power spectrum for which the consistency probability
$\x$ is maximized. 
This optimization problem, in which
all experiments are treated on an equal footing, will be seen to lead directly to
the asymmetric case above where the signal in one is
maximized given constraints from the others.
For definiteness, we will continue using the example with the LP, SP91
and COBE experiments.
As seen in 
Section~\ref{pindepsec2}, 
the source of the low consistency probabilities is that
$\slph$ is quite high when compared to $\ssph$ and $\scobeh$. Thus
it is fairly obvious that for the power spectrum that maximizes the
consistency probability, we will have $\slph > \expec{\slp}$, whereas
$\ssph < \expec{\ssp}$ and $\scobeh < \expec{\scobe}$, so we can neglect power
spectra that do not have this property. Let us first restrict ourselves to
the subset of these power spectra for which  $\clp = D$ and $\ccobe = E$, where
$D$ and $E$ are some constants.
Then these power spectra all predict the same probability distributions for 
$\slp$ and $\scobe$. The consistency probability $\eta$ is clearly maximized
by the power spectrum that maximizes $\expec{\ssp}$, and this will be a
linear combination of one or two delta functions as shown in 
Section~\ref{pindepsec4.1}. 
The
key point is that since the locations of these delta functions are
independent of $D$ and $E$ (within the range discussed in 4.1), the
infinite-dimensional optimization problem reduces to the following two
simple steps:

\begin{enumerate}

\item Solve for the optimal number of delta functions $m$ and their locations
$k_i$ as described in 
Section~\ref{pindepsec4.1}

\item Find the $m$ coefficients $p_i$ for which the power spectrum 
$P(k) = \sum_{i=1}^m p_i\delta(k-k_i)$ maximizes the consistency probability.

\end{enumerate}

\subsection{Power spectrum independent constraints on LP, SP91 and COBE}

When applying the above consistency test to the LP, SP91 and COBE
experiments, we obtain exactly the same consistency probability
as in 
Table~\ref{pindeptable3b}.
The reason for this is that the optimal 
normalization turns out to be zero. This will obviously change if the
LP error bars become smaller in the future. 
Thus dropping the CDM assumption does not improve the situation
at all, which indicates that main source of the
inconsistency must be something other than the CDM model.

In anticipation of future developments, 
consistency probabilities were also computed for a number of cases
with less noise in the LP experiment. 
Comparing only LP and SP91, the optimum power spectrum has a delta function at 
$k\approx (941\Mpc)^{-1}$.
When including all three experiments, treating the COBE and SP91
constraints as upper limits, 
the optimum power spectrum has a single delta
function at $k\approx (79\Mpc)^{-1}$, so 
the addition of COBE strengthens the constraint only
slightly, due to the
flatness of $\fsp$ in 
Figure~\ref{pindepfig4}. 
Interestingly, for all these cases with smaller LP error bars, 
consistency  probabilities were found to be almost as low when
allowing arbitrary power spectra as for the CDM case. This is again
attributable to the flatness of $\fsp$, since weighted averages of a
flat function are fairly independent of the shape of the weight
function (here the power spectrum).

\markboth{CHAPTER 7: POWER SPECTRUM INDEP{.} CONSTRAINTS}{7.5. DISCUSSION}
\section{Discussion}
\label{pindepsec5}

We have developed a formalism for testing multiple cosmological experiments for
consistency. 
As an example of an application, we have used it to place 
constraints on bulk flows of
galaxies using the COBE and SP91 measurements of fluctuations in the cosmic
microwave background.
It was found that taken at face value, 
the recent detection by Lauer and Postman of a bulk
flow of $689$ km/s on scales of $150h^{-1}$Mpc 
is inconsistent with SP91
within the framework of a CDM
model, at a significance level of about 95\%.
However, interestingly, this cannot be due solely to the CDM assumption,
since the LP result was shown to be inconsistent with COBE and
SP91 at the same significance level even when no assumptions whatsoever
were made about the power
spectrum. This leaves four possibilities:

\begin{enumerate}

\item The window functions are not accurate.

\item Something is wrong with the quoted signals
or error bars for at least one of the experiments,

\item The observed fluctuations cannot be explained within the framework 
of gravitational instability and the Sachs-Wolfe effect.

\item The random fields are not Gaussian,

\end{enumerate}

\noindent
Case (1) could be attributed to a number of effects:
If $\Omega\neq 1$, then both the calculation of the Sachs-Wolfe 
effect (which determines $\Wsp$ and $\Wc$) and the growth of
velocity perturbations (which determines $\Wlp$) are altered.
If the universe became reionized early enough to rescatter a significant 
fraction of all CBR photons, then small scale CBR anisotropies
were suppressed, which would lower $\Wsp$. A quantitative treatment of these 
two cases will be postponed to future work. 
Other possible causes of (1) include 
a significant fraction of the density 
perturbations being isocurvature (entropy) perturbations or 
tensor-mode perturbations
(gravity waves).
Apart from these uncertainties, we have made several 
simplifying assumptions about the window functions for LP and SP91.
To obtain more accurate consistency probabilities than those derived
in this chapter, a more accurate LP window function should be used
that incorporates the discreteness and the asymmetry of 
the sample of Abell clusters used. This can either be done 
analytically (Feldman \& Watkins 1993) or circumvented altogether 
by performing Monte-Carlo simulations like those of LP or 
SCO, but for the whole family of power spectra under consideration. 

As to case (2), there has been considerable debate about both the LP and 
the SP91 experiments.
A recent Monte-Carlo Simulation of LP by SCO
basically confirms the large error bars quoted by LP. As is evident from the
flatness of the LP curve in 
Figure~\ref{pindepfig3}, 
it will be impossible to make very strong
statements about inconsistency until future experiments produce smaller error
bars. 
With the SP91 experiment, a source of concern is the validity of using
only the highest of the four frequency channels to place limits, even though
it is fairly clear that the other three channels suffer from problems with
galactic contamination. The situation is made more disturbing by the fact that a
measurement by the balloon-borne MAX experiment (Gundersen {\etal} 1993) has
produced detections of degree-scale fluctuations that that are higher than
those seen by SP91, and also higher than another MAX measurement (Meinhold
{\etal} 1993). Other recent experiments that have detected greater fluctuations
include ARGO (de Bernardis {\etal} 1993), 
PYTHON (Dragovan {\etal} 1993) and MSAM (Cheng {\etal} 1993).
On the other hand, SP91 has been used only as an upper limit
in our treatment, by including only the Sachs-Wolfe effect
and neglecting both 
Doppler contributions
from peculiar motions of the surface of last scattering 
and intrinsic density fluctuations at the recombination epoch.
If these effects (which unfortunately depend strongly on parameters such as
$h$ and $\Omega_b$) where included, the resulting constraints 
would be stronger.

Case (3) might be expected if the universe 
underwent a late-time 
phase transition, since this could generate new large-scale 
fluctuations in an entirely non-gravitational manner.

In the light of the many caveats in categories (1) and (2),
the apparent 
inconsistency between LP and SP91 (Jaffe {\etal} 1993) 
is hardly a source of major concern at the present time, 
and it does not appear necessary to invoke (3) or (4).
However, we expect the testing formalism developed in this chapter 
to be able to provide many useful constraints in the future, as
more experimental data is accumulated and error bars become smaller.

\def\fheight{10.3cm} \def\fwidth{14.5cm}

\newpage
\vfill
\goodbreak

\begin{figure}[h]
\psfig{figure=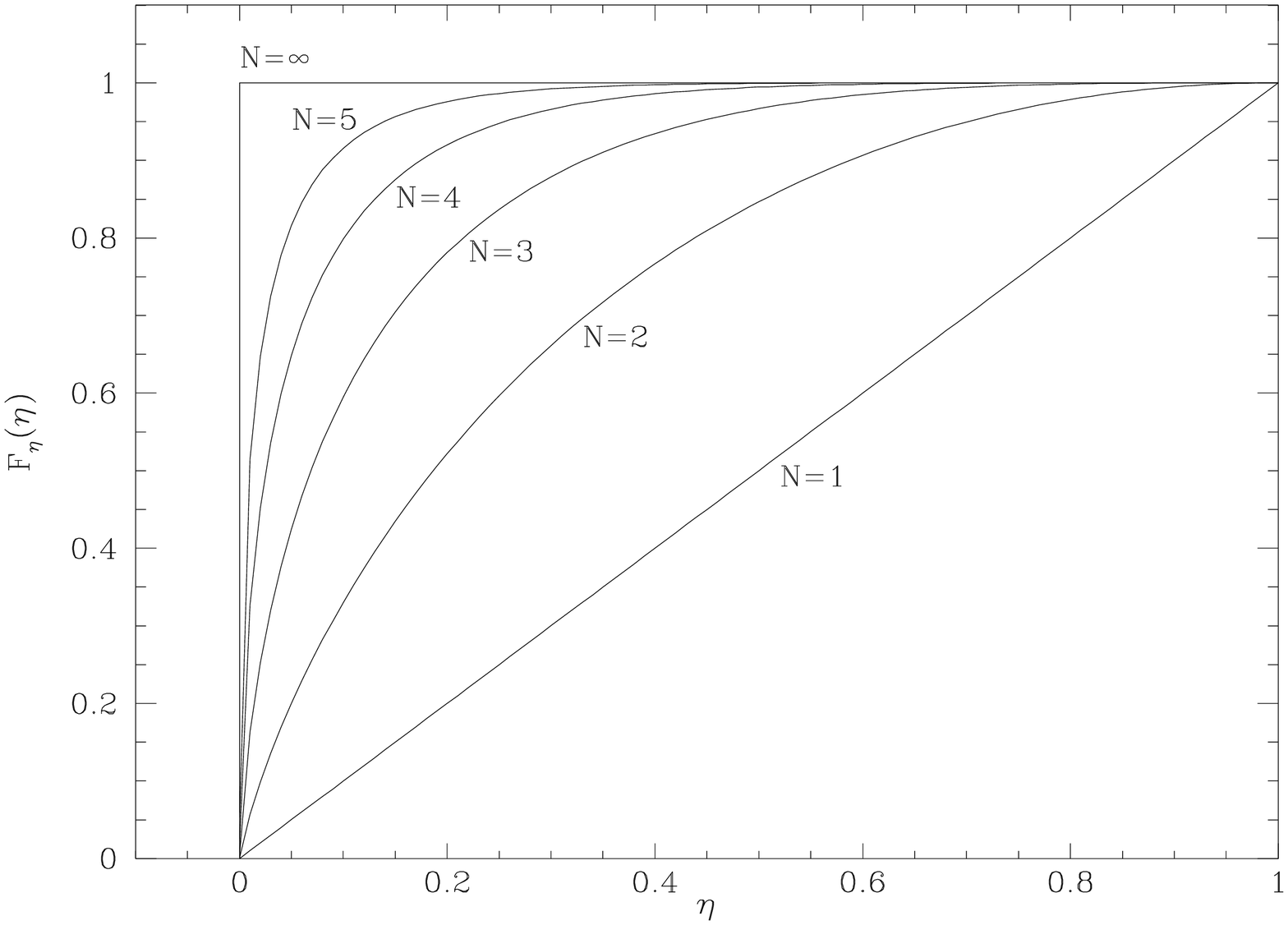,width=\fwidth,height=\fheight}
\caption{The function $F_{\eta}$.}
\label{pindepfig1}

\mycaption{The cumulative probability 
distribution for the goodness-of-fit parameter 
$\eta$ is plotted for a few different $n$-values.
}
\end{figure}
\vfill

\goodbreak
\begin{figure}[h]
\psfig{figure=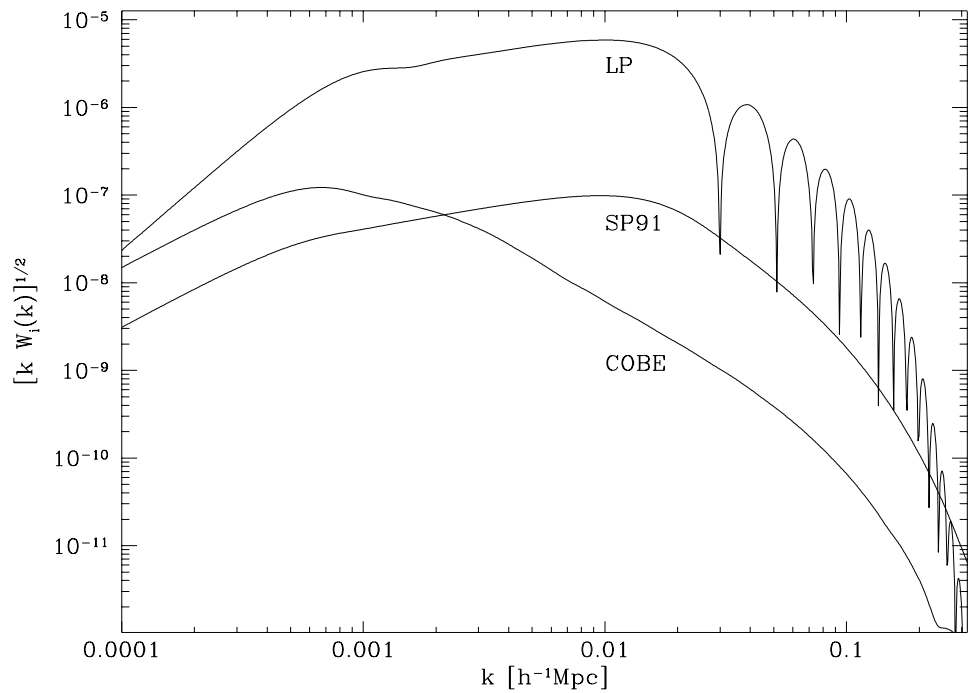,width=\fwidth,height=\fheight}
\nobreak
\caption{Window functions.}
\label{pindepfig2}

\mycaption{The window functions of the Lauer-Postman
bulk flow measurement (LP), the 
South Pole 1991 nine point scan (SP91),
and the COBE DMR $10^{\circ}$ pixel {r.m.s.} 
measurement (COBE) are plotted as a function of 
comoving wavenumber $k$.
}
\end{figure}
\vfill

\goodbreak
\begin{figure}[h]
\psfig{figure=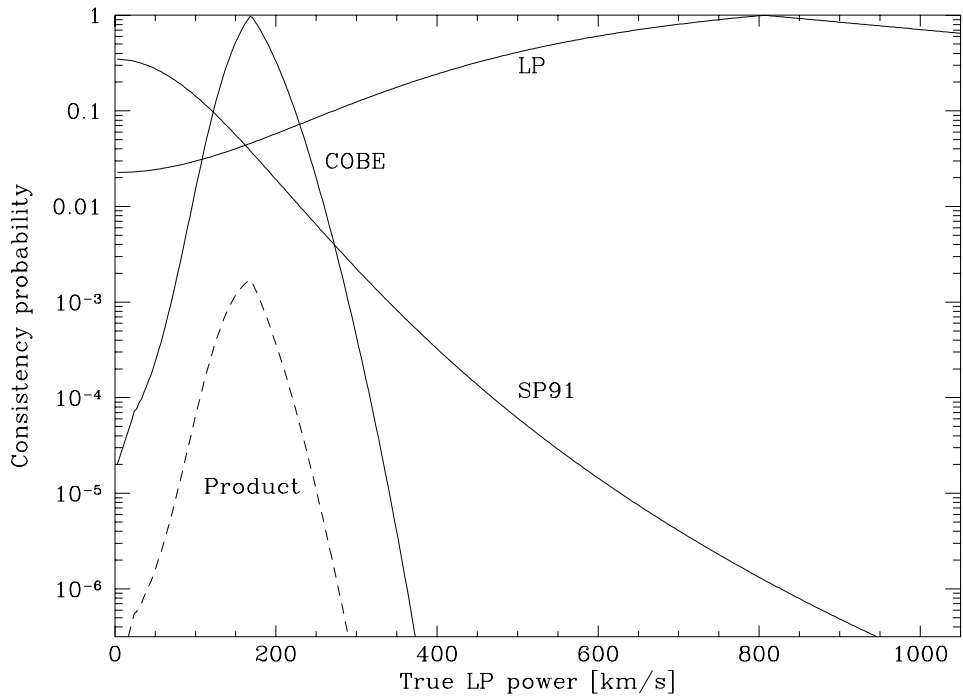,width=\fwidth,height=\fheight}
\nobreak
\caption{Consistency probabilities.}
\label{pindepfig3}

\mycaption{The probability that the the Lauer-Postman
bulk flow measurement (LP), the COBE DMR experiment and the 
South Pole 1991 nine point scan (SP91)
are consistent with CDM is plotted as a function of the normalization 
of the power spectrum. The normalization is expressed in terms of 
the expected bulk flow in a LP measurement.
The dashed line is the product of these three probabilities, and takes a 
maximum for a normalization corresponding to 168 km/s. 
}
\end{figure}
\vfill

\goodbreak
\begin{figure}[h]
\psfig{figure=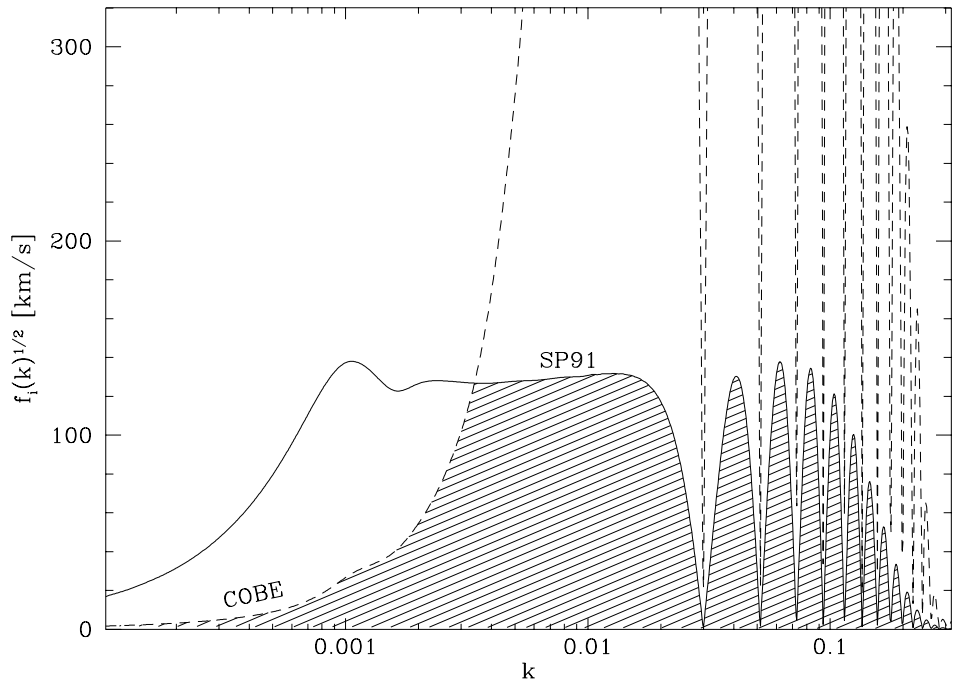,width=\fwidth,height=\fheight}
\caption{The best places to hide power.}
\label{pindepfig4}

\mycaption{The functions $f_{sp}$ (solid line) 
and $f_{cobe}$ (dashed line)
are plotted as a 
function of wavenumber $k$. 
The shaded region, {\it i.e.} the area lying beneath both curves,
constitutes the LP bulk flows that would be consistent with 
the SP91 and COBE experiments using the $N=1$ constraints only, when 
the power spectrum is a single delta function located at $k$.
}
\end{figure}
\vfill

\cleardoublepage

\pagestyle{headings}
\appendix
\chapter{The Efficiency Parameter $\fion$}
\label{fionappendix}

In this appendix, we will discuss the parameter $\fion$, and see that
it rarely drops below $30\%$. All the notation is that of
Chapter~\ref{reionchapter}. 
We will first discuss the thermal
evolution of intergalactic hydrogen exposed to a strong UV flux, and then
use the results to write down a differential equation for the
volume fraction of the universe that is ionized, subject to 
point sources of UV radiation that switch on at different times. 
We will see that photoionization is so
efficient within the ionized regions of the IGM that
quite a simple equation can be given for the expansion of 
the ionized regions. 

The evolution of IGM exposed to ionizing radiation has been discussed by many
authors. Important early work includes that of Arons \& McCray (1970),
Bergeron \& Salpeter (1970) and 
Arons \& Wingert (1972). 
The main novelty of the treatment that follows is that 
whereas previous treatments focus on late ($z<5$) epochs, when 
various simplifying approximations 
can be made because the recombination and Compton rates are low,
we are mainly interested in the case $50<z<150$.
We show that IGM exposed to a strong UV flux rapidly approaches
a quasistatic equilibrium state, where it is almost fully ionized and
the temperature is such that 
photoionization heating exactly balances Compton cooling.
This simplifies the calculations dramatically, since the entire thermal 
history of
the IGM can be summarized by a single function $\v(z)$, the volume 
fraction that is ionized. 
Thus a fraction $\v(z)$ is ionized and hot (with a temperature 
that depends only on $z$, not on when it became ionized), 
and a fraction $1-\v(z)$ is neutral and cold.

In the first section, we justify this approximation.
In the second section, we derive a differential equation for the
time-evolution of $\v$ as well as a useful analytic estimate
of $\fion$.

\section{Intergalactic Str\"omgren Spheres}

Let $\i$ denote the ionization fraction in a small, 
homogeneous volume of intergalactic hydrogen, {\ie}
$\i\equiv n_{HII}/(n_{HI}+n_{HII})$. 
($\i$ is not to be confused with $\v$, the volume fraction in
ionized bubbles.)
When this IGM is  
at temperature $T$, exposed to a
density of $\eta$ UV photons per proton, the
ionization fraction $\i$ evolves as follows: 
\beq{aIonizationEq}
{d\i\over d(-z)} = 
{1+z\over\soz}
\left[\lpi(1-\i) + \lci\i(1-\i) - \lrec^{(1)}\i^2\right],
\eeq
where $H_0^{-1}(1+z)^{-3}$ times the rates per baryon for
photoionization, collisional ionization and recombination are given by
\beq{aRateEq}
\cases{
\lpi \aet {1.04}{12} \left[h\Ob\uvsigfid\right]\eta,&\crr
\lci \aet{2.03}{4} h\Ob T_4^{1/2} e^{-15.8/T_4,}&\crr
\lrec^{(1)} \approx 0.717 h\Ob 
T_4^{-1/2}\left[1.808-0.5\ln T_4 + 0.187 T_4^{1/3}\right],&
}
\eeq
and $T_4\equiv T/10^4\K$. $\uvsigfid$ is the thermally averaged 
photoionization cross section in units of $10^{-18}\cm^2$, and has been
computed in 
Table~\ref{reiontable3} 
for various spectra 
using the differential cross section
from Osterbrock (1974). The collisional ionization rate is from Black (1981).
The recombination rate is the total rate to all hydrogenic levels
(Seaton 1959). 

Below we will see that in the ionized Str\"omgren bubbles that will
appear around the galaxies or quasars, the photoionization
rate is so much greater than the other rates that to a good
approximation, \eq{aIonizationEq} can be replaced by the
following simple model for the IGM:

$\bullet$ It is completely ionized ($\i=1$).

$\bullet$ When a neutral hydrogen atom is formed
through recombination, it is instantly photoionized again.

Thus the only unknown parameter is the IGM temperature, which
determines the recombination rate, which in turn equals the
photoionization rate and thus determines the rate of heating.

Let us investigate when this model is valid.
Near the perimeter of an ionized Str\"omgren sphere of radius $r$
surrounding a galaxy, the number of UV photons per proton is roughly 
$$\eta = {S_{uv}\over
4\pi r^2 c n},$$ 
where $S_{uv}$ is the rate at which UV photons leave the galaxy. 
For an O5 star, the photon flux above the Lyman limit is 
approximately $3.1\times 10^{49}\s^{-1}$ (Spitzer 1968), so if each 
$N$ solar masses of baryons in a galaxy leads to production of a UV
flux equivalent to that of an O5 star, then 
\beq{etaEq}
\eta \geq 0.77 {\fesc M_6\over h^2 r_1^2 N (1+z)^3},
\eeq
inside the sphere, where $r_1\equiv r/1\Mpc$ and
$M_6\equiv M/10^6\Ms$.
When a fraction $f_s$ of all matter has formed galaxies of a typical
total (baryonic and dark) mass $M$, then in the absence of strong
clustering, the typical separation between two galaxies is $$R =
\left({M\over\fs\rho}\right)^{1/3}\approx  \left({15\kpc\over 1+z}\right)
\left({M_6\over h^2\fs}\right)^{1/3},$$
where $M_6\equiv M/10^6\Ms$. Thus $r$ continues to increase until 
$r\approx R$, and spheres from neighboring galaxies begin to overlap.
We are interested in the regime where $z<150$. Substituting this and 
\eq{etaEq} into\eqnum{aRateEq}, we see that
$\lpi\gg\lci$ and $\lpi\gg\lrec$ for any reasonable parameter values.
Hence we can neglect collisional ionization in 
\eq{aIonizationEq}. Since $\lpi\gg 1$,
the photoionization timescale is much shorter than the Hubble
timescale, so \eq{aIonizationEq} will quickly approach a
quasistatic equilibrium solution where the recombination rate equals the
photoionization rate, {\ie}
$$\i\approx 1-{\lrec\over\lpi}\approx 1.$$
In conclusion, the simple $\i=1$ model is valid for all parameter
values in our regime of interest.

When a hydrogen atom gets ionized, the photoelectron acquires 
an average
kinetic energy of ${3\over 2} k\Tpi$, where $\Tpi$ is defined by
${3\over 2}k\Tpi = \euv - 13.6\eV$, and $\euv$ is the average energy of
the ionizing photons 
(see Table~\ref{reiontable3}).

Since the timescale for Coulomb
collisions is much shorter than any other timescales involved,
the electrons and protons rapidly thermalize, and we can always assume
that their velocity distribution is Maxwellian, corresponding to some 
well-defined temperature $T$.
Thus shortly after the 
hydrogen gets photoionized, after the electrons have 
transferred half of their energy to the protons, the plasma temperature is 
$T = {1\over 2} \Tpi$.

The net effect of a recombination and subsequent photoionization is to
remove the kinetic energy of the captured electron,
say ${3\over 2} kT\recfac(T)$, from 
the gas and replace it with
${3\over 2} k\Tpi$,
the kinetic energy 
of the new photoelectron.
Since the recombination cross section is approximately proportional
to $v^{-2}$, slower electrons are
more likely to get captured. Hence the mean energy of the captured electrons 
is slightly lower than ${3\over 2}kT$, {\ie} $\recfac(T)$ is slightly less
than unity (Osterbrock 1974). 
We compute $\recfac(T)$ using Seaton (1959). 
The complication that $\recfac(T)\neq 1$ turns out to 
be of only marginal importance: 
$\recfac(10^4\K)\approx 0.8$, which only 
raises the equilibrium temperatures calculated below by a few percent.

The higher the recombination
rate, the faster this effect will tend to push the temperature 
up towards
$\Tpi$. The two dominant cooling effects are Compton drag against the
microwave background photons and cooling due to the adiabatic expansion of
the universe. Line cooling from collisional excitations 
can be neglected, since the neutral fraction $1-\i \approx 0$.
 Combining these effects, we obtain the evolution equation
for the IGM inside of a Str\"omgren bubble:
\beq{aTeq}
{dT\over d(-z)} = -{2\over 1+z}T + 
{1+z\over\sqrt{1+\Oz z}}\left[\lcomp(T_{cbr}-T) +
{1\over 2}\lrec(T)[T_{cbr}-\eta_{rec}(T)T]\right]
\eeq
where
$$\lcomp = {4\pi^2\over 45}
\left({k T_{cbr}\over\hbar c}\right)^4
{\hbar\st\over H_0 m_e}(1+z)^{-3}
\approx 0.00418 h^{-1}(1+z)$$
is $(1+z)^{-3}$ times the Compton cooling rate per Hubble time
and $T_{cbr} = T_{cbr,0}(1+z)$.
We have taken $T_{cbr,0}\approx 2.726\K$ (Mather {\etal} 1994).
The factor of ${1\over 2}$ in front of the $\lrec$ term is due to the
fact that the photoelectrons share their acquired energy with the protons.
The average energy of
the ionizing photons is given by the spectrum $P(\nu)$ as
$\euv = h\expec{\nu}$,
where 
$$\expec{\nu} = 
{\izi P(\nu)\sigma(\nu) d\nu\over
 \izi \nu^{-1}P(\nu)\sigma(\nu) d\nu}.$$
Here the photoionization cross section
$\sigma(\nu)$ is given by
Osterbrock (1974). Note that, in contrast to certain nebula
calculations where all photons get absorbed sooner or later, the
spectrum should be weighted by the photoionization cross section. This
is because most photons never get absorbed in the Str\"omgren  regions
(only in the transition layer), and all that is relevant is the energy
distribution of those photons that do. $P(\nu)$ is the energy
distribution (W/Hz), not the number distribution which is proportional
to  $P(\nu)/\nu$.

The spectral parameters $\euv$ and $\Tpi$ are given in 
Table~\ref{reiontable3}
for some selected spectra. A Planck spectrum 
$P(\nu)\propto\nu^3/\left(e^{h\nu/kT}-1\right)$
gives quite a good prediction of $T^*$ for stars with surface temperatures
below $30,000\K$. For very hot stars, more realistic spectra (Vacca, 1993)
have a sharp break at the Lyman limit, and fall off much slower above it,
thus giving higher values of  $T^*$. As seen in 
Table~\ref{reiontable3},
an
extremely metal poor star of surface temperature $50,000\K$ gives roughly the
same $T^*$ as QSO radiation. 
The only stars that are likely to be relevant to
early photoionization scenarios are hot and short-lived ones,
since the universe is only about $10^7$ years
old at $z = 100$, and fainter stars would be unable to inject enough
energy in so short a time. Conceivably, less massive stars could play
a dominant role later on, thus lowering $T^*$. However, since they
radiate such a small fraction of their energy above the Lyman limit,
very large numbers would be needed, which could be difficult to
reconcile with the absence of observations of Population III stars
today. If black holes are the dominant UV source, the stellar spectra
of 
Table~\ref{reiontable3}
are obviously irrelevant. A power law spectrum
$P(\nu)\propto \nu^{-\alpha}$ with $\alpha=1$ fits observed QSO
spectra rather well in the vicinity of the Lyman limit (Cheney \&
Rowan-Robinson 1981; O'Brien {\etal} 1988), and is also consistent
with the standard model for black hole accretion. 

Numerical solutions to 
\eq{aTeq} are shown in 
Figure~\ref{reionfig7},
and it is
seen that the temperature evolution separates into three distinct phases.
In the first phase, 
the IGM is outside of the Str\"omgren regions, 
unexposed to UV radiation, and 
remains cold and neutral.
In the second phase, the IGM suddenly becomes ionized, and its temperature
instantly rises to ${1\over 2}\Tpi$. After this, Compton cooling rapidly 
reduces the temperature to a
quasi-equilibrium value of a few thousand K.
After this, in the third phase, $T$ changes only quite slowly,
and is approximately given by setting the expression
in square brackets in \eq{aTeq} equal to zero.
This quasi-equilibrium temperature is typically many times lower
than $T^*$, since Compton cooling is so efficient at the high 
redshifts involved.

\section{The Expansion of Str\"omgren Regions}

This rapid approach to quasi-equilibrium, where the IGM 
``loses its memory" of how long ago it became part of a 
Str\"omgren region,
enables us to construct a very simple model for
the ionization history of the universe.
At redshift $z$, a volume fraction $\v(z)$ of the universe is
completely ionized and typically has a temperature of a few thousand K.
The ionized part need not consist of non-overlapping
spheres; it can have any topology whatsoever. The remainder is cold and
neutral.

Between the ionized and neutral regions is a
relatively thin transition layer, where 
the IGM becomes photoionized and its temperature 
adjusts to the quasistatic value as in 
Figure~\ref{reionfig7}.
As this IGM becomes part of the hot and ionized
volume, the transition layer moves, and $\v(z)$ increases\footnote
{We are tacitly assuming that the UV luminosity of the galaxy that 
creates each Str\"omgren sphere never decreases.
Although obviously untrue, this is in fact an excellent approximation,
since these early dwarf-galaxies correspond to perturbations 
far out in the Gaussian tail.
Since $\fs(z)$ grows so dramatically as the redshift decreases and we
move from five sigma to four sigma to three sigma, {\it etc.}, almost 
all galaxies in existence at a given redshift 
are in fact very young, so that 
older ones that have begun to dim can be safely neglected.
}.

As long as $\v<1$, all UV photons produced are absorbed instantly to a
good approximation. Thus the rate at which UV photons are released
is the sum of the rate at which they are used to counterbalance
recombinations inside the hot bubbles and the rate at which they are
used to break new ground, to increase $\v$. Thus
$$\fupp{d\fs\over dt} = \alpha^{(2)}(T) n\v + {d\v\over dt},$$
where $\alpha^{(2)}(T)$ is the total recombination rate 
to all hydrogenic levels except the ground state\footnote
{A reionization directly to the ground state produces a UV photon that 
usually propagates uninterrupted through the 
highly ionized Str\"omgren region, and then ionizes another atom 
in the transition layer
between the
expanding Str\"omgren region and its cold and neutral surrounding.
Thus 
recombinations directly to the ground state 
were included in the above calculation of the quasi-equilibrium
temperature of the Str\"omgren bubbles, since 
the resulting UV photons could be considered lost from 
the latter.
Here, on the contrary, recombinations directly to the ground state
should not be included, since the UV photons they produce are not wasted
from an energetics point of view.}.
Changing the independent
variable to redshift and using \eq{fsEq}, 
we find that this becomes
\beq{ExpansionEq}
{d\v\over d(-z)} + \lrec^{(2)}{1+z\over\sqrt{1+\Omega_0 z}}\v = 
{2\over\sqrt\pi}\left({\fupp\over 1+\zvir}\right)
\exp\left[-\opzopzv^2\right].
\eeq
Here 
$$\lrec^{(2)} \approx 0.717 h\Ob 
T_4^{-1/2}\left[1.04-0.5\ln T_4 + 0.19 T_4^{1/2}\right]$$
is $H_0^{-1}(1+z)^{-3}$ times 
the total recombination rate per 
baryon to all hydrogenic levels except the ground
state. The fit is to the data of Spitzer (1968) and is accurate to
within $2\%$ for $30\K<T<64,000\K$.
$\lrec^{(2)}$ is to be evaluated at the quasi-equilibrium temperature 
$T(z)$ discussed above.

Using the values in 
Table~\ref{reiontable3}
for the pessimistic, middle-of-the-road and
optimistic estimates, the parameter $\fupp$ equals 
roughly 4, 190 and 24,000, 
respectively.

In the absence of photon waste through recombination, equation
\eq{ExpansionEq} would have the solution 
$\v^*(z) = \fupp\fs(z),$
so the ionization efficiency is
$$\fion(z) = \v(z)/\v^*(z).$$ 
Since \eq{ExpansionEq} is linear in $\v$ and the initial
data is $\v=0$ at some redshift, it is readily seen that the solution
$\v(z)$ is proportional to $\fupp$, the constant in front of the
source term. Combining these last two observations, we see that 
$\fion$ is independent of $\fupp$ and hence independent of the
poorly known parameters $\fmet$, $\fuv$ and $\fesc$.

Plots of $\fion(z)$ from numerical solutions 
of \eq{ExpansionEq} are shown in 
Figure~\ref{reionfig8}
for various
parameter values, and it is seen that the dependence on $z$ is generally
quite weak. Let us make use of this fact by substituting the 
{\it Ansatz} 
$\v(z) = \fion(z)\fupp\fs(z)$ into \eq{ExpansionEq}, 
and setting $\fion'(z)\approx 0$. 
Using \eq{fsEq} and an asymptotic
approximation for the error function, we obtain
$$\fion(z) \approx
{1\over 1 + 0.48{\lrec^{(2)}(1+\zvir)^2/\soz}},$$
independent of $\fupp$, which agrees to within $10\%$ with the numerical
solutions for all reasonable parameter values.
This expression highlights 
the connection between $\fion$ and the thermal evolution
of the Str\"omgren bubbles: 
Essentially, the higher the quasi-static temperature,
the lower the recombination rate $\lrec^{(2)}$, and
the higher $\fion$ becomes.

The value of $\fion$ relevant to computing the ionization redshift is 
obviously that where $z=\zion$. As we have seen, 
$\zion$ typically lies between $2\zvir$ and $3\zvir$. 
Substituting $T\approx 2,500\K$ into the expression for $\lrec$ and
taking $\Omega_0 \approx 1$ and $z=\zion\approx 2.5\zvir$, the above reduces to  
$$\fion \approx
{1\over 1 + 0.8 h\Ob(1+\zvir)^{3/2}},$$
so we see that $\fion$ will be of order unity
unless $\zvir\gg 15$ or $h\Ob\gg 0.03$.

\cleardoublepage
\chapter{Comparing Goodness-of-fit Parameters}
\label{goodnessappendix}

In this appendix, we compare the performances of 
the probability product and the likelihood product as goodness-of-fit
parameters. The notation is that of Chapter~\ref{pindepchapter}.

First of all, what do we mean by a goodness-of-fit parameter
$\eta$ being good?
That $\eta$ leads to the correct model being ruled out at some
confidence level $x$ cannot be held against it
--- by definition of significance
level, this happens a fraction $(1-x)$ of the time. 
Rather, the conventional criterion for rating goodness-of-fit 
parameters
is {\it rejection power}: given a model and a set of observations, 
one $\eta$ is said to be more powerful than another if 
it rejects the model at a higher level of significance.
An example of a very
stupid goodness-of-fit parameter, which in fact has the lowest rejection
power possible,
is a random variable $\eta$ drawn from a uniform distribution on
$[0,1]$, thus containing no information whatsoever about the model or the
observed data. Use of this parameter will reject the model at 95\% confidence
only 5\% of the time, even if the data is blatantly inconsistent with the model.

\section{Both $\elp$ and $\epp$ can be ``fooled"...}

Given a random variable $s_i$ with probability distribution $f_i$, we define 
the corresponding random variable for likelihood by 
$L_i = f_i(s_i)/f_{max},$
where we chose the normalization constant 
$f_{max} \equiv \max_x f_i(x)$ so that we will always have
$0\leq L_i\leq 1$.
Thus for $N$ experiments, the likelihood product 
$$\elp\equiv\prod_{i=1}^N L_i$$ 
will also be a random variable on the interval $[0,1]$.

It is easy to construct examples where either 
the probability product $\epp$ (as defined in 
Section~\ref{pindepsec2.1}) 
or the
likelihood product $\elp$ give very low rejection power.
The Achilles' heel of the probability product is multimodal 
distributions, where values near the mean are rather unlikely. 
For example, suppose $N=1$ and we have a double-humped distribution such as 
$$f_1(s) \propto s^2 e^{-s^2}.$$
If we observe a value $\sh_1 = 0$, then $\elp$
would reject the model with $100\%$ confidence whereas the 
probability product fails miserably, rejecting with $0\%$ confidence
since $\sh_1$ equals the mean.
The likelihood product, on the other hand, 
has the weakness that the
highest likelihood may be attained far out in the tail of the distribution.
Suppose for instance that $N=5$ and we have triangle distributions
$$f_i(s) = 2s$$
on the interval $0\leq s\leq 1$.
If all five observed values $\sh_i$ lie between $0.999$ and $1$, something is
clearly wrong with the model. 
Since $\epph > 0.004^5$, 
using \eq{FetaEq}
indeed rejects the model at a confidence level exceeding
$1-F_{\eta}(0.004^5) \approx 99.999997\%$. 
Yet the likelihood product is near its maximum value:
$\elph\geq 0.999^5 \approx 0.995$, which gives virtually no rejection power at
all.

\section{...but they usually give similar results.}

It is important to note that 
neither of the two examples above were particularly physical. 
The random variables arising from cosmic variance have Gaussian or 
chi-squared distributions, and the same tends to holds for the 
various experimental noise distributions.
Thus the probability distributions to which our goodness-of-fit
parameter is applied in this chapter are unimodal (which 
eliminates the first example) and taper off to zero smoothly
(which eliminates the second example). 
Hence for cosmological applications, 
goodness-of-fit parameters should not be rated by their performance with
such pathological distributions, but rather 
by their rejection power when applied to 
continuous, unimodal distributions. We will now compare the performance of
$\epp$ and $\elp$ for a few such cases. 

For a symmetric {\it exponential} distribution
$$f_i(s) = {1\over 2} e^{-|s|},$$
it is easy to see that the likelihood $L_i$ has a uniform distribution. 
This means that $\epp$ and $\elp$ will have identical distributions, for
arbitrary $N$. It is straightforward to show that the same holds for 
symmetric {\it triangle} distributions
$$f_i(s) = 1 - |s|.$$
A third case where 
$\epp$ and $\elp$ give identical results is when $N=1$ and $f$ is 
any smooth unimodal function.

For a {\it Gaussian} distribution
$$f_i(s) = (2\pi)^{-1/2} e^{-s^2/2},$$
we have
$$P(L_i < x) = 2\erfc\left[(-2\ln x)^{1/2}\right].$$
Although the probability distribution of $\elp$ appears not to be expressible
in terms of elementary functions for arbitrary $n$, 
it is easy to show that $\elp$ has a uniform distribution for 
the special case $N=2$. Thus the likelihood product gives rejection at a
confidence level of 
$$P(\elp<\elph) = e^{-(\sh^2_1 + \sh^2_2)/2}.$$
Comparing this with the corresponding confidence level based on 
$\epp$ shows a remarkable agreement between the two methods. Within the 
disc $\sh^2_1 + \sh^2_2 < 4$, 
over which $P(\eta<\hat\eta)$
varies with many orders of magnitude, the two methods
never differ by more than a factor of two. Thus, at worst, one may
yield a confidence level of say $99.98\%$ where the other yields $99.99\%$.
The probability product is stronger in slightly more than half of the 
$(\sh_1,\sh_2)$-plane, roughly for regions that are more than $20^{\circ}$
away from any of the coordinate axes. 

In conclusion, we have seen that for unimodal, continuous
probability distributions, the likelihood product and the probability product 
tend to give fairly similar --- in a few special cases even identical
--- results. 
Thus chosing one parameter over the
other is more a matter of personal preference than 
something that is likely to seriously affect any scientific conclusions. 
There is however one important practical consideration: 
The probability distribution  of $\elp$ depends on the
probability distributions of the random variables $s_i$. 
This means that, apart
from a few fortuitous special cases such as described above, it can
generally not be calculated analytically. Rather, it must be computed
numerically,  through numerical convolution or
Monte Carlo simulation.
The probability distribution for $\epp$, on the other
hand, is always known analytically, as given by 
\eq{FetaEq}, 
so the probability product is considerably simpler to use.

\cleardoublepage

\chapter{Window Functions}
\label{windowappendix}

\def\rhat{\hat{\bf r}}
All the notation in this appendix is defined 
in Chapter~\ref{pindepchapter}.
The results of CMB anisotropy experiments can be conveniently described 
by expanding the temperature fluctuation in spherical harmonics:
$$
{\Delta T\over T}(\rhat)=\sum_{l=2}^\infty\sum_{m=-l}^l a_{lm}Y_{lm}(\rhat).
$$
(The monopole and dipole anisotropies
have been removed from the above expression, since
they are unmeasurable.)
If the fluctuations are Gaussian, then each coefficient $a_{lm}$ is an
independent Gaussian random variable with zero mean (Bond \& Efstathiou
1987).  The statistical
properties of the fluctuations are then completely specified by the
variances of these quantities
$$
C_l\equiv \left<|a_{lm}|^2\right>.
$$
(The fact that the variances are independent of $m$ is an immediate 
consequence of spherical symmetry.)  Different CMB experiments are sensitive
to different linear combinations of the $C_l$'s:
\beq{TedEqOne}
S=\sum_{l=2}^\infty F_lC_l,
\eeq
where $S$ is the ensemble-averaged mean-square signal in a particular
experiment, and the ``filter function'' $F_l$ specifies the sensitivity
of the experiment on different angular scales.  The filter functions for 
COBE and SP91 are
\beq{TedEqTwo}
\cases{
F_l^{(cobe)}&$={(2l+1)\over 4\pi}e^{-\sigma_c^2(l+{1\over 2})^2}$,\crr
F_l^{(sp)}&$=4e^{-\sigma_s^2(l+{1\over 2})^2}\sum_{m=-l}^l H_0^2(\alpha m)$,
}
\eeq
where $H_0$ is a Struve function.
$\sigma_c=4.25^\circ$ and $\sigma_s=0.70^\circ$ are the {r.m.s.} beamwidths
for the two experiments, and $\alpha=1.5^\circ$ is the amplitude of the
beam chop (Bond {\etal} 1991; Dodelson \& Jubas 1993; White {\etal} 1993).

For Sachs-Wolfe fluctuations in a spatially flat Universe with the standard
ionization history, the angular power spectrum $C_l$ is related to the power
spectrum of the matter fluctuations in the following way
(Peebles 1984; Bond \& Efstathiou 1987):
\beq{TedEqThree}
C_l={8\over \pi\tau_0^4} \int_0^\infty dk P(k)\jbar_l^2(k).
\eeq
Here $\tau_0$ is the conformal time at the present epoch, and 
$$
\jbar_l(k)\equiv\int j_l(k\tau)V(\tau)\,d\tau,
$$
where $j_l$ is a spherical Bessel function.
The visibility function $V$ is the probability distribution for the conformal
time at which a random CMB photon was last scattered.  $\jbar_l(k)$ 
is therefore the average of $j_l(k\tau)$ over the last scattering surface. 
We have used the $V$ of Padmanabhan (1993).

We can combine equations\eqnum{TedEqOne},\eqnum{TedEqTwo}, 
and\eqnum{TedEqThree}
to get the window
functions for the two experiments:
$$\cases{
\Wc &$= {2\over\pi^2k^2\tau_0^4}\sum_{l=2}^{\infty}
\jbar_l^2(k)e^{-\sigma_c^2\left(l+{1\over 2}\right)^2}(2l+1)$\crr
\Wsp &$= {32\over\pi k^2\tau_0^4}\sum_{l=2}^{\infty}
\jbar_l^2(k)
e^{-\sigma_s^2\left(l+{1\over 2}\right)^2}
\sum_{m=-l}^l H_0^2(\alpha m)$\crr
}$$

The mean-square bulk flow inside of a sphere of radius $a$ is
(see, {\it e.g.}, Kolb \& Turner 1990)
\beq{TedEqFour}
\left<v^2\right>=\int dk P(k){18\over \pi^2\tau_0^2}\, {j_1^2(ka)\over (ka)^2}.
\eeq
However, we must make two corrections to this result before applying it
to the LP data.  This formula applies to a measurement of the bulk
flow within a sphere with an infinitely sharp boundary.  In reality,
errors in measuring distances cause the boundary of the spherical region
to be somewhat fuzzy.  If we assume that distance measurements are subject
to a fractional error $\epsilon$, then the window function must be
multiplied by $e^{-(\epsilon ka)^2}$.  We have taken $\epsilon=0.16$, 
the average value quoted by LP. It should be noted that this value varies
from galaxy to galaxy in the LP sample, due to the distance estimation 
technique used, and that a  more accurate window function that reflects 
the discrete locations of the Abell clusters used in the survey 
should take this into account.

The second correction has to do with the behavior of the window function
at small $k$.  Equation\eqnum{TedEqFour} applies to the velocity 
relative to the rest frame of the Universe.  The velocity measured by LP is
with respect to the CMB rest frame.  If there is an intrinsic CMB dipole
anisotropy, then these two reference frames differ.  Therefore, we must
include in \eq{TedEqOne}, 
a term corresponding to the intrinsic CMB dipole.  This correction was first
noticed by G\'orski (1991).
After applying both of these corrections, the LP window function is
$$
W_{lp}={18\over \pi^2\tau_0^2}\left({j_1(ka)\over ka}e^{-(\epsilon ka)^2}
-{\jbar_1(k\tau_0)\over k\tau_0}\right)^2.
$$

\cleardoublepage

\setcounter{secnumdepth}{-1}
\chapter{References}
\setcounter{secnumdepth}{2}

\rf Alcock, C. {\etal} 1993;Nature;365;621
% + Akerlof, C. W., Allsman, R. A., Axelrod, T. S. and others 1993;

\rf Alpher, R. A., Bethe, H. A. \& Gamow, G. 1948;Phys. Rev.;73;803

\rf Anninos, P. {\etal} 1991;ApJ;382;71
% Anninos, P.; Matzner, R.A.; Tuluie, R.; Centrella, J.
% Anisotropies of the cosmic background radiation in a 'hot' dark matter
% universe.

\rf Arnaud, M. {\etal} 1992;A\&A;254;49
  
\rf Arons. J. \& McCray, R. 1970; Ap. Lett.;5;123

\rf Arons, J. \& Wingert, D. W. 1972;ApJ;177;1

\rf Bardeen, J. M., Bond, J. R., Kaiser, N. \& Szalay, A. S.
1986;ApJ;304;15
 
\rf Bartlett, J. \& Stebbins, A. 1991;ApJ;371;8

\rf Bergeron, J. \& Salpeter, E. E. 1970;Ap. Lett.;7;115

\rf de Bernardis, P. {\etal} 1994;ApJ;422;L33
% ARGO.
% de Bernardis, P.; Aquilini, E.; Boscaleri, A.; De Petris, M.; and others.
% Degree-scale observations of cosmic microwave background anisotropies.

\rf Binggeli, B., Sandage, A. \& Tammann, G. A.
1988;ARA\&A;26;509
 
\rf Binney, J. 1977;ApJ;215;483

\rf Birkinshaw, M. \& Hughes, J. P. 1994;ApJ;420;33
%   A MEASUREMENT OF THE HUBBLE CONSTANT FROM THE X-RAY PROPERTIES AND THE
%   SUNYAEV-ZELDOVICH EFFECT OF ABELL 2218.
%   ASTROPHYSICAL JOURNAL, 1994 JAN 1, V420 N1:33-43.
  
\rf Black, J. 1981;MNRAS;197;553
 
\rf Blanchard, A., Valls-Gabaud, D. \&
Mamon, G. A. 1992;A\&A;264;365
 
\rf Blumenthal, G. R., Faber S. M., Primack, J. R., \&
Rees, M. J.1984;Nature;311;517

\rf Bond, J. R. \& Efstathiou, G. 1984;ApJ;285;L45
% CBR ANISOTROPIES IN A UNIVERSE DOMINATED BY CDM
% Give their famous fitting formula and tabulate a, b, c and nu for different models.
 
\rf Bond, J. R. \& Szalay, A. S. 1983;ApJ;274;443
 
\rf Brainerd, T. G. \& Villumsen, J. V. 1992;ApJ;394;409
 
\rf Bruhweiler, F. C., Gull, T. R., Kafatos, M., \&
Sofia, S. 1980;ApJ;238;L27

\rf Carlberg, R. G. \& Couchman, H. M. P. 1989;ApJ;340;47

\rf Carroll, S. M., Press, W. H. \& Turner, E. L. 1992;ARA\&A;30;499
%  THE COSMOLOGICAL CONSTANT.
%  ANNUAL REVIEW OF ASTRONOMY AND ASTROPHYSICS, 1992, V30:499-542.
 
\rf Cen, R., Gnedin, N. Y., Koffmann, L. A., \& Ostriker, J. P.
1992;ApJ;399;L11
 
\rf Cen, R., Ostriker, J. P. \& Peebles, P. J. E. 1993;ApJ;415;423
 
\rf Cheney, J. E. \& Rowan-Robinson, M. 1981;MNRAS;195;831
 
\rf Cheng, E. {\etal} 1994;ApJ;422;L37
% MSAM
 
\rf Cioffi, D. F., McKee, C. F., \&
Bertschinger, E. 1988;ApJ;334;252
 
\rf Conklin, E. K. 1969; Nature;222;971
%  Velocity of the Earth with respect to the cosmic background radiation.
%  Nature, 7 June 1969, vol.222, (no.5197):971-2.
 
\rf Couchman, H. M. P. 1985;MNRAS;214;137

\rf Couchman, H. M. P. \& Rees, M. J. 1986;MNRAS;221;53
 
\rf Cox, D. P. \& Smith, B. W. 1974;ApJ;189;L105

\rf Dalgarno, A. \& McCray, R. A. 1972;ARA\&A;10;375

\rf David, L. P., Arnaud, K. A., Forman, W. \& Jones, C. 1990;ApJ;356;32

\rf David, L. P., Forman, W. \& Jones, C. 1991;ApJ;369;121
 
\rf Davis, M., Summers, F. J. \& Schlegel, D. 1992;Nature;359;393

\rf Dekel, A. \& Silk, J. 1986;ApJ;303;39
 
\rn
Devlin, M. {\etal} 1992, in {\it Proc. NAS Colloquium on Physical
Cosmology}, Physics Reports (in press).
% MAX 

\rf Dicke, R. H., Peebles, P. J. E., Roll, P. G. \&
Wilkinson, D. T. 1965;ApJ;142;414

\rf Dodelson, S., Gyuk, G. \& Turner, M. S. 1994a;
Phys. Rev. Lett.;72;3754
% Fermilab preprint 93/236-A
% PRIMORDIAL NUCLEOSYNTHESIS WITH A DECAYING TAU NEUTRINO

\rf Dodelson, S., Gyuk, G. \& Turner, M. S. 1994b;
Phys. Rev. D;49;5068
% Fermilab preprint 94/026-A
% IS A MASSIVE TAU NEUTRINO JUST WHAT CDM NEEDS?
 
\rf Dodelson, S. \& Jubas, J.M. 1993; Phys. Rev. Lett.;70;2224

\rn Dolgov, A. D., Sazhin, M. V. \&
Zel'dovich,  Y. B. 1990, 
{\it Modern Cosmology},
Editions Frontieres, Gif-sur-Yvette, France.

\rf Donahue, M. \& Shull, J. M. 1991;ApJ;383;511
 
\rn Dragovan, M. {\etal} 1993, preprint.
% PYTHON 

\rn Edge, A. C. 1989, Ph.D. thesis, University of Leicester.
 
\rf Efstathiou, G., Bond, J. R. \& White, S. D. M. 1992;MNRAS;258;1P
% Argue that their old fitting formula works OK for most other models as
% well, if you just fudge \Gamma.
 
\rf Efstathiou, G., Frenk, C. S., White, S. D. M., \&
Davis, M. 1985;ApJ;57;241
 
\rf Efstathiou, G., Frenk, C. S., White, S. D. M., \&
Davis, M. 1988;MNRAS;235;715
 
\rf Efstathiou, G. \& Rees, M. J. 1988;MNRAS;230;5P
%  [High-redshift quasars in the Cold Dark Matter cosmogony].

\rn Feldman, H. A. \& Watkins, R. 1993, preprint.
 
\rf Feynman, R. P. 1939;Phys. Rev.;56;340

\rn Gamow, G. 1970, {\it My World Line}, Viking, New York.

\rf Gaier, T. {\etal} 1992;ApJ;398;L1
 
\rn Gelb, J. M. \& Bertschinger, E. 1992, preprint.
 
\rf Ginsburg, V. L. \& Ozernoi, L. M. 1965; Astron. Zh.;42;943
(Engl. transl. 196 Sov. Astron. AJ, 9, 726)

\rf Glanfield, J. R. 1966;MNRAS;131;271
 
\rf Gnedin, N. Y. \& Ostriker, J. P.1992;ApJ;400;1
% "GO"
 
\rf G\'orski, K. M. 1991;ApJ;370;L5
 
\rf G\'orski, K. M. 1992;ApJ;398;L5

\rf G\'orski, K. M., Stompor, R. \& Juszkiewicz, R. 1993;ApJ;410;L1
%  COLD DARK MATTER AND DEGREE-SCALE COSMIC MICROWAVE BACKGROUND ANISOTROPY
%  STATISTICS AFTER COBE.
 
\rf Gott, J. R. \& Rees, M. J. 1975;A\&A;45;365
 
\rf Gundersen, J.O. {\etal} 1993;ApJ;413;L1
% + Clapp, A.C., Devlin, M., Holmes, W., Fischer, M.L.,
% Meinhold, P.R., Lange, A.E., Lubin, P. M., Richards, P.L., \& Smoot, G.F.
 
\rf Gunn, J. E. \& Peterson, B. A. 1965;ApJ;142;1633

\rn Gyuk, G. \& Turner, M. S. 1994,
astro-ph/9403054 preprint.
% RELAXING THE BIG-BANG BOUND TO THE BARYON DENSITY

\rn Hancock, S. {\etal} 1994;Nature;367;333
% TENERIFE

\rn Hatsukade, I. 1989, thesis, Osaka University.
 
\rf Heckman, T. M., Armus, L., \&
Miley, G. K. 1990;ApJS;74;833
 
\rf Henry, P. S. 1971; Nature;231;516

\rf Holtzman, J. A. 1989;ApJS;71;1
 
\rf Hu, W., Scott, D. \& Silk,  J. 1994;Phys. Rev. D;49;648
% REIONIZATION AND COSMIC MICROWAVE BACKGROUND DISTORTIONS - A COMPLETE
% TREATMENT OF SECOND-ORDER COMPTON SCATTERING.
 
\rf Hu, W. \& Silk, J. 1993; Phys. Rev. D;48;485
% THERMALIZATION AND SPECTRAL DISTORTIONS OF THE COSMIC BACKGROUND RADIATION.

\rf Hubble, E. 1929;Proc. N. A. S.;15;168

\rf Hughes, J. P., Yamashita, K., Okumura, Y.,
Tsunemi, H., \& Matsuoka, M. 1988;ApJ;327;615

\rf Ikeuchi, S. 1981;Publ. Astr. Soc. Jpn.;33;211
 
\rf Ikeuchi, S. \& Ostriker, J. P. 1986;ApJ;301;522
 
\rf Ikeuchi, S. \& Turner, E. L 1991;ApJ;381;L1
%  THE EVOLUTION OF THE DIFFUSE COSMIC ULTRAVIOLET BACKGROUND CONSTRAINED BY
% HUBBLE SPACE TELESCOPE OBSERVATIONS OF 3C 273.

\rf Jaffe, A. H., Stebbins, A. \& Frieman, J. A. 1994;ApJ;420;9
%  MINIMAL MICROWAVE ANISOTROPY FROM PERTURBATIONS INDUCED AT LATE TIMES.
      
\rn Jubas, J. \& Dodelson, S. 1993, preprint.
 
\rn Juszkiewicz, R. 1993, private communication.
 
\rf Juszkiewicz, R., G\'orski, K. M., \& Silk, J.1987;ApJ;323;L1

\rf Kamionkowski, M. \& Spergel, D. N. 1994;ApJ;432;7
 
\rf Kamionkowski, M.,
Spergel, D. N. \& Sugiyama, N. 1994;ApJ;426;L57
 
\rf Kashlinsky 1992;ApJ;399;L1
 
\rf Klypin, A., Holtzman, J., primack, J. \& Reg\"os, E. 1993;ApJ;416;1
%     STRUCTURE FORMATION WITH COLD PLUS HOT DARK MATTER.
 
\rn Kolb, E. W. \& Turner, M. S. 1990, {\it The Early Universe}, Addison-Wesley.
 
\rf Kompaneets, A. S. 1957; Soviet Phys.--JETP;4;730

\rf Lauer, T. \& Postman, M. 1994;ApJ;425;418
 
\rf Lea, S. M., Mushotzky, R., \& Holt, S. 1982;ApJ;262;24

\rf Maddox, S. J., Efstathiou, G., Sutherland, W. J.
\& Loveday, J. 1990;MNRAS;242;43
 
\rf Malaney, R. A. \& Mathews, G. J. 1993;Phys. Rep.;229;145
% PROBING THE EARLY UNIVERSE - A REVIEW OF PRIMORDIAL NUCLEOSYNTHESIS BEYOND
% THE STANDARD BIG BANG.

\rf Mather J. C. {\etal} 1990;ApJ;354;L37
%  MATHER JC; CHENG ES; EPLEE RE; ISAACMAN RB; and others.
%  A PRELIMINARY MEASUREMENT OF THE COSMIC MICROWAVE BACKGROUND SPECTRUM BY
%  THE COSMIC-BACKGROUND-EXPLORER (COBE) SATELLITE.
  
\rf Mather {\etal} 1994;ApJ;420;439
% MEASUREMENT OF THE COSMIC MICROWAVE BACKGROUND SPECTRUM BY
% THE COBE FIRAS INSTRUMENT.
 
\rf Matsumoto, T. {\etal} 1988;ApJ;329;567
 
\rf McCray, R. \& Kafatos, M. 1987;ApJ;317;190
 
\rf McCray, R. \& Snow, T. P. Jr. 1979;ARA\&A;17;213
 
\rf McKee, C. F. \& Ostriker, J. P. 1977;ApJ;218;148

\rf Meinhold, P. R. \& Lubin, P. M. 1991;ApJ;370;L11
 
\rf Meinhold, P. R. {\etal} 1993; ApJ;409;L1
% + Clapp, A.C., Cottingham, D., Devlin, M., Fischer, M.L.,
% Gundersen, J.O., Holmes, W., Lange, A.E., Lubin, P.M.,
% Richards, P.L., \& Smoot, G.F.

\rf Michelson, A. A. \& 
Morley, E. W. 1887; Am. Journ. of Science;34;333
 
\rf Miralda-Escud\'e, J. \& Ostriker, J. P. 1990;ApJ;350;1

\rf Mushotzky, R. F. 1984;Physica Scripta;T7;157

\rn Narlikar,  J. V. 1993, {\it Introduction to Cosmology} (2nd ed.),
Cambridge U. P., Cambridge.

\rf O'Brien P.T., Wilson, R \& Gondhalekar, P. M 1988;MNRAS;233;801
 
\rn Osterbrock, D. E. 1974, {\it Astrophysics of Gaseous Nebulae},
Freeman, San Francisco.
 
\rn Ostriker, J. P. 1991, {\it Development of Large-Scale
Structure in the Universe}, Fermi Lecture Series,
Cambridge U. P., Cambridge.
 
\rf Ostriker, J. P. \& Cowie, C. F. 1981; ApJ;243;L127
 
\rf Ostriker, J. P. \& McKee, C. F. 1988;Rev. Mod. Phys.;60;1

\rf Oukbir, J. \& Blanchard, A. 1992;A\&A;262;L21

\rn Padmanabhan, T. 1993, {\it Structure Formation in the Universe},
Cambridge U. P., Cambridge.
 
\rf Peacock, J. A. \& Heavens, A. F. 1990;MNRAS;243;133

\rn Peacock, J. A., Heavens, A. E. \& Davies, A. T. 1990 (eds.),
{\it Physics of the Early Universe},
Scottish Universities' Summer School in Physics,
IOP Publishing, Bristol.
 
\rf Peacock, J. A. \& Dodds, S. J. 1994;MNRAS;267;1020

\rn Peebles, P. J. E. 1971, {\it Physical Cosmology},
Princeton U. P., Princeton.

\rn Peebles, P. J. E. 1980, {\it The large-scale structure of the universe},
Princeton U. P., Princeton.
% (Phillip James Edwin)
% The large-scale structure of the universe / by P. J. E. Peebles.  
 
\rf Peebles, P. J. E. 1984; ApJ;284;439
 
\rf Peebles, P. J. E. 1987; ApJ;315;L73
 
\rn Peebles, P. J. E 1993, {\it Principles of Physical Cosmology},
Princeton U. P., Princeton.
 
\rf Penzias, A. A. \& Wilson, R. W. 1965;ApJ;142;419
% Measurement of excess antenna temperature at 4080Mc/s
 
\rf Pettini, M., Boksenberg, A., \& Hunstead, R. 1990;ApJ;348;48
 
\rf Press, W. H. \& Schechter, P. 1974;ApJ;187;425

\rf Rees, M. J. \& Ostriker, J. P. 1977;MNRAS;179;541
 
\rf Rees, M. J.1986;MNRAS;218;25P
 
\rf Ressell, M. T. \& Turner, M. S. 1990;Comments on Astroph.;14;323
 
\rf Rothenflug, R. L., Vigroux, R., Mushotzky, R.,
\& Holt, S. 1984;ApJ;279;53
 
\rn Rubin, V. C. \& Coyne, G. V. 1988 (eds.),
{\it Large-scale motions in the universe : a Vatican study week},
Princeton U. P., Princeton.

\rf Sadoulet, B. \& Cronin, J. W. 1991;Physics Today;44;53
%  Sadoulet, B.; Cronin, J.W. 
%  Particle astrophysics. 
%  Physics Today, April 1991, vol.44, (no.4):53-7. 
%  Pub type:  General or Review. 
%  Abstract: The authors discuss various aspects of dark matter in cosmology, 
%  solar neutrinos, supernova neutrinos, gamma -rays detection, galactic 
%  neutrino sources, and cosmic ray primary particles. 
 
\rn Sanchez, F., Collados, M. \& Rebolo, R. (eds.) 1990,
{\it Observational and Physical Cosmology},
2nd Canary Islands Winter School on Astrophysics,
Cambridge U. P., Cambridge.
 
\rn Schlegel, D., Davis, M., Summers, F. \& Holtzman, J. 1993, preprint.

\rn Schuster, J. {\etal}, private communication. 

\rf Schwartz, J., Ostriker, J. P., \& Yahil, A. 1975;ApJ;202;1

\rf Seaton, M. 1959;MNRAS;119;84

\rn Sedov, L. I. 1959, {\it Similarity and dimensional methods in 
mechanics}, Academic, New York.
 
\rf Shaefer, R. K. \& Shafi, Q. 1992; Nature;359;199
 
\rf Shafi, Q. \& Stecker, F. W. 1984;Phys. Rev. Lett.;53;1292

\rf Shapiro, P. R. 1986;PASP;98;1014
 
\rf Shapiro, P. R. \& Giroux, M. L. 1987;ApJ;321;L107

\rn Shapiro, H. S. \& Tegmark, M. 1994, {\it Soc. Ind. App. Math. Rev.}, {\bf 36}.

\rf Sherman, R. D. 1980;ApJ;237;355
 
\rn Shuster, J. {\etal} 1993, private communication.
 
\rf Silk, J. 1977;ApJ;211;638
 
\rf Smith, M. S., Kawano, L. H. \& Malaney, R. A. 
1993;ApJS;85;219

\rf Smoot, G. F., Gorenstein, M. V. \& Muller, 
R. A. 1977; Phys. Rev. Lett.;39;898
% THE AIRCRAFT DIPOLE DETECTION
\rf Smoot, G. F. {\etal} 1992;ApJ;396;L1
% STRUCTURE IN THE COBE DMR 1ST YEAR MAPS

% \rf Spinrad, H. 1986 [Faint galaxies and cosmology].

\rn Spitzer, L. 1968, {\it Diffuse Matter in Space}, Wiley, New York.
 
\rf Stebbins, A. \& Silk, J. 1986;ApJ;300;1
 
\rf Steidel, C. C. 1990;ApJS;74;37
 
\rf Steidel, C. C. \& Sargent, W. L. W. 1987;ApJ;318;L11
 
\rn Strauss, M., Cen, R. \& Ostriker, J.P. 1993, preprint. 

\rf Subrahmanyan, R. {\etal} 1993;MNRAS;263;416
%     SUBRAHMANYAN R; EKERS RD; SINCLAIR M; SILK J.
%     A SEARCH FOR ARCMIN-SCALE ANISOTROPY IN THE COSMIC MICROWAVE BACKGROUND.

\rn Sugiyama, N. 1993, private communication.

\rf Sugiyama, N., Silk, J. \& Vittorio, N. 1993;ApJ;419;L1
% REIONIZATION AND COSMIC MICROWAVE ANISOTROPIES.
 
\rf Suto, Y., G\'orski, K. M. Juszkiewicz, R., \& Silk, J.1988; Nature;332;328

\rf Tegmark, M. 1993;Found. Phys. Lett.;6;571
%     APPARENT WAVE FUNCTION COLLAPSE CAUSED BY SCATTERING.
%     FOUNDATIONS OF PHYSICS LETTERS, 1993 DEC, V6 N6:571-590.
   
\rf Tegmark, M., Bunn, E. \& Hu, W. 1994;ApJ;434;1
%  CfPA preprint 93-th-40.
%  Power spectrum independent constraints on cosmological models

\rf Tegmark, M. \& Shapiro, H. S. 1994;Phys. Rev. E;50;2538
% gaussians.tex

\rn Tegmark, M. \& Silk, J. 1994, {\it ApJ}, {\bf 423}, 529; {\bf 434}, 395
% y.tex

\rn Tegmark, M., Silk, J. \& 
Blanchard, A. 1994, {\it ApJ}, {\bf 420}, 484; {\bf 434}, 395
%     ON THE INEVITABILITY OF REIONIZATION - IMPLICATIONS FOR COSMIC MICROWAVE
%     BACKGROUND FLUCTUATIONS.
%     ASTROPHYSICAL JOURNAL, 1994 JAN 10, V420 N2:484-496.
 
\rf Tegmark, M., Silk, J. \& Evrard, A. 1993;ApJ;417;54
%    LATE REIONIZATION BY SUPERNOVA-DRIVEN WINDS.
%    ASTROPHYSICAL JOURNAL, 1993 NOV 1, V417 N1:54-62.

\rf Tegmark, M. \& Yeh, L. W. 1994;Physica A;202;342
%    STEADY STATES OF HARMONIC OSCILLATOR CHAINS AND SHORTCOMINGS OF HARMONIC
%    HEAT BATHS.
%    PHYSICA A, 1994 JAN 1, V202 N1-2:342-362.
    
\rn Teresawa, N. 1992, preprint.
 
\rf Tomisaka, K., Habe, H, \& Ikeuchi, S. 1980;Progr. Theor.
Phys. (Japan);64;1587

\rn Vacca, W. 1993, private communication.
 
\rf Vishniac, E. 1987;ApJ;322;597
 
\rf Vittorio, N. \& Silk, J. 1984;ApJ;285;L39

\rf Vittorio, N. \& Silk, J. 1992; ApJ;385;L9
%  ANISOTROPIES OF THE COSMIC MICROWAVE BACKGROUND IN NONSTANDARD
%  COLD DARK MATTER MODELS.

\rf Walker, P. N. {\etal} 1991;ApJ;376;51
 
\rf Watson, R.A.,{\etal} 1992;Nature;357;660
% \rf Watson, R.A., Guti\'errez de la Cruz, C.M., Davies, R.D., Lasenby, A.N.,
% Rebolo, R., Beckman, J.E., \& Hancock, S.1992;Nature;357;660

\rf Watson, R.A. \& Guti\'errez de la
Cruz, C.M. 1993;ApJ;419;L5

\rf Webb, J. K., Barcons, X., Carswell, R. F., \&
Parnell, H. C. 1992;MNRAS;255;319
 
\rf Weaver, R., McCray, R, Castor, J.,
Shapiro, P., \& Moore, R. 1977;ApJ;218;377
 
\rn Weinberg, S. 1972, {\it Gravitation and Cosmology},
Wiley, New York.
 
\rf White, M., Krauss, L., \& Silk, J. 1993;ApJ;418;535

\rf White, S. D. M. \& Rees, M. J. 1978;MNRAS;183;341
%  Core condensation in heavy halos: a two-stage theory for galaxy formation
%  and clustering.
 
\rn White, M., Scott, D. and Silk, J. 1994, to appear in ARA\&A.
 
\rf White, S. D. M., \& Rees, M. J. 1986;MNRAS;183;341

\rf Woody, D. P. \& Richards, P. 1981;ApJ;248;18

% \rf Woosley  \& Weaver 1986

\rf Zel'dovich, Y., \& Sunyaev, R. 1969;Ap. Space Sci.;4;301

\end{document}